\documentclass[12pt, oneside]{book}
\usepackage{geometry} 
\usepackage{graphicx}
\usepackage{pxfonts}

\usepackage{tikz}
\usepackage{amsmath}
\usepackage{varwidth}
\usepackage{multirow}
\usepackage{rotating}
\usepackage{parskip}

\def\checkmark{\tikz\fill[scale=0.4](0,.35) -- (.25,0) -- (1,.7) -- (.25,.15) -- cycle;}
\usepackage[doublespacing]{setspace}

\title{
	{\Huge The Relation Between \\ Classical and Quantum Mechanics}\\
	{\vspace{40pt}\LARGE by Peter Taylor}\\
\vspace{40pt}}
	
\author{Foreword by \\ Simon Saunders, Oxford University}
\date{}

\begin{document}

\maketitle
\newpage
\addcontentsline{toc}{chapter}{Foreword}
{\center \section*{Foreword}}
\singlespacing

No part of Peter Taylor's {\em{The Relation between Classical and Quantum Mechanics}} was ever published in journals. It has been available in the Bodleian Library of the University of Oxford, as any other Oxford DPhil thesis, since 1984, but there it has lain unread and unknown. In the subsequent decades there have been several crucial advances in the understanding of the relation between classical and quantum, most notably in the field of decoherence theory, but there remain aspects to this relation that are far from certain: in inter-theory reduction, localisation, controls over approximations, and what I would loosely call the 'axiomatics' of quantum theory. 

Peter Taylor, in his Introduction, highlighted two of these foci: approximations and localisation. The first he addressed with the full rigour of the methods as recently perfected by Michael Reed and Barry Simons in their remarkable {\em{Methods of Modern Mathematical Physics}}, published in 4 volumes in the mid to late 1970s. The second he expressed in terms of a compactness condition on sets of pure states.  The two others listed are 'well-defined theories....as a necessary precursor to inter-theoretic reduction', and 'pure states: the assertion of realism in physics by employing pure states as primitive abstractions'. He addressed both by providing a new lattice-theoretic axiomatization of quantum mechanics, using ideas introduced by Veeravalli Varadarajan in his magisterial {\em{Geometry of Quantum Theory}}, published in 1968,  buttressed by an analysis of inter-theory reduction that is original and, after more than three decades, timely. The thesis does not solve the measurement problem, but it does not aspire to: it is concerned with the circumstances in which quantum Hamiltonians drive evolutions well-approximated by classical Hamiltonian flows, not those, as in measurement processes, that do not.

This theory of reduction is accompanied by an account of dimensional constants, and a detailed evaluation of the proposal, by the mathematical physicist Klaus Hepp in 1974, on a definition of theory-reduction (and in particular the classical limit of quantum mechanics) in terms of a family of quantum theories with decreasing magnitude of Planck's constant. As set out in Appendix 4.2, this is one of several sections that would in my view have merited a journal publication in its own right -- and that still does. As for the lattice-theoretic realism -- an ally of  the quantum logic approach to quantum foundations -- its time may yet come: the quantum information-theoretic approach to the axiomatisation of quantum theory, despite some initial successes, has languished, its prospects for realism dim. 

The ideas and methods here set out in lattice theory, theory reduction, dimensional analysis, and particle localisation, are impressive taken in isolation: much more so taken in unison.  The monograph is restricted to non-relativistic theory, but within that orbit combines philosophical scope, axiomatic method, and mathematical rigour, to an extent that I have seldom seen.  

Work of this calibre is not easily completed within the time-scale of a funded graduate degree. Peter Taylor left academia to begin a successful career in the London insurance markets in 1981. He was able to take out sufficient time to finish the thesis in 1984, but did no more with it. His premature death in November 2015, at the age of 61, has forever put to an end his efforts to clarify the foundations of quantum mechanics, but may mark the beginning of their influence, as only now made available to wider communities: in foundations of physics, mathematical physics, philosophy of science, and philosophy of physics. 

Thanks are due to Ian Nicol, Thomas M\"{o}ller-Nielsen, and David Shipley, for editing and resetting of the monograph in TeX. Without them its publication would not have been possible. 
\newline
\newline
Simon Saunders
\newline
Merton College, Oxford
\newline
June 2018

\newpage
\begin {center}
\doublespacing
\vspace{60pt}
	{\huge The Relation Between \\ Classical and Quantum Mechanics}\\
	{\vspace{14pt}\large Peter Taylor}\\
	{\large Magdalen College, Oxford}\\
	\vspace{80pt} {\includegraphics[scale=0.15]{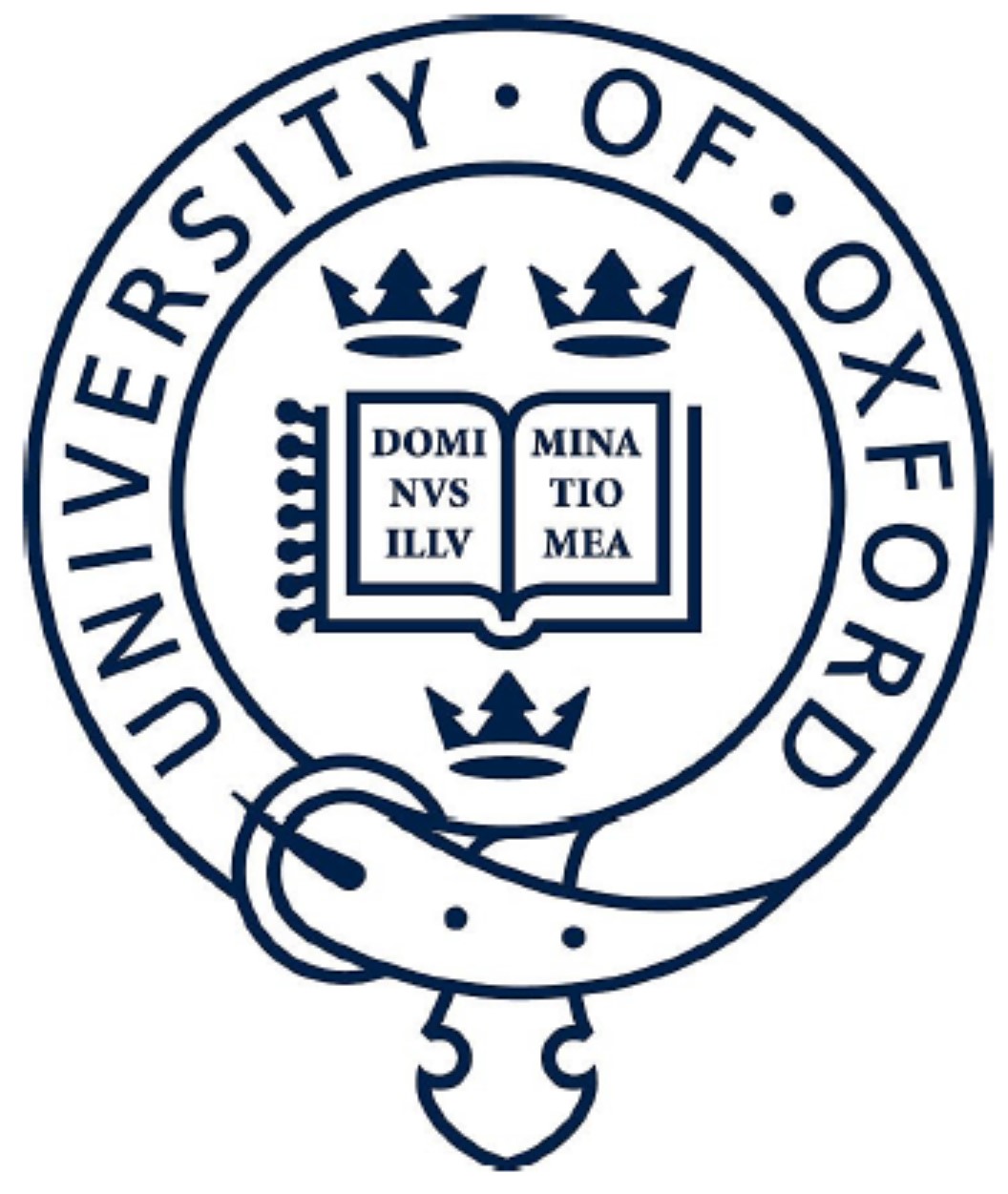}}
\vspace{40pt}

\author{Thesis submitted for the degree of \\ Doctor of Philosophy}
\date{Hilary Term, 1984}
\end {center}
\singlespacing

\newpage
\addcontentsline{toc}{chapter}{Abstract}
{\center \section*{Abstract}}
\singlespacing

This thesis examines the relation between classical and quantum mechanics from philosophical, mathematical and physical standpoints.

It first presents arguments in support of ``conjectural realism'' in scientific theories distinguished by explicit contextual structure and empirical testability; and it analyses intertheoretic reduction in terms of weakly equivalent theories over a domain of applicability.

Familiar formulations of classical and quantum mechanics are shown to follow from a general theory of mechanics based on pure states with an intrinsic probability structure. This theory is developed to the stage where theorems from quantum logic enable expression of the state geometry in Hilbert space. Quantum and classical mechanics are then elaborated and applied to subsystems and the measurement process. Consideration is also given to space-time geometry and the constraints this places on the dynamics.

Physics and Mathematics, it is argued, are growing apart; the inadequate treatment of approximations in general and localisation in quantum mechanics in particular are seen as contributing factors. In the description of systems, the link between localisation and lack of knowledge shows that quantum mechanics should reflect the domain of applicability. Restricting the class of states provides a means of achieving this goal. Localisation is then shown to have a mathematical expression in terms of compactness, which in turn is applied to yield a topological theory of bound and scattering states.

Finally, the thesis questions the validity of ``classical limits'' and ``quantisations'' in intertheoretic reduction, and demonstrates that a widely accepted classical limit does not constitute a proof of reduction. It proposes a procedure for determining whether classical and quantum mechanics are weakly equivalent over a domain of applicability, and concludes that, in this restricted sense, classical mechanics reduces to quantum mechanics.
\newpage

\addcontentsline{toc}{chapter}{Dedication}
{\center \section*{Dedication}}

This thesis is dedicated with love to my mother, in appreciation of all her encouragement and help over the years.
\newpage

\addcontentsline{toc}{chapter}{Acknowledgements}
{\center \section*{Acknowledgements}}
It is my pleasure to thank the many people whose help has guided me to completing this thesis.

My greatest thanks and deepest gratitude go to Professor Brian Davies for his encouragement, inspiration and supervision throughout the research. I am also grateful to Professor John Rowlinson and Dr. Keith Hannabuss for their supervisory support.

Of those who have kindly reviewed and discussed points, notably in Chapter 1, I would like to single out Dr. David Barry, Mr. Ian Nicol, Dr. Sean Keating and Dr. Greg Ezra. For her superb typing of this thesis I particularly wish to thank Mrs. Joan Bunn.

For her continual support and love my fondest thanks go to my dear wife, Anne.

I wish to thank the Science Research council and Magdalen College, Oxford, for financial support.

The responsibility for any errors or failings in the thesis is, of course, my own.

\tableofcontents \markboth{Contents}{Contents}
\newpage
\chapter*{Introduction}

\addcontentsline{toc}{chapter}{Introduction} \markboth{INTRODUCTION}{Introduction}
What is an atom? Simple, as everyone knows it is a small ball-bearing (the nucleus) orbited by even smaller ball-bearings (electrons). Further investigation casts doubt on the smaller ball-bearings; no matter, replace them by a cloud of energy subject to little jumps in excitation. Allow further that very little ball-bearings can behave like waves and that light waves are prone to behave like ball-bearings and the mental furniture of the pragmatic scientist is nearly complete. Know which equations to turn on and the theory works.

It is in the spirit of molecules as balls joined together by flexible sticks that this thesis is written. The simple fact remains that to understand chemistry needs only minor modifications to the classical mechanical picture. Yet, we are told, quantum theory is true and to quote Dirac's famous words from 1928:
\begin{quote}
``The underlying physical laws necessary for the mathematical theory of a larger part of physics and the whole of chemistry are thus completely known...''
\end{quote}
The quantum theory of atoms and molecules is remarkable. Mathematically abstruse, difficult to visualise, still a hub of controversy but successful - and with no serious contender in nearly sixty years. In short, quantum theory has revealed little but delivered much.

So here are two apparently conflicting views of chemistry; on the one hand a conceptual framework based on classical mechanics, on the other the mystery of quantum theory. Put another way, how can chemists have so few qualms in practising their science when, as highlighted by Primas (Pr 2), quantum mechanics is at odds with many of the chemist's assumptions?

The aim of this thesis is to reconcile the classical conceptual framework to the quantum reality by examining the relation between classical and quantum mechanics.

Four themes underlie the presentation of ideas in the thesis:
\begin{itemize}
\item {\underline{Well-defined theories}}: the full explication of theories as a necessary precursor to any demonstration of intertheoretic reduction.
\item {\underline{Pure states}}: the assertion of realism in physics by employing pure states as primitive abstractions.
\item {\underline{Approximations}}: the role of proved approximations in identifying theories over a certain domain.
\item {\underline{Localisation}}: the use of compact sets of pure states to express localisation.
\end{itemize}

In summary, Chapter 1 sets the scene by reviewing the nature of scientific theories and their interrelationship. Chapter 2 presents a self-contained axiomatic theory of mechanics which includes classical and quantum mechanics as special cases. Chapter 3 exploits the analogy between compactness and localisation and takes a new look at scattering theory. Finally, Chapter 4 brings these ideas together to provide a clear method for determining if classical mechanics reduces to quantum mechanics.

\chapter{The Structure of Scientific Theories} \markboth{Chapter 1}{The Structure of Scientific Theories}
This thesis investigates the relation between two individually successful and sophisticated theories, classical and quantum mechanics. Some basic questions pose themselves at the outset:
\begin{itemize}
\item What is a theory?
\item Why are theories important?
\item How are theories related?
\end{itemize}
This first Chapter examines these questions with the aim of providing a reasoned framework for the more specific topics which follow.

\newpage
\section{Abstractions and Understanding}
\begin{quote}
``{\em{And we extend our concept...as in spinning a thread we twist fibre upon fibre. And the strength of the thread does not reside in the fact that some fibre runs through its whole length, but in the overlapping of many fibres}}.''

\hspace{295pt}Wittgenstein
\end{quote}

Both as a methodology and a body of knowledge science is considered by many to provide our most profound understanding of the world. Yet what is understanding?

We understand or can claim to understand many things - words, sentences, poetry, politics, scientific theories, mathematics and so on. Each requires a level or type of understanding which may be precise or vague, shallow or profound, concerned with a particular aspect of a subject or the subject as a whole. Such diversity suggests a return to basics. These basics, the categories with which we distinguish and organise experience, will be termed `abstractions'. For example, a component of communication such as a gesture is understood by someone to the extent of its meaning to them and this will assuredly evoke that individual's experience. Immediately, this leads to not only a discussion of meaning but also the prospect that the `meaning' of an abstraction rests on people sharing the same experience. To avoid such connotations we shall abandon the word `meaning', with its suggestion of uniqueness and absolutism, and adopt instead the less emotive word `significance'. Take a simple abstraction - the name of a person. Although a person's name may evoke different experiences for each individual, a simple test demonstrates common understanding: one individual brings forward the person to whom he believes the name belongs and associates the relevant symbols to this person. There may be some temporary confusion but the response from other individuals will soon be a mimicry of the association or some conventional expression of agreement such as a nodding of heads.

Proper names have, in this way, primarily perceptual significance, yet they also admit of understanding through their relation to abstractions for which denotation is accepted or presumed. Thus, in the absence of the person, we could refer to a photograph or construct sentences such as ``Churchill was the Prime Minister of Great Britain during the Second World War'', passing the denotative buck. This leads us to distinguish two ways in which an abstraction attains significance: firstly, by an agreement on the denotation of individual experience, which, we call {\em{denotative}} significance, and secondly, by the relation to other abstractions through language conventions, which we call {\em{contextual}} significance.

The distinction between denotative and contextual significance is not, as might have been hoped at first sight, clear cut. Consider again an individual's experiences denoted by a proper name. These experiences inevitably contribute to the denotation of other abstractions and induce an association of abstractions facilitating, for example, their conjunction in a sentence of verbal communication. We do not, therefore, attribute meaning solely by denotation.

Contextual significance, on the other hand, yields more readily to analysis. By placing an abstraction in context we are identifying it as an element of a structure - a set of auxiliary abstractions bearing well-defined relations to one another. Particular contexts may be isolated by choosing particular combinations - or patterns - within such a set. This appeal to a reference structure can be viewed as an act of abstraction which may be implicit, as in metaphor, or explicit, as in the axiomatising of logical argument. Moreover, a variety of reference structures may be employed and the process of abstraction repeated. In summary, the contextual significance of an abstraction derives from the structure of which it is deemed to be a component. If we view the branches of pure mathematics as reference structures, (even though motivation for their formulation may well reside in features of the experienced world), meaning is derived solely through axioms, rules of inference and theorems of the structure. We therefore distinguish logical and mathematical abstractions, in the above sense, and call them {\em{contextual}}, whilst we term the others {\em{descriptive}}.

Contextual abstractions do not as they stand denote anything, though associations may be made to other structures (models) yielding interpretations of one string of symbols in terms of others. A large part of mathematical activity may be looked upon as the analysis of such interrelation of structures. Now suppose that the `model' for a set of symbols in a logical system is a set of words in verbal language. If, by reference to his accustomed usage (based on denotation), an individual accepts this association, the words derive enhanced contextual significance from the logical structure. But we should not conclude from this that there exists a fixed correspondence between sets of words and (strings of) contextual abstractions. In fact, we shall argue that the usefulness of descriptive abstractions resides in their non-allegiance to any such fixed mapping.

We began by considering abstractions as components of communication, whose denotative significance is determined by social agreement on the symbolisation of each individual's experience. What, then, of a society with only one member; what of the `personal understanding' of an individual? In discriminating, organising and inspiring various experiences, `personal' abstractions conform to the analysis given above. We can go further; the fact that any organism must interact with its environment requires that certain external stimuli will trigger a form of internal signal, which `abstracts' the stimulus and will, in turn, induce certain responses. Allowing that the organism is capable of storing signals, then it will naturally form an image of its environment. The individual act of abstraction associated with the formation of such an image may thus be considered a basic biological function, rather than a sophisticated facility of higher mammals. Yet it is only through communal abstractions that any personal understanding may be revealed. The mere expression of an idea does not guarantee that other people will understand it in the sense intended, for discussion and elaboration may be needed to make it comprehensible. However, any claim that a `personal understanding' is, in principle, inexpressible at once sets it beyond discussion and thereby also outside the scope of this chapter.

That experiences are distinctive enough to be abstracted by humanity en masse leads to the belief that there exists an independent objective world structured in accordance with the abstractions we use to describe these experiences. But it is neither necessary nor desirable to presuppose such existence, convenient though this proves in normal discourse. Instead `reality' and its `existence' can be treated as a conjecture, a point of view which will be elaborated shortly. By so abandoning Naive Realism and indeed any claim to the existence of a universal underlying `truth', the fundamental distinction between subjectivity and objectivity evaporates, to be replaced by a recognition that understanding is primarily interactive.

It is natural to suppose that the use of a communal abstraction - such as `apple' - implies a shared identity between certain experiences of different individuals. Yet, given the diversity of our perceptions, such an assumption is unwarranted. Can {\em{my}} experiences of `apple' ever be said to strictly coincide with anyone else's?  Although there is a loose identity of significance following from our conventional agreement on denotation, we must allow individual's experiences, and thereby their denotations and associations of abstractions, to differ.

For proper names the agreed denotation of distinct sets of experiences is usually unambiguous, so to this extent denotation is independent of context. But for most descriptive abstractions it is the context which determines denotation and this, in turn, induces strings of associations peculiar to each individual. Thus any strict demarcation between contextual and denotative significance is lost. It is not, perhaps, surprising that the further away from proper names one goes the greater the risk of ambiguity, and the greater the reliance upon context. The more diverse the denotation, the less applicable become either/or classifications as shown, for example, by descriptions of states of mind or emotions.  The net result is an inherent woolliness of meaning, standing in marked contrast to the categoric contextual significance imposed by symbolic logic.

We propose that the various compromises in the conflict between contextual precision and denotative woolliness are responsible for the diversity of understanding noted at the beginning of this Section. This should certainly not be taken as an approval of woolliness per se, since ambiguity is usually undesirable (especially when describing experiences), but two points deserve emphasis. Firstly, acceptance of the difference of each individual's experience entails an intrinsic imprecision in denotation of descriptive abstractions, and a diversity of their contextual significance. Secondly, the flexibility of usage of descriptive abstractions, and their evocation of various associations to each individual, facilitates the generation of opinions, conjectures and theories. This flexibility, far from being undesirable, is a characteristic of language responsible for its fertility.

A feature of human understanding following from these considerations is its reliance upon metaphor, that is, the (implicit) recognition of a reference structure common to two or more denotatively disparate sets of organised descriptive abstractions. Indeed, identification of such structures prompts the formulation of logical systems.

\newpage
\section{Appearance and Reality}
\begin{quote}
``{\em{The principle that everything is open to criticism (from which this principle itself is not exempt) leads to a simple solution of the sources of knowledge}}''.

\hspace{295pt} Popper
\end{quote}

It is common sense to view the world as comprising independently existing objects, with immediate perception merely our transient experience of their various aspects. Who, for instance, would seriously doubt that the furniture in a room continues to exist and remain organised when the light is switched off? Reality is ascribed, though usually uncritically, to a variety of abstractions; after all, are electrons and protons any more real than cups and saucers, or these more real than love and hate? On the other hand, we learn to distinguish appearance from reality: dreams, fairies and optical illusions all, in their different ways, occur as experiences yet they, or what they signify, fail to qualify as real.

To clarify the notion of reality, we must step away from existence in isolation. We shall call the set of mental data arising from experiences, coded and co-ordinated by abstraction and association, a {\em{world-picture}}. (It is not unreasonable to allow that some of these experiences, criteria for abstraction and patterns of association may be hereditary). The `reality' of an abstraction may be loosely defined as the status of this abstraction in the world-picture. Although such a definition does not prohibit one individual's reality from corresponding to another's illusion, the constraint of social existence in aligning world-pictures removes most confusion. Still, this does not amount to a claim of independent existence which is the chief assertion of Realism. It is the repeatable distinguishability of certain experiences and their conjunctions which make it natural to presume that just as abstractions denote and relate to other abstractions, so the experiences of immediate perception are but part of the denotation of independent entities bearing various relations to one another. The distinction between Idealism (as the doctrine of only accepting existence `in the mind') and Realism is this switch from world-picture to world. Never a clear distinction, it can be abandoned if we view a world-picture as a conjecture on the structure of experience, both that of the individual and, through the use of communal abstractions, that of others. Taking this view, which may be called `Conjectural Realism', there can be no absolute reality - or knowledge of that reality - hiding, as it were, behind the mask of appearance, only more or less adequate conjectures for co-ordinating experience. Through its relation to the rest of a world-picture, the adequacy of a conjecture may be assessed by subjecting it to criticism and tests. Such a `call and response' approach to epistemology will be examined in Section 1.3.

In connection with these conclusions, let us briefly consider two well-known philosophical problems:

\vspace{8pt}
{\bf{1) The Problem of Universals}}
\vspace{8pt}

In putting forward the doctrine that the objects we identify through perception are but the imperfect impressions on matter of universal `Ideas' - such as the universal `cat' - Plato claims to see beyond appearances to a world of ultimate truth. However, his arguments, and those for `Essentialism', are just elaborations of the argument for reality, namely:

We use abstractions to denote objects and attributes, but only perceive their aspects; these abstractions refer to something, therefore there exists an ultimate reality comprising the entities of abstraction which, due to human frailty, we cannot directly apprehend.

With its immediate appeal as an 'explanation' of our verbal categorisation of experience, the world of `Ideas' or `Essences' consists of whichever abstractions are deemed fit for immortality (irrespective of consistency), at the same time immunising itself against empirical criticism by reserving the right to reject as mere appearance the inconvenient `reality' apprehended through the senses. Accordingly, the theory of Universals is an unnecessary and unfalsifiable conjecture which is only a problem if we are gullible enough to accept it.

\vspace{8pt}
{\bf{2) The Problem of Induction}}
\vspace{8pt}

This is simply stated as the problem of justifying reasoning from singular empirical statements to general laws. The logical part of the problem is solved, following Popper, if we note that laws, as conjectures, may be refuted; that is, whilst no number of confirming instances can ever render a general law `true', just one falsifying instance makes it false. However, this recourse to the mathematical technique of disproof by counter-example does not entirely banish the `Problem of Induction', as it reveals two new difficulties: the first concerns the `truth' of singular empirical statements, and the second the relative importance we may attribute to non-false laws.

Even if we interpret empirical truth as `correspondence with the facts' and suppose a statement of the `facts' to be understood, these `facts' may still be denied; for example, the claimed experience could be disregarded as being an hallucination, fabrication or misinterpretation. For this reason, the sceptic requires independent corroboration before accepting `facts', and faced with such possible denials the most acceptable laws are those amenable to testing by repetition so that anyone in doubt may observe for himself the consistency of the `facts'.

Now suppose that there is a law for which there are no accepted falsifying instances or, as is more often the case, one that has been modified to exclude falsifications. There do not appear to be any explicit criteria for estimating the importance of such a non-false law but confidence in it will be influenced by its applicability under diverse circumstances, and how it accords with the rest of a world-picture. This is of particular interest when more than one law is in competition as an `explanation of the facts' - a case which will be considered in detail below.

\newpage
\section{The Structure of Scientific Theories}
\begin{quote}
``{\em{Theories put phenomena into systems}}''.

\hspace{275pt} N. R. Hanson
\end{quote}

In marked contrast to the confident Logical Positivist explication of the features required of a scientific theory, philosophers of science have more recently given up devising categoric distinctions between science and other forms of knowledge preferring, in Suppe's words (Su 1 p.618) ``The examination of historical and contemporary examples of actual scientific practice''. Just as any hope of characterising a generic `scientific theory' appears to founder on the entangled diversity of the varied collections of knowledge and method we call `science' so, similarly, the corpus of mathematical structures, computational recipes, iconic models, paradigms, experimental procedures and verbal associations to other theories constituting Quantum Theory, defeats isolation of what we usually suppose to be the Quantum Theory. However, this need not condemn us to the bland scepticism evinced in the following quotation from Achinstein (Ac 1 p.129):
\begin{quote}
``T is a theory, relative to the context if, and only if, T is a set of propositions that...is...not known to be true or to be false, but believed to be somewhat plausible, potentially explanatory, relatively fundamental, and somewhat integrated''.
\end{quote}

There are several readily identifiable characteristics of all sciences and, more than this, if we concentrate on analysing the claims made by a scientific theory - particularly one employing explicit logical or mathematical structures - we may distinguish and typify its major ingredients. We therefore propose the following three characteristics of science:
\begin{enumerate}
\item `Call-and-response' epistemology: the `call' being a conjecture on the occurrence and conjunction of certain distinguishable experiences (relating to the `reality' of our world-pictures by use of descriptive abstractions); the `response' being an arbitration on the validity of the conjecture by appeal to perception under conditions broadly specified by descriptive abstractions.
\item Explanatory: each conjecture of science constitutes part of a systematic classification and organisation of experience, (communally expressed through abstractions); implicitly, therefore, this system conforms to some logical or mathematical principles.
\item Predictive: novel conjectures may be deduced, thereby extending the explanatory capabilities of science.
\end{enumerate}
These characteristics are shared by the particular systematisations of knowledge and conjecture we call `scientific theories'. A theory is never a completely demarcated and static body of knowledge, and, whilst not quite all things to all men, different people will, according to their various needs and motives, emphasise different aspects. With a view to forming an opinion on how a scientific theory should be explicated, let us call attention to four components (or aspects) of its structure:
\begin{enumerate}
\item Fundamental Model: sets of logical or mathematical axioms, rules of inference, and theorems, often described - and derided - as the `formalism'.
\item Phenomenology: a body of experimental data, organised through correlations called `experimental laws', the description of which is based on ordinary language (communal abstractions). Loosely, the phenomenology constitutes the `facts'.
\item Co-ordinative Definitions: associations between the contextual abstractions of the fundamental models, and sets of descriptive abstractions in the rest of the world-picture (which may include other theories); co-ordinative definitions provide both the empirical interpretation of the fundamental models and the relation between the phenomenology and the fundamental models.
\item Recipes: formal rules, not necessarily derivable from the fundamental model, for going from one set of data to another. The data may be phenomenological or in the form of fundamental constants and parameters; a collection of recipes may be sufficiently coherent to qualify as a subtheory, employing some of the symbols of the main theory and with similar co-ordinative definitions, but need not be consistent with the fundamental model.
\end{enumerate}
In practice, any theory is a hotchpotch of these ingredients, with various alternatives for fundamental models, implicit dependencies on other theories, numerous subtheories and analogies, and recipes varying from algorithms through paradigms to overt experimental procedures. Just as the woolliness of descriptive abstractions makes their meaning difficult to pin down, so the chief culprits responsible for the confusion about what constitutes a scientific theory are the co-ordinative definitions. It is to these that we now look for resolution of the debate between the `Received View' and `Weltanschauung' analysis of scientific theories. By categorising the many philosophies of science in this way we are adopting the terminology of (Su 1). Familiarity with this reference will be assumed in what follows.

Let us first make a few remarks concerning the observational-theoretical distinction on the meaning of terms in a scientific theory. The motivation for dividing the descriptive abstractions of a scientific theory into `observational' and `theoretical' is to distinguish the names, attributes and relationships of objects available to direct observation, such as 'a red brick', from those which are not, such as `a wave function'. Although consideration of any list of scientific terms will reveal that there is no tenable sharp distinction between `observational' and `theoretical' in normal usage, adherents of the `Received View' proposed that a clean division could be effected which would retain, indeed reinforce, the scientific significance of the abstractions used in a theory. Allied to this is a reformulation of Kant's notion of `analytic' and `synthetic' sentences (propositions, judgements), where a sentence is called analytic if it is true because of its logical form and the explicitly defined meaning of its terms, or called synthetic if it is true due to its observed validity as a `fact'. Both distinctions break down because of, firstly, the elusiveness of `meaning' (other than logical) for analytic sentences and theoretical terms and, secondly, the problem of demarcating `observable' for synthetic sentences and observational terms. In view of our previous discussions of abstractions it is not surprising that any attempted enforcement of these distinctions leads to a highly artificial language with ad hoc meanings.

Implicit in the observational-theoretical distinction is the presumed existence of a `neutral observation language', that is, a theory-independent language describing the objects, and their attributes and relationships, observed by direct perception.

Proponents of the various `Weltanschauung' analyses deny the theory-in\-dependence of perception, arguing that science is part of a perspective on the world of experience and that the structure of scientific theories will be revealed by characterising their context within this perspective, in particular, by focusing attention on how science is actually done and evolves rather than what it is, or should be, as a finished product. However, the `Weltanschauung' soon becomes a metaphysical panacea for all philosophical ills, an intangible bag of paradigms, social attitudes, historical conditioning and individual dispositions. If we choose not to pursue the `Weltanschauung' we are still left to deal with the dependence of facts and observations on the theories which are supposed to describe them. Taken to the extreme, if the world is what we decide (or what our language constrains) it to be, how then can there be objective knowledge, and are we not forced to retreat into a subjective Idealism? The root of this difficulty with the `Received View' is, we propose, an unwarranted faith in Realism, resulting from the desire for a categoric distinction between theoretical description (reality) and its manifestation in perceptual terms (appearance). If the arguments of Section 1.1 are accepted, then we can do no better than employ descriptive abstractions whose contextual meaning derives from the logical systems to which they are assigned, so the `Weltanschauung' objection is justified but only insofar as it inevitably applies to all descriptive abstractions (and, therefore, co-ordinative definitions). As far as anything can be, ordinary language (and its technological elaborations) is a `neutral observation language' designed for unambiguously demarcating and relating communal experiences, with logic lending precision to its organisation. In conclusion, therefore, if the `observational-theoretical' distinction is replaced by a `descriptive-contextual' awareness, the main argument for the `Weltanschauung' evaporates.

A key feature of science is that it self-consciously turns the tables on everyday understanding, and begins with a logical or mathematical structure which has to be related to ordinary language through co-ordinative definitions involving descriptive abstractions. However, in providing denotative significance, the co-ordinative definitions cannot be perfect, if by perfect is meant logically precise, in the experiences they specify; they can only, at best, be unambiguous.

Muddled though a scientific theory may be in practice, the conjectural status of knowledge demands that the claims involved in a theory be made clear so that it can `stand up and be counted'. In terms of the components mentioned above this requires explication of, in particular, the fundamental model and co-ordinative definitions.

Switching attention from the structure of theories to their function in scientific enquiry brings out the primary role of recipes in understanding. After all, the acceptability of a theory is judged not so much by its aesthetic purity as by the adequacy of its canonical divisions, associations and predictions of phenomena. The doctrine of {\em{Instrumentalism}} espouses this hard-nosed attitude by viewing a scientific theory as a set of rules for:
\begin{enumerate}
\item Identifying certain features of experience (which we call `experimental categories').
\item Inferring one set of experimental categories from another.
\end{enumerate}
Instrumentalism approaches the world phenomenologically, with theories the instruments for dealing with experience, and knowledge the use of these instruments. In this way, questions concerning the 'reality' of theoretical terms, or their translatability into observational terms, become meaningless.

Modern science, with its proliferation of exotic mathematical structures which do not admit commonplace analogies and its use of involved experimental techniques, is considerably less amenable than ordinary experience to Naive Realism. Couple this with the many unresolved controversies over interpretation, notably in Quantum Mechanics, then Instrumentalism becomes the default epistemology for the practically-minded sceptic.

The Instrumentalist view of theories as mere `leading principles' undoubtedly characterises a substantial part of scientific practice, although deduction from conjectured universal laws, (together with singular statements), is also widely used. Both draw attention to the implicit logic of Instrumentalism, namely, that laws are transformed into rules of inference. To take Nagel's example, (Na 1 p.67):
\begin{quote}
``The conclusion that a given piece of wire {\em{a}} is a good electrical conductor can be derived from the two premises that {\em{a}} is copper and all copper is a good electrical conductor....However, that same conclusion can also be obtained from the single premise that {\em{a}} is copper if we accept as a principle of inference the rule that a statement of the form '{\em{x}} is a good electrical conductor' is derivable from a statement of the form `{\em{x}} is copper'.''
\end{quote}
Here the law (universal premise) ``all copper is a good electrical conductor'' is replaced by the rule of inference (universal conditional) ``For any {\em{x}}, if {\em{x}} is copper, then {\em{x}} is a good electrical conductor''. This example indicates a principal weakness of Instrumentalism: by reducing theories to lists of rules of identification and inference the unity of understanding accrued by the hypothetico-deductive view of theories is not merely confused, it is disavowed. Whilst many specific scientific claims are Instrumentalist, a theory is the unification of such diverse claims under an explanatory umbrella of deduction from explicit assertions. By denying that its rules are conclusions, Instrumentalism can avoid questioning the validity of these primitive theoretical assertions.

But to espouse Instrumentalism is not just to express oneself circumspectly, it is also to resurrect the observational-theoretical distinction since the experimental categories are presumed available to direct observation. If we accept that observational terms are theory-laden then, as a dogma on the exclusive 'reality' of experimental categories, Instrumentalism must be abandoned.

Finally in this Section, a word about the use of analogies in a theory. In Section 1.1 the important role of metaphor in human understanding was viewed as the recognition of a common logical structure in two or more denotatively disparate sets of organised descriptive abstractions. It should not be surprising, therefore, to find that metaphors are used in scientific theories where, being more extensive and explicit, they can be given the status of `analogies'. Analogies vary in precision from areas of ordinary experience through substantive (`iconic') models to detailed mathematical structures, and what is being analogised ranges in extent from parts of the phenomenology through recipes, or particular applications of a theory, to identifiable subtheories. By associating parts of a novel or complex or highly abstract theory to more familiar systems of knowledge, analogies became a key component in the development and understanding of the theory, but their function as heuristic and pedagogical aids should not overshadow the proper interpretation claimed for the theory. Analogies are allegorical, and as Erich Heller puts it when distinguishing between symbol and allegory (albeit referring to denotative significance with undertones of Universals):
\begin{quote}
``The symbol is what it represents; the allegory represents what, in itself, it is not.''
\end{quote}

\newpage
\section{Intertheoretic Reduction}
\begin{quote}
``{\em{Yet the postulate that lies at the root of every scientific enquiry, the act of faith which has always sustained scientists in their unwearying search for explanation, consists in the assertion that it must be possible - though perhaps at the heavy cost of ideas held for long and concepts of proved usefulness - to reach a synthetic view uniting all the partial theories suggested by the various groups of phenomena, and embracing them all despite their apparent contradictions}}.''

\hspace{275pt} Louis De Broglie
\end{quote}
By intertheoretic reduction, or reduction for short, we mean the doctrine that one theory (the `secondary theory') can be subsumed under another theory (the `primary theory'). To avoid a confusion about reduction present in the literature, we are here concerned with neither the historical circumstances of reductions nor, directly, the reasoning patterns actually used in scientific enquiry. Instead, this Section addresses the requirements which need to be satisfied in order that one theory or set of theories be considered a special case of another theory.

If, following Section 1.3, we accept that scientific theories are logically or mathematically organised sets of abstractions (descriptive through co-ordinative definitions) employed to explain diverse bodies of empirical fact, then the importance of reduction is evident if we interpret `empirical fact' to be the correlation of certain experiences through experimental laws and accord to it the status of a theory. Intertheoretic reduction is thus the natural extension, from a phenomenology to a distinct theory, of this explanatory unification. Since there is no ground other than the general success of science to suppose that some universal all-embracing theory lies just beyond the horizon, reduction should not be viewed as an inevitably true, even self-fulfilling, prophecy. Like any conjecture, a reduction has to be explicated and subjected to logical and empirical canons of validity.

It is often argued that in the hierarchical ordering of knowledge, which Reductionism purports to explain, there are `properties' and `objects' at each level of organisation in the hierarchy not deducible from the supposedly explanatory level above. Confining attention to the theories involved rather than invoking any Essentialism about entities or their attributes, this `Gestalt' view of emergent features, often called {\em{Holism}}, asserts that certain (sets of) descriptive abstractions in a secondary theory cannot be identified with or derived from combinations of terms in the primary theory. However, the arguments for Holism are plagued by confusion of the meaning of abstractions in the various theories so, although not rejecting it out of hand, we shall reserve Holism as a default in favour of considering reduction for theories in the mathematical sciences.

A necessary condition for reduction is that, for the circumstances corresponding to the experimental domain under consideration, the empirically relevant propositions of the secondary theory may be deduced from the fundamental model of the primary theory. We call this task the {\em{analytic problem of reduction}}. What constitutes a `deduction' has to be considered carefully: we take it to be the proof, in the fundamental model of the primary theory, of a set of propositions, denoted P(1), which can be identified, through certain criteria, with the empirically relevant propositions, denoted ERP(2), of the secondary theory. That is, we require a solution, which need not be unique, of the analytic problem of reduction, to specify:
\begin{enumerate}
\item Conditions of Deducibility, denoted CD: the mathematical conditions given by the theorems which deduce P(1) in the primary theory.
\item Identifications, denoted I: an association between the symbols, and combinations of symbols, in ERP(2) and in P(1). 
\item Criteria of Identity, denoted C(.,.): a set of criteria for according the propositions in P(1) equivalence to those identified from ERP(2).
\end{enumerate}
Denoting the fundamental models of the primary and secondary theories by M(1) and M(2), respectively, the relations between the various quantities may be illustrated by:
\vspace{4pt}

\includegraphics[scale=0.55]{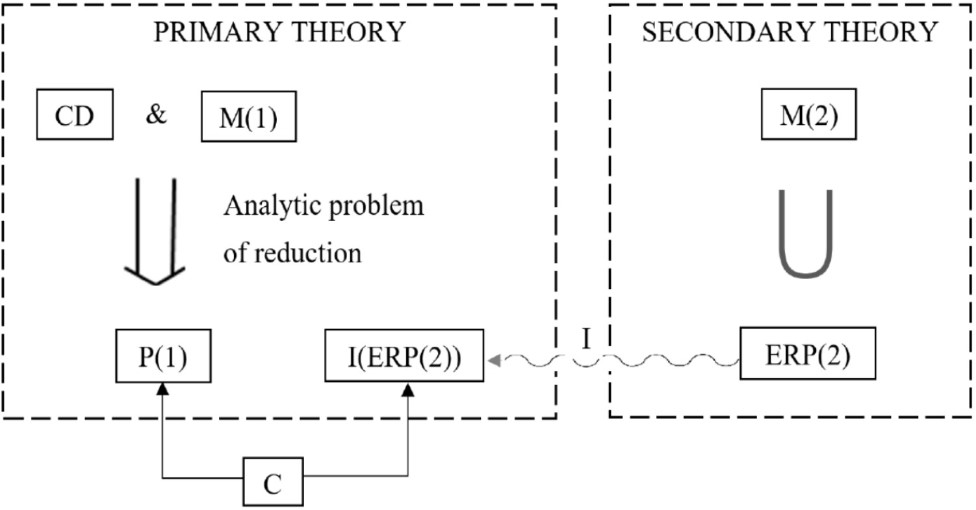}


{\bf{Figure 1.1: The Analytic Problem of Reduction}}

\vspace{16pt}
The analytic problem of reduction is solved if, for each b $\in$ ERP(2) there exists a $\in$ P(1) such that C(a, I(b)) are satisfied. For fixed fundamental models M(1) and M(2), and criteria of identity C, a solution is then the pair (CD, I).

An application of these principles is given in Section 4.2, to which the reader may turn for an example.

Now suppose a solution of the analytic problem of reduction has been found, then we propose that the specifications 1), 2), and 3) must satisfy the following three requirements, respectively, before the reduction of theories can be considered acceptable:
\begin{quote}
1') Applicability: the conditions of deducibility should include, when interpreted through the co-ordinative definitions, the circumstances appropriate to the application of the secondary theory.

2') Connectability: the identifications should not lead to a contradiction in the meaning - given by co-ordinative definitions - of descriptive abstractions in both theories, under the conditions of deducibility.

3') Indistinguishability: the criteria of identity should be consistent with the experimental resolutions in the domain of applicability of the secondary theory.
\end{quote}
Thus, if a reduction is acceptable, both the primary and secondary theories will accommodate the empirical facts equally well for the domain of applicability of the secondary theory. In such a case we call the choice of theoretical explanation {\em{weakly conventional}}, and the theories weakly conventional alternatives, where `weakly' signifies that one of the theories is primary with respect to the other. In practice, it is the secondary theory which is usually chosen for explaining its own domain, since the constraints imposed by the conditions of deducibility make the primary theory too cumbersome.

In the case where the fundamental models of two theories can be demonstrated to be equivalent - for trivial criteria of identity - we call the choice of theoretical explanation {\em{strongly conventional}}. An example of strong conventionality can be found in the Schr\"{o}dinger and Heisenberg pictures of Quantum Mechanics. This stronger form of Conventionalism does not, as some have thought, relegate all physical laws to the status of `concealed definitions'; rather, it determines which propositions can be taken, conventionally, as laws or as definitions.

In practice, it is rarely the case that if two or more theories account for the 'facts' they are, as they stand, demonstrably weakly conventional alternatives. All that can be said is that for the domain under consideration - typically an experiment or class of experiments - the theories are {\em{empirically equivalent}}. However, it may be possible to formulate a primary theory to which the empirically equivalent alternatives, restricted to the particular domain, reduce, where either the `facts' already constitute a phenomenology common to the theories, or they can be acceptably reinterpreted in the primary theory. We are drawing a distinction here between the `domain' and the `facts'; the former refers to the general ordinary language description of the experimental circumstances, whereas the latter includes singular empirical statements which may well be expressed in terminology peculiar to the theory in question.

Even if it is agreed that intertheoretic reduction, as described above, is a worthwhile ambition, its demonstration - if, indeed, it can be demonstrated for a pair of theories - is a major undertaking. Nevertheless, secondary theories have often been incorporated into primary theories and given the name `subtheories', so let us finally introduce some terminology for these in line with the discussion of reduction. Although, as part of a more extensive body of knowledge, the meanings of terms are inevitably modified, subtheories usually retain their own auxiliary symbols, hypotheses, co-ordinative definitions, analogies and recipes, and remain the principal explanatory tool for their, often well-demarcated, experimental domains. So suppose that a subtheory is sufficiently autonomous to be explicated separately - as far as any theory can be - from the full theory, then I shall call it an {\em{approximation}} if it is rigorously reducible to the full theory, and an {\em{idealisation}} if it is not. This distinction carries over to the various mathematical structures, often called `models', conjectured for circumstances covered, in principle, by the full theory but apparently too complicated to be amenable to direct analysis in terms of the fundamental model. In order that the mathematics be tractable, a model typically suppresses certain features, and `idealises' others, of the full theory. Whilst it is undoubtedly convenient to accept an idealisation or model as a subtheory in the fond hope that it is `really' an approximation, simply calling a lemon a peach does not make it taste sweet. If the ideal of a unified theoretical explanation, and with it the gain of greater understanding, is to be preserved, this act of faith must be replaced by an acceptable proof.

\chapter{A Theory of Mechanics} \markboth{Chapter 2}{Chapter 2: A Theory of Mechanics}
The main purpose of this Chapter is to state the theories of Classical and Quantum mechanics. Unfortunately, the two theories are usually formulated in quite different terms, both conceptually and mathematically. For this reason we devote considerable effort to determining a common foundation of the theories and, in particular, the extent to which they share the features of a more general theory of mechanics. Such a programme has been attempted before, but primarily from the point of view of quantum logic. For a recent review see (B $\&$ C 1). Although we draw heavily on these results our approach, and subsequent emphasis, is different. In essence, we adopt the `state of a system' as a primitive concept in mechanics.

Classical mechanics was based on the notion of a material object in independent possession of properties which it was the business of theory and experiment to uncover. Quantum Theory, rising out of the statistical mire of atomic phenomena, changed all that. Very small objects - or their theoretical counterparts - would not conform to the `classical' principles which governed everyday objects. The reality-status of theoretical terms, notably the state of a system, became obscure and contentious causing physicists to sound the retreat from Realism. Observed results - `what we know' - became the focus of attention and from this apparently secure footing evolved modern Quantum Theory as a theory of `observables'. Yet the status of observables, especially their relation to experimental results, is not clear.  Observables (or `propositions') we take to be theoretical quantities which represent measuring devices (or statements about measurements) with respect to a system. Three considerations motivate our abandonment of these quantities as primitive concepts:
\begin{enumerate}
\item Experiments and measurements are highly involved physical processes which are not in general amenable to simple analysis. Notable by their rarity are specific choices of observable for particular measurements and systems.
\item The observables that actually are specified do not represent actual instruments but are, instead, kinematic (or space-time) properties - for example, position, momentum, angular momentum, free particle energy. 
\item The `observables' approach does not correspond to scientific practice where it is invariably the state of a system which is taken to be fundamental.
\end{enumerate} 
The third of these is, perhaps, the most important - how, for example, does a chemist conceive of atoms and molecules if not as objects in particular states?

Our aim, therefore, is to re-establish Realism in mechanics - in the sense of Conjectural Realism using the state of a system - with observables, and what is observed, deduced rather than assumed.

Unlike subsequent Chapters, the mathematical development has been placed with the main text. The format is, as a consequence, somewhat monolithic but will, we hope, satisfy the more mathematically minded reader.

{\bf{Note}}: Since the Chapter was written there has appeared an up-to-date review of the logic of quantum mechanics by Beltrametti and Cassinelli (B $\&$ C 1). In the terminology of this review we have used a `transition-probability space' and fulfilled part of the programme they refer to (on p. 241) as
\begin{quote}
`` ... more a hope than an immediate possibility.''
\end{quote}
The review does, however, provide much of our development in Section 2.1 albeit with different terminology. The additional feature of our work which enables us to utilise Piron's Theorem (2.47), namely Axiom 4 on the existence of a `closest element', might provide a key to further research in this area.

\newpage

\section{Systems, Pure States, and Intrinsic Probabilities}
A hallmark of the experimental method is its careful selection of particular experiences from the diversity of those available. I shall call any such set of particular experiences a {\em{domain}}, its theoretical counterpart a {\em{system}}, and all the rest of experience the {\em{environment}}. Were it the case that domains could not be rationalised without reference to the environment then scientific explanation would be a tall order indeed; fortunately, however, some domains, and I shall call these {\em{isolated}}, may be rationalised irrespective of the condition of the environment, whilst for many others, which I shall call {\em{separated}}, the environment can be accommodated by employing only a few auxiliary quantities. Perhaps the most noteworthy feature of theories of mechanics, and the one responsible for their wide range of applicability, is their capacity for describing both domains and subdomains as separated. Not that the notion of a separated domain (or system) is without difficulties, especially in quantum mechanics; but let us for the moment assume domains (and systems) to be separated.

The fundamental notion to be elaborated in this Chapter is that of a {\em{pure state}} of a system - this we take to be a mathematical object which provides a complete description of the preparation or condition of the system, and we assume such a description is possible in the theory even if experimentally attainable only as some form of limit of operations. It is here that we are applying the philosophy of Conjectural Realism. We are conjecturing that we may think of the existence of a system's condition, just as we normally think of the state of an everyday object such as a chair. Throughout this Chapter a system will be denoted by $\Sigma$, and the set of pure states of $\Sigma$ by $S$.  Suppose, then, that $s \in S$ provides a complete description of a condition of a system. Although in a classical theory there would be no chance of any different pure state providing the description given by $s$, in quantum mechanics there is such a possibility, which may be expressed by saying that whilst a pure state is a complete description, it need not be an exclusive description. Accordingly, we introduce an {\em{intrinsic probability function}} $p_s$ associated to each pure state, where $p_s(s')$ is the probability that the description of the system by $s$ can be given by $s' \in S$. Loosely, a system in the pure state $s$ has a probability $p_s(s')$ of being in the pure state $s'$. The word `probability' in these motivating remarks may be cause for some discomfort; justifiably so, and we shall consider its interpretation - upon which the interpretation of mechanics depends - in Section 2.2. In this Section, `probability' should be viewed mathematically; elucidation, if needed, of any guiding verbal descriptions can be found in Section 2.2.

Whereas a system is a verbal, and necessarily rather nebulous, notion, pure states and intrinsic probability functions can be given contextual meanings:

A pure state is an element of a set $S$. To each $s \in S$ associate a positive function $p_s$, the intrinsic probability function, on $S$ which satisfies:

{\bf{2.1 Axiom 1}}

For each $s$, $s'$ $\in$ $S$:

$p_s(s')$ $\leq$ 1 \hspace{10pt} with equality iff $s = s'$.

{\bf{2.2 Axiom 2}}

$p_s(s')$ =  $p_{s'}(s)$ \hspace{10pt} $\forall$ $s$, $s'$ $\in$ $S$.

An immediate consequence of Axiom 1 is:

{\bf{2.3 Lemma}}

$p_s$ = $p_{s'}$ $\Leftrightarrow$ $s = s'$ \hspace{10pt} $\forall$ $s$, $s'$ $\in$ $S$.

{\bf{Proof}}

$\Leftarrow$ is obvious, so suppose $\Rightarrow$ is false. Then $\exists$$s$, $s'$ $\in$ $S$ s.t.:

$p_s$ = $p_{s'}$ $\nRightarrow$ $s = s'$.

$p_s(t)$ = $p_{s'}(t)$ $\forall$$t$ $\in$ $S$, so in particular $p_s(s)$ = $p_{s'}(s)$, then by Axiom 1 $s = s'$, which is a contradiction and proves the Lemma.

Let $2^{S}$ denote the power set of $S$, then define: 

{\bf{2.4 Definition}}

Let $T$ $\in$ $2$$^{S}$, then the {\em{annihilator set}} of $T$ is defined as the set:

$T^{\bot}$ $\equiv$ $\{r$ $\in$ $S$ $|$ $p_{r}(t)$ = 0 \hspace{10pt} $\forall$ $t \in T$$\}$.

{\bf{2.5 Definition}}

Let $T$ $\in$ 2$^{S}$, then the {\em{superposition set}} of $T$ is defined as the set:

$\bar{T}$ $\equiv$ $\{$$t$ $\in$ $S$ $|$ $p_{t}(r)$ = 0 \hspace{10pt} $\forall$ $r$ $\in$ $T$$^{\bot}$$\}$ $\equiv$$T$$^{\bot}$$^{\bot}$ .

The next Lemma summarises some simple properties of annihilator and superposition sets. The usual notation of set theory is employed.

{\bf{2.6 Lemma}}

Let $T$, $R$ $\in$ 2$^{S}$, then:

(i) $T^{\bot}$ = $\overline{T}^{\bot}$ = $T^{\overline{\bot}}$

(ii) $T \subseteq$ $\overline{T}$ = $\overline{\overline{T}}$

(iii) $T \subseteq$ $R$ $\Rightarrow$ $R^{\bot}$ $\subseteq$ $T^{\bot}$ $\Leftrightarrow$ $\bar{T}$ $\subseteq$ $\bar{R}$

(iv) $T^{\bot}$ $\cup$ $R^{\bot}$ $\subseteq$ $(T \cap R)^{\bot}$

(v) $(T \cup R)^{\bot}$ $\subseteq$ $T^{\bot}$ $\cap$ $R^{\bot}$.

{\bf{Proof}}

Note first that the inclusion in (ii) is obvious.

(i) First equality: let $x$ $\in$ $\overline{T}^{\bot}$ then by Definition 2.4 $p_{x}(t)$ = 0 $\forall t$ $\in$ $\overline{T}$ hence $p_{x}(t)$ = 0 $\forall t$ $\in$ $T$ which makes $x$ $\in$ $T^{\bot}$; conversely, let $x$ $\in$ $T^{\bot}$ then by Definition 2.5 $p_{x}(t)$ = 0 $\forall t$ $\in$ $\overline{T}$ so that $x$ $\in$ $\overline{T}^{\bot}$. The second equality is trivial.

(ii) Clearly $\overline{T}$ $\subseteq$ $\overline{\overline{T}}$, so let $x$ $\in$ $\overline{\overline{T}}$ then $p_{x}(u) = 0$ $\forall u$ $\in$ ($\overline{T}$)$^{\bot}$ = $T^{\bot}$ by (i).

(iii) First implication is obvious, and the second follows from the first using (i).

(iv) Let $x \in (T^{\bot} \cup R^{\bot})$ then either $p_{x}(y) = 0$ $\forall y \in T$ or $p_{x}(z) = 0$ $\forall z \in R$ (or both), hence $p_{x}(w) = 0$ $\forall w \in T \cap R$ and so $x \in {(T \cap R)}^{\bot}$.

(v) Let $x \in {(T \cup R)}^{\bot}$ then $p_{x}(u) = 0$ $\forall u \in T \cup R$ so $x \in T^{\bot}$ and $x \in R^{\bot}$ hence $x \in T^{\bot} \cap R^{\bot}$.
\vspace{8pt}

{\bf{2.7 Proposition}}

(i) $\overline{T \cap R} \subseteq \overline{\overline{T} \cap \overline{R}} = \overline{T} \cap \overline{R}$

(ii) $\overline{T} \cup \overline{R} \subseteq \overline{T \cup R} = \overline{\overline{T} \cup R} = \overline{T \cup \overline{R}}$

\vspace{8pt}

{\bf{Proof}}

(i) The inclusion follows immediately from (ii) of Lemma 2.6. From (iv) and (iii) of Lemma 2.6 we obtain $\overline{\overline{T} \cap \overline{R}} \subseteq (\overline{T}^{\bot} \cup \overline{R}^{\bot})^{\bot}$, but by (i) and (v) of Lemma 2.6 we also have $(\overline{T}^{\bot} \cup \overline{R}^{\bot})^{\bot} = (T^{\bot} \cup R^{\bot})^{\bot} \subseteq \overline{T} \cap \overline{R}$. Clearly $\overline{T} \cap \overline{R} \subseteq \overline{\overline{T} \cap \overline{R}}$, so (i) of the Proposition is proved.

(ii) To prove the inclusion notice that $\overline{T} \cup \overline{R} = T^{\bot \bot} \cup R^{\bot \bot} \subseteq (T^{\bot} \cap R^{\bot})^{\bot}$ by (iv) of Lemma 2.6. But applying (v) and (ii) of Lemma 2.6 gives $(T^{\bot} \cap R^{\bot})^{\bot} \subseteq (T \cup R)^{\bot \bot}$. This proves the inclusion and leaves us to prove only that $\overline{\overline{T} \cup R} \subseteq \overline{T \cup R}$. But obviously $\overline{\overline{T} \cup R} \subseteq \overline{\overline{T} \cup \overline{R}}$, which, with the inclusion gives $\overline{\overline{T} \cup R} \subseteq \overline{\overline{T \cup R}} = \overline{T \cup R}$ by (ii) of Lemma 2.6.

\vspace{8pt}
{\bf{2.8 Proposition}}
\begin{quote}
(i) $(T \cup R)^{\bot} = T^{\bot} \cap R^{\bot}$

(ii) $(\overline{T} \cap \overline{R})^{\bot} = \overline{T^{\bot} \cup R^{\bot}}$.
\end{quote}

\vspace{8pt}

{\bf{Proof}}

(i) By (v) of Lemma 2.6 it is sufficient to prove $T^{\bot} \cap R^{\bot} \subseteq (T \cup R)^{\bot}$. Using various combinations of the foregoing results we obtain:
\begin{quote}
$T^{\bot} \cap  R^{\bot} = \overline{T^{\bot} \cap  R^{\bot}} = (T^{\bot} \cap R^{\bot})^{\bot \bot} \subseteq (\overline{T} \cup \overline{R})^{\bot} = (\overline{T \cup R})^{\bot} = (T \cup R)^{\bot}$.
\end{quote}
(ii) $\subseteq$ : $(\overline{T} \cap \overline{R})^{\bot} = (T^{\bot \bot} \cap R^{\bot \bot})^{\bot} \subseteq (T^{\bot} \cup R^{\bot})^{\bot \bot}$.

$\supseteq$ : $\overline{T^{\bot} \cup R^{\bot}} = \overline{\overline{T}^{\bot} \cup \overline{R}^{\bot}} \subseteq \overline{(\overline{T} \cap \overline{R})^{\bot}} = (\overline{T} \cap \overline{R})^{\bot}$.

\vspace{8pt}

{\bf{2.9 Definition}}

Let $T, R \in 2^{S}$, then $T$ and $R$ will be said to be {\em{orthogonal}}, denoted $T \bot R$, if:
\begin{quote}
$p_{t}(r) = 0$ \hspace{8pt} $\forall t \in T$ and $\forall r \in R$.
\end{quote}

\vspace{8pt}

{\bf{2.10 Lemma}}
Let $T, R \in 2^{S}$, then the following are equivalent:
\begin{quote}
(i) $T \bot R$

(ii) $T \cap R^{\bot} = T$

(iii) $T^{\bot} \cap R = R$

(iv) $R \subseteq T^{\bot}$

(v) $T \subseteq R^{\bot}$

(vi) $\overline{T} \bot \overline{R}$.
\end{quote}

\vspace{8pt}

{\bf{Proof}}

All are trivial except for (vi). Clearly if  $\overline{T} \bot \overline{R}$ then $T \bot R$, so suppose $T \bot R$. Using the other equivalences we get: $T \bot R \Leftrightarrow T \subseteq R^{\bot} \Rightarrow \overline{R} \subseteq T^{\bot} \Leftrightarrow T \bot \overline{R} \Rightarrow \overline{T} \bot \overline{R}$.

\vspace{8pt}
{\bf{2.11 Proposition}}

Let $Q, T, R \in 2^{S}$, then:
\begin{quote}
$(Q \bot T$ and $Q \bot R)$ $\Leftrightarrow$ $Q\bot(\overline{T \cup R}) \Leftrightarrow Q\bot(T \cup R)$.
\end{quote}

\vspace{8pt}

{\bf{Proof}}

($Q \bot T$ $\&$ $Q \bot R) \Leftrightarrow (Q \subseteq T^{\bot}$ $\&$ $Q \subseteq R^{\bot}) \Leftrightarrow Q \subseteq T^{\bot} \cap R^{\bot} \Leftrightarrow Q \subseteq (T \cup R)^{\bot} \Leftrightarrow Q \bot(\overline{T \cup R}) \Leftrightarrow Q\bot(T \cup R)$.

\vspace{8pt}

{\bf{2.12 Definition}}

Let the empty set, (which is an element of $2^{S}$ but not of $S$), be denoted by $\O$, then the annihilator of $\O$ is defined as:
\begin{quote}
$\O^{\bot} = S$.
\end{quote}
\vspace{8pt}

{\bf{2.13 Proposition}}
\begin{quote}
(i) $S = \overline{S}$

(ii) $S^{\bot} = \O = \overline{\O}$

(iii) Let $T \in 2^{S}$ then $\overline{T} \cap T^{\bot} = \O$

(iv) Let $T \in 2^{S}$ then $\overline{\overline{T} \cup T^{\bot}} = S$.
\end{quote}

\vspace{8pt}

{\bf{Proof}}
\begin{quote}
(i) $\overline{S} = S^{\bot \bot} = \O^{\bot} = S$.

(ii) Let $x \in S^{\bot}$, then $p_{x}(s) = 0$ $\forall s \in S$. But by Axiom 1 this means that $x \not\in S$, hence that $S^{\bot} = \O$. For the second equality we have $\overline{\O} = \O^{\bot \bot} = S^{\bot} = \O$.

(iii) Let $x \in \overline{T} \cap T^{\bot}$, then $p_{x}(t) = 0$ $\forall t \in \overline{T}$, but $x \in \overline{T}$ so $x \not\in S$ hence $\overline{T} \cap T^{\bot} = \O$.

(iv)$\overline{\overline{T} \cup T^{\bot}} = \overline{T^{\bot \bot} \cup T^{\bot}} = (T^{\bot} \cap \overline{T})^{\bot} = \O^{\bot} = S$.
\end{quote}

\vspace{8pt}

{\bf{2.14 Definition}}

The {\em{set of superposition sets}} is defined as:
\begin{quote}
$L_{S}$ $\equiv$ $\{ T \in 2^{S}$ $|$ $T = \overline{T}\}$.
\end{quote}

$L_{S}$ is a poset under the ordering relation of set inclusion.

\vspace{8pt}

{\bf{2.15 Definition}}

Let $T, R \in 2^{S}$, then their {\em{join}} is the subset of $S$ defined as:
\begin{quote}
$T \vee R \equiv \overline{\overline{T} \cup \overline{R}} = \overline{T \cup R}$.
\end{quote}

\vspace{8pt}

{\bf{2.16 Definition}}

Let $T, R \in 2^{S}$, then their {\em{meet}} is the subset of $S$ defined as:

$T \wedge R \equiv \overline{\overline{T} \cap \overline{R}} = \overline{T} \cap \overline{R}$.

\vspace{8pt}

{\bf{2.17 Theorem}}

$(L_{S}, \vee, \wedge, \bot)$ is an orthocomplemented complete lattice.

\vspace{8pt}

{\bf{Proof}}

Under the operations $\vee$ and $\wedge$, $L_{S}$ is clearly a lattice with zero element $\O$ and unit element $S$. It is also complete since the join and meet of an arbitrary family of elements of the lattice can be defined from their set-theoretic counterparts: Let $I$ be any index set, then:
\begin{quote}
$\underset{i \in I}{\vee} R_{i} = \overline{\underset{i \in I}{\cup}R_{i}}$ and $\underset{i \in I}{\wedge} R_{i} = \underset{i \in I}{\cap} R_{i}$.
\end{quote}
The mapping:
\begin{quote}
$\bot$ : $L_{S} \rightarrow L_{S}$; \hspace{10pt} $R \rightarrow R^{\bot}$
\end{quote}
taking each superposition set to its annihilator set is clearly an automorphism of $L_{S}$ which is involutive and satisfies:
\begin{quote}
$R \subseteq T \Rightarrow T^{\bot} \subseteq R^{\bot}$; \hspace{8pt} $R^{\bot \bot} = R$; \hspace{8pt} $R \cap R^{\bot} = \O$
\end{quote}
and is therefore an orthocomplementation of $L_{S}$.

\vspace{8pt}

{\bf{2.18 Corollary}}

Let $Q, T, R \in 2^{S}$, then:
\begin{quote}
(i) $ R \vee (T \wedge Q) \subseteq (R \vee T) \wedge (R \vee Q)$

(ii) $(R \wedge T) \vee (R \wedge Q) \subseteq R \wedge (T \vee Q)$.
\end{quote}
\vspace{8pt}
\newpage
{\bf{Proof}}

These are well-known properties of any lattice. Proof in our case is easy using the distributivity of the set operations:
\begin{quote}
(i) $R \vee (T \wedge Q) = \overline{R \cup (T \cap Q)} = \overline{(R \cup T) \cap (R \cup Q)} \subseteq \overline{(R \cup T)} \cap \overline{(R \cup Q)}$.

(ii) $(R \wedge T) \vee (R \wedge Q) = \overline{(R \cap T) \cup (R \cap Q)} = \overline{R \cap (T \cup Q)} \subseteq \overline{R} \cap \overline{(T \cup Q)}$.
\end{quote}
A lattice of superposition sets is all very well but there is as yet no guarantee that the pure states are even in $L_{S}$, nor any explicit justification for distinguishing $L_{S}$ from other possible collections of subsets of $S$. To satisfy these criticisms another condition has to be placed on the intrinsic probability functions.

If, for some $s$, there exists an $s' \not= s$ such that $p_{s}(s') > 0$ then there is an obvious problem interpreting sums of intrinsic probability functions since, for example $p_{s}(s) + p_{s}(s') > 1$. It would seem, then, that the number $p_{s_{1}}(s) + p_{s_{2}}(s)$ for $s, s_{1}, s_{2} \in S$ cannot be interpreted as the probability that $s$ is in $s_{1} \cup s_{2}$. A similar difficulty arises in the usual theory of probability for: $p_{\Delta_{1}}(s) + p_{\Delta_{2}}(s)$ if $\Delta_{2} \not\subseteq S \smallsetminus \Delta_{1}$ where $\Delta_{1}, \Delta_{2}$ are measurable, which is overcome by simply requiring $\Delta_{2} \subseteq S \smallsetminus \Delta_{1} \equiv \Delta_{1}^{c}$. Clearly, the problem arises from the possibility of $s_{1}$ being $s_{2}$ and vice-versa, so it is natural to require $s_{1} \bot s_{2}$. Thus, for $s_{1} \bot s_{2}$, we look for a set $s_{1} \bigcirc s_{2}$, say, and a function
\begin{quote}
$p_{s_{1} \bigcirc s_{2}}$ : $S \rightarrow [0,1]$
\end{quote}
expressible in terms of the $p_{s_{i}}$ such that:
\begin{quote}
$p_{s_{1} \bigcirc s_{2}}(s) = 1$ iff $s \in s_{1} \bigcirc s_{2} \in 2^{S}$
\end{quote}
where $p_{s_{1} \bigcirc s_{2}}(s)$ is to be interpreted as the probability that $s$ is in the subset $s_{1} \bigcirc s_{2}$. That such sets exist follows from the following conditions on the intrinsic probability functions:

\vspace{8pt}

{\bf{2.19 Axiom 3}}

Each $p_{s}$ can be uniquely extended to a function on $2^{S}$ satisfying, for $R, T, Q \in 2^{S}$:
\begin{quote}
$R \bot T \Rightarrow p_{s}(R) + p_{s}(T) + p_{s}((R \cup T)^{\bot}) = 1$
\end{quote}
and
\begin{quote}
$R \subseteq Q \Rightarrow p_{s}(R) \leq p_{s}(Q)$.
\end{quote}

From now on I shall assume Axiom 3 is satisfied (in addition to Axioms 1 and 2). Immediately we have:

\vspace{8pt}

{\bf{2.20 Lemma}}

For each $s \in S$ and $R, T \in 2^{S}$:
\begin{quote}
(i) $p_{s}(\O) = 0$, $p_{s}(S) = 1$, and $0 \leq p_{s}(R) \leq 1$

(ii) $p_{s}(R) + p_{s}(R^{\bot}) = 1$

(iii) $p_{s}(R) = p_{s}(\overline{R})$

(iv) $R \bot T \Rightarrow p_{s}(R \cup T) = p_{s}(R \vee T) = p_{s}(R) + p_{s}(T)$.
\end{quote}

\vspace{8pt}

{\bf{Proof}}

(i) By Axiom 3 we have: $p_{s}(\O) \leq p_{s}(R) \leq p_{s}(S)$ $\forall R \in 2^{S}$, so in particular $p_{s}(t) \leq p_{s}(S)$ $\forall t \in S$, hence $p_{s}(S) = 1$, which implies $p_{s}(\O) = 1$.

(ii) Put $T = \O$ in Axiom 3 and use (i).

(iii) By (ii): $p_{s}(R^{\bot}) + p_{s}(R^{\bot \bot}) = 1$, hence $p_{s}(\overline{R}) = p_{s}(R)$.

(iv) By (ii): $p_{s}(R \cup T) + p_{s}(( R \cup T)^{\bot}) = 1 = p_{s}((R \cup T)^{\bot}) + p_{s}(R \vee T)$ hence result.

(iv) and (iii) are the key results of Lemma 2.20, interpretable as: the probability of a pure state being in a subset $R$ or a subset $T$ of $S$, where $R$ and $T$ are orthogonal, is the same as its probability of being in their join; and the probability of it being in a subset is the same as the probability of being in the superposition set of the subset.

\vspace{8pt}

{\bf{2.21 Proposition}}

Let $R, Q \in 2^{S}$ and $t \in S$, then:

\begin{quote}
(i) $p_{t}(R) = 0 \Leftrightarrow t \bot R$

(ii) $\overline{R} = \{ u \in S$ $|$ $ p_{u}(R) = 1 \}$

(iii) $p_{s}(R) = p_{s}(Q)$ $\forall s \in S \Leftrightarrow \overline{R} = \overline{Q}$.
\end{quote}

\vspace{8pt}

{\bf{Proof}}

(i) $\Rightarrow$ : suppose false, then $\exists r \in R$ s.t. $p_{t}(r) > 0$, but $p_{t}(r) \leq p_{t}(R) = 0$, hence $R \bot t$ which provides a contradiction.

$\Leftarrow$: If $t \bot R$ then $R \subseteq t^{\bot}$, so $p_{t}(R) \leq p_{t} (t^{\bot})$. But $p_{t}(t) + p_{t}(t^{\bot}) = 1$, so $p_{t}(t^{\bot}) = 0$ and hence $p_{t}(R) = 0$.

(ii) Define $\underline{R} = \{ u \in S$ $|$ $ p_{u}(R) = 1 \}$.

If $r \in \overline{R}$ then $1 = p_{r}(r) \leq p_{t}(\overline{R}) = p_{r}(R)$, so $p_{r}(R) = 1$ and $r \in \underline{R}$.

If $u \in \underline{R}$ then $p_{u}(R) = 1$, hence $p_{u}(R^{\bot}) = 0$. But then, by (i): $u \bot R^{\bot}$ so $u \subseteq R^{\bot \bot} = \overline{R}$.

(iii) Obviously $R = Q \Rightarrow p_{s}(R) = p_{s}(Q)$ by Axiom 3. For the other implication let $r \in \overline{R}$. Since $p_{r}(R) = 1$ then $p_{r}(Q) = 1$ so by (ii) we conclude that $r \in \overline{Q}$. Similarly for the proof of $\overline{Q} \subseteq \overline{R}$.

\vspace{8pt}

{\bf{2.22 Corollary}}

$L_{S}$ is an atomic lattice.

\vspace{8pt}

{\bf{Proof}}

It is clearly sufficient to show that each $s \in S$ is an element of $L_{S}$, that is, $s = \overline{s}$. But by Proposition 2.21 (ii):
\begin{quote}
$\overline{s} = \{ u$ $|$ $ p_{u}(s) =1 \} = s$ by Axiom 1.
\end{quote}

Stepping aside from the general development for a moment, the next Proposition gives an interesting condition on the elements of a superposition set:

\vspace{8pt}

{\bf{2.23 Proposition}}

Let $R \in 2^{S}$ then:
\begin{quote}
$\overline{R} = \{ u \in S$ $|$ $ p_{s}(u) \leq p_{s}(R)$ $\forall s \in S \}$.
\end{quote}

\vspace{8pt}

{\bf{Proof}}

Define $ R^{0} \equiv \{ u$ $|$ $ p_{s}(u) \leq p_{s}(R)$ $\forall s \in S \}$.

If $u \in \overline{R}$ then $p_{s}(u) \leq p_{s}(R)$ $\forall s$, so $u \in R^{0}$. Let $u \in R^{0}$ and suppose $u \not\in \overline{R}$. Clearly $\exists t \in R^{\bot}$ s.t. $p_{t}(u) > 0$ (if there didn't then $u$ would be in $R^{\bot \bot} = \overline{R})$. But $p_{t}(u) \leq p_{t}(R) = 0$ which is a contradiction and proves the proposition.

\vspace{8pt}

{\bf{2.24 Definition}}

Let $R, T \in 2^{S}$. If $R \subseteq T$ then their {\em{difference}} is defined as:
\begin{quote}
$\overline{T} - \overline{R} \equiv \{ u \in S$ $|$ $ p_{s}(u) \leq p_{s}(T) - p_{s}(R)$ $\forall s \in S \}$.
\end{quote}

\vspace{8pt}

{\bf{2.25 Proposition}}:

For each $s \in S$:
\begin{quote}
$p_{s}(\overline{T} - \overline{R}) = p_{s}(T) - p_{s}(R) = p_{s}(R^{\bot} \wedge \overline{T})$.
\end{quote}

\vspace{8pt}

{\bf{Proof}}

If $R \subseteq T$ then clearly $\overline{R} \bot T^{\bot}$.

Since: $p_{s}(R^{\bot} \wedge T) + p_{s}((R^{\bot} \wedge T)^{\bot}) = 1$ then:
\begin{quote}
$p_{s}(R^{\bot} \wedge T) = 1 - p_{s}(R \vee T^{\bot}) = p_{s}(T) - p_{s}(R)$.
\end{quote}
Applying Proposition 2.23 then we conclude that $\overline{T} - \overline{R} = R^{\bot} \wedge T$, and the result then follows from Proposition 2.21 (iii).

\vspace{8pt}

{\bf{2.26 Lemma}}

Let $R, T \in 2^{S}$. If $R \subseteq T$ then:
\begin{quote}
(i) $\overline{T} = (R^{\bot} \wedge T) \vee R$

(ii) $\overline{R} = (T^{\bot} \vee R) \wedge T$.
\end{quote}

\vspace{8pt}

{\bf{Proof}}

(i) Clearly $R \bot (R^{\bot} \wedge T)$ so:
\begin{quote}
$p_{s}((R^{\bot} \wedge T) \vee R) = p_{s} (R^{\bot} \wedge T) + p_{s}(R) = p_{s}(T)$

$\Leftrightarrow$ $\overline{T} = (R^{\bot} \wedge T) \vee R$ by Proposition 2.21 (iii).

\end{quote}
(ii) $p_{s}((R \vee T^{\bot}) \wedge T) = 1 - p_{s} ((( R \vee T^{\bot}) \wedge T)^{\bot}) = 1 - p_{s}((R^{\bot} \wedge T) \vee T^{\bot})$ clearly  $(R^{\bot} \wedge T) \bot T^{\bot}$ hence:
\begin{quote}
$p_{s}((R \vee T^{\bot}) \wedge T) = 1 - p_{s}(R^{\bot} \wedge T) - p_{s}(T^{\bot}) = p_{s}(R)$

$\Leftrightarrow$ $\overline{R} = (R \vee T^{\bot}) \wedge T$ by Proposition 2.21 (iii).
\end{quote}
For $R = \overline{R}$ and $T = \overline{T}$, (i) of Lemma 2.26 is known as {\em{weak modularity}}, whilst (ii) is sometimes called {\em{orthomodularity}}; the more familiar orthomodularity condition (which is, as is well known, equivalent to (i) or (ii)) is contained in the following Proposition, which strengthens Corollary 2.18:

\vspace{8pt}

{\bf{2.26 Proposition}}

Let $Q, R, T \in 2^{S}$. If $Q \bot R$ and $R \subseteq T$ then:
\begin{quote}
$(Q \vee R) \wedge T = (Q \wedge T) \vee R$.

\end{quote}

\vspace{8pt}
\newpage
{\bf{Proof}}

Call $Y = (Q \wedge T) \vee R$ and $Z = (Q \vee R) \wedge T$. By Corollary 2.18, we have $Y \subseteq Z$. Form $Z - Y = Y^{\bot} \wedge Z$. Clearly: $Z - Y \subseteq Z \subseteq Q \vee R$ and $Z - Y \subseteq Y^{\bot} = (Q \wedge T)^{\bot} \wedge R^{\bot} \subseteq R^{\bot}$, hence: $Z-Y \subseteq (Q \vee R) \wedge R^{\bot} = Q$ by Lemma 2.26 (ii) (since $Q \subseteq R^{\bot}$). Also: $Z - Y \subseteq Z \subseteq T$ so $Z - Y \subseteq Q \wedge T \subseteq Y$. But $Z-Y \subseteq Y^{\bot}$ so $Z - Y = \O$ and $Z = Y$. Putting this result together with Corollary 2.22 and Theorem 2.17 gives:

\vspace{8pt}

{\bf{2.27 Theorem}}

$L_{S}$ is a complete orthomodular atomic lattice.

The terminology `superposition set' arises from the next definition:

\vspace{8pt}

{\bf{2.28 Definition}}

Let $R \in 2^{S}$, then $t \in S$ is said to be a {\em{superposition}} of elements of $R$ if:

\begin{quote}
$p_{s}(R) = 0 \Rightarrow p_{t}(s) = 0$, $s \in S$
\end{quote}

{\bf{2.29 Lemma}}

Let $R \in 2^{S}$ then:

\begin{quote}
$p_{t}(s) = 0$ \hspace{5pt} $\forall s \in R^{\bot}$ $\Leftrightarrow$ $t \in \overline{R}$.
\end{quote}

\vspace{8pt}

{\bf{Proof}}

$\Leftarrow$ is obvious from $p_{t}(s) \leq p_{t}(R^{\bot}) = 0$. Suppose $\Rightarrow$ is false, then $\exists t \not\in \overline{R}$ such that $p_{t}(s) = 0$ $\forall s \in R^{\bot}$, but if $t \not \in \overline{R} = (R^{\bot})^{\bot}$ then $\exists u \in R^{\bot}$ such that $p_{t}(u) > 0$.

Thus we conclude (as was, perhaps, obvious) that any state which is a superposition of states in $R$ is in $\overline{R}$, that is, the set of states of superposition of $R$ is precisely the superposition set of $R$.

The `Superposition Principle' familiar from quantum mechanics is of a rather different nature, and may be formulated as:

\vspace{8pt}

{\bf{2.30 Definition}}

A subset $R$ of $S$ is said to satisfy:
\begin{quote}
(a) The {\em{Weak Superposition Principle}} (WSP) if, for some pair of elements $r_{1}, r_{2} \in R$ there exists a distinct $r_{3} \in R$ such that:
\begin{quote}
$ r_{1} \vee r_{2} = r_{3} \vee r_{2} = r_{1} \vee r_{3}$.
\end{quote}
(b) The {\em{Strong Superposition Principle}} (SSP) if, for every pair of elements $r_{1}, r_{2} \in R$ there exists a distinct $r_{3} \in R$ such that:
\begin{quote}
$ r_{1} \vee r_{2} = r_{3} \vee r_{2} = r_{1} \vee r_{3}$.
\end{quote}
Clearly SSP $\Rightarrow$ WSP. SSP is rather too strong a condition for the deduction of useful results from its negation, so I treat WSP first:
\end{quote}

\vspace{8pt}

{\bf{2.31 Lemma}}

Let $R, T \in 2^{S}$ with $R \subseteq T$ and $\overline{R} \not= \overline{T}$, then there exists $t \in \overline{T}$ with $t \bot \overline{R}$ such that: $t \vee \overline{R} \subseteq \overline{T}$.

\vspace{8pt}

{\bf{Proof}}

Form $R^{\bot} \wedge \overline{T}$; by Lemma 2.26 this must be non-empty (for otherwise $\overline{T} = \overline{R}$), so choose any $t \in R^{\bot} \wedge \overline{T}$ to satisfy the Lemma.

\vspace{8pt}

{\bf{2.32 Lemma}}

Let $s_{1}, s_{2} \in S$ with $s_{1} \bot s_{2}$. For any $t_{1}, t_{2} \in s_{1} \vee s_{2}$ with $t_{1} \not= t_{2}$ then: $t_{1} \vee t_{2} = s_{1} \vee s_{2}$.

\vspace{8pt}

{\bf{Proof}}

We first prove that if $t_{1} \bot t_{2}$ then $t_{1} \vee t_{2} = s_{1} \vee s_{2}$. Clearly $t_{1} \vee t_{2} \subseteq s_{1} \vee s_{2}$. Suppose not equal, then by Lemma 2.31 $\exists t_{3}$ such that $t_{3} \bot (t_{1} \vee t_{2})$ and $t_{1} \vee t_{2} \vee t_{3} \subseteq s_{1} \vee s_{2}$. But then:
\begin{quote}
$p_{s} (s_{1} \vee s_{2}) \geq p_{s}(t_{1}) + p_{s}(t_{2}) + p_{s}(t_{3})$.
\end{quote}
In particular, putting $s = s_{1}$ and $s = s_{2}$ and adding them together gives:
\begin{quote}
$ 2 \geq p_{t_{1}}(s_{1}) + p_{t_{1}}(s_{2}) + p_{t_{2}}(s_{1}) + p_{t_{2}}(s_{2}) + p_{t_{3}}(s_{1}) + p_{t_{3}}(s_{2})$
\end{quote}
But then, since $s_{1} \bot s_{2}$, the right hand side of the inequality is:
\begin{quote}
$p_{t_{1}} (s_{1} \vee s_{2}) + p_{t_{2}} (s_{1} \vee s_{2}) + p_{t_{3}} (s_{1} \vee s_{2}) = 3$
\end{quote}
which is a contradiction, hence $t_{3} = \O$ and $t_{1} \bot t_{2} \Rightarrow t_{1} \vee t_{2} = s_{1} \vee s_{2}$. Now suppose $t_{1} \not\perp t_{2}$. Then, by Lemma 2.31, $\exists t_{4}$ s.t. $t_{1} \bot t_{4}$ and:
\begin{quote}
$t_{1} \vee t_{4} \subseteq t_{1} \vee t_{2} \subseteq s_{1} \vee s_{2}$
\end{quote}
but we have just shown that for such a $t_{4}, t_{1}$ : $t_{1} \vee t_{4} = s_{1} \vee s_{2}$.

Unfortunately, the converse of Lemma 2.32 does not seem to hold; that is, if $s_{1} \not\perp s_{2}$ then it is not necessarily the case that there exist $t_{1}, t_{2} \in s_{1} \vee s_{2}$ with $t_{1} \bot t_{2}$ such that $t_{1} \vee t_{2} = s_{1} \vee s_{2}$. Thus we cannot deduce the `size' of $s_{1} \vee s_{2}$ (we could, for example, have $s_{1} \vee s_{2} = S$). Hence the next definition:

\vspace{8pt}

{\bf{2.33 Definition}}

A set $R \in 2^{S}$ will be said to be {\em{covered}} if, for any $r_{1}, r_{2} \in \overline{R}$ there exist (not necessarily distinct) $r_{3}, r_{4} \in \overline{R}$ with $r_{3} \bot r_{4}$ such that $r_{1} \vee r_{2} \subseteq r_{3} \vee r_{4}$.

\vspace{8pt}

{\bf{2.34 Proposition}}

Let $R \in 2^{S}$ be covered, and let $r_{1}, r_{2} \in R$ with $r_{1} \not= r_{2}$, then:
\begin{quote}
(i) For any distinct $r_{3}, r_{4} \in r_{1} \vee r_{2}$ : $r_{1} \vee r_{2} = r_{3} \vee r_{4}$

(ii) For any distinct $r \in r_{1} \vee r_{2}$ : $r \vee r_{1} = r \vee r_{2} = r_{1} \vee r_{2}$.
\end{quote}

\vspace{8pt}

{\bf{Proof}}

Immediate from Lemma 2.32 and Definition 2.33.

\vspace{8pt}

{\bf{2.35 Proposition}}

Let $R \in 2^{S}$ be covered, then the following are equivalent:
\begin{quote}
(i) $\overline{R}$ does not satisfy the Weak Superposition Principle.

(ii) For any $r_{1}, r_{2} \in \overline{R}$ : $r_{1} \vee r_{2} = r_{1} \cup r_{2}$.

(iii) For each $r \in \overline{R}$ : $p_{r}(r') = 0$ \hspace{5pt} $\forall r' \in \overline{R} \smallsetminus r$.
\end{quote}

\vspace{8pt}

{\bf{Proof}}

(ii) $\Rightarrow$ (i) is obvious. The others are proved by contradiction:

(i) $\Rightarrow$ (ii): Suppose false, then $\exists r_{3} \in \overline{R}$ s.t. $r_{1} \cup r_{2} \cup r_{3} \subseteq r_{1} \vee r_{2}$. But then, by Proposition 2.34 (ii): $r_{1} \vee r_{3} = r_{2} \vee r_{3} =  r_{1} \vee r_{2}$, so WSP is satisfied.

(ii) $\Rightarrow$ (iii): Suppose false, then $\exists r_{1}, r_{2} \in \overline{R}$ s.t. $p_{r_{2}}(r_{1}) > 0$, so, by Lemma 2.31, $\exists t \in r_{1} \vee r_{2}$ with $t \bot r_{1}$.

(iii) $\Rightarrow$ (ii): Suppose false, then $\exists r'' \in \overline{R}$ s.t. $r \cup r' \cup r'' \subseteq r \vee r'$, hence by (iii) and Axiom 3:
\begin{quote}
$p_{s} (r \vee r') \geq p_{s} (r) + p_{s}(r') + p_{s}(r'')$.
\end{quote}
But if $p_{r}(r') = 0$ \hspace{5pt} $\forall r, r' \in \overline{R}$ with $r \not= r'$, then:
\begin{quote}
$p_{s}(r \cup r') = p_{s} (r) + p_{s} (r') = p_{s} (r \vee r')$,
\end{quote}
hence $p_{s}(r'') = 0$ and $r'' = \O$. Note that (ii) $\Leftrightarrow$ (iii) irrespective of whether $R$ is covered.

\vspace{8pt}

{\bf{2.36 Theorem}}

Let $R \in 2^{S}$ be covered, then the following are equivalent:
\begin{quote}
(i) $(L_{\overline{R}}, \vee, \wedge)$ is a Boolean lattice.

(ii) $\overline{R}$ does not satisfy the Weak Superposition Principle.

(iii) $L_{\overline{R}} = 2^{\overline{R}}$.
\end{quote}

\vspace{8pt}

{\bf{Proof}}

(ii) $\Rightarrow$ (iii): If WSP is not satisfied, then by Proposition 2.35: $r_{1} \vee r_{2} = r_{1} \cup r_{2}$ \hspace{4pt} $\forall r_{1}, r{_2} \in \overline{R}$ which implies and is implied by: $L_{\overline{R}} = 2^{\overline{R}}$ with $\vee \equiv \cup$ and $\wedge \equiv \cap$.

(iii) $\Rightarrow$ (i): $2^{\overline{R}}$ is obviously Boolean.

(i) $\Rightarrow$ (ii): Suppose false, but then, for distinct $r_{1}, r_{2}, r_{3} \in \overline{R}$ satisfying $r_{3} \in r_{1} \vee r_{2}$, the distributive law:
\begin{quote}
$r_{3} \wedge (r_{1} \vee r_{2}) = (r_{3} \wedge r_{1}) \vee (r_{3} \wedge r_{2})$
\end{quote}
implies that $r_{3} = \O$.

Noting, from Proposition 2.23, that:
\begin{quote}
$\overline{R} = \{ u \in S$ $|$ $ p_{s}(u) \leq p_{s}(R)$ \hspace{5pt} $\forall s \in S \}$
\end{quote}
we are led to a condition on intrinsic probability functions, holding in both classical and quantum mechanics, which is sufficient for $S$ to be covered:

\vspace{8pt}

{\bf{2.37 Proposition}}

If, for each $R \in L_{S}$, $s \in S$, there exists (a not necessarily unique) $u \in R$ such that:
\begin{quote}
$p_{s}(u) = p_{s}(R)$
\end{quote}
then $S$ is covered.

\vspace{8pt}

{\bf{Proof}}

Suppose false, then for some $s_{1}, s_{2} \in S$ with $s_{1} \not\perp s_{2}$ \hspace{3pt} $\not\exists s_{3}, s_{4}$ with $s_{3} \bot s_{4}$ such that $s_{1} \vee s_{2} \subseteq s_{3} \vee s_{4}$. But, from: 
\begin{quote}
$p_{s_{1}}(s_{2}) + p_{s_{1}}({s_{2}}^{\bot}) = 1$
\end{quote}
there exists by hypothesis $u \in {s_{2}}^{\bot}$ such that:
\begin{quote}
$p_{s_{1}}(s_{2}) + p_{s_{1}}(u) = 1$
\end{quote}
Hence $s_{1} \in s_{2} \vee u$, so, with $s_{2} \in s_{2} \vee u$ we have:
\begin{quote}
$s_{1} \vee s_{2} \subseteq s_{2} \vee u$
\end{quote}
which is a contradiction and proves the Proposition.

As will be shown shortly, the condition in Proposition 2.37 is sufficient for us to draw far-reaching conclusions about $L_{S}$ if SSP is satisfied. Consequently, we shall elevate it to an Axiom:

\vspace{8pt}

{\bf{2.38 Axiom 4}}

For each $R \in L_{S}$ and $s \in S$ there exists (a not necessarily unique) $u \in R$ such that:
\begin{quote}
$p_{s}(u) = p_{s}(R)$.
\end{quote}

Axiom 4 will be assumed to hold for remainder of this Section. Notice that the uniqueness of our extension of the intrinsic probability functions from $S$ to $L_{S}$ is now guaranteed since we have by definition that:
\begin{quote}
$p_{s}(R) = \underset{r \in \overline{R}}{max}$ $p_{s}(r)$.
\end{quote}
Axiom 4 can be viewed as providing `closest elements', for which reason we shall term it a `completeness condition'. In order to eventually identify the lattice of superposition sets as a projective geometry, let us now make precise the notion of `size' alluded to earlier.

\vspace{8pt}

{\bf{2.39 Definition}}

A {\em{partition}} of $R \in L_{S}$ is any set $\{r_{i}\}$ of mutually orthogonal elements of $R$ such that:
\begin{quote}
$R = \vee_{i}r_{i}$.
\end{quote}

\vspace{8pt}

{\bf{2.40 Lemma}}

Let $\{r_{i}\}$ $_{i=1}^{m}$ and $\{t_{i}\}$ $_{i=1}^{n}$ be any two finite partitions of $R \in L_{S}$, then:
\begin{quote}
$m = n$.
\end{quote}
\newpage
{\bf{Proof}}

Construct the array:

\hspace{85pt} $p_{r_{1}}(t_{1})$ \hspace{13pt} . \hspace{13pt}  . \hspace{13pt}  . \hspace{13pt}  . \hspace{13pt} . \hspace{13pt} . \hspace{13pt} $p_{r_{m}}(t_{1})$

\hspace{95pt} . \hspace{185pt} .

\hspace{95pt} . \hspace{185pt} .

\hspace{95pt} . \hspace{185pt} .

\hspace{95pt} . \hspace{185pt} .

\hspace{95pt} . \hspace{185pt} .

\hspace{85pt} $p_{r_{1}}(t_{n})$ \hspace{13pt} . \hspace{13pt}  . \hspace{13pt}  . \hspace{13pt}  . \hspace{13pt} . \hspace{13pt} . \hspace{13pt}$p_{r_{m}}(t_{n})$
\vspace{8pt}
\newline
then each row and each column sums to 1. Also, the sum of all the summed rows adds up to $n$, and the sum of all the summed columns adds up to $m$. But these two must be equal, hence $m = n$.

\vspace{8pt}

{\bf{2.41 Definition}}

Define the function:
\begin{quote}
$d$ : $L_{S} \rightarrow \mathbb{Z}^{+}\cup\{\infty\}$
\end{quote}
by: $d(R) = m$ where  $\{r_{i}\}$ $_{i=1}^{m}$ is any finite partition of $R$.

\hspace{25.5pt} = $\infty$ otherwise (that is, if no finite partition exists).

If $d(R) < \infty$ then $R$ will be said to be {\em{finite}}.

It is obvious from this Definition that if $R \bot T$ then $d(R \vee T) = d(R) + d(T)$, and if $R \subseteq T$ then $d(R) \leq d(T)$.

\vspace{8pt}

{\bf{2.42 Lemma}}

If $R \in L_{S}$ be finite, then for any $t \not\in R$:
\begin{quote}
$d(R \vee t) = d(R) + 1$.
\end{quote}

\vspace{8pt}

{\bf{Proof}}

By Axiom 4 $\exists u \in R^{\bot}$ such that $p_{t}(R) + p_{t}(u) = 1$. Hence $R \vee t \subseteq R \vee u$ and $d(R \vee t) \leq d(R) +1$. Since $t \not\in R$ then by Lemma 2.31 $\exists v \in R^{\bot}$ such that $R \vee v \subseteq R \vee t$. Hence $d(R) + 1 \leq d(R +t)$.

\vspace{8pt}

{\bf{2.43 Lemma}}

Let $R \in L_{S}$ be finite. Let $k \in S$ with $k \not\in R$. If:
\begin{quote}
$s \in R \vee k$
\end{quote}
then there exists $z \in R$ such that $s \in z \vee k$.

{\bf{Proof}}

Define an orthogonal complement $^{*}$ in $(R \vee k)$ by: 
\begin{quote}
For $Q \in (R \vee k)$ then $Q^{*} = (R \vee k) \wedge (Q^{\bot})$.
\end{quote}
Notice that if we call $d(R \vee k) = N$ then $d(Q^{*}) = N - d(Q)$. If $s \in R$ or $s = k$ the Lemma is trivial, so suppose $s \not\in R$ and $s \not= k$. We claim that $(k \vee s) \wedge R$ is non-empty. To see this suppose that it is empty. Form: $Y = (k \vee s)^{*}$ ; $x = R^{*}$, then:
\begin{quote}
$d(Y) = d(R \vee k) - d(k \vee s) = N - 2$ by Lemma 2.42 and $d(x) = 1$.
\end{quote}
So if $Y^{*} \wedge x^{*} = (Y \vee x)^{*} = \O$, then:
\begin{quote}
$N = d(Y \vee x) \leq d(Y) + 1 = N - 1$
\end{quote}
hence $Y^{*} \wedge x^{*} = \O$.

Let $z \in (k \vee s) \wedge R$, then by Proposition 2.34: $k \vee z = k \vee s$ (since $z \in R$ and $k \not\in R$), so $s \in k \vee z$ as required.

\vspace{8pt}

{\bf{2.44 Lemma}}

Let $R, T, \in L_{S}$ be finite, then for each $s \in R \vee T$ there exists (not necessarily unique) $r \in R$ and $t \in T$ such that:
\begin{quote}
$s \in r \vee t$.
\end{quote} 

\vspace{8pt}

{\bf{Proof}}

The result obviously holds if $R$ and $T$ are atoms. The general proof will be by induction. Suppose Lemma 2.44 is satisfied for $R_{0}, T_{0} \in L_{S}$ then it is sufficient to show that it is also satisfied for $R_{0}$, $(T_0 \vee k)$ where $k \not \in R_{0} \vee T_{0}$.

Let $s \in R_{0} \vee T_{0} \vee k$, then by Lemma 2.43 $\exists z \in R_{0} \vee T_{0}$ s.t. $s \in k \vee z$. But, by hypothesis, $z \in r \vee t$ for some $r \in R_{0}$, $t \in T_{0}$, hence:
\begin{quote}
$s \in k \vee r \vee t$.
\end{quote}
Applying Lemma 2.43 again: $\exists y \in k \vee t$ s.t. $s \in r \vee y$, as required.

\vspace{8pt}

{\bf{2.45 Proposition}}

Let $R, T, Q \in L_{S}$ be finite with $Q \subseteq R$, then:
\begin{quote}
$R \wedge (T \vee Q) = (R \wedge T) \vee Q$.
\end{quote}

\vspace{8pt}

{\bf{Proof}}

$\supseteq$ is trivial; for $\subseteq$ let $s \in R \wedge (T \vee Q)$, then by Lemma 2.44 there exist $t \in T$ and $q \in Q$ such that $s \in t \vee q$, so, by Proposition 2.34, either $s \vee q = t \vee q$ or $s = q$. If the former, then $t \vee q \subseteq R$ so that $t \in R$, but then $t \in R \wedge T$ and hence $s \in (R \wedge T) \vee Q$. If the latter, then $s \in (R \wedge T) \vee Q$ trivially.

Thus the set of all finite superposition sets (and $S$) constitutes a {\em{modular}} sublattice of $L_{S}$. This modularity will be used to prove a well-known representation theorem, but first a few definitions:

\vspace{8pt}

{\bf{2.46 Definition}}

Let $H$ be a vector space over a division ring (i.e. a skew field) $\mathbb{D}$. Let $\theta$ be an involutive anti-automorphism of $\mathbb{D}$ (i.e. $\theta^{2} = 1$ and $\theta (d_{1} + d_{2}d_{3}) = \theta (d_{1}) + \theta(d_{3})\theta(d_{2})$. Let $\langle \cdot, \cdot \rangle$ be a $\mathbb{D}$-valued, symmetric (i.e. $\langle x_{1}, x_{2} \rangle = \theta (\langle x_{2}, x_{1} \rangle)$), definite (i.e. $\langle x, x \rangle = 0 \Leftrightarrow x = 0$), $\theta$-bilinear (i.e. $\langle d_{1}x_{1}, d_{2}x_{2} \rangle = \theta(d_{1}) \langle x_{1}, x_{2} \rangle d_{2}$) form on $H \times H$. Then the quadruple $(H, \mathbb{D}, \theta, \langle \cdot, \cdot \rangle)$, or, loosely, just $H$, will be called {\em{Hilbertian}} if and only if:
\begin{quote}
$H = M^{0} \oplus M^{00}$ for every $M \in 2^{H}$

where: $M^{0} \equiv \{ x \in H$ $|$ $\langle m, x \rangle = 0$ \hspace{4pt} $\forall m \in M \}$.
\end{quote}
$M \in 2^{H}$ will be called $\langle \cdot, \cdot \rangle${\em{-closed}} if and only if $M = M^{00}$.

\vspace{8pt}

{\bf{2.47 Theorem}} (Piron)

Let $d(S) \geq 4$, then the following are equivalent:
\begin{quote}
(i) $S$ satisfies the Strong Superposition Principle.

(ii) There exists a Hilbertian quadruple $(H, \mathbb{D}, \theta, \langle \cdot, \cdot \rangle)$ such that $L_{S}$ is isomorphic to the lattice of all $\langle \cdot, \cdot \rangle$-closed linear manifolds of $H$.
\end{quote}

\vspace{8pt}

{\bf{Proof}}

Provided we can show that $L_{S}$ is a {\em{complete projective logic}}  (defined on p. 176 of (Va 1)), we can use the proof of Theorem 7.40 of (Va 1) which, it should be noted, does {\em{not}} depend on $H$ being finite-dimensional. In Lemma 2.57 below we show that SSP is equivalent to {\em{irreducibility}}, so, since $L_{S}$ is complete and atomic by construction, and by Lemma 2.31 and Proposition 2.34 we have that for any $s \in S$ and $R \in L_{S}$ with $\O \not= R \not= S$ $\exists r \in R$ and $t \in R^{\bot}$ such that $s \in r \vee t$, then it remains to verify that if $R \not= \O$ is the lattice sum of a finite number of atoms, then the set: $\{ T \in L_{S}$ $|$ $\O \subseteq T \subseteq R \}$ is a complemented modular lattice. But it is clearly a sublattice of $L_{S}$ and, by Proposition 2.45, modular. A complement is $T^{*} = R \wedge T^{\bot}$.

\vspace{8pt}

{\bf{2.48 Remarks}}

The division ring $\mathbb{D}$ is determined up to isomorphism by the distinct elements in any `line' (i.e. set of the form $s_{1} \vee s_{2}$) of $L_{S}$, addition and multiplication in $\mathbb{D}$ being defined by means of certain special and general projectivities, respectively, of the line (see (Va 1) p. 86 for details). The involutive anti-automorphism $\theta$ of $\mathbb{D}$ arises directly from the orthocomplementation on $L_{S}$. Conditions for the vector space $H$ to be a Hilbert space are provided by:

\vspace{8pt}

{\bf{2.49 Corollary}}

Let $d(S) \geq 4$; let $\mathbb{D}$ be one of $\mathbb{R}$ (reals), $\mathbb{C}$ (complex numbers), $\mathbb{Q}$ (quaternions), and let $\theta$ be continuous (which is only a restriction for $\mathbb{D} = \mathbb{C}$), then the following are equivalent:
\begin{quote}
(i) $S$ satisfies the Strong Superposition Principle.

(ii) $L_{S}$ is isomorphic to the set of all closed linear manifolds of a Hilbert space over $\mathbb{D}$.
\end{quote}

\vspace{8pt}

{\bf{Proof}}

Use Theorem 2.47 above together with Lemma 7.42 (which proves that $H$ is complete) of (Va 1). 

The final task of this Section is to demonstrate, following Jauch (Ja 1), that a general $S$ may be decomposed into the union of a collection of superposition sets each of which satisfies SSP. Noting that if $s_{1} \vee s_{2}$ satisfies WSP it also satisfies SSP, we define:

\vspace{8pt}

{\bf{2.50 Definition}}

Let $s_{1}, s_{2} \in S$. $s_{1}$ will be said to be {\em{perspective}} to $s_{2}$ if $s_{1} \vee s_{2}$ satisfies WSP. Each $s \in S$ is defined to be perspective to itself.

\vspace{8pt}

\newpage

{\bf{2.51 Lemma}}

Perspectivity is an equivalence relation on $S$.

\vspace{8pt}

{\bf{Proof}} 

Reflexivity is taken care of in the definition, and symmetry is obvious. For transitivity we need to show that if $s_{1} \vee s_{2}$ and $s_{2} \vee s_{3}$ satisfy WSP then so does $s_{1} \vee s_{3}$. If $s_{3} \in s_{1} \vee s_{2}$ then the result is trivial, so suppose that $s_{3} \not\in s_{1} \vee s_{2}$. By hypothesis there exist distinct $t \in s_{1} \vee s_{2}$ and $r \in s_{2} \vee s_{3}$. Repeating the dimensionality argument used in the proof of Lemma 2.43, it is immediate that:
\begin{quote}
$(s_{1} \vee s_{3}) \wedge (r \vee t) \not= \O$.
\end{quote}
In fact $d((s_{1} \vee s_{3}) \wedge (r \vee t)) = 1$ so the required element to satisfy WSP for $s_{1} \vee s_{3}$ is $(s_{1} \vee s_{3}) \wedge (r \vee t)$.

\vspace{8pt}

{\bf{2.52 Definition}}

Let $R, T \in L_{S}$. $R$ will be said to be {\em{compatible}} with $T$, written $R \leftrightarrow T$, iff:
\begin{quote}
$(R - (R \wedge T)) \bot T$.
\end{quote}
The {\em{centre}}, $Z_{S}$, of $L_{S}$ is then defined to be the set:
\begin{quote}
$Z_{S} \equiv \{ R \in L_{S}$ $|$ $R \leftrightarrow T$ \hspace{3pt} $\forall T \in L_{S} \}$.
\end{quote}
The following Lemmas are mostly well-known; we include Proofs for the sake of completeness.

\vspace{8pt}

{\bf{2.53 Lemma}}

Let $R, T \in L_{S}$, then the following are equivalent:
\begin{quote}
(i) $R \leftrightarrow T$

(ii) $T \leftrightarrow R$

(iii) $R \leftrightarrow T^{\bot}$

(iv) $R$ and $T$ generate an orthocomplemented Boolean sublattice of $L_{S}$. We also have that:
\begin{quote}
$R \subseteq T \Rightarrow R \leftrightarrow T$, and $R \subseteq T^{\bot} \Rightarrow R \leftrightarrow T$.
\end{quote}
\end{quote}

{\bf{Proof}}

(i) $\Leftrightarrow$ (ii): $T \bot (R \wedge (R \wedge T)^{\bot})$ $\&$ $(R \wedge T)^{\bot}\bot(R\wedge T) \Rightarrow T \wedge (R \wedge T)^{\bot} \bot (R \wedge (R \wedge T)^{\bot}) \vee (R \wedge T) \Leftrightarrow T \wedge (R \wedge T)^{\bot} \bot R$. 

(i) $\Leftrightarrow$ (iii): Sufficient to prove that $R \bot (T^{\bot} \wedge (R \wedge T^{\bot})^{\bot})$: $R \bot R^{\bot}$ $\&$ $R \wedge (R \wedge T)^{\bot} \bot T \Rightarrow (R \wedge (R \wedge T)^{\bot}) \bot (R^{\bot} \vee T)$, but $(R \wedge T) \bot T^{\bot}$, so $((R \wedge (R \wedge T)^{\bot}) \vee (R \wedge T)) \bot (T^{\bot} \wedge (R^{\bot} \vee T)) \Leftrightarrow R \bot (T^{\bot} \wedge (R^{\bot} \vee T)) \Leftrightarrow R \bot (T^{\bot} \wedge (R \wedge T^{\bot})^{\bot})$.

(i) $\Rightarrow$ (iv): Form the superposition sets:

$R_{1} = R - (R \wedge T)$; $R_{2} = R \wedge T$; $R_{3} = T - (R \wedge T)$; $R_{4} = (R \vee T)^{\bot}$. Clearly, $S = R_{1} \vee R_{2} \vee R_{3} \vee R_{4}$, and, by (i), $R_{i} \bot R_{j}$ for $i \not= j$. Hence, defining the set:
\begin{quote}
$B (R, T) \equiv \{ \underset{j \in J}{\vee} R_{j}$ $|$ $\forall J \in 2^{\{1, 2, 3, 4\}}\}$
\end{quote}
we see that $B (R,T)$ is an orthocomplemented sublattice of $L_{S}$; it is trivial to use orthogonality to prove that the distributive laws hold in $B (R,T)$, hence $B (R, T)$ is Boolean, and clearly the smallest such containing $R$ and $T$.

(iv) $\Rightarrow$ (i): Let $B$ be any Boolean sublattice of $L_{S}$ containing $R$ and $T$ for which $\bot$ is an orthocomplementation. Then $R - (R \wedge T)$, $T$ and $T^{\bot}$ are in $B$, so:

\begin{quote}
$R - (R \wedge T) = (R - (R \wedge T)) \wedge (T \vee T^{\bot})$

\hspace{59pt} $= ((R - (R \wedge T)) \wedge T) \vee ((R - (R \wedge T)) \wedge T^{\bot})$

\hspace{59pt} $= (R - (R \wedge T)) \wedge T^{\bot} \subseteq T^{\bot}$.
\end{quote} 

\vspace{8pt}

{\bf{2.54 Corollary}}

$Z_{S}$ is an orthocomplemented Boolean sublattice of $L_{S}$.

\vspace{8pt}

{\bf{Proof}}

Immediate from Lemma 2.53.

\vspace{8pt}

{\bf{2.55 Lemma}}

Let $T \in L_{S}$. Let $\{R_{i}\}$, $i \in I$ for any index set $I$, be any subset of $L_{S}$, then $T \leftrightarrow R_{i}$ $\forall i \in I$ implies that:
\begin{quote}
$T \leftrightarrow  \underset{i \in I}{\vee}R_{i}$ and $T \leftrightarrow  \underset{i \in I}{\wedge}R_{i}$.
\end{quote}

{\bf{Proof}}

Call $R = \underset{i \in I}{\vee}R_{i}$. By Lemma 2.53 it is sufficient to prove $R \leftrightarrow T$. Now \\ $T \wedge R_{i} \subseteq T \wedge R$, so $T - (R \wedge T) \subseteq T - (T \wedge R_{i})$ $\forall i \in I$, but then, since $T \leftrightarrow R_{i}$: $T - (T \wedge R) \subseteq R_{i}^{\bot}$ $\forall i \in I$, hence: $T - (T \wedge R) \subseteq \underset{i \in I}{\wedge}{R_{i}}^{\bot} = R^{\bot}$.

It is now a simple matter to make the desired decomposition: denote the perspectivity equivalence classes of $S$ by $Q_{i}$, $i \in I$ for some index set $I$, then, by construction, each $Q_{i}$ satisfies SSP. Moreover, we have:

\vspace{8pt}

{\bf{2.56 Lemma}}

Let $\{Q_{i}\}_{i \in I}$ be the perspectivity equivalence classes of $S$, then:
\begin{quote}
(i) $S = \underset{i \in I}{\vee}Q_{i}$ and $Q_{i} \bot Q_{j}$ $\forall i, j \in I$ with $i \not= j$.

(ii) Each $Q_{i}$ satisfies SSP, and each $Q_{i} \in Z_{S}$.
\end{quote}

\vspace{8pt}

{\bf{Proof}}

(i) Clearly $S = \underset{i \in I}{\vee}Q_{i}$ since every $s \in S$ is in one of the $Q_{i}$. Suppose $Q_{i} \not\perp Q_{j}$ for some $i, j$ where $i \not= j$, then there exist $r \in Q_{i}$ and $t \in Q_{j}$ such that $p_{r}(t) > 0$. But then, by Lemma 2.31, $r \vee t$ satisfies WSP, so $r$ and $t$ are perspective which contradicts the definition of the $Q_{i}$ as distinct perspectivity equivalence classes.

(ii) Each $Q_{i}$ satisfies SSP by transitivity of perspectivity. To prove that $Q_{i} \in Z_{S}$ we have to show that $Q_{i} \leftrightarrow T$ $\forall T \in L_{S}$. Write $T = \vee_{j} T_{j}$ where $T_{j} = T \wedge Q_{j}$, then evidently $Q_{i} \leftrightarrow T_{j}$ $\forall j \in I$ (from orthogonality), so by Lemma 2.55: $Q_{i} \leftrightarrow T$.

\vspace{8pt}

{\bf{2.57 Lemma}}

Let $R \in L_{S}$, then the following are equivalent:
\begin{quote}
(i) $R$ satisfies SSP.

(ii) $Z_{R} = \{\O, R\}$.
\end{quote}

\vspace{8pt}

{\bf{Proof}}

(i) $\Rightarrow$ (ii): Suppose false, then $\exists Q \in Z_{R}$, $Q \not= \O$ or $R$. So, in particular, $Q \leftrightarrow r$ $\forall r \in R$. But $Q \leftrightarrow r$ iff either $r \bot Q$ or $r \in Q$ (by definition). Pick any $t \in R - Q$, and form $t \vee Q$, then we have a contradiction if there exists $r \in Q \vee t$ such that $r \not\in Q$ and $r \not\perp Q$. To find such an element, pick any $q \in Q$ and form $q \vee t$, then by SSP $\exists$ distinct $r \in q \vee t$ with $r \not\perp q$ and $r \not\perp t$. Now $r \not\in Q$ since $r \not\perp t$, so it remains to prove that $r \not\perp Q$. Suppose that $r \perp Q$, then $r \in (Q \vee t) -Q$, hence $\exists u \in r \vee t$ such that $u \bot t$, but then $u \bot (Q \vee t)$ which is a contradiction. Hence $r \not\perp Q$ and we are done.

(ii) $\Rightarrow$ (i): Suppose false, then $\exists r_{1}, r_{2} \in R$ such that $\not\exists$ distinct $t \in r_{1} \vee r_{2}$. Hence $r_{1} \vee r_{2} = r_{1} \cup r_{2}$ and $r_{1} \bot r_{2}$. But if $[r_{1}]$ and $[r_{2}]$ denote the perspectivity equivalence classes of $r_{1}$ and $r_{2}$, respectively, then $[r_{1}] \wedge [r_{2}] = \O$ (since if it did not then $r_{1} \vee r_{2}$ would satisfy WSP). But now, by Lemma 2.56 (ii), $[r_{1}]$ and $[r_{2}]$ are in the centre of $R$, which is a contradiction.

\vspace{8pt}

{\bf{2.58 Proposition}}

$Z_{S}$ is an orthocomplemented Boolean atomic sublattice of $L_{S}$. The atoms of $Z_{S}$ are precisely the perspectivity equivalence classes in $S$.

\vspace{8pt}

{\bf{Proof}}

By Lemma 2.55 and Corollary 2.54 it is clearly sufficient to prove that $\{Q_{i}\}_{i \in I}$ are atoms of $Z_{S}$. Let $Q \in Z_{S}$ with $Q \subseteq Q_{i}$. Since $Q_{i}$ satisfies SSP then, by Lemma 2.57, $Z_{Q_{i}} = \{ \O, Q_{i}\}$, but $Q_{i} \subseteq S \Rightarrow L_{Q_{i}} \subseteq L_{S}$ (if not, then $\exists T \in L_{Q_{i}}$ s.t. $\overline{T} \smallsetminus T \not= \O$, but $x \in \overline{T} \Rightarrow p_{x}(y) = 0$ $\forall y \in T^{\bot} \Rightarrow p_{x}(y) = 0$ $\forall y \in R - T \Rightarrow x \in T)$, so $Q = \O$.

\vspace{8pt}

{\bf{2.59 Theorem}}

$S$ can be uniquely written as the union of mutually orthogonal superposition sets, $\{Q_{i}\}$, where:
\begin{quote}
(i) Each $Q_{i}$ satisfies SSP.

(ii) The $Q_{i}$ are the atoms of $Z_{S}$, which is an orthocomplemented Boolean atomic sublattice of $L_{S}$.
\end{quote}
The $Q_{i}$ are the perspectivity equivalence classes in $S$.

\vspace{8pt}

{\bf{Proof}}

From the preceding results; uniqueness follows by our construction of the perspectivity equivalence classes.

\vspace{8pt}

{\bf{Remarks}}

Theorems 2.47 and 2.59 can be combined to yield a powerful representation result for any theory of mechanics whose pure states satisfy Axioms 1 to 4: if $d(Q_{i}) \geq 4$ $\forall i$, then $L_{S}$ is associated to a {\em{vector bundle}} over the set of perspectivity equivalence classes, though the dimensions of, and division rings associated to, different fibres can, in general, be different.

\newpage

\section{Probability in Mechanics}
{\bf{(a) Interpretation}}

Interpretations of `probability' are diverse and controversial; however, to an extent which we will make clear, we consider the choice to be irrelevant; thus we will outline the chief contenders, leaving the reader to decide which, if any, he prefers, and concentrate instead upon the status of the various types of probability that occur in mechanics.

`Probability' arises in any theory which involves statistical assertions. We shall throughout distinguish the probability functions used for describing the condition of a system from the probability statements which assert the results of measurements associated with the system. We start with the former, of which there are two types in our theory of mechanics:
\begin{enumerate}
\item {\bf{Intrinsic Probability}} which expresses the non-exclusiveness of descriptions (pure states) of a system.
\item {\bf{Avoidable Probability}} which expresses an incompleteness in the description of a system.
\end{enumerate}
The reason for this terminology is that avoidable probability can be minimised by a careful {\em{state preparation}} procedure, whereas both give rise to probability statements for the possible results of a {\em{measurement}}. In classical mechanics the pure state descriptions are exclusive, so that only avoidable probabilities are non-trivial;  moreover, no distinction between state preparation and measurement
need be made. Despite these simplifications there is still a problem, which we shall consider later, concerning the interpretation of this avoidable probability in classical mechanics.

Let us call the mathematical object that describes, even if incompletely, the preparation or condition of a system the {\em{statistical state}} of the system. A statistical state will be taken to be some form of `probability function' on the set of pure states, and every pure state will be identifiable as a statistical process through the association $s \leftrightarrow p_{s}$. We shall suppose that probability functions in the fundamental model lead to probability statements of the form:
\begin{quote}
The probability that the statistical state $v$ gives a value of the `observable' $A$ in the range $\Delta \subseteq\mathbb{R}$ is the number $Pr(v, A, \Delta) \in [0, 1]$.
\end{quote}
The `observable' $A$ is that mathematical object in the fundamental model which is taken to represent the measurement procedure, yielding numbers or small ranges of numbers as results, in the experiment to which the probability statement implicitly refers. These probability statements are {\em{assertions}} on the outcome of measurements, based upon the theory of the system, which are to be compared with the results of one or a sequence of experiments. As such, it is irrelevant whether the experiments are performed before or after the assertions signified by the function $Pr$ are made. We shall attribute to these assertions the same status as the probabilities asserted for results of a game of chance (such as the throwing of an initially symmetric, but not indestructible, die - for example, one made of sugar), `idealised' only to the extent of making explicit the set of conjectures constituting the theory of the game. As an assertive device, a probability statement is open to empirical comparison with the statistical frequencies obtained by repetition of the experiment, although its validation or not depends upon the credence given to some statistical test of this comparison - thus, for example, if a die yielded a hundred consecutive sixes we could, for this system, produce a number expressing our confidence in the validity of the usual assumption of randomness in the theory of die throwing. Although the significance of probability statements as far as assertion is concerned is non-problematic, the basis of the assertion - the choice of state - receives different emphasis according to which of the following two general views of the wider significance of probability statements is adopted (the terminology is due to Scheibe - see (Sc 1)):
\begin{enumerate} 

\item {\bf{Epistemic}}: the probability statement signifies the amount of knowledge (or lack of knowledge) about an individual case.
\item {\bf{Statistical}}: the probability statement denotes the relative frequencies of components of a hypothetical infinite ensemble of individual cases.
\end{enumerate}

This divergence of opinion on the significance of probability can be attributed to the impossibility of strictly verifying or falsifying probability statements by means of a single or, indeed, any finite number of experiments - which is why ``All Horse Players Die Broke"! Although many intermediate positions may be held, the epistemic and statistical views can be considered to be aspects of, respectively, the following two extreme interpretations of probability:
\begin{enumerate}
\item {\bf{Subjective}}: probability `does not exist'; rather, it is invented to accommodate uncertainty about some domain of experience, and expresses, in particular, each person's knowledge and ignorance concerning an individual event. A probability statement is then an assertion of a person's degree of rational belief.
\item {\bf{Objective}}: probabilities `exist' as the limiting relative frequencies of occurrence of particular events in a sequence of repeatable experiments, and are thereby empirically testable; a probability statement is an assertion about these relative frequencies in an infinite `ensemble' of individual experiments. (For elaboration of the objective interpretation of probability see, for example, (Pr 1)).
\end{enumerate}

Recalling that in Chapter 1 we argued that understanding is neither subjective nor objective, but interactive, it is, perhaps, clear why we may reject both of these extreme interpretations. In the subjective interpretation, the grounds for `rational belief' are suppressed by the expedient of using personal opinion, whilst in the objective interpretation the theoretical basis of probabilistic assertion is attributed to a hypothetical but at the same time empirically predetermined ensemble. But if the grounds for probabilistic assertion are explicated in the form of a theory about the domain under consideration, the question of subjectivity or objectivity becomes irrelevant. lt is for just this reason that the practical application of probability, especially in quantum mechanics, is unaffected by the controversy which rages over its `meaning'. This should not, however, be taken as showing that either interpretation is `wrong', logically or otherwise, merely that they are unnecessary if the theory has been explicated.

Operational arguments concerning the measurement of continuous parameters may be readily advanced for the necessity of an incomplete specification of the condition of a system in classical mechanics. This incompleteness appears, for everyday magnitudes, to be avoidable to any required degree by improving the precision of the state preparations (measurements) involved. A virtue of this necessity is made in classical statistical mechanics, where the large number of degrees of freedom and the limited information available combine to enforce an incomplete description - although it should be noted that the system, and its state, no longer refer to a point particle, but to an infinite collection of point particles. These classical examples have familiar epistemic and statistical interpretations, and are often taken uncritically as visualisable bases for the interpretation of probability in quantum mechanics. However, caution should be exercised: the chief dangers in adopting an interpretation of probability for quantum mechanics lie, firstly, in the aspiration that it provides `reasons' for the occurrence of probabilities, and, secondly, in the application of the interpretation not only to probability statements but also to the statistical and pure states which give rise to the statements. Failure to recognise these dangers leads to the unnecessary intrusion of classical analogies into quantum theory, generates endless confusion about supposed `existence' of various mathematical objects in the theory - witness the many conflicting discussions of the epistemological significance of Heisenberg's uncertainty principle - and can obliterate the important distinction of pure from other statistical states (for discussion of this point see Section 2.6b).

\vspace{8pt}
{\bf{(b) Technical Problems}}

In probability theory one usually starts by specifying a {\em{space of alternatives}}, that is, a set of possible `elementary events' or `distinguishable outcomes'. Note that the space of alternatives is not necessarily identifiable with the set of possible results obtained by some measurement; however, the alternatives should be in some way distinguishable by experimental procedures (even if only in some limiting sense). In this approach, then, the space of alternatives is a set, denoted $S$, of points, each point being an elementary event. $S$ could, for example, be the set of all intervals of the form ($a_{n}, a_{n+1}$] in $\mathbb{R}$, where $a_{n+1}$ - $a_{n}$ = $\epsilon > 0$; $n \in \mathbb{Z}$, with $a_{0} = 0$, say. In the next Section we shall consider an extension of this simple notion of a space of alternatives which, by switching the emphasis away from elementary events to distinguishable outcomes, considers the alternatives to be a certain collection of subsets of a set; the points of the set need not then be alternatives. For the moment, however, let me suppose that the `distinguishable outcomes' are points (atoms) in $S$.

The technical problems arise in trying to define functions which provide the probabilities of the various alternatives. The problems are essentially concerned with finding a suitable extension of the case where $S$ is finite and the `probability functions' are (finitely) additive in the sense that, for distinct alternatives  $\{a_{i}\}$$_{i=1}^{N}$, the probability of $a_{1}$ or $a_{2}$ or ... $a_{N}$ is the sum of the probabilities of the $a_{i}$. In particular, the probability functions should be defined on some domain $D_{S} \subseteq 2^{s}$, take values in the interval [0,1], and be at least finitely additive on any finite partition of $S$ for which the partition sets lie in $D_{S}$. The obvious choice for this extension to general $S$ is to define the functions on $2^{S}$ and require them to be finitely additive, but this would restrict one to some form of Riemann integral in subsequent analysis, with its problems of integrability and convergence. An alternative approach, due to Kolmogorov, is to allow the probability functions to be countably additive and defined on a $\sigma$-algebra, $M_{S}$ say, generated from sets in $S$. Although the assumption of countable additivity is difficult to justify on operational grounds, it has the virtue of making available the abstract integration methods associated with the Lebesgue theory, and we will accept it on these, admittedly rather suspect, grounds of mathematical convenience. Of more concern is the choice of $\sigma$-algebra $M_{S}$. In particular, when $S$ is uncountable, the choice of $M_{S} = 2^{S}$ yields only a restricted number of countably additive functions, whereas the choice $M_{S} =$ the $\sigma-algebra$ generated by the points of $S$ is, intuitively, too `small'. It seems that in order to find a suitable candidate for $M_{S}$ lying between these two, it is necessary to look beyond the probabilistic aspects. We leave the reader to decide whether or not he finds the following argument convincing: suppose there is a topology on $S$ providing a criterion for the `closeness' of elementary events, then to specify `how many are how close' it is desirable for neighbourhoods to be measurable, which is accomplished by choosing $M_{S}$ to be the smallest $\sigma$-algebra on $S$ containing the open sets. This choice is obviously convenient mathematically, but where does the topology come from? In classical mechanics the answer is clear: the topology is determined by the geometry of space and time (see Chapter 3 for details). Generalising, we offer the prescription that a topology for the set of values - for example, real numbers - of random variables can be used to determine $M_{S} = D_{S}$ by requiring the random variables to be $M_{S}$-measurable (relative to the $\sigma$-algebra generated by the topology on the set of values).

The above remarks should be borne in mind in the next Sections, where the theory of probability will be extended, following Mackey (Ma 1) to include orthocomplemented lattices of elementary events.

\newpage

\section{A Fundamental Model for Mechanics with Intrinsic Probability}
Although an axiomatic formulation of a theory suffers from the drawback of choosing particular axioms from other sets of equivalent or more or less restrictive axioms, it has the virtue of isolating (what the axiomatiser considers to be) the essential assumptions of the theory. Accordingly, the fundamental model of mechanics detailed below will provide an axiomatic basis from which it will be possible to derive familiar formulations of classical and quantum mechanics as special cases, and in so doing will emphasise the features common to both theories. Consideration is given in the following two Sections to the additional assumptions required for classical and quantum mechanics, respectively. With Section 2.7, where spatio-temporal notions are expressed in terms of mechanics, the programme of determining the common ground of classical and quantum mechanics is completed and the way made clear for resolution of their major differences. This resolution, in the sense of a well-defined intertheoretic reduction, is the content of Chapter 4.

\vspace{8pt}
{\bf{(a) Motivation of Definitions}}

In Section 2.1, a pure state was defined to be a complete description of the condition or preparation of a system, but what if it is not practical to fully specify the preparation procedure, or otherwise determine the condition of the system? Such a circumstance is familiar even from Newtonian mechanics, where the idealisation that the condition of a system can be described by a finite set of real numbers is not empirically attainable (this point will be considered in some detail later on). The basic problem is to describe the condition of the system in a manner that reflects the experimental limitations of state preparation. By viewing these limitations as generating uncertainty (or ignorance), resort can be made to probabilistic notions, and, as in Section 2.2, {\em{statistical states}} are introduced as a species of probability function on the set of pure states. To make this more precise, consider first the case where there are a finite number of pure states, then a function $v$ from $S$ to $[0, 1]$ will be said to be a probability function if, for any partition $\{s_{i}\}$ of $S$ (see Definition 2), it satisfies:
\begin{quote}
$\sum_{i} v(s_{i}) = 1$ (finite additivity)
\end{quote}
Clearly, each such $v$ can be uniquely extended to a function on $L_{S}$ by defining: $v(R) = \sum_{i} v(r_{i})$ for $R \in L_{S}$ where $\{r_{i}\}$ is any partition of $R$. The use of the orthomodular lattice structure of $L_{S}$ rather than that of $2^{S}$ follows from the selection, by intrinsic probability functions, of superposition sets as those subsets of $S$ to which probabilities may be consistently assigned; broadly, therefore, we are requiring that statistical states are no more discerning than pure states in their assignment of probabilities to subsets of $S$. Although we should be wary of interpretations, the following brief glossary may be useful: pure states are `elementary events'; superposition sets are `events' ; $v(R)$ is the probability that the system described by $v$ is in (or is describable by) a pure state belonging to the superposition set $R$. Obviously each pure state can, as its associated intrinsic probability function, be considered to be a statistical state.

To motivate the general definition, consider first two important special cases:
\begin{quote}
(i) $S$ {\em{is countably partitioned}}: If every partition of $S$ (and hence of any superposition set in $S$) has only a countable number of elements, then a probability function $v$ is defined to be any function from $S$ into $[0, 1]$ satisfying:
\begin{quote}
$\sum_{i} v (s_{i}) = 1$
\end{quote}
for every partition $\{s_{i}\}$ of $S$.

(ii) $L_{S}$ {\em{is Boolean}}: This is just the case considered in Section 2.2(b); if $S$ is uncountable, a probability function $v$ is defined to be any countably additive function from some fixed $\sigma$-algebra $D_{s}$, constructed from $S$, into $[0, 1]$ and satisfying $v(S) = 1$.

\end{quote}

Thus, for the general case, we are led to propose that some $\sigma$-complete orthocomplemented lattice $D_{S}$ must be constructed from $L_{S}$. {\em{Probability function}} (or {\em{measure}}) is then defined to be any function $v$ from $D_{S}$ into $[0, 1]$ such that, for any countable collection $\{R_{i}\}$ of mutually orthogonal sets in $D_{S}$ which satisfies $v_{i}R_{i} = S$:
\begin{quote}
$\sum_{i} v(R_{i}) = 1$ (countable additivity on $D_{S}$)
\end{quote}

In view of the countable additivity demanded for probability functions (and hence for statistical states), it is appropriate to extend Axiom 2.3 by requiring the intrinsic probability functions to be defined and countably additive on $D_{S}$. Now this may not be possible for an arbitrary $D_{S}$ (for example, if we make the choice for $D_{S}$ discussed in the next paragraph), and thereby provides a useful condition for $D_{S}$ to be suitable. In order that the intrinsic probability functions be at least defined on $D_{S}$, we shall eventually require $D_{S} \subseteq L_{S}$; but first, however, an example where this is not true:

The problem of choosing $D_{S}$ is most notable in classical mechanics, for which $L_{S} = 2^{S}$. Considering, for simplicity, the case of $S \approx \mathbb{R}^{2n}$, a popular candidate for $D_{S}$ is the quotient algebra of Borel sets in $\mathbb{R}^{2n}$ modulo Lebesgue-null sets; let us denote this choice by $Pop (S)$. Notice first that $Pop (S) \not\subseteq L_{S}$, and second that pure states are not statistical states. The basic idea is that $Pop (S)$ represents the limitation to experimental determination of pure states; indeed, Primas (Pr 1) terms the associated statistical states {\em{epistemic}}, whilst the `inaccessible' pure states he calls {\em{ontic}}. There appear to be three criteria for choosing $Pop (S)$; two - the existence of a natural topology and canonical measure on $S$ - are mathematical, and the third - the experimental inaccessibility of real numbers - is epistemological. The mathematical criteria have no obvious counterparts in the general case, but some appreciation (albeit unsympathetic!) of the epistemological criterion can be obtained from the notion of a property of a system:

A {\em{property}} of a system may be simply thought of as a labelling of pure states by numbers. Properties serve primarily two purposes: firstly, that of keeping numerical track of certain features of pure and statistical states, and secondly, that of a mathematical representation of measuring instruments (in the sense that these associate numbers to states) - if used for this latter purpose, we shall call a property an {\em{observable}}. The range of possible results of an experiment for a given measuring instrument we shall term the {\em{set of values}} of the observable associated to the instrument. In general, the set of values could be any set equipped with a suitable structure such that both set and structure are deemed appropriate to the measuring instrument's display of results, but let us for simplicity restrict the possibilities for the set to the real number line or subsets thereof; of more importance is the `suitable structure', and this will be considered below. Although an observable strictly refers only to a particular measuring instrument in the particular experiment under consideration, once determined it is available for use as a bookkeeping device in domains not involving the instrument (provided the set of pure states is the same). Under these circumstances we will still use the term `observable', even though as a property its measurement significance is only a potentiality. Not every property need be an observable, indeed, it is by no means evident that observables can be found at all. However, we shall so define properties as to accommodate any `reasonable' measuring instruments, namely, those instruments which yield, for each statistical or pure state of the system, probability measures on their respective sets of values. No general prescription will be given for determining particular observables from particular experimental arrangements - this is a matter either for ad hoc conjecture or for analysis of the measuring process (see Section 2.6).

To define a property we need to find an appropriate labelling of pure states by numbers; let us start by supposing that a property $P$ is a function from $E(\mathbb{R})$ into $2^{S}$, where the set $E(\mathbb{R})$, constructed from $\mathbb{R}$, represents the set of values with some `suitable structure'. With the interpretation that if $s \in P(E)$ then $s$ has the property $P$ with a value in $E \in E(\mathbb{R})$, then, if $s'$ is another pure state, $s'$ has a probability $p_{s'}(P(E))$ of having the property $P$ with a value in $E$. Thus we look for a $P$ such that $p_{s}(P(\cdot))$ is a probability measure on $\mathbb{R}$ for each pure state $s \in S$. Given that the probabilistic aspects of the theory are committed to countable additivity, it is clear that the `suitable structure' of the set of values should include a $\sigma$-algebra in the set of values. The key question, then, is which $\sigma$-algebra in $\mathbb{R}$ should we choose? It seems not unreasonable to utilise the topology of $\mathbb{R}$, and this leads us to consider again the criteria for the `popular choice' of $D_{S}$ in classical mechanics referred to earlier; if the arguments put forward by proponents of the popular choice are accepted, the $\sigma$-algebra should surely be the quotient algebra of Lebesgue-Borel sets modulo Lebesgue-null sets. However, this choice would not only prove an embarrassment for quantum mechanics (where point spectra are notably useful) but it is also, in our opinion, epistemologically unsound. We devise instruments to present results as finite strings of digits; the results might be in a directly numerical, such as binary, form, or in an indirectly numerical form, such as a graph, from which the numerical quantities can be derived. The length of the string of digits depends upon the pre-specified degree of precision, and each measurement may be viewed as a `call-and-response', the call being a specified precision, and the response the measured result. This `call-and-response' view could, by itself, motivate the inclusion of each interval and each real number in the set of values, but in fact our choice has already been determined by the assumption of countable additivity. Each result, being a finite string of digits, is an interval in $\mathbb{R}$ with rational end-points, so, if the theory is to include the wide variety of experimentally attainable precision, there are only two alternatives available for the set of values:
\begin{quote}
(1) Accept countable additivity; the appropriate $\sigma$-algebra is then that generated by all intervals in $\mathbb{R}$ with rational end-points. Hence it is the $\sigma$-algebra of Borel sets in $\mathbb{R}$, denoted $B(\mathbb{R})$ , and includes, in particular, each real number.

(2) Reject countable additivity; the set of values then simply consists of all intervals in $\mathbb{R}$ with rational end-points.
\end{quote}
(The intervals can be taken to be open, half-open or closed depending on one's preference. Note that it makes no difference if, in (1) or (2), we replace $\mathbb{R}$ by some open interval in $\mathbb{R}$ appropriate to the range of a particular measuring instrument, since changes of scale lead us back to $\mathbb{R}$).

Naturally, we adopt alternative (1) and, accordingly, reject the `popular choice' for $D_{S}$ in classical mechanics. So, with a clear conscience, we will assume that $D_{S}$ includes the pure states. Thus, if $C_{S}$ denotes the $\sigma$-complete sublattice of $L_{S}$ generated by all the points of $S$, the requirement on $D_{S}$ is:
\begin{quote}
$C_{S} \subseteq D_{S} \subseteq L_{S}$
\end{quote}
Having chosen the domain of a property to be $B(\mathbb{R})$, consider now its range. In view of Lemma 2.20, the range should be contained in $L_{S}$, but since it is desired that statistical as well as pure states give rise to probability measures, the range should be contained in $D_{S}$. Overall, therefore, we look for functions:
\begin{quote}
$P : B(\mathbb{R}) \rightarrow D_{S}$; $\Delta \rightarrow P(\Delta)$
\end{quote}
such that $p_{S}(P(\cdot))$ is a countably additive probability function on $B(\mathbb{R})$ for each $s \in S$. This now determines the function $P$ completely:

{\bf{2.60 Proposition}}

Let each intrinsic probability function be countably additive, then $p_{S}(P(\cdot))$ is a countably additive probability function on $B(\mathbb{R})$ for each $s \in S$ if and only if:

(i) $P(\O) = \O$ and $P(\mathbb{R}) = S$ 

(ii) For any countable collection $\{\Delta_{i}\}$ of mutually disjoint elements of $B(\mathbb{R})$ then:
\begin{quote}
$P(\Delta_{i})$ $\bot$ $P(\Delta_{j})$ for $i \not= j$, and $P(U_{i} \Delta_{i}) = v_{i} P(\Delta_{i})$.
\end{quote}
{\bf{Proof}}

$\Rightarrow$ : (i): $p_{S}(P(\O)) = 0$, $\forall s \in S \Leftrightarrow P(\O) = \O$

\hspace{37pt} $p_{S}(P(\mathbb{R})) = 1$, $\forall s \in S \Leftrightarrow P(\mathbb{R}) = S$.

(ii): first prove that if $\Delta_{1} \subseteq \Delta_{2}$ then $P(\Delta_{1}) \subseteq P(\Delta_{2})$: since $p_{s}(P(\cdot))$ is a probability measure, then $\Delta_{1} \subseteq \Delta_{2}$ implies that  $p_{s}(P(\Delta_{1})) \leq p_{s}(P(\Delta_{2}))$ $\forall s \in S$, so by Proposition 2.23, $u \in P(\Delta_{1})$ implies that $u \in P(\Delta_{2})$, which is the required result. Now, since $p_{s}(P(\cdot))$ is additive, then for $i \not= j$ we have:
\begin{quote}
$p_{s}(P(\Delta_{i} \cup \Delta_{j})) =  p_{s}(P(\Delta_{i})) + p_{s}(P(\Delta_{j}))$
\end{quote}
but, since $P(\Delta_{i}) \subseteq P (\Delta_{i} \cup \Delta_{j})$ then we also have, by Proposition 2.25:
\begin{quote}
$p_{s}(P(\Delta_{i} \cup \Delta_{j})) =  p_{s}(P(\Delta_{i})) + p_{s}(P(\Delta_{i})^{\bot} \wedge P(\Delta_{i} \cup \Delta_{j}))  $
\end{quote}
hence, by Proposition 2.21 (iii):
\begin{quote}
$P(\Delta_{j}) = P(\Delta_{i})^{\bot} \wedge P(\Delta_{i} \cup \Delta_{j})$
\end{quote}
so that $P(\Delta_{i})$ $\bot$ $P(\Delta_{j})$.

From the countable additivity of $p_{s}(P(\cdot))$, we have:
\begin{quote}
$p_{s}(P(\cup_{i} \Delta_{i})) = \sum_{i}p_{s}(P(\Delta_{i}))$
\end{quote}
but we have just seen that the $\{P(\Delta_{i})\}$ are mutually orthogonal, so from the countable additivity assumed for each $p_{s}$ we conclude:
\begin{quote}
$\sum_{i}p_{s}(P(\Delta_{i})) = p_{s}(v_{i}P(\Delta_{i}))$
\end{quote}
$\Leftarrow$: If $P$ satisfies (i) and (ii) then it is immediate that $p_{s}(P(\cdot))$ is a probability measure on $B(\mathbb{R})$.

A function from $B(\mathbb{R})$ to $D_{S}$ which satisfies (i) and (ii) of Proposition 2.60 is said to be a $D_{S}$-{\em{valued measure on}} $\mathbb{R}$ (for the Borel $\sigma$-algebra on $\mathbb{R}$), so we define a property to be a $D_{S}$-valued measure on $\mathbb{R}$.

Now for each property $P$ and each statistical state $v$, the function $v^{P}$ on $B(\mathbb{R})$ given by:
\begin{quote}
$v^{P}(\Delta) = v(P(\Delta))$
\end{quote}
is clearly a countably additive probability measure on $\mathbb{R}$. $v^{P}(\Delta)$ may be given the interpretation of the probability that the system described by $v$ will yield a value in $\Delta$ of the property $P$. Notice that each pure state is a statistical state by the identification $v \equiv p_{s}$. Since $v^{P}$ is a countably additive measure, we can define its mean value for any Borel set; if $\mathbb{P}$ denotes the set of all properties, and $\mathbb{V}$ the set of all statistical states, then we introduce the {\em{expected value functional}}, $E$, for the system as:
\begin{quote}
$E: B(\mathbb{R}) \times \mathbb{P} \times \mathbb{V} \times \mathbb{R}$; $(\Delta, P, v) \rightarrow E(\Delta, P, v)$
\end{quote}
where: $E(\Delta, P, v) = \int_{\Delta} x dv^{P}(x)$ will be called the {\em{expected value}} of the property $P$ in the state $v$ for the Borel set $\Delta$, and it has the usual probabilistic significance. We define $\mathbb{R} = \mathbb{R} \cup \{ \phi \}$ where $\phi$ where is a null result assigned whenever the integral is not defined.

Some properties are obtained by defining, for $R \in L_{S}$:

\[
   P_{R}(\Delta)=\left\{
               \begin{array}{ll}
                 R \hspace{4pt} \text{if} \hspace{4pt} \{1\} \subseteq \Delta; \text{and} \hspace{4pt} R^{\bot} \hspace{4pt} \text{if} \hspace{4pt} \{0\} \subseteq \Delta  \\
                S \hspace{4pt} \text{if} \hspace{4pt} \{0\} \cup \{1\} \subseteq \Delta \\
                \O \hspace{4pt} \text{otherwise},
               \end{array}
             \right.
 \]
which be loosely interpreted as the property of the system being in a pure state contained in $R$. Special cases are the {\em{identical property}}, $\prod = P_{S}$, and the pure state properties, $\{P_{s}\}_{s \in S}$. It does not seem possible, however, to find a property corresponding to each statistical state.

Finally, some further remarks concerning $D_{S}$: besides being $D_{S}$-valued measures, properties are also $\sigma$-homomorphisms of Boolean $\sigma$-algebras (see (Va 1) p.12 for definition), from which we conclude that $Ran(P)$ is a Boolean $\sigma$-complete sublattice of $L_{S}$ for each property $P$. Now $B (\mathbb{R})$ is the `largest' of all separable $\sigma$-algebras in the sense of the following Proposition:

{\bf{2.61 Proposition}}

Let $B$ be any separable Boolean $\sigma$-algebra, then there exists a $\sigma$-homomorphism, $h$, from $B(\mathbb{R})$ to $B$ such that:

\begin{quote}
$B = Ran(h)$.

\end{quote}
{\bf{Proof}}

This is the second part of Theorem 1.6 (i) in (Va 1).

{\bf{2.61 Corollary}}

Let $B$ be a Boolean $\sigma$-complete sublattice of $L_{S}$, then in order that there exists a property $P$ such that:

\begin{quote}
$B = Ran (P)$

\end{quote}
it is necessary and sufficient that $B$ is separable.

{\bf{Proof}}

Immediate from Proposition 2.61 and the previous remarks.

From these results it seems not unreasonable to require $D_{S}$ to be separable in the sense of the following definition:

{\bf{2.62 Definition}}

A $\sigma$-complete orthocomplemented lattice $L$ will be said to be {\em{separable}} if, for each Boolean $\sigma$-complete sublattice, $B$, of $L$ there exists a countably generated Boolean $\sigma$-complete sublattice, $B'$, of $L$ such that $B \subseteq B'$.

Notice that this generalises to lattices the usual definition for $\sigma$-algebras of separable as countably generated. A stronger type of separability for lattices is adopted by some authors (see, for example, (Ja 1) or (Va 1)), namely, the requirement that every Boolean $\sigma$-complete sublattice is countably generated; however, this latter definition has the disadvantage of excluding the usual theory of classical mechanics, where $D_{S}$ is the $\sigma$-algebra of Borel sets in phase space. To see this, consider $B(\mathbb{R})$: although $B(\mathbb{R})$ is countably generated, the collection of all sets with countably many elements together with the complements of these sets is a $\sigma$-algebra contained in $B(\mathbb{R})$ which is not, however, countably generated. (I thank Dr. E. B. Davies for bringing this example to my attention).

\vspace{16pt}

{\bf{(b) The Fundamental Model}}

The definitions of Section 2.1 are retained, but in place of Axioms 1 to 4 there is the following fundamental definition:

{\bf{2.63 Definition}}

A set $S$ will be said to be a {\em{set of pure states}} of a system $\sum$ if, for each $s \in S$, there exists a positive function $p_{s}$ on $S$, called the {\em{intrinsic probability function}} of $s$, satisfying:
\begin{quote}
(1) (i) $p_{s}(s') \leq 1$, $\forall s' \in S$ with equality iff $s = s'$.

\hspace{13pt} (ii) $p_{s}(s') = p_{s'}(s)$, $\forall s' \in S$.

\end{quote}
(2) $p_{s}(R) =$ $_{r \in \overline{R}}^{max}$$p_{s} (r)$ exists for each $R \in 2^{S}$. ($\overline{R}$ defined in Definition 2.5).

(3) For any countable collection $\{R_{i}\}$ of mutually orthogonal elements of $2^{S}$ such that $v_{i}R_{i} = S$ then: $\sum_{i}p_{s}(R_{i}) = 1$.

{\bf{2.64 Remarks}}

(a) (2) is the extension of $p_{s}$ from the set of all states to the set of all superposition sets, and can be viewed as providing the `closest elements' to a state in a superposition set.

(b) (1) and (3) are generalised probability axioms. Note that although we have required countable additivity over $2^{S}$ (or, equivalently, over $L_{S}$), this could be weakened to hold only over some $D_{S}$ where $D_{S}$ is defined - independently of (3) - in the next definition.

{\bf{2.65 Definition}}

$D_{S}$ is any separable $\sigma$-complete orthocomplemented sublattice of $L_{S}$ which contains the points of $S$. 

(By {\em{orthocomplemented sublattice}} we mean one with the inherited orthocomplementation and lattice operations).

Choose $D_{S}$, then:

{\bf{2.66 Definition}}

A {\em{statistical state}}, $v$, of $\sum$ is any probability measure on $D_{S}$; that is, a function:
\begin{quote}
$v: D_{S} \rightarrow [0,1]$; $R \rightarrow v(R)$
\end{quote}
satisfying, for any countable collection $\{R_{i}\}$ of mutually orthogonal elements of $D_{S}$ such that $v_{i}R_{i} = S$:
\begin{quote}
$\sum_{i}v(R_{i}) = 1$.
\end{quote}
The set of all statistical states will be denoted $\mathbb{V}$.

{\bf{2.67 Remarks}}

(a) Clearly the set of all statistical states is a convex set.

(b) Each pure state is a statistical state by the identification $s \equiv p_{s}$.

(c) Two auxiliary conditions, neither of which follow from the above definition, are desirable on statistical states in order that $v(R)$ may be consistently viewed as the probability that $\sum$ is in (or is described by) a pure state in $R$. These are:
\begin{enumerate}
\item $v(s) = 1$ for some $s \in S$ implies that $v = p_{s}$.
\item For any countable collection $\{T_{i}\}$ of elements of $D_{S}$ such that $v(T_{i}) = 1$, $\forall i$, then:
\begin{quote}
$v(\wedge_{i} T_{i}) = 1$
\end{quote}
It will not be necessary to impose these conditions in classical or quantum mechanics, since, as shown in Sections 2.4 $\&$ 2.5, in these cases they are satisfied for every statistical state of Definition
2.65.
\end{enumerate}
(d) It is an interesting question as to whether or not the set of pure states coincides with the set of all extreme points of the convex set $W$. Now if condition (1) of Remark (c) is satisfied, then it is easy to show that every pure state is extreme, but to show that every extreme statistical state is pure is harder: it is, however, true under either of the following conditions:
\begin{quote}
(i) If $D_{S}$ is Boolean.

(ii) If conditions (1) and (2) of Remark (c) are satisfied and every Boolean $\sigma$-complete sublattice of $D_{S}$ is separable.
\end{quote}
(Both proofs are straightforward and use separability, although (ii) also needs the Axiom of Choice (Hausdorff's Maximality Principle)). A more general proof, however, still seems to require extra conditions on either $S$ or $D_{S}$.

{\bf{2.68 Definition}}

A {\em{property}}, $P$, of $\sum$ is any $D_{S}$-valued measure on the $\sigma$-algebra, $B(\mathbb{R})$, of Borel sets in $\mathbb{R}$; that is, a function:
\begin{quote}
$P: B(\mathbb{R}) \rightarrow D_{S}$; $\Delta \rightarrow P(\Delta)$
\end{quote}
satisfying, for any countable collection $\{\Delta_{i}\}$ of mutually disjoint elements of $B(\mathbb{R})$ such that $\cup_{i}\Delta_{i} = \mathbb{R}$:
\begin{enumerate}
\item $P(\Delta_{i})$ $\bot$ $P(\Delta_{j})$ for $i \not= j$
\item $v_{i}P(\Delta_{i}) = S$.
\end{enumerate}
The set of all properties will be denoted $\mathbb{P}$.

{\bf{2.69 Definition}}

The {\em{expected value functional}}, $E$, of $\sum$ is the function:
\begin{quote}
$E: B(\mathbb{R}) \times \mathbb{P} \times \mathbb{V} \rightarrow \mathbb{R}$; $(\Delta, P, v) \rightarrow E(\Delta, P, v)$
\end{quote}
where: $E(\Delta, P, v) = \int_{\Delta} x dv^{P}(x)$ is called the {\em{expected value}} of the property $P$ in the state $v$ for the Borel set $\Delta$, and $v^{P}$ is the measure on $B(\mathbb{R})$ given by: $v^{P}(\Delta') = v(P(\Delta'))$.

Clearly each $v^{P}$ is a probability measure on $B(\mathbb{R})$, but notice there is no guarantee that $E(\Delta, P, v)$ is finite. Following Mackey ((Ma 1) p. 69) the set:
\begin{quote}
$J_{P} = \{ I$ $ |$ $ v^{P}(I) = 0$, $\forall v \in \mathbb{V}$; $I$ an open interval in $\mathbb{R}\}$
\end{quote}
is open and contains every open set $\Delta$ in $\mathbb{R}$ satisfying $v^{P}(\Delta) = 0$. The set: $Sp(P) = \mathbb{R} \smallsetminus J_{P}$, called the {\em{spectrum}} of $P$, is a closed set, and $P$ is said to be {\em{bounded}} iff $Sp(P)$ is bounded; it is then obvious that if $P$ is bounded then the expected value of $P$ is always finite, and similarly if the expected value of $P$ is indeterminate for some state and Borel set, then $P$ is unbounded.

The following definitions and results are of interest in expressing space-time geometry in the general theory of mechanics:

{\bf{2.70 Definition}}

Let $S$ and $T$ be sets of pure states, then a {\em{morphism}}, $\alpha$, from $S$ to $T$ is any mapping:
\begin{quote}
$\alpha: S \rightarrow T$; $s \rightarrow \alpha(s)$
\end{quote}
such that:
\begin{quote}
(1) $\alpha$ is one-to-one and onto.

(2) $p_{s_{1}}(s_{2})$ = $p_{\alpha(s_{1})}(\alpha(s_{2}))$, $\forall s_{1}, s_{2} \in S$
\end{quote}
If $S = T$ then $\alpha$ will be called an {\em{automorphism}}.

{\bf{2.71 Lemma}}

Let $\{R_{i}\}$ be any collection of elements of $L_{S}$. If $\alpha$ is a morphism, then:
\begin{quote}
(i) The inverse, $\alpha^{-1}: T \rightarrow S$, defined by $\alpha^{-1}(\alpha(s)) = s$, $\forall s \in S$, is a morphism.

(ii) $p_{s}(R) = p_{\alpha(s)}(\alpha(R))$

(iii) $\alpha(R) \in L_{S}$

(iv) $\alpha(R^{\bot}) = (\alpha(R))^{\bot}$

(v) $R_{i} \subseteq R_{j} \Leftrightarrow \alpha(R_{i}) \subseteq \alpha(R_{j})$

(vi) $\alpha(v_{i}R_{i}) = v_{i} \alpha (R_{i})$ and $\alpha(\wedge_{i}R_{i}) = \wedge_{i} \alpha(R_{i})$

(vii) $R_{i} \leftrightarrow R_{j} \Leftrightarrow \alpha(R_{i}) \leftrightarrow \alpha(R_{j})$

(viii) $\alpha$ preserves perspectivity equivalence classes.
\end{quote}
{\bf{Proof}}

(i) is trivial since $\alpha$ is a bijection.

(ii): $p_{s}(R) = _{r \in R}^{max} p_{s}(r) =_{r \in R}^{max} p_{\alpha(s)}(\alpha(r)) \leq p_{\alpha(s)}(\alpha(R))$ from which the required result follows.

(iii) By definition of $\alpha(R)$ and from Proposition 2.23 we have:
\begin{quote}
$\alpha(R) = \{ \alpha(r) \in T$ $|$ $p_{s}(r) \leq p_{s}(R)$, $\forall s \in S \}$

\hspace{23.5pt} $= \{ u \in T$ $|$ $p_{t}(u) \leq p_{\alpha^{-1}(t)}(R)$, $\forall t \in T \}$

\hspace{23.5pt} $= \{ u \in T$ $|$ $p_{t}(u) \leq p_{t}(\alpha(R))$, $\forall t \in T \} = \overline{\alpha(R)}$

\end{quote}
(iv): 
\begin{quote}
$(\alpha(R))^{\bot} = \{ u \in T$ $|$ $p_{u}(v) = 0$, $\forall v \in \alpha(R) \}$

\hspace{23.5pt} $= \{ u \in T$ $|$ $p_{\alpha^{-1}(u)}(r) = 0$, $\forall r \in R \}$

\hspace{23.5pt} $= \{ \alpha(x) \in T$ $|$ $p_{x}(r) = 0$, $\forall r \in R \} = \alpha(R^{\bot})$

\end{quote}
(v): From (i) it is obvious that $\alpha(\alpha^{-1}(t)) = t$ from which we conclude that if $R_{i} \subseteq R_{j}$ then $t \in \alpha(R_{i}) \Rightarrow t \in \alpha (R_{j})$.

(vi) Now $\alpha(R_{i}) \subseteq \alpha(v_{i}R_{i})$, $\forall i$, so, since we are, by (iii), dealing only with superposition sets, we have: $v_{i} \alpha (R_{i}) \subseteq \alpha (v_{i} R_{i})$ which, with (iv), is clearly sufficient.

(vii) and (viii) are immediate from the other results.

{\bf{2.72 Remark}}

As might have been expected, a morphism preserves all the lattice structure of the superposition sets; however, although it takes a given $D_{S}$ into some $D_{T}$, there is no guarantee that it gives a bijection between two pre-specified D's. Consequently, when $D_{S}$ and $D_{T}$ are specified, we shall call a morphism $D${\em{-bimeasurable}} if it provides a bijection between $D_{S}$ and $D_{T}$.

From Lemma 2.71 it is clear that a morphism not only effects a permutation of perspectivity classes but also induces an isomorphism between the lattices of superposition sets associated to each perspectivity equivalence class. Denoting the perspectivity equivalence classes of $S$ by $\{Q_{i}\}_{i \in I}$, where $I$ is an index set, then provided $d(Q_{i}) \geq 4$, $\forall i$, we have from Theorem 2.47 that each $L_{Q_{i}}$ is isomorphic to the lattice, denoted $L(H_{i}, \mathbb{D}_{i})$, of $\langle \cdot, \cdot \rangle$-closed linear manifolds of a Hilbertian vector space $H_{i}$. Hence a morphism effects an isomorphism between the $L(H_{i}, \mathbb{D}_{i})$.

{\bf{2.73 Definition}}

By a {\em{semilinear transformation}}, $F$, between any two vector spaces $H_{1}$ and $H_{2}$ over, respectively, the division rings $\mathbb{D}_{1}$ and $\mathbb{D}_{2}$, we shall mean the pair:

(1) An isomorphism: $f: \mathbb{D}_{1} \rightarrow \mathbb{D}_{2}$; $d \rightarrow d^{f}$

(2) An f-linear isomorphism, $F$, from $H_{1}$ to $H_{2}$; that is, a bijection such that for any $x_{1}, x_{2} \in H_{1}$ and $d \in \mathbb{D}_{1}$:
\begin{quote}
$F (x_{1} + dx_{2}) = Fx_{1} + d^{f}Fx_{2}$.
\end{quote}

Now let $H_{1}$ and $H_{2}$ be Hilbertian, and let $F$ be a semilinear transformation between $H_{1}$ and $H_{2}$, then we define the mapping $\xi_{F}$ from $L(H_{1}, \mathbb{D}_{1})$, (the lattice of $\langle \cdot, \cdot \rangle$-closed linear manifolds of $H_{1}$), into $2^{H_{2}}$ by:
\begin{quote}
$\xi_{F}(M) = \{ F(x)$ $|$ $x \in M \}$ for each $M \in L(H_{1}, \mathbb{D}_{1})$.
\end{quote}
The next Proposition relates isomorphisms of lattices of closed linear manifolds to semilinear transformations of the underlying vector spaces:

{\bf{2.74 Proposition}}

Let $H_{1}$ and $H_{2}$ be Hilbertian vector spaces of dimension $\geq 3$.

(i) If $\xi$ is any isomorphism $L(H_{1}, \mathbb{D}_{1})$ to $L(H_{2}, \mathbb{D}_{2})$ then there exists a semilinear transformation, $F$, between $H_{1}$ and $H_{2}$ such that:
\begin{quote}
$\xi = \xi_{F}$.
\end{quote}
Moreover, if $F'$ is another semilinear transformation between $H_{1}$ and $H_{2}$, then the following are equivalent:
\begin{quote}
(a) $\xi_{F} = \xi_{F'}$.

(b) There exists $0 \not= c \in \mathbb{D}_{2}$ such that: $d^{f'} = cd^{f}c^{-1}$, $\forall d \in \mathbb{D}_{1}$, and $F'x = cFx$, $\forall x \in H_{1}$.
\end{quote}
(ii) If $F$ is a semilinear transformation between $H_{1}$ and $H_{2}$, then the following are equivalent:
\begin{quote}
(a) $\xi_{F}$ is an isomorphism from $L(H_{1}, \mathbb{D}_{1})$ to $L(H_{2}, \mathbb{D}_{2})$.

(b) $\xi_{F}(M^{0}) = (\xi_{F}(M))^{0}$, $\forall M \in L(H_{1}, \mathbb{D}_{1})$.

(c) There exists $0 \not= k \in \mathbb{D}_{2}$ such that: $\langle Fx_{1}, Fx_{2} \rangle = \langle x_{1}, x_{2} \rangle^{f}k$, $\forall x_{1}, x_{2} \in H_{1}$.

\end{quote}

{\bf{Proof}}

The proof is based upon results for finite-dimensional, in particular 3-dimensional, subspaces:

(i) Let $V$ be any finite-dimensional sunspace of $H_{1}$, then by the ``Conversely, if...'' part of Theorem 3.1 of (Va 1) there exists a semilinear transformation $F$ such that $\xi(V) = \xi_{F}(V)$. It is easy to show that $F$ is a bijection and that $f$ is independent of $V$. Since $\xi$ is an isomorphism from $L(H_{1}, \mathbb{D}_{1})$ to $L(H_{2}, \mathbb{D}_{2})$, then for any $M \in L(H_{1}, \mathbb{D}_{1})$ we have:
\begin{quote}
$\xi (M) = \{ \xi(x)$ $|$ $x \in M \}$ $=$ $\{ \xi_{F}(x)$ $|$ $x \in M \}$ $=$ $\{ F(x)$ $|$ $x \in M \}$ $=$ $\xi_{F}(M)$.
\end{quote}
Now let $F'$ be another semilinear transformation:

(a) $\Rightarrow$ (b): $\xi_{F} = \xi_{F'}$ $\Rightarrow$ $\xi_{F}(V) = \xi_{F'}(V)$ for every finite-dimensional subspace $V$; hence, by Lemma 3.15 of (Va 1) we have the required result holding for each $V$, and it is easy to show that $c$ must be independent of $V$.

(b) $\Rightarrow$ (a): obvious.

(ii) (a) $\Rightarrow$ (b) is obvious.

(b) $\Rightarrow$ (a): $\xi_{F}$ is, by supposition, a mapping from $L(H_{1}, \mathbb{D}_{1})$ into $L(H_{2}, \mathbb{D}_{2})$; to see that it is a bijection, note that $F$ is a bijection so that $\xi_{F^{-1}}$ is well-defined from $L(H_{2}, \mathbb{D}_{2})$ into $2^{H_{1}}$, but then: $\xi_{F}((\xi_{F^{-1}}(P))^{0}) = P^{0}$, $\forall P \in L(H_{2}, \mathbb{D}_{2})$, so, since $\xi_{F^{-1}}(\xi_{F}(M)) = M$, then: $\xi_{F^{-1}}(\xi_{F}((\xi_{F^{-1}}(P))^{0})) = \xi_{F^{-1}}(P^{0}) = (\xi_{F^{-1}}(P))^{0}$, which makes $\xi_{F^{-1}}$ a mapping into $L(H_{1}, \mathbb{D}_{1})$, hence $\xi_{F}$ is a bijection. We can mimic the proof of Lemma 2.21 (vi) to show that $\xi_{F}$ is a lattice isomorphism by noting that $M \subseteq N \Rightarrow \xi_{F}(M) \subseteq \xi_{F}(N)$.

(c) $\Rightarrow$ (b): Mimic the proof of Lemma 2.71 (iv).

(a) $\Rightarrow$ (c): In particular, $\xi_{F}$ is then an isomorphism for any finite-dimensional subspace $V$, so the result follows by applying Lemma 4.8 of (Va 1) and noting that $k$ must be independent of $V$.

\vspace{8pt}
{\bf{2.75 Remarks}}

(a) This is essentially the fundamental theorem of projective geometry.

(b) In (Ja 1) p. 144, Jauch incorrectly omits conditions (b) or (c) in (ii).

(c) If $\mathbb{R}$ is a subfield of $\mathbb{D}$, then $\mathbb{D}$ is one of $\mathbb{R}$, $\mathbb{C}$, or $\mathbb{Q}$ as is well-known. It is then easy to show (see pp. 168-169 of (Va 1)) that $f$ is continuous (whether or not the associated Hilbert spaces are separable). The continuous isomorphisms of these division rings are well-known, (continuity is a restriction only for $\mathbb{C}$), and allows us to state:
\begin{enumerate}
\item $\mathbb{D} = \mathbb{R}$: $f$ is the identity and $F$ is linear. (Note: $\theta$ can only be the identity).

\item $\mathbb{D} = \mathbb{C}$: either $f$ is the identity and $F$ is linear, or $f$ is complex conjugation and $F$ is conjugate linear. (Note: if $\theta$ is continuous, it can only be complex conjugation).

\item $\mathbb{D} = \mathbb{Q}$: $f$ is an inner automorphism and $F$ is linear. (Note: $\theta$ can only be canonical conjugation).
\end{enumerate}
(Proofs of these assertions can be found in (Va 1) pp. 45-49 $\&$ 62-65).

Notice also that if $\mathbb{D} = \mathbb{R}$, $\mathbb{C}$ or $\mathbb{Q}$, then $F$ is a bounded semilinear operator on the associated Hilbert spaces.

Combining Proposition 2.74 with the remarks preceding it, we obtain:

\vspace{8pt}
{\bf{2.75 Theorem}} (Wigner)

Let $\{Q_{i}\}_{i \in I}$, $I$ a fixed index set, denote the perspectivity equivalence classes of $S$. Let $d(Q_{i}) \geq 4$, $\forall i$. For each $Q_{i}$ let $J_{i}$ denote a fixed isomorphism from $Q_{i}$ to the Hilbertian vector space $H_{i}$ associated to $Q_{i}$ by Theorem 2.47. Let $\alpha$ be any mapping of $S$ into itself, then the following are equivalent:
\begin{quote}
(i) $\alpha$ is an automorphism.

(ii) There exist:

(a) A permutation, also denoted $\alpha$, of perspectivity equivalence classes such that $Q_{\alpha(i)} = \alpha(Q_{i})$ for each $i \in I$.

(b) For each $i \in I$, a semilinear transformation $F_{i}$ between $H_{i}$ and $H_{\alpha(i)}$ such that $F_{i}(M^{0}) = (F_{i}(M))^{0}$, $\forall M \in L(H_{i}, \mathbb{D}_{i})$, and $J_{{\alpha (i)}^{\circ \alpha}}\circ {J_{i}}^{-1} = \xi_{F_{i}}$.
\end{quote}

\vspace{8pt}

{\bf{2.76 Definition}}

Let $G$ be a group, then a {\em{realisation}}, $\alpha$, of $G$ in $\sum$ is defined to be a mapping:
\begin{quote}
$\alpha: G \rightarrow Aut(S)$; $g \rightarrow \alpha_{g}$
\end{quote}
such that: ($e$ is the identity of $G$):
\begin{quote}
(i) $\alpha_{e} = id_{s}$ (the identical automorphism of $S$).

(ii) $\alpha_{g_{2}} \circ \alpha_{g_{1}} = \alpha_{g_{2}g_{1}}$, $\forall g_{1}, g_{2} \in G$.
\end{quote}
A realisation $\alpha$ will be called {\em{irreducible}} if and only if, for $R \in L_{S}$:
\begin{quote}
$\alpha_{g}(R) = R$, $\forall g \in G \Rightarrow R = \O$ or $S$.
\end{quote}
If $\alpha$ is an irreducible realisation of $G$ in $\sum$, then $\sum$ will be called an {\em{elementary system}} with respect to $G$.

\vspace{8pt}

{\bf{2.77 Remarks}}

(a) As defined, $\alpha$ is a left action of $G$ on $S$; right actions may also be defined similarly, but we shall not need them.

(b) If $\alpha$ is an irreducible realisation, then we conclude from Theorem 2.75 and Proposition 2.74 that:
\begin{quote}
(i) $\alpha$ acts transitively on the atoms of $Z_{S}$. 

(ii) All the Hilbertian vector spaces $H_{i}$ associated with the perspectivity equivalence classes $Q_{i}$ are isomorphic.

(iii) $\alpha$ provides a homomorphism from $G$ into the group of isomorphisms of $\mathbb{D}$. 
\end{quote}
(c) In Section 2.7 we will describe the elementary systems when $G$ is the Galilei group; as such, $G$ describes all spatio-temporally distinct observers, and it will be argued that each associated elementary system:
\begin{quote}
(i) expresses spatio-temporal notions in mechanical terms by defining a `free particle' and a set of spatio-temporal (kinematic) properties;

(ii) determines a possible set of pure states of a system which is interacting with an `external field' or other systems.
\end{quote}
(d) Extra conditions will be placed on a realisation if $G$ is a topological or Lie group; also, if $D_{S} \not= L_{S}$, then a realisation will be required to be D-bimeasurable.

\vspace{16pt}

{\bf{2.78 Definition}}

A {\em{flow}}, $F$, on $\sum$ is any mapping:
\begin{quote}
$F: (a,b) \rightarrow Aut(S)$; $\tau \rightarrow F_{\tau}$
\end{quote}
where $(a,b)$ is some open interval in $\mathbb{R}$.

For each flow $F$, we define an associated {\em{propagator}}, $\mathbb{F}$ by:
\begin{quote}
$\mathbb{F}(\tau_{2}, \tau_{1}) \equiv {F_{\tau_{2}} \circ F_{\tau_{1}}}^{-1}$.
\end{quote}
$\mathbb{F}$ is thereby a mapping from $(a,b) \times (a,b)$ into $Aut(S)$ such that:
\begin{quote}
$\mathbb{F}(\tau_{2}, \tau_{1}) = \mathbb{F}(\tau_{2}, \tau) \circ \mathbb{F}(\tau, \tau_{1})$, $\forall \tau_{1}, \tau_{2}$, $\tau \in (a,b)$.
\end{quote}

\vspace{8pt}

{\bf{2.79 Remarks}}

(a) In taking the flow rather than the propagator to be fundamental we are implicitly choosing a base time; thus, for a fixed $t_{0} \in (a,b)$ and propagator $\mathbb{F}$ we may define the flow $F^{t_{0}}$ by: 
\begin{quote}
$F^{t_{0}}_{\tau} \equiv \mathbb{F}(\tau, t_{0})$.
\end{quote}
(b) The idea of a flow is that for each $s \in S$ the function:
\begin{quote}
$\tau \rightarrow F_{\tau}(s)$
\end{quote}
is a one-parameter {\em{curve}} in $S$ which describes the evolution of the pure state $s$. Since we shall be viewing this evolution to be a feature of the system which is independent of any particular observer's spatio-temporal description, the parameter $\tau$ will be called the {\em{proper time}} of the system.

(c) As defined, the curves arising from a flow need not be in any sense `continuous'; extra conditions restricting flows to those that are suitably `continuous' will, however, arise naturally from the structure of $S$ in the special cases of classical and quantum mechanics.

(d) Any flow on $(a,b)$ may be extended to a flow on $\mathbb{R}$; this will also be possible when the flows are required to be `continuous'.

\vspace{8pt}

{\bf{2.80 Definition}}

Let $F$ be a flow on a system $\sum$. A group $G$ will be said to be a {\em{symmetry}} of the pair $(\sum, F)$ iff there exists a realisation $\alpha$ of $G$ on $\sum$ such that:
\begin{quote}
(i) $\alpha$ is an injection (that is, $\alpha$ is {\em{faithful}});

(ii) For each $\tau \in (a,b)$ we have:
\begin{quote}
$\alpha_{g} \circ F_{\tau} = F_{\tau} \circ \alpha_{g}$, $\forall g \in G$.
\end{quote}
\end{quote}
Although {\em{subsystems}} are of considerable  in mechanics, it is somewhat artificial to deal with them in the general theory, and as they will, anyhow, only be of interest to us in quantum mechanics, we reserve appropriate definitions and discussion for Sections 2.5 $\&$ 2.6.

\newpage

\section{Classical Mechanics}
Suppose that $\sum$ is an elementary system with respect to some group $G$, then from the preceding Sections we have that $L_{S}$ is determined (up to isomorphism) once we specify:
\begin{quote}
(i) A Hilbertian quadruple $(H, \mathbb{D}, \theta,\langle \cdot, \cdot \rangle)$, and

(ii) A set $S'$ (the set of atoms of $Z_{S}$).
\end{quote}

On the other hand, the theory of mechanics on $\sum$ requires us also to specify the set $D_{S}$. It is evident that this latter specification is a problem only if:
\begin{quote}
(i) $H$ is not countably partitioned, (for example, if $H$ is a non-separable Hilbert space), and/or

(ii) $S'$ is uncountable.
\end{quote}

In the next Section we shall consider $H$, but for the present Section we make any one of the following equivalent assumptions:
\begin{quote}
(1a) The intrinsic probability functions are all trivial.

(1b) $S = S'$

(1c) $L_{S} = Z_{S}$

(1d) $L_{S} = 2^{S}$

(1e) The Hilbertian vector spaces $H_{i}$ are all trivial.

(1f) $S$ does not satisfy WSP.
\end{quote}
It could be argued that in any non-trivial theory of mechanics there should exist a `free flow' on $\sum$ such that the automorphisms $\{ F_{\tau} \}_{\tau \in (a,b)}$ are all distinct, from which it follows that $S$ should be uncountable. Hence we are faced with the problem of choosing a suitable $\sigma$-algebra $D_{S}$ in $2^{S}$. For this purpose we make the assumption:
\begin{quote}
(2) $S$ is a (smoothly differentiable) manifold.
\end{quote}

The obvious candidate for $D_{S}$ is then the Borel $\sigma$-algebra generated by the topology on $S$. But where does this manifold structure come from? To this question we have no answer other than that it arises from the manifold structure of space-time which is imposed upon $S$ by means of a realisation of a group of (given) space-time transformations. Assumptions (1) and (2) are, however, insufficient for us to conclude that the resultant $\Sigma$ is a `classical mechanical system' in the usual sense; for this we also require that:
\begin{quote}
(3) There exists a symplectic structure, (that is, a closed non-degenerate 2-form, $\omega$) on $S$.
\end{quote}

Symplectic structures usually arise in connection with cotangent bundles, but in the absence of a `configuration manifold' there does not appear to be any simple justification of assumption (3) which, like (2), will therefore be treated as ad hoc.

With these remarks in mind, let us proceed to the definitions; throughout, `C' will denote `classical' so that, for example, `C-system' should be read as `classical system'.

\vspace{8pt}
{\bf{2.81 Definition}}

A system $\sum$ will be called a {\em{C-system}} if and only if the set of pure states of $\sum$ is the set of points of a symplectic manifold. 

$D_{S}$ is defined to be the $\sigma$-algebra, $B(S)$, of {\em{Borel sets}} in {\em{S}}.

Thus, for a C-system, $\mathbb{V}$ is the set of all probability measures on $S$, and the extreme points of $\mathbb{V}$ are thereby just the Dirac measures of mass one on $S$ and may be identified with the pure states. $\mathbb{P}$ is the set of all $B(S)$-valued measures on $\mathbb{R}$ .

\vspace{8pt}

{\bf{2.82 Proposition}}

Let $\sum$ be a C-system. Let $A$ be any function from $B(\mathbb{R})$ into $B(S)$, then the following are equivalent:
\begin{quote}
(i) $A$ is a property.

(ii) $A = f^{-1}$ for some real-valued Borel function $f$ on $S$.
\end{quote}
\vspace{8pt}

{\bf{Proof}}

(ii) $\Rightarrow$ (i) is easy. For (i) $\Rightarrow$ (ii) the construction of the Borel function $f$ requires some analysis, and this is provided in the proof of Theorem 1.4 of (Va 1).

Thus, $\mathbb{P}$ may be identified with the real-valued Borel functions on $S$.

\vspace{8pt}

{\bf{2.83 Definition}}

A {\em{C-morphism}}, $\alpha$, between C-systems $\sum_{1}$ and $\sum_{2}$ is a symplectomorphism, that is, a diffeomorphism, $\alpha$, from $S_{1}$ to $S_{2}$ such that:
\begin{quote}
$\alpha^{*}\omega_{2} = \omega_{1}$
\end{quote}
(where $\omega_{1}$ and $\omega_{2}$ are the symplectic forms on $S_{1}$ and $S_{2}$ respectively).

If $S_{1} = S_{2}$ then the group of all symplectomorphisms will be denoted $Aut(S_{1}, \omega)$.

Thus the C-morphisms are just the morphisms which preserve the extra (symplectic) structure we have imposed on the pure states of C-systems. Since $S$ is a manifold it is of particular interest to consider the Lie transformation groups on $S$:

\vspace{8pt}

{\bf{2.84 Definition}}

Let $G$ be a Lie group, and $\sum$ a C-system. A {\em{C-realisation}}, $\alpha$, of $G$ in $\sum$ is any mapping:
\begin{quote}
$\alpha: G \rightarrow Aut (S, \omega)$; $g \rightarrow \alpha_{g}$
\end{quote}
such that:
\begin{quote}
(i) $\alpha_{e} = id_{S}$.

(ii) $\alpha_{g_{2}} \circ \alpha_{g_{1}} = \alpha_{g_{2}g_{1}}$, $\forall g_{1}, g_{2} \in G$.

(iii) $g \rightarrow a_{g}(s)$ is smooth for each $s \in S$.
\end{quote}
If $G$ is a lie group with Lie Algebra $\Im$, let:
\begin{quote}
$d\alpha: \Im \rightarrow LHV(S)$; $A \rightarrow d\alpha(A)$
\end{quote}
where $d\alpha(A) \rvert_{S} = \frac{d}{dt} \rvert_{t=0}$ \hspace{2pt} $\alpha_{exp(tA)}(s)$ \hspace{25pt} $(\in T_{s}S$).

$d\alpha$ defines a representation of $\Im$ in the Lie algebra of locally Hamiltonian vector fields on $S$. It is readily verified that:
\begin{quote}
$[d\alpha(A), d\alpha(B)] = d\alpha ([A,B])$.
\end{quote}
The range of $d\alpha$ is thus the Lie algebra of generators of the transformations on $S$ given by the realisation $\alpha$ of $G$. Since the generators are locally Hamiltonian vector fields, they are each associated locally to a smooth function. To make this association global, we introduce:

\vspace{8pt}

{\bf{2.85 Definition}}

A C-realisation, $\alpha$, of a Lie group $G$ on $\sum$ will be called {\em{strict}} iff:
\begin{quote}
$Ran(d\alpha) \subseteq HV(S)$
\end{quote}
where $HV(S)$ denotes the Hamiltonian vector fields on $S$.

Denoting the smooth functions on $S$ by $C(S)$, (note: $C(S) \subset \mathbb{P}$), then $C(S)$ is a Lie algebra under the {\em{Poisson bracket}}, $\{\cdot, \cdot \}$, where:
\begin{quote}
$\{ f, g \} \equiv \omega (X_{f}, X_{g})$
\end{quote}
and each Hamiltonian vector field $X_{f}$ is determined by $f \in C(S)$ by:
\begin{quote}
$\omega (\cdot, X_{f}) = df(\cdot)$.
\end{quote}
Note, however, that each $X_{f}$ only determines $f$ up to an additive constant.

If $\alpha$ is a strict C-realisation of a Lie group $G$ in $\sum$, then any linear mapping:
\begin{quote}
$\lambda: \Im \rightarrow C(S)$; $A \rightarrow \lambda(A)$
\end{quote}
such that:
\begin{quote}
$X_{\lambda(A)} = d\alpha(A)$, $\forall A \in \Im$
\end{quote}
defines a representation of the Lie algebra $\Im$ of $G$ in $HV(S)$. It is readily verified that:
\begin{quote}
$X_{\{\lambda(A), \lambda(B) \}} = [X_{\lambda(A)}, X_{\lambda(B)}]$
\end{quote}
In general, however, $\lambda$ is not a representation of $\Im$ in $C(S)$ since:
\begin{quote}
$\sigma(A,B) \equiv \lambda([A,B]) - \{\lambda(A), \lambda(B)\}$
\end{quote}
is not zero, but defines a {\em{multiplier}} on $\Im$. It is by analysing the possible linear mappings of the above type that all the C-elementary systems with respect to the Galilei group may be determined, where:

\vspace{8pt}

{\bf{2.86 Definition}}

A {\em{C-elementary system}} of a Lie group $G$ is any C-system $\sum$ such that there exists an irreducible strict C-realisation of $G$ in $\sum$.

Notice that a C-realisation of a group $G$ is irreducible if and only if the action of $G$ is transitive on $S$.

\vspace{8pt}

{\bf{2.87 Definition}}

A {\em{C-flow}} $F$, on a C-system $\sum$ is any mapping:
\begin{quote}
$F: (a,b) \rightarrow Aut (S, \omega)$; $\tau \rightarrow F_{\tau}$
\end{quote}
for which there exists a smooth function:
\begin{quote}
$h: (a,b) \times S \rightarrow \mathbb{R}$; $(\tau, s) \rightarrow h_{\tau}(s)$
\end{quote}
such that: $\frac{d}{dt} \rvert_{t= \tau}F_{t}(s) = X_{h_{\tau}} \rvert F_{\tau}(s)$.

The {\em{C-propagator}} $\mathbb{F}$ associated to a C-flow $F$ is defined by:
\begin{quote}
$\mathbb{F}(\tau_{2}, \tau_{1}) = {F_{\tau_{2}} \circ F_{\tau_{1}}}^{-1}$
\end{quote}
and any function $h$ as above will be said to {\em{generate}} $\mathbb{F}$.

\vspace{8pt}

{\bf{2.88 Remarks}}

(a) Since first order ordinary differential equations admit local solutions it is easy to see that a vector field on $S$ determines a local flow; if $U_{t}^{\tau}$ denotes this flow for the vector field $X_{h}$, then the condition in Definition 2.87 is equivalent to the requirement:
\begin{quote}
$\frac{d}{dt} \rvert_{t= \tau}(\mathbb{F}(t, \tau)(s) - U_{t}^{\tau}(s)) =0$ for each $s \in S$.
\end{quote}
Thus the condition may be viewed as following from an anticipation that any flow on a C-system will, for small proper times, have similar features to the `free flow' (see Section 2.7) determined by a realisation of the time translation subgroup of the Galilei group which, by the preceding remarks, will be generated by a Hamiltonian vector field.

(b) It is readily verified (for example on p. 562 of (LS 1)) that Definition 2.87 is a {\em{contact structure}} (in the sense used in (AM 1)), so we may take over the results of the time-dependent Hamilton-Jacobi theory given in Chapter 5 of (AM 1). Note: $\tau \rightarrow F_{\tau}(s)$ is smooth.

\newpage

\section{Quantum Mechanics}
This Section considers the special case where $Z_{S}$ is trivial; that is, we make either of the following equivalent assumptions:
\begin{quote}
(1a) $Z_{S} = \{\O, S\}$

(1b) $S$ satisfies SSP.
\end{quote}
Hence, when $d(S) \geq 4$, $L_{S}$ is associated to a Hilbertian quadruple $(H, \mathbb{D}, \theta,\langle \cdot, \cdot \rangle)$. Quantum mechanics is considerably less ad hoc than classical mechanics in that it is only necessary to make additional assumptions relating to this quadruple. We start with the requirement:
\begin{quote}
(2) $\mathbb{D} = \mathbb{C}$
\end{quote}
The justification of this choice for the division ring is not, however, clear. On the one hand, $\mathbb{D}$ is constructed from the distinct elements in each line in $L_{S}$ so it might be possible to show that $\mathbb{R}$ is a subfield of $\mathbb{D}$ if, for each $s \in s_{1} \vee s_{2}$ and each number $x \in [0, 1]$ there exists $s' \in s_{1} \vee s_{2}$ such that $p_{s}(s') = x$, ($s_{1} \not= s_{2}$). The non-uniqueness of such $s'$ might then determine $\mathbb{D}$ completely. On the other hand, $\mathbb{D}$ must possess sufficient structure for $L_{S}$ to admit the automorphisms required by an irreducible realisation of the Galilei group, (in this connection, see (Jo 1)), and for a realisation to be non-trivial we would expect $\mathbb{D}$ at least to include $\mathbb{R}$ .

Anyhow, given that the complex numbers have been chosen for $\mathbb{D}$, then from Corollary 2.49 and the remarks following Proposition 2.61 we have that $\theta$ must be complex conjugation, $H$ be a Hilbert space, and $\langle \cdot, \cdot \rangle$ be the usual inner product on $H$.

The final assumption, which determines $D_{S}$ as $L_{S}$, but is otherwise unjustified, is:
\begin{quote}
(3) $H$ is separable.
\end{quote}

Assumptions (1), (2), and (3) now determine the intrinsic probability functions completely:

\vspace{8pt}

{\bf{2.89 Proposition}}

Let $S$ be a set of pure states for which the assumptions (1), (2), and (3) above hold. Suppose $d(S) \geq 4$, and fix an isomorphism $J$, from $L_{S}$ onto $L(H, \mathbb{C})$. Then:
\begin{quote}
$p_{s}(s') = Tr[P_{J(s)}P_{J(s')}]$, \hspace{10pt} $\forall s, s' \in S$
\end{quote}
where $P_{T}$ denotes the orthogonal projection onto $T \in L(H, \mathbb{C})$.

\vspace{8pt}

{\bf{Proof}}

Each $p_{s}$ is, by definition, a probability measure on $L_{S}$, so, by Gleason's Theorem, (Proposition 2.91 below), there exists, for each $s \in S$, a positive operator, $B_{s}$ say, with trace one on $H$ such that:
\begin{quote}
$p_{s}(R) = Tr[B_{s}B_{J(R)}]$, \hspace{10pt} $\forall R \in L_{S}$
\end{quote}
By the spectral theorem, $B_{s} = \sum_{n} c_{n}^{s} p_{n}^{s}$ for some countable set $\{c_{n}^{s}\}$ of numbers with $c_{n}^{s} \geq 0$ and $\sum_{n} m_{n}^{s} c_{n}^{s} = 1$, where the multiplicity $m_{n}^{s} = dim$ $Ran(P_{n}^{s})$, and where $\{P_{n}^{s}\}$ is a countable set of mutually orthogonal projections on $H$ with $\sum_{n} P_{n}^{s} = \prod$. Now, since $s \not\in R \Rightarrow Tr[P_{J(R)}P_{J(s)}] < 1$, it follows that if $J(s) \not\in Ran(P_{n}^{s})$ for any $n$, then $p_{s}(s) < \sum_{n} m_{n}^{s} c_{n}^{s} = 1$; hence $J(s) \in Ran(P_{n}^{s})$ for some $n$. But then, from $p_{s}(s') = 1 \Leftrightarrow s = s'$, we can only have $B_{s} = P_{n}^{s} = P_{J(s)}$, whence result.

For the definitions which follow, `Q' will always denote `Quantum'.

\vspace{8pt}

{\bf{2.90 Definition}}

A system $\sum$ will be a called a {\em{Q-system}} if and only if the pure states of $\sum$ are the rays of a separable complex Hilbert space.

Thus, for a Q-system, $L_{S}$ is identified with $L(H, \mathbb{C})$, where $H$ is separable. Each pure state $s$ is a ray, (that is, a one-dimensional manifold), in $H$; so, if $\psi_{s}$ denotes any unit vector in $s$, and $P_{R}$ denotes the (orthogonal) projection onto the closed linear manifold $R \in L_{S}$, we have:
\begin{quote}
$p_{s}(s') = Tr[P_{s} P_{s'}] = | \langle \psi_{s}, \psi_{s'} \rangle |^{2}$
\end{quote}
Our requirements on $D_{S}$ determine that $D_{S} = L_{S}$ for a Q-system. It is possible to characterise the convex set, $\mathbb{V}$, of statistical states of a Q-system as precisely the convex set, $J^{+}(H)_{1}$, of positive trace-class operators of unit trace in $H$.

\vspace{8pt}

{\bf{2.91 Proposition}} (Gleason)

Let $\sum$ be a Q-system, with $d(s) \geq 3$, then there exists a unique convex isomorphism:
\begin{quote}
$\rho: \mathbb{V} \rightarrow J^{+}(H)_{1}$; \hspace{10pt} $v \rightarrow \rho_{v}$
\end{quote} 
such that: $v(R) = Tr[\rho_{v} P_{R}]$, $\forall R \in L_{S}$.

\vspace{8pt}
\newpage
{\bf{Proof}}

See, for example, the article by Jost in (Jo 1).

\vspace{8pt}

{\bf{2.92 Remarks}}

(a) That the above result also holds for $\mathbb{D} = \mathbb{R}$ or $\mathbb{Q}$ is evident from the proof for $\mathbb{C}$ (see (Va 1) Chapter 7.2).

(b) It is now easy to show (see Proposition 2.18) that for each $s \in S$, where we denote $\rho_{p_{s}}$ by $\rho_{s}$, then:
\begin{quote}
$\rho_{s} = p_{s}$.
\end{quote}
Thus, we may identify $\mathbb{V}$ with $J^{+}(H)_{1}$; it is clear that the extreme points of $\mathbb{V}$ are just the intrinsic probability functions, and may therefore be placed in one-to-one correspondence with the pure states. The following result is well-known from the spectral theorem:

\vspace{8pt}

{\bf{2.93 Proposition}}

Let $\sum$ be a Q-system. Then for each $v \in \mathbb{V}$ there exists a unique countable set $\{c_{n}^{v}\}$ of distinct positive numbers, and a unique countable set $\{P_{n}^{v}\}$ of mutually orthogonal projection operators on $H$ with $\sum_{n} P_{n}^{v} = \prod$ such that:
\begin{quote}
$\rho_{v} = \sum_{n=1} c_{n}^{v} P_{n}^{v}$
\end{quote}
(the sum converging in trace norm). Moreover, for each $n$ with $c_{n}^{v} > 0$, $P_{n}^{v}$ has a finite multiplicity, $m_{n}^{v}$, and $\sum_{n} c_{n}^{v} m_{n}^{v} = 1$.

In particular, therefore, we can find, for each $v \in \mathbb{V}$, a complete orthonormal set $\{\psi_{n}^{v}\}$ of vectors in $H$ such that:
\begin{quote}
$\rho_{v} = \sum_{n} c_{n}^{v} | \psi_{n}^{v} \rangle \langle \psi_{n}^{v} |$
\end{quote}
although unless the multiplicities are each equal to 1, the $c_{n}^{v}$ need to be repeated. The set of vectors need not, of course, be unique.

For the properties, $\mathbb{P}$, we can use the spectral theorem to prove the following characterisation:

\vspace{8pt}

{\bf{2.94 Proposition}}

Let $\sum$ be a Q-system. Let P be any function from $B(\mathbb{R})$ into $L_{S}$, then the following are equivalent:
\begin{quote}
(i) $P$ is a property.

(ii) $P$ is a projection-valued measure.

(iii) There exists a self-adjoint operator, $A$, on $H$ such that:
\begin{quote}
$A = \int_{\mathbb{R}}xdP(x)$
\end{quote}
(where the integral is defined in the sense that:
\begin{quote}
$\langle \psi, A\O \rangle = \int_{\mathbb{R}}xd \langle \psi, P(x) \O \rangle$, \hspace{10pt} $\forall \psi \in H$, $\forall \O \in D(A)$
\end{quote}
where $D(A)$ is the domain of $A$).
\end{quote}
Hence $\mathbb{P}$ may be identified with the set of all self-adjoint operators on $H$, and to each property $P$ we will associate the unique self-adjoint operator $A$ given by (iii) of Proposition 2.94.

We immediately have the following formula for expected values:
\begin{quote}
$E(\Delta, P, v) = \int_{\Delta} xdTr[P(x)\rho_{v}]$
\end{quote}
so, if $|P|$ denotes the property corresponding to the positive operator $|A_{P}|$, and if $E(\mathbb{R}, |P|, v) <  \infty$, then:
\begin{quote}
$E(\mathbb{R}, P, v) = Tr [A_{P} \rho_{v}]$.
\end{quote}
From the remarks following Proposition 2.61 we have that for each automorphism, $\alpha$, of a Q-system there is an operator $U$ on $H$ such that:
\begin{quote}
$\alpha = \xi_{U}$
\end{quote}
and where $U$ is a semilinear transformation of one of the following two types, $(\psi, \O \in H)$:
\begin{quote}
(i) Unitary: $U(\lambda \psi + \O) = \lambda U \psi + U \O$ and $\langle \psi, \O \rangle = \langle U \psi, U \O \rangle$

(ii) Antiunitary: $U(\lambda \psi + \O) = \overline{\lambda} U \psi + U \O$ and $\langle \psi, \O \rangle = \overline{\langle U \psi, U \O \rangle}$.
\end{quote}
Moreover, if $U'$ is any other operator on $H$ such that:
\begin{quote}
$U' = cU$ for some $c \in \mathbb{T} = \{ c \in \mathbb{C}$ $\rvert$ $|c| = 1 \}$
\end{quote}
then we also have $\alpha = \xi_{U'}$; and, conversely, if $\alpha = \xi_{V}$ for some operator $V$ on $H$, then $c \in \mathbb{T}$ exists such that $V = cU$. Clearly, if $W$ is any operator on $H$ such that $\alpha^{2} = \xi_{W}$, then $W$ must be unitary.

With these remarks in mind, we define:

\vspace{8pt}
\newpage
{\bf{2.95 Definition}}

Let $\sum$ be a Q-system and $G$ a Borel group, then a {\em{Q-realisation}}, $\alpha$, of $G$ is any mapping:
\begin{quote}
$\alpha: G \rightarrow Aut(S)$; \hspace{10pt} $g \rightarrow \alpha_{g}$
\end{quote}
such that:
\begin{quote}
(i) $\alpha_{e} = id_{S}$

(ii) $\alpha_{g_{2}} \circ \alpha_{g_{1}} = \alpha_{g_{2}g_{1}}$, $\forall g_{1}, g_{2} \in G$.

(iii) $g \rightarrow p_{s}(\alpha_{g}(s'))$ is (Borel) measurable $\forall s, s' \in S$.
\end{quote}

Recall that a {\em{projective representation}}, $U$, of a second countable locally compact group $G$ in a separable complex Hilbert space $H$ is any weakly measurable mapping:
\begin{quote}
$U: G \rightarrow U(H)$; \hspace{10pt} $g \rightarrow U_{g}$
\end{quote}
into the group $U(H)$ of unitary operators on $H$. $U$ satisfies 
\begin{quote}
$U_{g_{2}} U_{g_{1}} = \sigma (g_{2}, g_{1}) U_{g_{2}g_{1}}$
\end{quote}
where the {\em{multiplier}}, $\sigma$, is some Borel mapping:
\begin{quote}
$\sigma: G \times G \rightarrow \mathbb{T}$.
\end{quote}
The multiplier satisfies the cocycle conditions:
\begin{quote}
(i) $\sigma (g_{3}, g_{2}g_{1})\sigma(g_{2},g_{1}) = \sigma(g_{3}, g_{2})\sigma(g_{3}g_{2},g_{1})$, \hspace{7pt} $\forall g_{1}, g_{2}, g_{3} \in G$

(ii) $\sigma (g, e) = \sigma(e,g) = 1$ \hspace{10pt} $\forall g \in G$.
\end{quote}

\vspace{8pt}

{\bf{2.96 Proposition}}

Let $\sum$ be a Q-system, $G$ a connected Lie group, and $\alpha$ any mapping from $G$ into $Aut(S)$, then the following are equivalent:
\begin{quote}
(i) $\alpha$ is a Q-realisation.

(ii) There exists a projective representation, $U$, of $G$ such that:
\begin{quote}
$\alpha_{g} = \xi_{U{g}}$ \hspace{10pt} $\forall g \in G$.
\end{quote}
\end{quote}

\vspace{8pt}

{\bf{Proof}}

(ii) $\Rightarrow$ (i) is trivial. For (i) $\Rightarrow$ (ii) let $F_{g}$ be any operator on $H$ such that $\alpha_{g} = \xi_{F_{g}}$. Since $G$ is a connected Lie group, then, for $g$ sufficiently close to the identity, there must exist $g'$ such that $g'^{2} = g$ and hence that $\alpha_{g} = {\alpha_{g'}}^{2}$. $F_{g}$ is therefore unitary in the neighbourhood of the identity, and the group property allows us to conclude this globally. We can now use Corollary 10.2 and Theorem 10.5 of (Va 2), where it is proved that if $G$ is a second countable locally compact group then for any mapping $F$ from $G$ into $U(H)$ such that $F_{g_{2}}F_{g_{1}} = \omega (g_{2}, g_{1})F_{g_{2}g_{1}}$ for some set of numbers $\omega(g_{2}, g_{1}) \in \mathbb{T}$, and $g \rightarrow {| \langle \psi, F_{g} \O \rangle |}^{2}$ is Borel $\forall \psi, \O \in H$, there exists a projective representation $U$ of $G$ in $H$ satisfying: ${| \langle \psi, U_{g} \O \rangle |}^{2} = {| \langle \psi, F_{g} \O \rangle |}^{2}$ $\forall g \in G$. It is trivial to find such a mapping $F$ for which $\alpha_{g} = \xi_{F_{g}}$ $\forall g \in G$, and we conclude that the projective representation $U$ satisfies $\alpha_{g} = \xi_{U_{g}}$ $\forall g \in G$ which proves the Proposition.

By analysing projective representations all the Q-elementary systems with respect to the Galilei group can be determined, where:
\
\vspace{8pt}

{\bf{2.97 Definition}}

A {\em{Q-elementary system}} of a Lie group $G$ is any Q-system $\sum$ together with an irreducible Q-realisation of $G$ in $\sum$.

The multiplier group of $\mathbb{R}$ is trivial, so a Q-realisation of $\mathbb{R}$ is some weakly measurable one-parameter group of unitary operators $\{U_{t}\}$. By the theorems of von Neumann and Stone concerning such groups, we conclude that $U_{t} = exp (-ith)$ for some self-adjoint operator $h$. Since `free flows' will be determined by the realisations of the time translation subgroup of the Galilei group, we define:

\vspace{8pt}

{\bf{2.98 Definition}}

A {\em{Q-flow}}, $F$, on a Q-system $\sum$ is any mapping:
\begin{quote}
$F: (a,b) \rightarrow U(H)$; \hspace{8pt} $\tau \rightarrow F_{\tau}$
\end{quote}
such that, for each $\tau \in (a,b)$, there exists a self-adjoint operator $h_{\tau}$ which satisfies:
\begin{quote}
$\frac{d}{dt}\rvert_{t= \tau}F_{t}(\psi) = -ih_{\tau}F_{\tau}(\psi)$ for each $\psi \in {F_{\tau}}^{*}(D(h_{\tau}))$
\end{quote}

The {\em{Q-propagator}} $\mathbb{F}$ associated to a Q-flow $F$ is defined by:
\begin{quote}
$\mathbb{F} (\tau_{2}, \tau_{1}) = {F_{\tau_{2}}F_{\tau_{1}}}^{*}$
\end{quote}
and the set of operators $\{h_{\tau}\}$ associated to $F$ will be said to {\em{generate}} $\mathbb{F}$.

\vspace{8pt}

{\bf{2.99 Remarks}}

(a) By the differentiation in the above Definition is meant the strong derivative, so that the condition is:
\begin{quote}
$_{t  \downarrow \tau}^{lim} | | \frac{1}{ t- \tau} (F_{t}\psi - F_{\tau}\psi) + ih_{\tau}F_{\tau}\psi || = 0$ for each $\psi \in {F_{\tau}}^{*}(D(h_{\tau}))$

\end{quote}
which we shall write as:

\begin{quote}

$s - \frac{d}{dt}\rvert_{t = \tau} F_{t} = -ih_{\tau}F_{\tau}$.

\end{quote}

(b) Denote the unitary group with infinitesimal generator $h_{\tau}$ by \\ $U_{t}^{\tau} \equiv exp(-i(t - \tau) h_{\tau})$. By Stone's Theorem we have:
\begin{quote}
(i) $U_{t}^{\tau}$ is strongly continuous

(ii) $s - \frac{d}{dt}\rvert_{t = \tau} U_{t}^{\tau} = -ih_{\tau}$.
\end{quote}

(c) It is a simple matter to use differentiability of $F_{t}$ to show that:
\begin{quote}
(i) $\mathbb{F}(t, s)$ is jointly strongly continuous with respect to $t$ and $s$.

\end{quote}

From (a) and (b) it is evident that:
\begin{quote}
(ii) $s - \frac{d}{dt}\rvert_{t = \tau} (\mathbb{F} (t, \tau) - U_{t}^{\tau}) = 0$.
\end{quote}
This expression is analogous to that in Remarks 2.88 (a), so we may interpret the condition in Definition 2.99 as a requirement that any Q-flow will be, for small properties, similar to a `free flow'.

\newpage

\section{The Measurement Process}

{\bf{(a) Experiments}}

In Section 2.1 a system was taken to be a theoretical representation of a domain of experience; we will now be more specific about what constitutes a domain. By an {\em{experiment}} will be meant some procedure, implemented by means of apparatus (however rudimentary), for analysing a domain. We contend that associated to any experiment is the following three-fold division of a domain: (see Figure 2.1):
\begin{enumerate}
\item {\bf{State Preparation}}: a selection from all available experience by means of some piece of apparatus.
\item {\bf{Interaction}}: a controlled change of the domain selected by (1). Typically, an interaction involves an auxiliary domain which, combined with the selected domain, is separable in the sense of Section 2.1; their joint change we call an {\em{evolution}}.
\item {\bf{Measurement}}: an assignment, by means of some further apparatus, of results, usually expressed in numerical form, to the domain selected by (1) and evolved under (2).
\end{enumerate}If both the state preparation and the interaction are trivial, then the experiment is just an (unanalysed) measurement, so the substance of the claim is that if a measurement can be analysed, then such a division may be effected. In practice, the possibility of this division is usually assumed, albeit only tacitly.

\vspace{16pt} 

\fbox{State Preparation} \hspace{3pt} $\rightarrow$ \hspace{3pt} \fbox{Interaction} \hspace{3pt} $\rightarrow$ \hspace{3pt} \fbox{Measurement} \hspace{3pt} $\rightarrow$ \hspace{3pt} \fbox{Results}

\hspace{44pt} $\updownarrow$ \hspace{97pt} $\updownarrow$ \hspace{90pt} $\updownarrow$ \hspace{45pt}

\hspace{15pt} \fbox{\begin{varwidth}{\textwidth}
\centering
Selection  \\ Apparatus
\end{varwidth}} \hspace{43pt} \fbox{\begin{varwidth}{\textwidth}
\centering
Auxiliary  \\ Domain
\end{varwidth}} \hspace{29pt} \fbox{\begin{varwidth}{\textwidth}
\centering
Measurement  \\ Apparatus
\end{varwidth}}

\vspace{17pt}

\hspace{15pt} Numerical \hspace{52pt} Numerical \hspace{39pt} Numerical
\\
\indent
\hspace{19pt} Controls \hspace{62pt} Controls \hspace{50pt} Controls

\vspace{7pt} 
{\bf{Figure 2.1: The Division of a Domain in a Typical Experiment}}

\vspace{8pt}

The state preparation, interaction and measurement are identified by specifications which usually include numerical controls of the apparatus associated to the selection, the auxiliary domain and the measurement, respectively.

We may now construct a theory of mechanics for the domain of an experiment by making the Co-ordinative Definitions listed in Table 2.1. 

\begin{center}
   \begin{tabular}{ | p{7.5cm} | p{8.5cm} |}
   \hline
   {\bf{\textsc{Experimental Notion}}} & {\bf{\textsc{Abstraction In The Fundamental Model}}} \\ \hline
   Domain of Experiment & System $\Sigma^{1}$ \\ \hline
    Auxiliary Domain & System $\Sigma^{2}$ \\ \hline
    Condition of a Domain & (Statistical) State of a system \\ \hline
    Condition of a Domain for fixed State Preparation & State, $v^{1}$, of $\Sigma^{1}$  \\ \hline
    Interaction & Automorphism of combined system $\Sigma^{1} + \Sigma^{2}$ \\ \hline
    Numerical Control of a Measurement & Element, $\Delta$, of set of values $B(\mathbb{R})$ \\ \hline
    Measurement & Expected Value functional $E(\Delta, P,  \cdot)$ associated to an observable $^{\cdot} P$ of $\Sigma^{1}$. \\ 
    \hline

   \end{tabular}
\end{center}

{\bf{Table 2.1: Co-ordinative Definitions in a Theory of Mechanics}}

\vspace{8pt}
{\bf{Notes on the Co-ordinative Definitions}}:
\begin{quote}
(a) The State Preparation and Interaction are fixed by means of numerical controls associated to their respective apparatuses.

(b) For convenience, we also use the term `Interaction' for the automorphism associated to an Interaction. This automorphism of $\sum^{1}$ + $\sum^{2}$ often (indeed always in the case of external fields - see Section 2.7) admits a weakly conventional alternative description as an automorphism of $\sum^{1}$ (see below).

(c) Again for convenience, we shall frequently just refer to the observable property $P$ as the abstraction associated to a measurement. The use of the expected value functional follows from the construction, based on probabilistic notions, of properties and statistical states (see Sections 2.2 $\&$ 2.3a)).

(d) It is not the case, in general, that the observable $P$ is independent of the Interaction.
\end{quote}
The aims and uses of a theoretical explanation of an experiment are diverse and depend upon the `knowledge' available. For example, it might provide a (perhaps previously unnoticed) {\em{correlation}} of results; a {\em{prediction}} of possible results from knowledge of an interaction, the observable and the prepared states; an {\em{observable determination}} from known results, interaction and prepared states; a {\em{probe}} of the interaction from known results, observable and states; or a {\em{state determination}} from known results, observable and interaction. Subsuming correlations under predictions, these alternatives are laid out in Table 2.2 below, where a `$\checkmark$' denotes that information for the column is known, and a `?' denotes that information for the column is derived:

\begin{center}
   \begin{tabular}{| p{2.4cm} | p{2.2cm} | p{2.5cm} | p{2.5cm} | p{2.7cm} |}

  \cline{2-5}
      \multicolumn{1}{c|}{} & Prepared \textsc{States} & \textsc{Interaction} & \textsc{Observable} & Results of \textsc{Measurement} \\ \hline
   \textsc{Prediction} of Results & $\hspace{26pt} \checkmark$ & $\hspace{26pt} \checkmark$ & $\hspace{26pt} \checkmark$ &\hspace{30pt}  ? \\ \hline
    \textsc{Observable} \textsc{Determination} & $\hspace{26pt} \checkmark$ & $\hspace{26pt} \checkmark$ &\hspace{26pt}  ? & $\hspace{30pt} \checkmark$ \\ \hline
    \textsc{Probe} of \textsc{Interaction} & $\hspace{26pt} \checkmark$ &\hspace{26pt}  ? & $\hspace{26pt} \checkmark$ & $\hspace{30pt} \checkmark$ \\ \hline
    \textsc{State Determination} &\hspace{26pt}  ? & $\hspace{26pt} \checkmark$ & $\hspace{26pt} \checkmark$ & $\hspace{30pt} \checkmark$ \\ \hline

   \end{tabular}
\end{center}

{\bf{Table 2.2: Some Uses for a Theoretical Explanation of an Experiment}}

\vspace{8pt}

It is most important, however, to note that the uses given in Table 2.2 do not, in general, provide sufficient information for the theoretical quantities, (state, interaction or observable), to be determined uniquely. In the ideal of arbitrarily precise results, they are still only determined up to an equivalence relation; all the elements in the equivalence class are then strongly conventional alternatives for that experiment. It is also worth remarking that prescriptions, derived from the theory, for computing these `equivalence classes' are not, in general, available, although a significant exception is the determination of the interaction in (the `inverse problem' of) scattering theory.

A further use of a theoretical explanation is to provide a {\em{Calibration}} of the numerical controls associated to the specification of the apparatus. (A Calibration need not, of course, involve the theory directly; it could be simply a correlation of results with the specifications of the apparatus). Thus, for example, the set of states calculated by a state determination may be placed in a correspondence with numerical controls associated to the selection apparatus, such as oven temperatures, slit widths, magnetic fields, or number of children of the experimenter. Whichever of these controls can be varied without altering the results may then be ignored.

The remarks of the last two paragraphs lead us to enquire about the source of the `knowledge' assumed to be available in the `$\checkmark$'s of Table 2.2. How might one know the prepared states, interactions, or observables? There appear to be two basic alternatives:
\begin{enumerate}
\item {\bf{Theoretical Assumptions}}: certain quantities in the theory are fixed. The status of the theoretical explanation is then: ``If so-and-so is the case, then such-and-such is the consequence''. In a prediction of results, for example, the assumptions might be progressively for: parameters of the system such as mass and electric charge; the various elementary systems comprising $\sum^{1}$; explicit forms of the electro-magnetic potentials relating to the elementary systems; and, finally, everything else bar the results to be predicted and some simply parameterised family of states of one of the elementary systems.
\item {\bf{Evidence from other Experiments}}: a certain piece of apparatus has, by a number of other experiments, been demonstrated to be associated to a particular state preparation, interaction or observable for the same, or similar, systems. Thus, for example, a photographic plate is found to detect electrons in a manner comparable to the expected value, ranging over (macroscopically) small sets of values, of the position observable (itself provided by the spatio-temporal analysis of elementary systems).
\end{enumerate}

Usually a combination of these alternatives is used, but it is the evidence from other experiments we wish to pursue, since this evidence not only justifies many of the theoretical assumptions, but also suggests an analysis of the measurement process. Before proceeding along this line, note that there may be alternative theories, (or recipes associated to a theory, or subtheories with varying proportions of theoretical assumptions), available to provide a `theoretical explanation' of an experiment. Supposing that each of these is in accord with the `facts', then, in the terminology of Chapter 1, they constitute a set of empirically equivalent theories for the domain under consideration.

In a given experiment, the particular division of the domain into state preparation, interaction and measurement typically depends upon three factors: the quantity, or set of quantities, that is of interest in the experiment; the apparatus that is available to assist in the enquiry; and how well understood is the functioning of each piece of apparatus. Experiments are often designed to investigate an interaction, so, with the division thereby enforced, we are led to consider the apparatus associated to state preparation and measurement. But any analysis of the apparatus necessarily invokes auxiliary theories and experiments - for example, components of the apparatus have usually been checked and calibrated in `independent' experiments involving standard samples, fields or detectors. It is just this inter-relation of various theories and experiments which makes Scientific understanding so comprehensive whilst at the same time contributing to the `measurement problem of quantum theory'. To circumvent this apparent dilemma, consider how the selection apparatus, for example, came to be designed, used, or understood in the first place. It might have been on the basis of other experiments involving well-understood interactions and measurements - for example, with the interaction stage set to `free flow', and with the measurement stage a set of diffraction gratings and photographic plates. Or, it could be that the selection apparatus started as an experiment itself, but, once having determined the evolved states, with the measurement stage subsequently replaced by a `filter' which allows this characterised portion of the evolved domain to evolve `freely' thereafter. To ascertain the effect, if any, of the filter, one could either perform further measurements or analyse the filter as an experiment in its own right. Overall, therefore, we contend that the state selection and measurement apparatuses can each, in turn, be subdivided into state preparation, interaction and measurement stages (see Figure 2.2). The first possibility mentioned above for the state preparation is then the special case where this subdivision corresponds to another set of experiments and the State Determination of Table 2.2.

\vspace{16pt} 

{\includegraphics[scale=0.35]{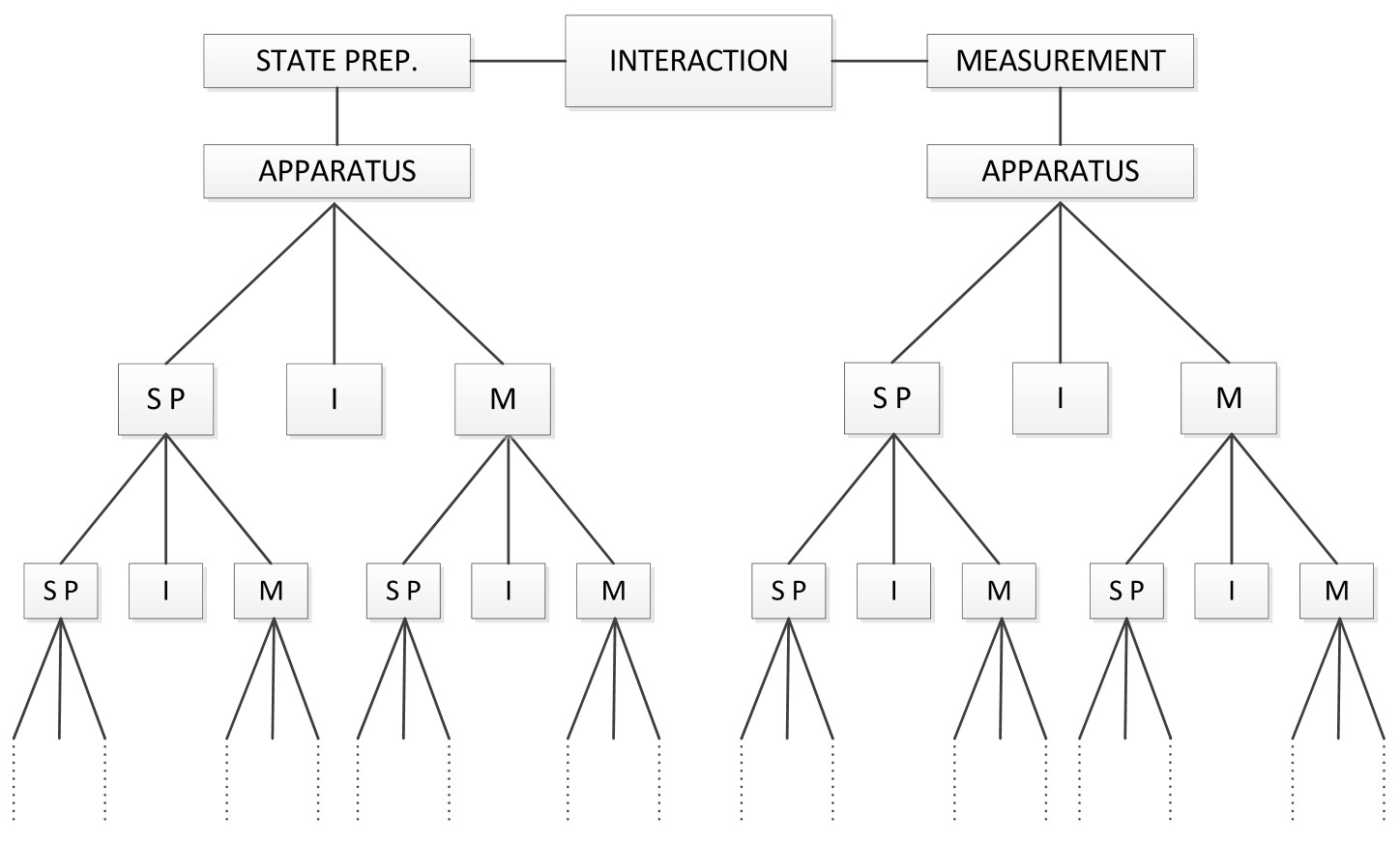}

{\bf{Figure 2.2: The Hierarchy of Experiments}}

\vspace{16pt} 

The hierarchy of experiments given by repeated subdivision, or correspondence with other experiments, is an idealisation since the domains of each experiment will differ somewhat. Consequently, it either terminates rapidly with a `well understood' experiment which is not further analysed, or else diverges to include the whole of Physics! Well, the former is at least the hope, although in practice it seems that one should read `unanalysable' for `well-understood'. For the measurement stage the `unanalysable' terminating step is usually an irreversible system with a macroscopic manifestation.

Let us at this point make a few remarks concerning experiments on systems described by classical mechanics:

The chief feature of classical mechanical systems, and a feature which might, indeed, have been anticipated from the fundamental model, is that their probabilistic aspects arise solely from limited information about the state preparation. A number of other circumstances then conspire to trivialise the analysis of experiments given above. Most notable of these is that the interactions associated to each of the state preparation and the measurement apparatus often have a negligible effect on the state of the system. Typically the auxiliary system for these interactions involves light rays. In many cases, therefore, filters and irreversible measurements are redundant, as is the hierarchy of experiments, since the possibility of repeated, non-perturbing, `measurement/preparations' allows values of properties - in particular, the {\em{position}} and its variation with time - for the state of an individual system to be determined to arbitrary precision (for everyday magnitudes). Although these values are strictly only intervals whose points are indistinguishable by everyday standards, it is customary to describe the system by a pure state. Avoidable probability is non-trivial if we consider, for example, a beam of particles prepared by firing a blunderbuss, or, alternatively, repeatedly firing a revolver, towards a collimating device together with a shutter which is opened for a certain time interval and then closed (both at fixed times from the firing of the blunderbuss or revolver). From results concerning the subsequent positions and velocities of the component particles, and knowledge of the forces - gravitational, Coriolis and so on - acting on the system, statistical states for the system after preparation could be computed and used for prediction of results if the experiment were repeated under the same conditions. In classical mechanics, therefore, statistical states may be viewed as ensembles of pure states.

\vspace{8pt}
{\bf{(b) Subsystems in Mechanics}}

Earlier in this Chapter the notion of a `domain' was introduced as a distinguishable set of experiences. Much of our understanding can be viewed in terms of the identification, and characterisation under various circumstances, of such domains. This approach is clearly evident in our compartmentalisation of the everyday material world. Thinking, in particular, of man-made objects another feature is notable, namely the hierarchical structure of the compartmentalisation so that, for example, we talk of a car, the various `systems' within it, and the working components within these systems.

Given that domains and subdomains are recognised, the theoretical task is to analyse the corresponding systems and subsystems. Here are a number of points to bear in mind concerning subsystems:
\begin{enumerate}
\item There are basically two types of subsystem: those in which the subsystem is analysed
\begin{quote}
(a) as a separated system (the rest of the system being an `environment')

(b) in conjunction with other subsystems as part of the overall system.
\end{quote}
\item Subsystems can arise in
 \begin{quote}
(a) Breaking down a system

(b) Building up a system.
\end{quote}
\item The identification of and benefit accrued from analysis using a subsystem will generally depend upon the condition of the system. Different subsystems may be appropriate to different conditions of the system.
\item A fundamental model may be directly applicable to more than one level in the subsystem hierarchy.
\item As a measure of the diversity of subsystems, consider some examples from Chemistry: there are fundamental particles (nuclei, electrons and, sometimes, photons), atoms, molecules, functional groups of atoms in molecules, liquids; molecules in various environments (e.g. lattice, polar solvent, non-polar solvent, gaseous), macromolecules, liquid crystals, liquids, various crystal lattices, metals and so on. Theoretical chemists generally model systems as `small perturbations' of subsystems. The identification of the relevant subsystems rarely results from mathematical introspection, mathematics being notably insensitive to orders of magnitude, rather it rests on an appeal to some visualisable classical analogue. For example, consider the mathematically similar cases of the helium atom and the hydrogen molecule-ion which are analysed on the basis of classical analogues for heavy nuclei orbited by light electrons (this example is from Primas in (Pr 1)).
 \end{enumerate}
The relations between systems and subsystems will be needed to describe state preparations, interactions and measurements. Throughout, the emphasis will be on Quantum Mechanics, though we start by considering the general theory of mechanics.

Recall that the pure states of a system were taken to be the elements of a set $S$ with an intrinsic probability structure. The general theory of subsystems is complicated by the need to identity, for given component subsystems, the intrinsic probability structure relating different subsystems. We have the basic, but insubstantial definition:

{\bf{2.100 Definition}}

Let $\sum_{1}$ and $\sum_{2}$ be systems, then the {\em{composite system}}, denoted by $\sum$ = $\sum_{1}$ $\times$ $\sum_{2}$ has a set of pure states, denoted $S$ = $S_{1}$ $\otimes$ $S_{2}$, generated by the intrinsic probability structure from the Cartesian product of the state spaces $S_{1}$ and $S_{2}$ of the {\em{component subsystems}} $\sum_{1}$ and $\sum_{2}$.

Where does this definition come from, and what does `generated by' mean? Given an intrinsic probability structure relating $S_{1}$ and $S_{2}$, the state space of the composite system must be consistent with the lattice description of the superposition sets, which requires the pure states to be the atoms in a lattice which contains $L_{S_{1}}$ $\times$ $L_{S_{2}}$. Complicated though this may appear, Theorem 2.59 allows us to break up the general case into composition of systems for classical and quantum mechanics. The important point is that the pure states of a composite system need not be just the elements of the Cartesian product of the pure states of the component subsystems.

There are a number of important results which hold in both Classical and Quantum Mechanics which cannot be conveniently proved in the general theory, so for the moment we pass to Classical Mechanics.

In Classical Mechanics a pure state in the composite system determines, and is determined by, pure states of the component system:

{\bf{2.101 Definition}}

Let $\sum_{1}$ and $\sum_{2}$ be C-systems, then the composite system has a pure state space: 
\begin{quote}
$S$ = $S_{1}$ $\times$ $S_{2}$
\end{quote}
that is, the Cartesian product of the state spaces $S_{1}$ and $S_{2}$.

For statistical states the relevant spaces are the Banach spaces $M(S_{i})$ of real (signed) Borel measures on the Borel spaces $S_{i}$, with the states being elements of $M_{1}(S_{i})^{+}$, that is, the measures of mass one on the cone $M(S_{i})^{+}$ of positive measures.

If $\rho \in \mathbb{V}$ $(S_{1} \times S_{2})$ $\equiv$ $M_{1}$ $(S_{1} \times S_{2})^{+}$ then the `partial state' of $\rho$ in the system $\Sigma_{1}$, say, is naturally defined as the restriction of the measure to $S_{1}$ which may be written as:
\begin{quote}
($PT_{1} (\rho)) (R) = \rho (R \times S_{2})$, $\forall R \in D_{S_{1}}$
\end{quote}
Note that $PT_{i}$ is an affine map from $\mathbb{V}$ $(S_{1} \times S_{2})$ onto $\mathbb{V}$($S_{1}$) which will be called the {\em{partial trace}}.

If $\rho_{1} \in \mathbb{V} (S_{1})$ and $\rho_{2} \in \mathbb{V} (S_{2})$ then, as is well known, there is a {\em{unique}} measure, denoted $\rho_{1}$ $\otimes$ $\rho_{2}$ such that:
\begin{quote}
$\rho_{1}$ $\otimes$ $\rho_{2}$ ($R \times T$) = $\rho_{1} (R) \otimes \rho_{2} (T)$, $\forall R \in D_{S_{1}}$, $T \in D_{S_{2}}$
\end{quote}

Notice, however, that this does not imply that there is a unique statistical state for the composite system such that its partial states are $\rho_{1}$ and $\rho_{2}$. As will be proved in Proposition 2.109 this non-uniqueness follows from convexity and is by no means peculiar to Quantum Mechanics. In the sense that various states of the composite system have the same partial states this feature allows for `correlations' of the states of the subsystems.

Let $\alpha$ be an automorphism of $S$, then its `restriction' to $S_{i}$ determines an automorphism $\alpha_{i}$, where ($i = 1, 2$):
\begin{quote}
$\alpha_{i}: S_{i} \rightarrow S_{i}$; \hspace{10pt} $s_{i} \rightarrow \alpha_{i}(s_{1}, s_{2}) =  (\alpha(s_{1}, s_{2}))_{i}$
\end{quote}
If $F$ is a C-flow on $\sum$, then the {\em{reduced dynamics}} on $\sum_{i}$ is simply the restriction of automorphisms $F_{t}$ to $S_{i}$. It should be noted that whilst the reduced dynamics is a C-flow, it will generally be generated by a time-dependent Hamiltonian even though the full dynamics could be generated by a time-independent Hamiltonian.

For C-systems with symmetry, reduction of the state space is often possible, allowing `separation' of motions. However, the `subsystems' do not necessarily correspond to different material entities but rather to symmetry aspects of the motion. For further details, consult (AM 1) p. 298.

For Quantum Mechanics the relation between a system and its subsystems is more subtle. No longer is it the case that the set of pure states of the composite system is given by the Cartesian product of the pure states of the component subsystems. Indeed, the most remarkable feature of Quantum Mechanics is that:
\begin{quote}
A complete description of the composite system does not entail a complete description of each component subsystem.
\end{quote}
It is important to be clear that it is this, and not some woolly notion of the `whole being greater than the sum of its parts', which distinguishes Quantum from Classical subsystems. Results to support these contentions will be provided after we have elaborated the mathematical side of the Quantum theory of subsystems.

Suppose $\sum_{1}$ and $\sum_{2}$ are two Q-systems with associated Hilbert spaces $H_{1}$ and $H_{2}$, and lattices of superposition sets $L_{1}$ and $L_{2}$. let $L_{S}$ be the lattice of the composite system. By Theorem 2.59 the centre is either trivial, in which case the Strong Superposition principle holds throughout $S$, or non-trivial. Considering the second possibility first, it is easy to see (cf. (Va 1) Section 8.2) that $L_{S}$ is the direct union $L_{1} \times L_{2}$. This case is often described by saying that a {\em{superselection}} rule operates between $S_{1}$ and $S_{2}$. Whether superselection rules need to be invoked depends on one's viewpoint. For example, the Bargmann `mass superselection rule', referring to the inequivalent projective representations of the Galilei group (See Section 2.7), can be considered a Superselection rule if we view all non-relativistic (non-zero mass) particles as different states of the same particle. On the other hand, it can also be considered a criterion for different particles. Supposing now that the centre is trivial, then we look for a Hilbert space $H$ such that $L_{1} \times L_{2} \subseteq L_{H}$. The smallest such candidate is the Hilbert space generated by the algebraic tensor products $\{\psi \otimes \phi \}$, $\psi \in H_{1}$, $\phi \in H_{2}$, which is the {\em{tensor product}} of the Hilbert spaces. Note, however, that application of the Pauli principle can restrict the lattice $L_{S}$ of superposition sets of the composite system to sublattices of $L_{H}$ with appropriate symmetry under permutations. We do not, however, pursue this case here. Thus we are led to:

{\bf{2.102 Definition}}

Let $\sum_{1}$ and $\sum_{2}$ be Q-systems, then the composite system has a state space representable by the rays of the Hilbert space 
\begin{quote}
$H$ = $H_{1}$ $\otimes$ $H_{2}$
\end{quote}
where $H_{1}$ and $H_{2}$ are the Hilbert spaces representing the state spaces $S_{1}$ and $S_{2}$.

Recalling that the statistical states of a Q-system are representable by the convex set $J_{1}(H)^{+}$ of positive trace-class operators with trace one, we are led to the partial state of $\rho$ in the system $\sum_{1}$, say, as $PT_{1}(\rho)$ = $Tr_{H_{2}}[\rho]$. Notice that this definition follows from the abstract definition:
\begin{quote}
($PT_{1}(\rho)$)$(R)$ = $\rho (R \otimes H_{2})$
\end{quote}
where $\rho$ is the probability measure on the lattice of superposition sets of $S$, and $R$ is any superposition set in $H_{1}$.

As a basis for treating the relationship between states of systems and subsystems in Quantum Mechanics it is useful to review tensor products and partial traces in some detail.

We assume the reader is familiar with the construction of the tensor product $H_{1} \otimes H_{2}$ of two Hilbert spaces, $H_{1}$ and $H_{2}$, as the completed space of (conjugate) bilinear functionals on $H_{1} \times H_{2}$, where Riesz' Theorem guarantees uniqueness. There are no unexpected difficulties in defining tensor products of densely defined operators on $H_{1} \otimes H_{2}$.

Recalling $L(H) = J(H)^{*}$, the Partial Trace may be defined as:

{\bf{2.103 Definition}}

Let $\{H_{i}\}$ be Hilbert spaces, ($i = 1, ..., N$), then the {\em{Partial Trace}} $PT_{i}$ is an affine contraction
\begin{quote}
$PT_{i}$: $J(\otimes_{j}H_{j}$) $\rightarrow$ $J(H_{i})$
\end{quote}
determined by the condition:
\begin{quote}
$Tr[\prod \otimes .... \otimes \prod \otimes A \otimes \prod \otimes ... \otimes \prod \rho] = Tr [A$ $PT_{i} (\rho)]$
\end{quote}
for all $A$ $\in L (H_{i})$, $\rho \in J(\otimes_{i}H_{i})$. 

By means of the isometries between $H_{1} \otimes H_{2}$ and $HS ({H_{2}}^{*}, H_{1})$ we may view any $\Phi \in H_{1} \otimes H_{2}$ as a Hilbert-Schmidt operator from ${H_{2}}^{*}$ to $H_{1}$ or, alternatively, as a conjugate linear map from $H_{2}$ to $H_{1}$ for which we have the inner product
\begin{quote}
$\langle \Phi, \Psi \rangle = Tr_{H_{1}}[\Phi^{\ddagger}\Psi]$
\end{quote}
where, if $\theta_{1} \in H_{1}$, $\theta_{2} \in H_{2}$, the `conjugate adjoint' is defined by:
\begin{quote}
$\langle \theta_{1}, \Phi \theta_{2} \rangle = \langle \theta_{2}, \Phi^{\ddagger}\theta_{1} \rangle$
\end{quote}
Note that if $\Phi = \phi_{1} \otimes \phi_{2}$, then
\begin{quote}
$\Phi (\theta_{2}) = \langle \theta_{2}, \phi_{2} \rangle \phi_{1}$

 $\Phi^{\ddagger} (\theta_{1}) = \langle \theta_{1}, \phi_{1} \rangle \phi_{2}$
\end{quote}

This is Jauch's approach in (Ja 1) and leads to:

{\bf{2.104 Proposition}}

Let $\Phi \in H_{1} \otimes H_{2}$, then:
\begin{quote}
$PT_{1} (| \Phi \rangle \langle \Phi |) = \Phi \Phi^{\ddagger}$

 $PT_{2} (| \Phi \rangle \langle \Phi |) = \Phi^{\ddagger} \Phi$
\end{quote}
{\bf{Proof}}

See (Ja 1) p. 181 (note: the calculation is much tricker if $HS({H_{2}}^{*}, H_{1})$ is used).

We can now use this result to obtain from (Ja 1), the {\em{normal form}} of the partial states of a pure state $\Phi \in H_{1} \otimes H_{2}$:

{\bf{2.105 Proposition}}:

Let $\Phi \in H_{1} \otimes H_{2}$, then there exist orthonormal systems $\{\phi$$_{n}^{1}\}$, $\{\phi$$_{m}^{2}\}$ in $H_{1}$ and $H_{2}$ respectively, and positive numbers $\{a_{r}\}$ with $\sum_{r}a_{r} = 1$ such that:
\begin{quote}
$\Phi = \sum_{r} \sqrt{a_{r}} \phi_{r}^{1} \otimes \phi_{r}^{2}$
\end{quote} 
and $(i = 1, 2)$:

 $PT_{i} (| \Phi \rangle \langle \Phi |) = \sum_{r}a_{r} |\phi_{r}^{i} \rangle \langle \phi_{r}^{i}|$

{\bf{Proof}}

See Jauch (Ja 1) p. 182. An immediate Corollary is:

{\bf{2.106 Corollary}}  (`Schr\"{o}dinger's Non-invariance Theorem')

Let $\Phi \in H_{1} \otimes H_{2}$, and let $\{\psi_{n}\}$ be any orthonormal system in $H_{1}$. Define the normalised vectors $\{\theta_{m}\}$ in $H_{2}$ by:
\begin{quote}
$\Phi = \sum_{n} c_{n} \psi_{n} \otimes \theta_{n}$
\end{quote}
then the following are equivalent:
\begin{quote}
(i) $\{\theta_{n}\}$ is an orthonormal set.

(ii) $\{\psi_{n}\}$ and $\{|c_{n}|^{2}\}$ solve the eigenvalue problem:

$\Phi \Phi^{\ddagger} \psi$ = $\lambda \psi$
\end{quote}
{\bf{2.107 Remarks}}

The above Corollary was considered important for two reasons, both depending upon von Neumann's theory of measurement for their significance:
\begin{enumerate}
\item Suppose $\psi_{r}$ was held to be the state of $\sum_{1}$ (corresponding to $H_{1}$) then, if $\Phi$ was known to be the state of $\sum_{1} \times \sum_{2}$ (corresponding to $H_{1} \otimes H_{2}$), the state of $\sum_{2}$ after `measuring' $\psi_{r}$ would be $\theta_{r}$. Thus, if $\{\psi_{r}\}$ were the eigenvectors of an `observable' $A$, say (as would be the case in von Neumann's theory), $\theta_{r}$ would determine which observables could be `measured' in conjunction with $\sum_{2}$.
\item The uniqueness of the decomposition of $\Phi$ evidently depends upon the $|c_{n}|^{2}$ coefficients, and so, therefore, does the set of `compatible observables' in $\sum_{2}$. Thus, if the $|c_{n}|^{2}$ are all different, the $\{\psi_{r}\}$ and $\{\theta_{r}\}$ are uniquely determined,  whereas at the other extreme if all the $|c_{n}|^{2}$ are the same then the $\{\psi_{r}\}$ and $\{\theta_{r}\}$ may be chosen freely.
\end{enumerate}
We now aim to make precise the difference in status of subsystems between Classical and Quantum Mechanics. As a summary of the position so far we have:

\vspace{8pt}
{\bf{2.108 Proposition}}

Let $\sum = \sum_{1} \times \sum_{2}$ be a composite C- or Q-system, then:
\begin{quote}
(a) The statistical state spaces are convex sets whose extreme points are the pure states.

(b) The Partial Traces are affine and onto

(c) If $\rho_{1} \in \mathbb{V} (S_{1})$ and $\rho_{2} \in \mathbb{V} (S_{2})$ then there exists a {\em{unique}} $\rho \in \mathbb{V} (S_{1} \otimes S_{2})$, denoted $\rho_{1} \otimes \rho_{2}$ such that:
\begin{quote}
$\rho_{1} \otimes \rho_{2} (R \otimes T) = \rho_{1} (R) \rho_{2} (T)$, $\forall R \in D_{{S}_{1}}$,  $T \in D_{{S}_{2}}$
\end{quote}
where `$R \otimes T$' denotes the Cartesian product and tensor product of superposition sets for C- and Q-systems, respectively.
\end{quote}
{\bf{Proof}}

(a) is treated in Remarks 2.67 (see also sections 2.4 and 2.5).

(b) is readily demonstrated from the definitions of Partial Trace above.

(c) for C-systems is a well-known measure-theoretic result. For Q-systems the result is almost trivial since it amounts to:
\begin{quote}
$\langle \phi, \rho_{1} \otimes \rho_{2} \phi \rangle = \langle \phi, \rho \phi \rangle$, $\forall \phi \in H \Leftrightarrow \rho = \rho_{1} \otimes \rho_{2}$
\end{quote}
which is evidently true.

The following Proposition summarises the relation between uniqueness of the composite state and purity of the states involved:

{\bf{2.109 Proposition}}

Let $\sum = \sum_{1} \times \sum_{2}$ be a composite C- or Q-system. Let $\rho_{1} \in \mathbb{V} (S_{1})$ and $\rho_{2} \in \mathbb{V}(S_{2})$.
\begin{quote}
(a) The following are equivalent:
\begin{quote}
(i) There exists a unique $\rho \in \mathbb{V}$ such that $PT_{i}(\rho) = \rho_{i}$, $i = 1, 2$

(ii) $\rho_{1}$ or $\rho_{2}$ is pure.
\end{quote}
(b) If $\rho_{1}$ and $\rho_{2}$ are both pure then there exists a unique $\rho \in \mathbb{V} (S)$ such that $PT_{i} (\rho) = \rho_{i}$ and $\rho$ is pure.
\end{quote}
{\bf{Proof}}

(a) (i) $\Rightarrow$ (ii). Suppose false, then both $\rho_{1}$ and $\rho_{2}$ are non-extreme and we can find $0 < c < 1$ and statistical states $\mu_{i}$, $\lambda_{i}$ $i = 1, 2$ such that:

\begin{quote}
$\rho_{i} = c\mu_{i} + (1-c) \lambda_{i}$.
\end{quote}
Then not only does $PT_{i} (\rho_{1} \otimes \rho_{2}) = \rho_{i}$ but (by definition of $\otimes$ in Proposition 2.108) convexity allows also:
\begin{quote}
$PT_{i} (c\mu_{1} \otimes \mu_{2} + (1 - c) \lambda _{1} \otimes \lambda_{2}) = \rho_{i}$
\end{quote}
which contradicts the hypothesis.

(ii) $\Rightarrow$ (i). The proof for C-systems, which can probably be extended to Q-systems (although we shall use a different approach), employs the identity:
\begin{quote}
$R \times S_{2} = R \times (T \cup T^{\bot}) = (R \times T) \cup (R \times T^{\bot})$.
\end{quote}
We suppose that (ii) $\Rightarrow$ (i) is false. Hence there exists $\rho' \neq \rho_{1} \otimes \rho_{2}$ such that $PT_{i} (\rho') = \rho_{i}$. It follows that there exist Borel sets $R \in D_{{S}_{1}}$ and $T \in D_{{S}_{2}}$ such that
\begin{quote}
$\rho'(R \times T) \neq \rho_{1} \otimes \rho_{2} (R \times T)$.
\end{quote}
But if $\rho_{2}$, say, is pure then there exists $q \in S_{2}$ such that $\rho_{2} = \delta_{q}$, the Dirac measure at $q$. Whence:
\begin{quote}
$\rho_{1} \otimes \rho_{2} (R \times T) = \rho_{1} \otimes \rho_{2} (R \times S)$ =
\begin{quote}
- $\rho_{1} (R)$ if $q \in T$.

- 0 otherwise (i.e. if $q \in T^{\bot}$).
\end{quote}
\end{quote}
Hence, if $q \in T^{\bot}$ then $0 < \rho' (R \times T) \leq \rho' (S_{1} \times T) = \rho_{2} (T) = 0$; if $q \in T$ then using the identity above we obtain
\begin{quote}
$0 < \rho' (R \times T^{\bot}) \leq \rho'(S_{1} \times T^{\bot}) = \rho_{2}(T^{\bot}) = 0$.
\end{quote}
Either way there is a contradiction. (Note that the strict inequalities follow from the assumption $\rho' (R \times T) \neq \rho_{1} \otimes \rho_{2} (R \times T))$.

For Q-systems it is simplest to use Proposition 2.105: let $\rho$ be any state such that $PT_{i}(\rho) = \rho_{i}$, $i = 1,2$. Then $\rho$ can be written as:
\begin{quote}
$\rho = \sum_{n} c_{n} |\Phi_{n}\rangle \langle \Phi_{n}|$ 
\end{quote}
However, using Proposition 2.105 on each $|\Phi_{n}\rangle \langle \Phi_{n}|$ and letting $\rho_{2}$, say, be pure and equal to $|\alpha \rangle \langle \alpha |$, then there exist $\psi_{n} \in H_{1}$ such that $\Phi_{n} = \psi_{n} \otimes \alpha$ whence

\begin{quote}
$\rho = (\sum_{n} c_{n} |\psi_{n}\rangle \langle \psi_{n}|$) $\otimes$ $|\alpha \rangle \langle \alpha |$
\end{quote}
which determines $\rho$ uniquely.

(b) All we need to show here is that $\rho$ is pure. This follows from convexity of the state spaces and the fact that $PT_{i}$ is affine.

{\bf{2.110 Remarks}}

(1) We have proved that correlations are possible in both C- and Q-systems and, moreover, found necessary and sufficient conditions that the state of the composite system be uniquely determined by the states of the component subsystems.

(2) Part (b) shows that a complete description ($\equiv$ pure state) of the component subsystems entails a complete description of the composite system contrary to what seems to be claimed by some authors.

(3) We shall shortly consider how knowledge of the state of one of the component subsystems and the state of the composite system allows us to infer the state of the other component subsystem. The mathematics will be trivial, but the claim of knowledge of the state of the component subsystem will be seen to be the source of all the confusion surrounding the EPR `paradox' and the `holistic' nature of quantum theory.

It remains to delineate the difference between C- and Q-mechanics in their treatment of subsystems. The two theories differ on the `heredity' of the completeness of a description of a system. Thus, whilst it is true for both theories that if the states of the component subsystems are pure then the state of the composite system is also pure, the converse implication fails, in general, for Quantum Mechanics. Precisely, we have

{\bf{2.111 Proposition}}

Let $\sum = \sum _{1} \otimes \sum_{2}$ be a composite C- or Q-system.

(i) If $\rho_{1}$ and $\rho_{2}$ are pure states of the component systems then the state $\rho_{1} \otimes \rho_{2}$ of the composite system is:

(a) Pure

(b) The only state, $\rho$, such that $PT_{i} (\rho) = \rho_{i}$, $i = 1, 2$

(ii) If $\rho$ is any pure state of the composite system, then $PT_{i}(\rho)$ will always be pure states only if $\sum$ is a C-system.

{\bf{Proof}}

(i): follows from Proposition 2.109.

(ii): that it is true for C-systems follows from consideration of Dirac measures. To find a counterexample for Quantum Mechanics we need only choose  $\rho = |\Psi \rangle \langle \Psi |$ with $\Psi = \psi_{1} \otimes \phi_{1} + \psi_{2} \otimes \phi_{2}$ and apply Proposition 2.105.

\newpage

\section{Geometry and Mechanics}
Having set up the theories of mechanics as abstract state geometries, we now turn to the incorporation of space-time geometry into these theories. The results are well-known so this Section will be merely a brief review for the sake of completeness.

We start with space-time itself. Space and time we view as parameters used in an individual's description of the world. As show by Levy-Leblond (LL 1) the structure of space-time is determined up to a constant by the following three hypotheses:
\begin{enumerate}
\item Space and time are homogeneous in any reference frame.
\item Space and time are isotropic in any reference frame.
\item Reference frames are related by a group structure.
\end{enumerate}
Broadly speaking, the first two express the assumption of a Euclidean reference frame by any observer, whilst the last requires that observers can talk consistently to one another. If causality is also demanded, the case where the constant is negative is excluded. This leaves only two types of structure - space-time either supports Galilean transformations (when the constant is zero) or Lorentz transformations (when it is positive). The constant can, if we wish, be identified as the (reciprocal of the) speed of light.

That we can make such hypotheses - is space-time real? - is allowed provided that we adopt a {\em{conventionalist}} view of geometry. For a discussion of special and General Relativity in these terms see, for example, (Ro 1).

We have thus arrived at a relativity group acting on space-time. Our aim is to express this space-time structure in the theory of mechanics. The first step is to look for representations of the relativity group in the state spaces, in particular, to find {\em{elementary systems}} (Definitions 2.86 and 2.97) in classical and quantum mechanics. This programme has already been carried out - for a review of the Galilean case from a `geometric quantisation' standpoint see Bez (Be 1).

Suppose then that we have a state space $S$ and an irreducible representation $V$ of a relativity group $G$ defined in space-time $X$. So what? Well, we can use this information to determine properties with a space-time interpretation, namely, the configuration and momentum kinematic properties associated with the state space. To see what this entails consider two space-time frames linked by a relativity group transformation. Adopting the `passive' view of (objective) space-time and states, let $\Delta$ be a portion of space-time as viewed from the first frame, with $g(\Delta)$ the same portion of space-time but as viewed from the second frame. Similarly, the description $s$ of the state in the first frame is, in the second frame, given by $V(g)s$. For a property, $P$, to be a {\em{configuration kinematic property}} we require that it be defined on space-time, $X$, and give rise to expected values independent of frame; that is, we require covariance:

$P_{g(\Delta)}$$(V(g) s) = P_{\Delta}(s)$, 

or, equivalently:

\hspace{27pt} $P_{g(\Delta)}(s)$ =  $P_{\Delta}$ $(V(g^{-1}) s)$.

This defines what is known in group theory as a {\em{system of imprimitivity}}. For transitive group actions, such as we have here, the quantum (Hilbert space) systems of imprimitivity are fully characterised - see, for example, Chapter IX of (Va 2).

The group structure allows us to go further and identify {\em{momentum kinematic properties}} as the generators of one-parameter subgroups (symmetries).

At this stage we abandon our development to merely summarise the key points from a very extensive literature on the subject:
\begin{enumerate}
\item An elementary system has all the features of a free particle, in particular, a `rest-mass' parameter.
\item Configuration kinematic properties are, in the Galilean quantum case, the familiar position operators and time parameter. Embarrassingly, position does not appear so conveniently in the Lorentz case (see, for example, (Va 2) p. 236).
\item Momentum kinematic properties in the quantum case are familiar operators such as:
\begin{itemize}
\item linear momentum (generating space translations)
\item angular momentum (generating space rotations)
\item free Hamiltonian (generating time translations).
\end{itemize}
It should be noted that though these properties may be, and are, used to describe particles evolving under general flows, their significance rests with the free (`straight-line') particle. Under the conventionalist view of geometry alluded to above, dynamics can be thought of as a theory of deviations from straight-line (free particle) motion.
\item Projective (ray) representations are involved in both classical and quantum mechanics (cf. the discussions after Definition 2.85 and 2.96 above). As Levy-Leblond demonstrated (LL 2) this leads to intrinsic spin appearing for classical and quantum elementary systems.
\item By imposing a limited Galilean `covariance' condition on particles undergoing general flows, the one-particle Hamiltonian, $h$, is constrained to be of the form:
\begin{quote}
$h = \frac{1}{2m} (p - A)^{2} + V$
\end{quote}
where $A$ and $V$ commute with position. See, for example, (Va 1) p.206. However, the status of this limited covariance condition is unclear and does not seem to be applicable to the much more demanding Lorentz case.
\end{enumerate}

\chapter{Approximation and Localisation} \markboth{Chapter 3}{Chapter 3: Approximation and Localisation}
The previous two Chapters laid out fundamental principles for the theories of classical and quantum mechanics. We now turn to more everyday concerns of the physicist and chemist - the use of approximations. With the fundamental models usually too intractable to provide a working basis for applications, useful results are mostly obtained through various levels of idealisation. For example, Griffith ((Gr 1) Section 5.6), in discussing the Hamiltonian for atoms in an external magnetic field, makes the following assumptions (pp 128-130):
\begin{enumerate}
\item ``We now discuss an atom in a constant external magnetic field...''
\item ``We neglect small effects, such as nuclear hyperfine structure...''
\item ``Neglecting th?se latter...'' (i.e. magnetic interactions between the orbital and spin magnetic moments of pairs of electrons)
\item ``The last term of (5.50) is quite negligible compared with the other terms depending on H, the ratio between them being about 2.5 $\times$ $10^{-5} n^{-2}$ for an electron in an $n\ell$ orbital of hydrogen.''
\item ``The second [diamagnetic] term is very small and for atoms not in $S$ states is quite negligible compared with the paramagnetic part.''
\item ``In weak fields we regard $H_{1}$ as a perturbation small (energies of the order of 1 $cm^{-1}$) compared with the separation between levels of a term...''
\item ``In deriving (5.56) we have neglected the matrix elements of $L_{Z}$ between states of different $J$. In other words we have supposed those matrix elements small compared with the multiplet splitting between levels. This condition is satisfied in practice for most atoms even for macroscopically strong magnetic fields.''
\end{enumerate}
It is little wonder that mathematicians, faced with such a sequence of unproved assertions, prefer to ruminate on more fundamental matters! A common thread does, however, run through the successive idealisations of the Hamiltonian made by Griffith. It is the principle that for the states of interest to the physicist certain parts of the full Hamiltonian are, in an unspecified sense, negligible. Moreover, these states are somehow related to the low-energy localised states of a corresponding `unperturbed' system. For instance, though few physicists would quibble with the idealisation of a `constant' external magnetic field, this confidence is not based on any mathematical proof but rather on a `physical' view that the spatial localisation of wave functions around an atom is several orders of magnitude lower than fluctuations in the external magnetic field.

The key question motivating this Chapter is: Can we justify the `physical' view? This will lead us to a new approach to analysing the idealisations in quantum mechanics which are based on classical analogues.

\newpage

\section{The Physical Perspective}
{\bf{(1) The Divergence of Mathematics from Physics}}

It is often said that physics is becoming more mathematical. Certainly theories are nowadays couched in more abstract language and numerical methods play an increasing role both in analysing experimental results and in pursuing specific consequences of a formalism. Yet despite this apparent communality of purpose we shall argue that mathematics and physics are uneasy bedfellows, paying each other lipservice as they pursue their separate ends.

A pervasive feature of mathematical physics is {\em{modelling}} - the formulation of a `physical' theory, problem or circumstance in terms of a well-defined symbolic (contextual) structure, its model. Whether it be a fundamental investigation, such as that into the existence of quantum field theories, or a specific problem,  say the spatial decay of an eigenfunction, the method is a four-fold process:
\begin{enumerate}
\item Select the physics of interest.
\item Abstract the physics into a model.
\item Derive results within the model.
\item Apply these results back to the physics.
\end{enumerate}
But what of the unity, the hypothetico-deductive umbrella, required of a scientific theory in Chapter 1? To conform to a grand scheme, particular models should evidently be special cases of some fundamental model. This is met in practice by the choice of mathematical structure in which the model sits. However, indiscriminate application of a conventionally accepted mathematics may ignore the full conditions of the `physics of interest' so that the model reflects not the problem in hand but rather some other, mathematically more convenient, problem.

Consider, for example, the electronic spectrum of a hydrogen atom in an external magnetic or electric field. The popular model for this physics, to be found in any introductory text on quantum mechanics, is the finite-dimensional spectral theory of the Hamiltonian operator for an electron in a Coulomb potential in a constant magnetic or electric field. Harmless enough, perhaps, until we reflect that a different problem has been modelled, namely the properties of a charged particle in a Coulomb potential in a constant field over all space and time. It is unclear why this should be relevant to the physics of interest, especially when more sophisticated spectral theory for the electric field embarrassingly reveals a continuous rather than a discrete spectrum. The model using simple spectral theory `works', but we're not sure why.

From the example it seems reasonable to propose that a model should either reflect the conditions of the physics or demonstrably `coincide' with the fundamental model for the physics of interest. Either way, we would expect physics to look to mathematics for expression of its conditions.

For our example, the implicit conditions of physics include:
\begin{quote}
(a) The electromagnetic field is of sufficiently large wavelength to be taken as spatially constant over the electronic states of the atom (`electric dipole transitions').

(b1) Electronic states (wave functions) are sufficiently localised for variations in the external electric or magnetic field to be ignored.

(b2) The external electric or magnetic field is of small magnitude relative to the Coulomb potential around the nucleus.
\end{quote}

Whereas one might have expected conditions (a) and (b) to find expression within one mathematical structure, what happens in practice is rather different. These conditions - as `approximations' - generate reformulations of the problem in different mathematical terms.

In our example the rationale appears to be along the lines:
\begin{enumerate}
\item Condition (a) contributes to the demonstration (see, e.g. (Gr 1) p. 49) that provided the timescale of the interaction is short compared to the `natural lifetime' (whatever this is!) of the ground state, then the electromagnetic field induces transitions between states with maximum probability when these states are (certain) eigenfunctions of the original Hamiltonian.

{\bf{Conclusion}}: To analyse the electromagnetic spectrum of an atom or molecule, use the mathematical spectrum of the relevant unperturbed Hamiltonian.

\item Condition (b1) facilitates modification of the Hamiltonian by a simple extra term for which the external field is a constant. The spectrum of this modified Hamiltonian can, using condition (b2), be analysed by applying the perturbation theory of operators in a finite-dimensional vector space.

{\bf{Conclusion}}: The relevant mathematical structure is a finite-dimensional vector space based on low-energy eigenstates of the original Hamiltonian.
\end{enumerate}
Proofs to support this kind of reasoning are notable by their absence. Physicists, to whom such assumptions are many and frequent, treat mathematics as their tool not master and dismiss the use of different models with that sleight of hand known as `physical' reasoning. Nor do mathematicians have much to offer, pursuing consequences within a model rather than derivations of one model from another, in essence because mathematicians view empirical results as numerical values not physical magnitudes.

\vspace{8pt}
{\bf{(2) Physical Approximation and the Use of Limits}}

A scientific theory eventually makes contact with the empirical world through measurement of events. The central feature of this contact is the ``acceptable error'' within which theory explains the facts. Thus the confirmation of a theory does not rest on a coincidence of real numbers but on agreement of predictions within ranges of error (or intervals of imprecision). It seems reasonable, therefore, to propose that for a given acceptable error two theories ``agree'' provided their predictions are within this error. In the terminology of Chapter 1 such theories are {\em{weakly equivalent}}.

In more detail, let $Pred(T)$$|_{D}$ denote the predicted magnitude of a physical event, $D$, according to a theory $T$, then:

{\bf{3.1 Definition (Criterion for Physical Approximation)}}:

Two theories $T_1$ and $T_2$ will be said to be weakly equivalent for the event $D$ subject to an acceptable error $\epsilon$ if and only if:
\begin{quote} 
$|Pred(T_{1})$$|_{D}$ - $Pred(T_{2})$$|_{D}$ $|$ $<$ $\epsilon$.
\end{quote}

This is a pointwise or ``eventwise'' approximation of one theory or model by another, but can be readily extended to a set of events or circumstances by requiring uniform equivalence (over the set with respect to $\epsilon$).

With this background, what techniques do mathematicians bring to bear on approximations? The typical mathematical approach employs the notion of a {\em{limit}}. This is a powerful but demanding requirement whereby a family of objects can satisfy, in an ordered way, any request for closeness. As a simple example, a one-parameter family $\{p_{\lambda}\}$ of points in a metric space converges to a point, $p$, in the space provided:
\begin{quote}
For each $\epsilon > 0$, $\exists \lambda (\epsilon)$ such that:
\begin{quote}
$d(p_{\lambda}, p) < \epsilon$, $\forall \lambda \leq \lambda (\epsilon)$.
\end{quote}
\end{quote}
With the usual metric topology of the real numbers this leads to the disturbing:

{\bf{3.2 Lemma}} (formal):
A limit is neither necessary nor sufficient to satisfy our criterion for physical approximation (3.1).

\vspace{7pt}
{\bf{Proof}}:

Non-necessity: assume two predictions satisfy the criterion (3.1 above), then clearly we do not need to require $\epsilon \rightarrow 0$ as some $\lambda \rightarrow 0$.

Non-sufficiency: suppose a limit exists for some parameterisation, $\lambda$. Ask for coincidence within $\epsilon$, then although we know a suitable $\lambda (\epsilon)$ exists we do not know {\em{which one}}, that is, we don't know the actual events or conditions or circumstances under which the physical approximation holds.

Thus, approximation by a limit is of little use unless something is known about the rate of convergence. In particular, we need to know how the parameter, $\lambda$, determines the error, $\epsilon$; that is, how the error, $\epsilon$, depends on the parameter, $\lambda$. For example, if $p_{\lambda} = p + \lambda$ (positive numbers) then the condition:
\begin{quote}
$d(p_{\lambda}, p) \leq \lambda$
\end{quote}
tells us how to choose the parameter $\lambda$ in order to be within an acceptable error $\epsilon$:
\begin{quote}
$\lambda < \epsilon \Rightarrow d(p_{\lambda}, p) < \epsilon$.
\end{quote}

To support a physical approximation, therefore, the abstract existence of a limit (soft analysis) needs to be augmented by a concrete estimate of the convergence (hard analysis).

The state of affairs in practice is typically even worse. Not only have very few applicable hard estimates been proved to date but the abstract limit itself may not exist. Resort is then made to {\em{asymptotic approximation}}, where a function, $f(\lambda$) say, is said to be asymptotically approximated by an asymptotic expansion $\sum a_{n} \phi_{n} (\lambda)$, (where $\{\phi_{n}(\lambda)\}$ is an asymptotic series, e.g. $\{\lambda^{n}\}$) if, for each $N$:
\begin{quote}
For each $\epsilon > 0$, $\exists \lambda(\epsilon)$ such that:
\begin{quote}
$| f(\lambda) - \sum_{n=1}^{N} a_{n}\phi_{n}(\lambda)|$ $< \epsilon$ $| \phi_{N}(\lambda)|$ \hspace{2pt} $\forall$ $|\lambda |$ $\leq \lambda(\epsilon)$.
\end{quote}
\end{quote}

Asymptotic expansions provide a popular method of analysing physical problems, yet the definition is so weak that it does not help at all in meeting the criterion of a physical approximation. To quote Reed $\&$ Simon (R $\&$ S XII p. 26):
\begin{quote}
``Saying that $f$ has a certain asymptotic series gives us no information about the value of $f(z)$ for some fixed nonzero value of $z$. We know that $f(z)$ is well approximated by $a_{0}$ + $a_{1}z$ as $z$ gets ``small'' but the definition says nothing about how small is ``small''.''
\end{quote}
After considering an example, Reed $\&$ Simon conclude:
\begin{quote}
``Thus, we see the typical behaviour of wandering near the right answer for a while (and not even that near!) and then going wild.''
\end{quote}
In the present author's opinion the use of asymptotic approximation is a contrivance with no basis in the physics. Moreover, the need for asymptotics reveals that an inadequate mathematical model is under analysis. For example, the Stark effect requires asymptotic approximation because of two unjustifiable features of the mathematical model - the spatial behaviour of the electric field at infinity and the consideration of an infinite time problem. We therefore put forward:

{\bf{3.3 Conjecture}}
\begin{quote}

Asymptotic approximations occur whenever the full conditions (contingencies) of the physical problem have not been taken into account.
\end{quote}
Reflection on the nature of mathematical limits reveals a deeper malaise in mathematical models. Whereas physics is concerned with magnitudes, mathematics - including the call-and-response in limits - deals with numerical values. The result is that models and approximations contain no internal representation of physical magnitudes and take the same form whether representing, say, high or low energies, macroscopic or atomic distances. This feature is, of course, an advantage for all-embracing fundamental models, but shows up as a major deficiency in analysing physical systems whose behaviour varies according to order of magnitude.

What can be done? Our approach will be to build physical conditions and weak equivalence into the analysis according to the following two principles:
\begin{quote}
{\bf{(a) Build magnitudes into models}} - Represent physical magnitudes within the model. Here we require more than just the basic Galilean parameters of mass and scalar/vector potentials. Although these may suffice for eigenvalue problems they cannot handle initial conditions or durations which reflect spatial and temporal orders of magnitude. One way to do this is to base a model on bounded {\em{ranges}} of magnitudes, representing the ranges of experimental conditions.

{\bf{(b) Analyse approximations as comparisons of models}} - view approximations as possible alternative descriptions, with respect to acceptable errors, for a range of physical conditions. The success of an approximation may then be evaluated by call-and-response using physical magnitudes. Typically here one would consider the difference between two predictions of some relevant property (e.g. energy level), with an estimate for this difference as a function of physical parameters.
\end{quote}

\vspace{8pt}

\newpage

{\bf{(3) Localisation}}

A notable feature of the everyday world is the localisation of objects in space and time. Indeed, so fundamental is the idea of localisation that the mathematical language of classical mechanics - differential geometry - may be developed from the notion of a point executing a trajectory.

By contrast, the states (wave functions) of Quantum Mechanics are delocalised. Even free-particle states of compact support immediately become delocalised as is demonstrated in Proposition 3M.1.

As a matter of practical fact, however, experiments are conducted in a localised environment with the condition of the rest of the world irrelevant. For example, in a molecular beam experiment molecules are fired through an electric or magnetic field whose value outside the cylinder of the `classical' trajectory does not affect the outcome.

Experimental necessity is thus an embarrassment to the fundamental model of Quantum Mechanics. With knowledge of potentials over the full range of delocalisation being unachievable it is essential that the theory accommodates localised behaviour. We therefore require for Quantum Mechanics that:
\begin{quote}
(a) Localisation be well-defined within the theory.

(b) The theory can demonstrate that behaviour of a localised particle is independent of reasonable potential fluctuations outside a macroscopic region over which an experimenter has control or knowledge. By `independent' here we mean a weak equivalence relative to some acceptable error.
\end{quote}

\vspace{8pt}

{\bf{(4) Compact Sets and Phase Space Localisation}}

What is localisation in Quantum Mechanics? Some likely requirements are:
\begin{itemize}
\item localisation is a possible attribute of a set of states;
\item any one state (and hence any finite collection of states) is localised to some degree;
\item finite time evolution preserves localisation;
\item localised states are bounded in position.
\end{itemize}
An obvious criterion for localisation is compact support in position space yet, by Proposition 3M.1, this is too strict to be useful. Loosely, localisation could be said to be a finiteness in position which, by the requirement that it be preserved under evolution, is also a finiteness in momentum.

Mathematically, notions of `finiteness' and `boundedness' come together in the definition of compactness. It is, perhaps, no surprise to find that compactness is essentially a position-momentum localisation. This is demonstrated in Proposition 3M.3 where it is shown that a collection of wave functions is compact if and only if the wave functions get uniformly small (tending to zero) as position and momentum get large (tending to infinity). In that theorem the `largeness' of position and momentum is governed by two functions, $F(Q)$ and $G(P)$ respectively, which are strictly positive tending to infinity as position and momentum tend to infinity. Besides these conditions, the functions are very general indeed, the only other requirement being that they be measurable. For instance, they might both become infinite outside bounded intervals in position and momentum, although in this case no wave function (except the null wave function) meets the localisation requirement! Sobolev spaces can be viewed as special cases, $F$ becoming infinite outside a bounded spatial region with $G$ given by $P^{2}$.

In proposition 3M.4 the quadratic forms $F$ and $G$ are re-expressed in operator terms and in Proposition 3M.5 sufficient conditions provided in order to define the operator sum $F + G$. Whether these conditions are necessary we have not been able to determine. Nevertheless, Corollary 3M.6 provides an operator version of Proposition 3M.3.

For our main result, Theorem 3M.9, it is necessary to define carefully the inverse of an operator. This is done in Lemma 3M.7, and in Lemma 3M.8 it is shown how for {\em{Hilbert spaces}} compact operators take bounded sets onto compact, not just precompact, sets. Theorem 3M.9 itself brings together previous results to provide comprehensive criteria for compactness of a set in Hilbert space.

Returning from the mathematical development we see that compact sets satisfy all of the ``likely requirements'' given at the start of this sub-section. Indeed, compactness has many more qualifications to recommend it. Quoting from Sutherland (Su 2):
\begin{quote}
``(1) It allows us to pass from the local to the global...

(2) The second answer has been very well expressed by Hewitt (1960). Hewitt remarks that compactness is a substitute for finiteness, appropriate to the analysis of continuity. More explicitly, he points out that many statements about function $f: A \rightarrow B$ are:
\begin{quote}
(i) true and trivial if $A$ is a finite set,

(ii) true for continuous $f$ when $A$ is a compact space,

(iii) false, or very hard to prove, even for continuous $f$, when $A$ is non-compact.''
\end{quote}
\end{quote}
From our point of view, not only can compactness be viewed as the mathematical expression of localisation but in representing boundedness in general provides the relevant type of object in which to formulate physical theories, in accord with the first of our principles set out at the end of Section 3.1.2. above.

We now turn to an ``application'' - a reformulation of bound and scattering states in terms of compactness.

\newpage

\section{Topological Bound and Scattering States}
In this Section the theory of compact sets in Hilbert space is applied to develop a classically motivated ``phase space'' approach to bound and scattering states in quantum mechanics.

Most of the results are not new but their conceptual significance will, I hope, be demystified by the context in which they are presented. In particular, the following `topological' criteria for the spectral subspaces of a Hamiltonian are presented:
\begin{quote}
pure point = evolution contained in a compact region

continuous = zero average time spent in any compact region

absolutely continuous = finite transit time across any compact region.
\end{quote}

Questions of existence and completeness of wave operators are not attacked in this section as we only make passing reference to comparison dynamics. There are good reasons for this omission. Although `compactness' conditions are usually employed at some stage in the mathematical theory of comparison dynamics scattering, these conditions do not directly reflect the physical problem under consideration, which has position and momentum playing essentially different roles. Typically, (`potential scattering'), the comparison evolution is free evolution, generated by the Hamiltonian $\frac{P^{2}}{2m}$, and the evolution under analysis is generated by a Hamiltonian of the form $\frac{P^{2}}{2m} + V(Q)$, where the potential $V$ gets asymptotically small as its argument - distance from the scattering centre - gets large. We need not conclude, however, that definitions using `phase-space' compactness are irrelevant. Far from it, as shown by our results relating to the spectral subspaces of the Hamiltonian. Indeed, we might hazard the view that some of the fundamental mathematical problems in scattering theory arise from reconciling the asymmetric (in phase-space) problem with the symmetric definitions. That phase-space ideas are useful in scattering theory has been amply demonstrated by Enss' work (see, e.g. (RS 3) X1.17), although his methods bear little relation to those in this Section. For a recent review see also reference (Pe 1).

\vspace{8pt}

{\bf{1. Classical Ideas}}

We suppose that we are dealing with a classical evolution $U_{t}$ in a phase space of states $S$. Our starting point is to develop local definitions of {\em{bound states}}, {\em{transit time}} and {\em{average stay}}.

Let $\Omega$ be a compact region of phase space, and $\Delta T$ a compact interval of time, then:

(a) The {\em{bound states}} $B_{\Delta T}^{\Omega}$ of an evolution $U_{t}$ are those states which remain in $\Omega$ during $\Delta T$:
\begin{quote}
$B_{\Delta T}^{\Omega} \equiv \{\alpha \in S$ $|$ $U_{t}\alpha \in \Omega$ \hspace{3pt} $\forall t \in \Delta T\}$.
\end{quote}

(b) The {\em{transit time}} $\tau_{\Delta T}^{\Omega} (\alpha)$ of a state $\alpha$ across the region $\Omega$ in the interval $\Delta T$ is the time spent in $\alpha$ by $\Omega$:
\begin{quote}
$\tau_{\Delta T}^{\Omega} (\alpha) \equiv \int_{\Delta T}p^{\Omega}(\alpha, t) dt$
\end{quote}
where:
\begin{quote}
$p^{\Omega}(\alpha, t) = 1$ if $U_{t}\alpha \in \Omega$

\hspace{38pt} $= 0$ otherwise.
\end{quote}

(c) The {\em{average stay}} $\mu_{\Delta T}^{\Omega}(\alpha)$ of a state $\alpha$ in the region $\Omega$ for the interval $\Delta T$ is the mean of the transit time:
\begin{quote}
$\mu_{\Delta T}^{\Omega}(\alpha) \equiv \frac{1}{m(\Delta T)} \int_{\Delta T}p^{\Omega}(\alpha, t) dt = \frac{\tau_{\Delta T}^{\Omega} (\alpha)}{m(\Delta T)}$
\end{quote}
where $m(\Delta T)$ is the Lebesgue measure of $\Delta T$.

Although in reality we may strictly only talk about regions and time intervals under our control and, therefore, bounded, it seems to be the case that notions of bound and scattering states are independent of the region and time interval provided that they are big enough. In anticipation that the definitions we shall make will be non-trivial, let us try to extend the `finite theory' definitions to arbitrarily large regions and time intervals.

Considering bound states first, we note that continuity of the evolution implies:
\begin{quote}
$\Delta T$ compact $\Rightarrow \{U_{t}\alpha \}_{t \in \Delta T}$ contained in compact set in $S$.
\end{quote}
Accordingly, we introduce the future (+) and past (-) bound states for a region $\Omega$ as:
\begin{quote}
$B_{\pm}^{\Omega}$ $\equiv$ $\{\alpha \in S$ $|$ $U_{t}\alpha \in \Omega$ \hspace{3pt} $\forall t \in \mathbb{R}^{\pm}\}$.
\end{quote}
Extending to arbitrary regions of phase space we obtain:
\begin{quote}
$B_{\pm}$ $\equiv$ $\{\alpha \in S$ $|$ $\exists$ compact $\Omega$ with $\{U_{t}\alpha \}_{t \in \mathbb{R}^{\pm}} \subseteq \Omega \}$.
\end{quote}

Non-bound states cannot be termed `scattering' as they might return to a compact region $\Omega$, albeit intermittently. Thus we introduce the scattering states of $\Omega$, $S_{\pm}^{\Omega}$ as those states which leave $\Omega$ forever:
\begin{quote}
$S_{\pm}^{\Omega} \equiv \{\alpha \in S$ $|$ $\exists$ const $< \infty$ with $U_{t}\alpha \not\in \Omega$ \hspace{4pt} $\forall \pm t >$ const$\}$
\end{quote}
and extend to arbitrary regions by:
\begin{quote}
$S_{\pm} \equiv \{\alpha \in S$ $|$ for each compact $\Omega$ $\exists$ const $< \infty$ with $\Omega \cap \{U_{t} \alpha \}_{\pm t > const} = \O \}$.
\end{quote}

The states which are neither contained in a compact set, nor fully escape from all compact sets we shall call exceptional $E_{\pm}$:
\begin{quote}
$E_{\pm} \equiv B_{\pm}^{c} \cap S_{\pm}^{c}$.
\end{quote}

For the future and the past we have categorised states as bound, exceptional, or scattering. Equivalent, less abstract, definitions may be provided in terms of phase-space boundedness or transit times and average stay:

(a) Bound and scattering states in terms of phase-space boundedness.

Introduce the phase-space norm $||.||_{S}$ where, for $\alpha = (x, p)$, $|| \alpha ||_{S}^{2} = x^{2} + p^{2}$. It is shown in Proposition 3M.10 that the bound states are those which are uniformly norm-bounded, and the scattering states those whose phase-space distance from any fixed point, for example the origin, tends to infinity.

(b) Scattering states in terms of transit times.

We first extend the finite definition to the limits for future (+) and past (-) transit times of a state $\alpha$ across a region $\Omega$:
\begin{quote}
$\tau_{+}^{\Omega}(\alpha) \equiv \underset{T \rightarrow \infty}{lim} \int_{0}^{T} p^{\Omega}(\alpha, t) dt$; \\ $\tau_{-}^{\Omega}(\alpha) \equiv \underset{T \rightarrow \infty}{lim} \int_{-T}^{0} p^{\Omega}(\alpha, t) dt$
\end{quote}
where we allow the limit to infinity.

It is shown in Proposition 3M.11 that for reasonable evolutions the scattering states are precisely those with finite transit times across any compact region. By `reasonable' is meant, as in the proof, that the phase-space velocity, $\dot{\alpha_{t}}$, be bounded over any compact region in phase space. Equivalently, the phase-space gradient of the Hamiltonian needs to be bounded over any compact region. The actual condition used was that the Hamiltonian be infinitely differentiable on $S$ (a familiar requirement). It is possible that the transit time criterion for scattering states holds {\em{almost everywhere}} on phase space for a much wider class of evolutions.

(c) Average stay.

We extend the finite definition of average stay to the limits for future (+) and past (-) stays by:
\begin{quote}
$\mu_{+}(\alpha) \equiv \underset{T \rightarrow \infty}{lim} \frac{1}{T} \int_{0}^{T} p^{\Omega}(\alpha, t) dt$; \\ 

$\mu_{-}^{\Omega}(\alpha) \equiv \underset{T \rightarrow \infty}{lim} \frac{1}{T} \int_{-T}^{0} p^{\Omega}(\alpha, t) dt$
\end{quote}
where the limits exist.

By Birkhoff's theorem (see e.g. (Ha 1)) the limits exist almost everywhere. From our definitions we have that:
\begin{quote}
$\alpha \in S _{\pm} \Rightarrow \mu_{\pm}^{\Omega}(\alpha) = 0$ \hspace{3pt} $\forall$ compact $\Omega$.

$\alpha \in B _{\pm} \Rightarrow \mu_{\pm}^{\Omega}(\alpha) = 1$ for some compact $\Omega$.
\end{quote}
This leaves the exceptional states $E_{\pm}$. Von Neumann's ergodic theorem (See (RS 1) Section II.5) tells us that $ \mu_{\pm}^{\Omega}(\alpha)$ are invariant under $L^{2}$ functions (with support in $\Omega$), which leads us to {\em{conjecture}} that {\em{almost everywhere}} (i.e., except possibly on a set of Liouville measure zero):
\begin{quote}
$\alpha \in E _{\pm} \Rightarrow \mu_{\pm}^{\Omega}(\alpha) = 0$ \hspace{3pt} $\forall$ compact $\Omega$?
\end{quote}
Consideration of possible exceptional trajectories indicates that provided the phase space velocity is uniformly bounded on $S$, the only states not to have a zero average stay for a compact set are those which leave the compact set increasingly infrequently, going increasingly further away each time. It is to be hoped that for most evolutions the set of such trajectories is of Lebesgue measure zero.

\vspace{8pt}

{\bf{2. Classical No-Capture theorem}}

There is an easy result (Schwarzschild's theorem - see Proposition 3M.12) which says that capture by or escape from a compact set is impossible.

Precisely, we have that for invertible evolutions ($U_{t}^{-1} \equiv U_{-t}$) and for compact $\Omega$:
\begin{quote}
$B_{+}^{\Omega} = B_{-}^{\Omega}$ \hspace{4pt} a.e.
\end{quote}
That is, the two sets agree except possibly on a set of Liouville measure zero (a.e. $\equiv$ almost everywhere). Denoting the a.e. equivalence class of a set, $X$, say, by $\underline{X}$ we conclude that 
\begin{quote}
$\underline{B}_{+} = \underline{B}_{-} \equiv \underline{B}$.
\end{quote}

The proof of Proposition 3M.12 depends, once again, on the (a.e.) invariance of the Liouville measure under evolutions. This result, together with our remarks at the end of Section 2.1, indicate that a useful definition might be {\em{total average stay}}:
\begin{quote}
$\mu^{\Omega}(\alpha) = \underset{T \rightarrow \infty}{lim} \frac{1}{2T} \int_{-T}^{T} p^{\Omega}(\alpha, t) dt$.
\end{quote}

In Figure 3.1 the various results concerning our definitions are summarised.

\vspace{12pt}

{\bf{Figure 3.1 - Definitions of Bound, Scattering and Exceptional States in Classical Mechanics}}

\vspace{12pt}

\includegraphics[scale=0.45]{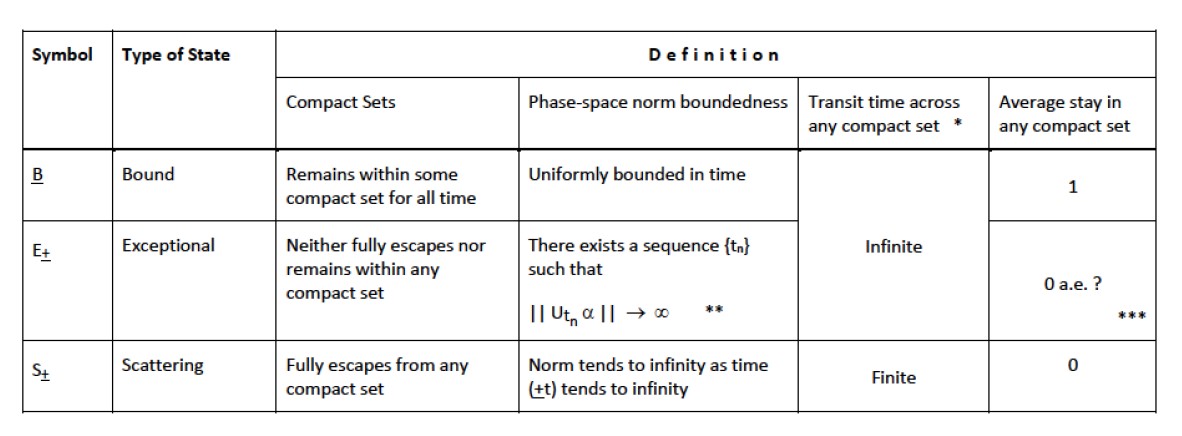}

* For evolutions with bound phase-space velocity over any compact region

** Follows from Proposition 3M.10

*** Conjecture

\vspace{24pt}

Figure 3.2, on the next page, illustrates, for different evolutions, examples of members from the sets $\underline{B}$, $E_{+}$ and $S_{+}$. It should be noted that our definition of scattering states includes those which disappear into a singularity - this is the price we pay for a uniform treatment in phase space. Such states need to be eliminated in normal scattering theory.

\newpage

\hspace{-2cm} \includegraphics[scale=0.33]{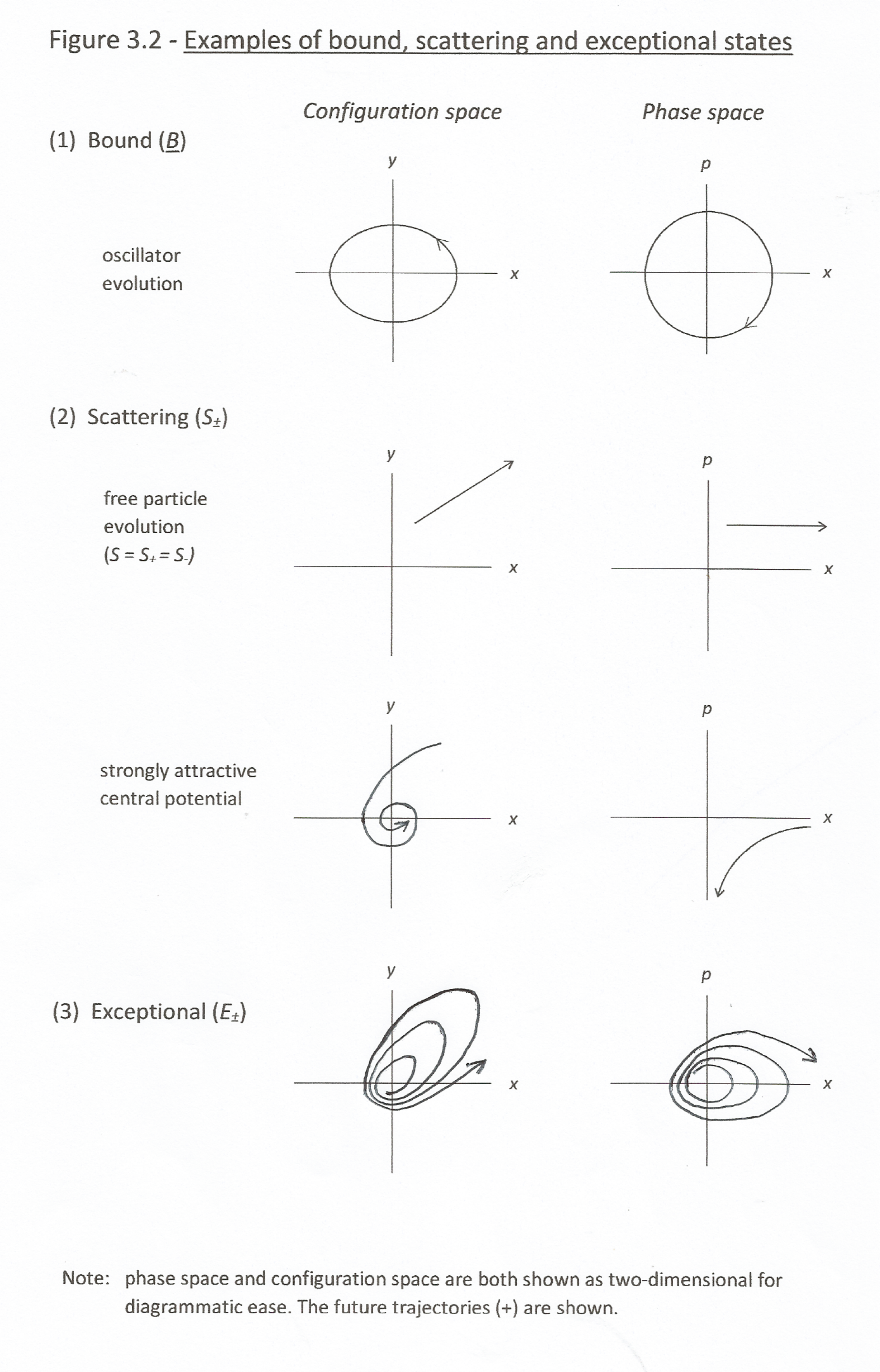}


\newpage

{\bf{3. General Questions}}

\vspace{8pt}

{\bf{(a) Geometric}}

(i) {\em{Exceptional States}} - for which evolutions does $E_{\pm} = \O$?

(ii) {\em{Geometric Asymptotic Completeness}} - for which evolutions are the past (incoming) and future (outgoing) scattering states the same, that is, when does $S_{+}(U) = S_{-}(U)$?

\vspace{8pt}

{\bf{(b) Comparison Dynamics}}

If $V_{t}$ is another, standard evolution the V-asymptotic states for the evolution $U_{t}$ are states $\alpha \in S$ such that there exists $\alpha' \in S$ with:
\begin{quote}
$U_{t}\alpha \thicksim V_{t}\alpha'$ as $t \rightarrow \pm \infty$.
\end{quote}
Specifically we define the V-asymptotic states of $U_{t}$ as those in the sets:
\begin{quote}
$S_{\pm}(U, V)$ $\equiv$ $\{\alpha \in S$ $|$ $\alpha = \underset{t \rightarrow \pm \infty}{lim} U_{t}^{-1}V_{t}\alpha'$ for some $\alpha' \in S\}$.
\end{quote}

(i) {\em{Existence of Wave operators}} - under what evolutions and for which states do wave operators $W_{\pm}$ exist, where:
\begin{quote}
$W_{\pm} (U,V) \alpha \equiv \underset{t \rightarrow \pm \infty}{lim} U_{t}^{-1}V_{t}\alpha$?
\end{quote}

(ii) {\em{Asymptotic Completeness for Comparison Dynamics}} - for which evolutions are the incoming and outgoing V-asymptotic states the same:
\begin{quote}
$S_{+} (U,V) = S_{-} (U,V)$?
\end{quote}

\vspace{8pt}

{\bf{(c) Asymptotic Completeness for Geometric and Comparison Dynamics}}

Combining (a) (ii) and (b) (ii): for which evolutions $U_{t}$ and standard evolutions $V_{t}$ does:
\begin{quote}
$S_{+} (U) = S_{-} (U) = S_{+} (U,V) = S_{-} (U,V)$?
\end{quote}

\vspace{8pt}

{\bf{(d) Relation to the Finite Theory}}

The theory involves limits for arbitrarily large quantities explicitly in time, implicitly in phase space. It is natural to ask how rapidly the limits are approached and, in particular are there `many' states for which the finite theory categories do not agree with the limiting categories, where $\Omega$, $\Delta T$, $\tau_{\Delta T}^{\Omega}$ and $\mu_{\Delta T}^{\Omega}$ are of some laboratory order of magnitude?

\vspace{8pt}

{\bf{(e) Remarks on the above questions}}

The geometric questions (a) will be dealt with in the quantum mechanical case by relating our geometric definitions to the spectral subspaces of the Hamiltonian.

The comparison dynamics questions (b) are a large field of study in themselves. Much is known in the quantum mechanical case where $V_{t}$ is the free evolution - see, for example, (AJS) or (RS 3).

There is a close relation between the geometric and comparison dynamics scattering states (c) by virtue of our answers to (a) and the textbook results for (b). In the classical case a geometric result is supplied (Proposition 3M.14) for fully scattering comparison evolutions. Specifically if $S_{\pm}(V) = S$ (i.e. all states are scattering states for $V_{t}$) then:
\begin{quote}
$S_{\pm}(U,V) \subseteq B(U)^{c}$
\end{quote}
that is, the V-asymptotic states of $U_{t}$ are not bound states. The proof is trivial.

Finally, the questions raised in (d) concerning the real-life applicability of the limiting theory deserve considerable attention. However, this is a difficult area and beyond the scope of this thesis. Nonetheless the related problem of localisation occupies much attention in the last Chapter.

\vspace{8pt}

{\bf{4. Quantum Mechanics - The Finite Theory}}

\vspace{8pt}

{\bf{(a) The use of compact sets in Hilbert Space}}

As shown in the first Section of this Chapter there is a close connection between compact sets in Hilbert space and position/momentum boundedness. It is well-known that there is no `phase space' projection operator so that compact operators provide the starting point for expressing phase-space boundedness in quantum mechanics. The choice of compact operator depends on the problem under consideration, each compact operator providing its own expression of boundedness just as would a choice of compact region in the classical case. To determine how the restrictions which determine a compact set, $\omega$, of vectors in Hilbert space $H$ should be formulated we note from Section 1 that the following are equivalent:
\begin{quote}
(i) $\omega \subseteq H$ is precompact

(ii) There exists a positive compact operator, $\Omega$ with $0 \leq \Omega \leq 1$ and a constant $R < \infty$ such that:
\begin{quote}
$\omega \subseteq$ range of $\Omega$ acting on the $R$-ball of $H$.
\end{quote}
(iii) With $\Omega$ and $R$ as in (ii), and $\Omega^{-1}$ defined as in Section 1:
\begin{quote}
$\omega \subseteq \{ \psi \in Ran(\Omega)$ $|$ $\langle \Omega ^{- \frac{1}{2}} \psi, \Omega ^{- \frac{1}{2}} \psi \rangle \leq R^{2} \}$ 

\hspace{9pt} $\equiv \{ \psi \in Ran(\Omega)$ $|$ $|| \Omega^{-1} \psi || \leq R \}$.
\end{quote}
Note that this last set is the R-ball of the Hilbert space \\ $H_{\Omega} \equiv (Ran(\Omega)$, $||.||_{\Omega})$, where  $|| \psi ||_{\Omega} \equiv  || \Omega ^{-1} \psi ||$, which is compact in the norm topology of $Ker({\Omega})^{\bot}$.
\end{quote}
An example of formulation (iii) is the choice $\Omega^{-1} = P^{2} + Q^{2}$ where $R$ can be loosely interpreted as a comparison oscillator energy bound.

In what follows we shall define the principal geometric quantities of bound state, transit time and average stay in terms of compact operators and compact sets in Hilbert space. The choice of Hilbert space vectors rather than rays (pure states) is both for analytic convenience and to tie up with existing results. The theory in terms of pure states will be presented in Section 3.2.6 below.

\vspace{8pt}

{\bf{(b) Bound vectors, transit time and average stay}}

We suppose throughout that an evolution $U_{t}$ is a one-parameter family of unitary operators in Hilbert space. At this stage we do not require $U_{t}$ to be a (strongly continuous) one-parameter group.

Let $\omega \subseteq H$ be a compact set, let $\Omega$ be a positive compact operator in $H$ and let $\Delta T$ be a compact interval of time then:

(i) The {\em{bound vectors}} $B_{\Delta T}^{\omega}$ of an evolution $U_{t}$ are those vectors which remain in $\omega$ during $\Delta T$:
\begin{quote}
$B_{\Delta T}^{\omega} \equiv \{ \psi \in H$ $|$ $U_{t}\psi \in \omega$ \hspace{3pt} $\forall t \in \Delta T \}.$
\end{quote}
For the case where $\omega = Ran_{R}(\Omega)$ where $Ran_{R}(\Omega)$ is the range of $\Omega$ acting on the R-ball in $H$ we may equivalently define:
\begin{quote}
$B_{\Delta T}^{\Omega_{R}} \equiv \{ \psi \in H$ $|$ $U_{t}\psi \in Ran(\Omega)$ $\&$ $|| \Omega^{-1} U_{t} \psi || \leq R$ \hspace{3pt} $\forall t \in \Delta T \}.$
\end{quote}
From now on, the condition that $U_{t}\psi \in Ran(\Omega)$ will be assumed since we may always define $|| \Omega^{-1} \phi ||$ as a limit (which may be infinity) by the spectral theorem for positive operators. Note that if $Ker({\Omega}) \not= 0$ then $\Omega^{-1}\phi \in Ker(\Omega)^{\bot}$. 

(ii) {\bf{Transit time}}

From Chapter 2 we have that if $R$ is a superposition set and $s$ a pure state then the probability of $s$ being ``in'' $R$ is given by $p_{s}(R)$. In Hilbert space terms this becomes: if $P_{R}$ is the orthogonal projection onto a closed linear manifold $R \subseteq H$ then if $\psi \in H$ with $|| \psi ||$ $= 1$, the transition probability of $\psi$ with respect to $R$ is:
\begin{quote}
$p_{s}(R) = ||P_{R}\psi ||^{2} = \langle \psi, P_{R} \psi \rangle$
\end{quote}
The classical function $p^{\Omega}(\alpha, t)$ of Section 3.2.1 can be interpreted as the probability that $U_{t}\alpha$ is in $\Omega$ (it takes the values 1 or 0). Accordingly we define the {\em{transit time}} $\tau_{\Delta T}^{R}(\psi)$ of a vector $\psi$ across the closed linear manifold $R$ in the interval $\Delta T$ as:
\begin{quote}
$\tau_{\Delta T}^{R}(\psi) \equiv \int_{\Delta T} \langle U_{t} \psi, P_{R} U_{t} \psi \rangle dt$
\end{quote}
Although there is no satisfactory way to talk of a transit time across a compact set in Hilbert space (since the probabilistic interpretation requires closed linear manifolds), consider an operator which can be written as a sum of mutually orthogonal projections:
\begin{quote}
$\Omega = \sum_{n} \lambda_{n} P_{R_{n}}$ where $\lambda_{n} > 0$
\end{quote}
then the {\em{transit time}} $\tau_{\Delta T}^{\Omega}(\psi)$ of a vector $\psi$ with respect to operator $\Omega$ in the interval $\Delta T$ can be defined as the sum of transit times:
\begin{quote}
$\tau_{\Delta T}^{\Omega}(\psi) \equiv \sum_{n} \lambda_{n} \tau_{\Delta T}^{R_{n}}(\psi)$

\hspace{35pt} $= \int_{\Delta T} \langle U_{t} \psi, \Omega U_{t} \psi \rangle dt$
\end{quote}
where the $\{ \lambda_{n} \}$ are bounded. By the spectral theorem the transit time can thus be defined with respect to any positive bounded operator. For example, we could choose the coherent state POV measure $A(\Omega)$ of (Da 1) Theorem 5.2 for a compact region, $\Omega$, in coherent state phase-space.

{\bf{ (iii) Average Stay}}

By the arguments used in (b) we define the {\em{average stay}} $\mu_{\Delta T}^{R}(\psi)$ of a vector $\psi$ in the closed linear manifold during $\Delta T$ as:
\begin{quote}
$\mu_{\Delta T}^{R}(\psi) \equiv \frac{1}{m(\Delta T)} \int_{\Delta T} \langle U_{t} \psi, P_{R} U_{t} \psi \rangle dt$
\end{quote}
and the {\em{average stay}} $\mu_{\Delta T}^{\Omega}(\psi)$ of a vector $\psi$ with respect to a positive bounded operator $\Omega$ during $\Delta T$ as:
\begin{quote}
$\mu_{\Delta T}^{\Omega}(\psi) \equiv  \frac{1}{m(\Delta T)} \int_{\Delta T} \langle U_{t} \psi, \Omega U_{t} \psi \rangle dt$
\end{quote}
where $m(\Delta T)$ is the Lebesgue measure of $\Delta T$.

\vspace{8pt}
\newpage
{\bf{5. Quantum Mechanics - Bound Vectors}}

\vspace{8pt}

{\bf{(a) Basic Definitions and Properties}}

As in Section 3.2.1 we first extend the finite theory to arbitrarily large times and introduce future (+) and past (-) bound states for a compact set $\omega$ as:
\begin{quote}
$B_{\pm}^{\omega} \equiv \{ \psi \in H$ $|$ $U_{t}\psi \in \omega$ \hspace{3pt} $\forall t \in \mathbb{R}^{\pm} \}$.
\end{quote}
Extending next to arbitrarily large compact sets:
\begin{quote}
$B_{\pm} \equiv \{ \psi \in H$ $|$ $\exists$ compact $\omega$ with $\{ U_{t} \psi \}_{t \in \mathbb{R}^{\pm}} \subseteq \omega \}$.
\end{quote}
It is shown in Proposition 3M.15 that $B_{\pm}$ can be alternatively defined as:
\begin{quote}
$B_{\pm} \equiv \{ \psi \in H$ $|$ $\exists$ positive compact operator $\Omega$ with $|| \Omega^{-1} U_{t} \psi || \leq 1$ $\forall t \in R^{\pm} \}$.
\end{quote}
Notice that in this definition we could equally well require $0 \leq \Omega \leq 1$ with $\{ \Omega^{-1} U_{t} \psi \}$ uniformly bounded.

Thus, as in the classical case:
\begin{quote}
$B_{\pm} = \underset{\omega \hspace{1.5pt} compact}{\bigcup}$ \hspace{2pt} $B_{\pm} = \underset{\Omega \hspace{1.5pt} compact}{\bigcup} B_{\pm}^{\Omega_{R}}$

\hspace{97pt} $R < \infty$
\end{quote}

Proposition 3M.16 demonstrates that $B_{\pm}$ are linear manifolds. We shall see later (3M.17 and 3M18) that $B_{+} = B_{-}$ and is, in fact, closed.

\vspace{8pt}

{\bf{(b) Remarks on Other `Geometric' Definitions}}

The popular `geometric' definition of bound states (see e.g. (AJS) p.262) utilises a family $\{ F_{r} \}$ of projections satisfying $s - \underset{r \rightarrow \infty}{lim} F_{r} = 1$. The bound states are defined as:
\begin{quote}
$M_{0}^{\{F_{r} \}} = \{ \psi \in H$ $|$ $\underset{ r \rightarrow \infty}{lim}$ $\underset{t \in \mathbb{R}}{sup}$ $|| (1 - F_{r}) U_{t} \psi || = 0 \}$.
\end{quote}
The choice for $F_{r}$ is geometric as it is taken to be the projection associated with the r-ball in position space. This definition suffers, however, from three drawbacks:
\begin{quote}
(i) There is no `finite' version.

(ii) The set of bound states depends on the choice of the family $\{ F_{r} \}$.

(iii) When the position space projections are used it is necessary to require some `compactness' condition in order to relate this geometric definition to the usual spectral definition as the pure point subspace. In particular the condition is of the form that $F_{r}(h + i)^{-n}$ be compact where $h$ is the Hamiltonian. Such compactness conditions, which occur throughout the `position space' geometric theory, can be understood as follows. The Hamiltonian is usually of the form $P^{2} + V$ with the potential $V$ at our disposal in applying conditions. Let $B$ be a positive bounded operator of the form $(A(Q) + 1)^{-1}$ where $A(x) \rightarrow \infty$ as $| x | \rightarrow \infty$. Suppose that the Hamiltonian is semibounded with, say, $-1 \in \rho (h)$. Formally we arrive at:
\begin{quote}
$B(h + 1)^{-1} = (A + 1)^{-1} (h + 1)^{-1} = (F(Q) + P^{2}A(Q) + P^{2})^{-1}$

\hspace{51.5pt} $ = (A + 1)^{-1} (G + 1)^{-\frac{1}{2}} . (G + 1)^{\frac{1}{2}}(h +1)^{-\frac{1}{2}} . (h +1)^{-\frac{1}{2}}$
\end{quote}
where $G \equiv G(P) = P^{2}$. Now $A(x) \rightarrow \infty$ as $x \rightarrow \infty$ so $B(x) \rightarrow 0$ as $x \rightarrow \infty$; similarly $G(k) \rightarrow \infty$ as $k \rightarrow \infty$ so that $(G(k) + 1)^{-\frac{1}{2}} \rightarrow 0$ as $k \rightarrow \infty$. Together these show, by a well-known argument (see e.g. (AJS 1) Lemma 7.6), that $B (G + 1)^{-\frac{1}{2}}$ is compact. Since $0 \leq G \leq h$ then $(G + 1)^{\frac{1}{2}}(h +1)^{-\frac{1}{2}}$ is bounded, and with $(h +1)^{-\frac{1}{2}}$ bounded by definition the result follows.

The purpose of this manipulation was to indicate how the $P^{2}$ in $h$ contributed momentum space (un)boundedness and $B$ (plus, possibly, $V$) contributed position space boundedness to make $B(h + 1)^{-1}$ compact.
\end{quote}

\vspace{8pt}

{\bf{(c) Quantum No-Capture and Poincar\'{e} Recurrence Theorems}}

The analogue to Schwarzschild's No-Capture theorem (Section 3.2.2) is proved in Proposition 3M.17, and states that:
\begin{quote}
$B_{+} = B_{-}$
\end{quote}
The proof used is directly analogous to the classical case, although the idea was taken from Chernoff (Ch 1). Again we require $U_{t}^{-1} = U_{-t}$ $\forall t$. In place of the invariance of  phase space volumes (Liouville measure) under evolutions we use the fact that an isometry of a compact metric space to itself which is into is also onto.

{\em{From now on I shall assume that $U_{t}^{-1} = U_{-t}$ and call $B \equiv B_{+} = B_{-}$}}.

Also from Chernoff we may lift the idea for proving a quantum mechanical version of Poincar\'{e}'s recurrence theorem. Namely that if $\psi$ is a bound vector then $U_{t}\psi$ will return arbitrarily close to $\psi$ at some later time. This result is proved in Proposition 3M.18 by using the simple fact that a compact metric space is sequentially compact (i.e., every sequence has a converging subsequence). We also require that $U_{t}$ be a one-parameter group.

\vspace{8pt}

{\bf{(d) Bound Vectors and Eigenvectors of the evolution}}

Suppose $\phi$ is an eigenvector of $U_{t}$, then for $\Omega$ compact:
\begin{quote}
$\langle U_{t} \phi, \Omega^{-1} U_{t} \phi \rangle = \langle \phi, \Omega^{-1} \phi \rangle $
\end{quote}
which is equal to 1 for the choice $\Omega = | \phi \rangle \langle \phi |$. Similarly if $\phi \in$ {\em{lin}} (eigenvectors of $U_{t}$), where `{\em{lin}}' denotes the finite linear span, we can always find a compact $\Omega$ such that the above holds.

It follows from this reasoning that:
\begin{quote}
$\underset{t \in \mathbb{R}}{\bigcap} lin$ (eigenvectors of $U_{t}$) $\subseteq B$.
\end{quote}
For simplicity let $U_{t} = e^{-iht}$ where the Hamiltonian, $h$, is self adjoint. In Proposition 3M.19 it is shown that not only are all finite linear combinations of eigenvectors of $h$ contained in $B$ but so are all infinite linear combinations. That is:
\begin{quote}
$clin$ (eigenvectors of $h$) $\subseteq B$
\end{quote}
where `$clin$' denotes closed linear span.

This confirms the popular analogy between eigenvectors and closed orbits. In fact, the converse is also true as shown by Theorem 3M.20, namely that every bound vector is composed from eigenvectors. In summary:
\begin{quote}
$B = clin$ (eigenvectors of $h$).
\end{quote}
$B$ is thus, as promised, closed and actually identical to the pure point spectral subspace $H_{pp}(h)$ of $h$.

The idea behind Theorem 3M.20 is as follows:

We first show that if there is a non-zero invariant compact set in $H$ then there exists an eigenvector of $h$. To do this a fixed point theorem for compact sets is used. Next, we project a bound vector into the continuous spectral subspace $(H_{pp}(h)^{\bot})$ of $h$ and repeat the argument to conclude that the component of any bound vector in the continuous spectral subspace is zero.

\vspace{8pt}

{\bf{(e) Analogues to Birkhoff's Theorem and the Ergodic Theorem}}

In classical statistical mechanics, {\em{Birkhoff's theorem}} (see, e.g. (Kh 1) Ch. 2) says, in the notation of Section 3.2.1, that if $\Omega$ is any invariant finite-volume region of phase space and $F \in L^{1}(\Omega, ds)$ then: 
\begin{quote}
$\hat{F}(\alpha) \equiv \underset{T \rightarrow \infty}{lim} \frac{1}{2T} \int_{-T}^{T} F (U_{t} \alpha) dt$ \hspace{3pt} exists a.e.
\end{quote}

The first part of Proposition 3M.21 provides a quantum mechanical analogue for operators which are bounded on the bound vectors: for $U_{t} = e^{-iht}$:
\begin{quote}
$\hat{F}(\psi) \equiv \underset{T \rightarrow \infty}{lim} \frac{1}{2T} \int_{-T}^{T} \langle U_{t} \psi, F U_{t} \psi \rangle dt$ \hspace{3pt} exists for $\psi \in B$.
\end{quote}
The Ergodic theorem in classical statistical mechanics says that if $\Omega$ has no invariant subsets of non-zero measure then:
\begin{quote}
$\hat{F}(\alpha) = \overline{F} \equiv \frac{1}{m_{L}(\Omega)} \int_{\Omega} F (\alpha) d \alpha$ \hspace{3pt} a.e.
\end{quote}
(see (Kh 1) p. 29). That is, the time average is a constant $\overline{F}$, independent of the state $\alpha$ in $\Omega$.

The second part of Proposition 3M.21 provides a partial result along these lines for quantum mechanics, namely that if $\psi \in B$ then:
\begin{quote}
$\hat{F}(\psi) = Tr [F\rho]$
\end{quote}
where
\begin{quote}
$\rho = \sum_{n}P_{n} | \psi \rangle \langle \psi | P_{n}$
\end{quote}
$P_{n}$ being the projections onto the eigenspaces $L_{n}$ of $h$.

This result (cf. Lemma 5.7 in (Da 2)) indicates that the notion of `ergodic' states for quantum mechanics depends strongly on the behaviour of the operator $F$ with respect to the spectral projections of $h$ - note that we could write:
\begin{quote}
$\hat{F}(\psi) = Tr[ \overline{F} | \psi \rangle \langle \psi |]$ where $\overline{F} \equiv \sum_{n}P_{n} F P_{n}$.
\end{quote}

{\bf{6. Quantum Mechanics - Bound States}}

To present the theory in terms of bound states it will suffice to set the scene as the results essentially carry over from bound vectors.

The continuous mapping:
\begin{quote}
$j$: $H \rightarrow J^{+}(H)$ ; $\psi \rightarrow | \psi \rangle \langle \psi |$
\end{quote}
takes the unit ball of $H$ onto the extreme points of $J_{1}^{+}(H)$, the positive trace class operators with unit trace. By Gleason's theorem these are identifiable as the pure states of the quantum system whose projective geometry is described by the closed linear manifolds of $H$.

Compactness is preserved by $j$ so we define the bound statistical states $\mathbb{B}^{\Omega_{R}}$ corresponding to the bound vectors $B^{\Omega_{R}}$ by:
\begin{quote}
$\mathbb{B}^{\Omega_{R}} = \overline{co}(j(B_{1}^{\Omega_{R}}))$
\end{quote}
where $B_{1}^{\Omega_{R}}$ are the unit vectors of  $B^{\Omega_{R}}$ and $\overline{co}$ denotes closed convex null. Evidently $\mathbb{B}^{\Omega_{R}}$ is compact and contained in the compact set (see (Ch 1) Prop 2.2 for proof of compactness):
\begin{quote}
$\{ \rho \in J_{1}^{+}(H)$ $|$ $Tr[U_{t} \rho U_{t}^{*} \Omega ^{-2}] \leq R^{2}$ \hspace{2pt} $\forall t \}$
\end{quote}
where $\Omega ^{-2}$ is understood in the sense of Section 1.

It is interesting that extreme points of this set need not be pure states, a possibility which could have implications for the theory of measurement when the set of `physical' states is taken to be compact. We do not, however, pursue this idea further here.

\vspace{8pt}

{\bf{7. Quantum Mechanics - Average Stays}}

Let us first extend the definition of average stay in the finite theory of Section 3.2.4 to arbitrarily large times:

The {\em{average stay}} $\mu^{\Omega}(\psi)$ of a vector $\psi$ with respect to a {\em{positive}} bounded operator $\Omega$ is given by the formula:
\begin{quote}
$\mu^{\Omega}(\psi) \equiv \underset{T \rightarrow \infty}{lim} \frac{1}{2T} \int_{-T}^{T} \langle U_{t} \psi, \Omega U_{t} \psi \rangle dt$.
\end{quote}
We have already seen from Proposition 3M.20 that this limit exists for $\psi \in B$. Proposition 3M.22 shows that for compact operators it exists for all vectors in $H$. Moreover, the bound and non-bound vectors of a Hamiltonian evolution $U_{t} = e^{-iht}$ can be characterised in terms of the average stay with respect to compact operators:
\begin{quote}
$B = \{ \psi \in H$ $|$ $\mu^{\Omega}(\psi) > 0$ for some compact $\Omega \}$

$B^{\bot} = \{ \psi \in H$ $|$ $\mu^{\Omega}(\psi) = 0$ for all compact $\Omega \}$.
\end{quote}

The proof uses Proposition 3M.21 and the RAGE theorem ((RS 3) Th. XI. 115) or Wiener's theorem ((RS 3) Th. XI. 114).

We could, in fact, have considered average stays with respect to vectors or finite dimensional linear manifolds - the `compactness' just gives the fullest expression of average stay with respect to (phase-space) bounded region.

\vspace{8pt}
\newpage
{\bf{8. Quantum Mechanics - Transit Times and Scattering States}}

We extend the definition of transit times given in Section 3.2.4 to arbitrarily large future (+) and past (-) times:

The {\em{transit time}} $\tau_{\pm}^{\Omega}(\psi)$ of a vector $\psi$ with respect to a positive bounded operator $\Omega$ is given by the formulae:
\begin{quote}
$\tau_{+}^{\Omega}(\psi) \equiv \underset{T \rightarrow \infty}{lim} \int_{0}^{T} \langle U_{t} \psi, \Omega U_{t} \psi \rangle dt$

$\tau_{-}^{\Omega}(\psi) \equiv \underset{T \rightarrow \infty}{lim} \int_{-T}^{0} \langle U_{t} \psi, \Omega U_{t} \psi \rangle dt$.
\end{quote}

The limits obviously exist if the integrals are uniformly bounded - in any other case we shall set the transit times to $\infty$.

The future (+) and past (-) sets of {\em{scattering vectors}} $S_{\pm}$ are defined as in the classical case by:
\begin{quote}
$S_{\pm} \equiv \{ \psi \in H$ $|$ $\tau_{\pm}^{\Omega}(\psi) < \infty$ for all compact $\Omega \}$.
\end{quote}
If we define the {\em{total transit time}} $\tau^{\Omega}(\psi)$ as:
\begin{quote}
$\tau^{\Omega}(\psi) \equiv \tau_{+}^{\Omega}(\psi) + \tau_{-}^{\Omega}(\psi) = \underset{T \rightarrow \infty}{lim} \int_{-T}^{T} \langle U_{t} \psi, \Omega U_{t} \psi \rangle dt$
\end{quote}
and the set of {\em{total scattering vectors}} $S$ by:
\begin{quote}
$S \equiv \{ \psi \in H$ $|$ $\tau^{\Omega}(\psi) < \infty$ for all compact $\Omega \} = S_{+} \cap S_{-}$
\end{quote}
then Proposition 3M.23 provides a characterisation of $S$ for Hamiltonian evolutions $U_{t} = e^{-iht}$ as:
\begin{quote}
$\overline{S} = H_{ac}(h)$
\end{quote}
that is, the scattering vectors with finite total transit time are dense in the absolutely continuous spectral subspace $H_{ac}(h)$ of the Hamiltonian $h$.

We may now collect together the relations between our geometric definitions and the spectral subspaces of the Hamiltonian $h$ when $U_{t} = e^{-iht}$.

Call the set of {\em{exceptional vectors}} $E$ those which are neither total scattering nor bound in the sense that:
\begin{quote}
$E = S^{\bot} \cap B^{\bot}$
\end{quote}
then we have:
\begin{quote}
$H = B \oplus E \oplus \overline{S}$.
\end{quote}
If $h$ is a Hamiltonian with $H_{pp}(h)$, $H_{sc}(h)$ and $H_{ac}(h)$ denoting its pure point, singular continuous and absolutely continuous spectral subspaces, then we know ((RS 1) Th. VII. 4):
\begin{quote}
$H = H_{pp}(h) \oplus H_{sc}(h) \oplus H_{ac}(h)$.
\end{quote}
Our results (see Corollary 3M.24) allow us to characterise the spectral subspaces by vectors with the following geometric features:
\begin{quote}
(a) Bound in some compact set for all time:
\begin{quote}
$B = H_{pp}(h)$
\end{quote}
(b) Zero average stay w.r.t. any compact operator:
\begin{quote}
$B^{\bot} = H_{sc}(h) \oplus H_{ac}(h)$
\end{quote}
(c) Exceptional: 
\begin{quote}
$E = H_{sc}(h)$
\end{quote}
(d) Finite transit time w.r.t. any compact operator:
\begin{quote}
$\overline{S} = H_{ac}(h)$.
\end{quote}
\end{quote}
Thus, the categories of Fig 3.1 remain essentially valid in the quantum case.

\vspace{8pt}

{\bf{9. Quantum Mechanics - Comparison Dynamics}}

What little we can say about comparison dynamics is summarised in Proposition 3M.26. This result mimics Proposition 3M.14 and we define, as for the classical case, the {\em{V-asymptotic}} vectors $S_{\pm}(U, V)$ of an evolution $U_{t}$ with respect to a comparison evolution $V_{t}$ by:
\begin{quote}
$S_{\pm}(U,V) \equiv \{ \psi \in H$ $|$ $\psi = \underset{t \rightarrow \pm \infty}{lim}U_{t}^{-1}V_{t}\psi'$ for some $\psi' \in H \}$.
\end{quote}
$S_{\pm}(U,V)$ are better known as the ranges of the wave operators $W_{\pm}(U,V)$ where:
\begin{quote}
$W_{\pm}(U,V) = \underset{t \rightarrow \pm \infty}{s-lim}{U_{t}}^{*}V_{t}$.
\end{quote}
The questions raised in Section 3.2.3 concerning the relation between the geometric theory and comparison dynamics remain open. (See, however, the remarks at the very start of Section 3.2).

The proof of Proposition 3M.26, which asserts that:
\begin{quote}
$S_{\pm}(U,V) \subseteq B(U)^{\bot} \equiv E(U) \oplus \overline{S}(U)$
\end{quote}
uses a set of vectors $D_{\pm}$ which includes scattering vectors but not necessarily all exceptional vectors. The definition and properties of $D_{\pm}$ are given in Lemma 3M.25.

\newpage

\addcontentsline{toc}{section}{Mathematics Section of Chapter 3}
\section*{Mathematics Section of Chapter 3}
This section develops the various mathematical results to support the main text of Chapter 3.

{\bf{3.1 The Physical Perspective}}

{\bf{3. Localisation}}

{\bf{3M.1 Proposition}}

If $U_{t}$ denotes the free-particle evolution in quantum mechanics and $\psi(x)$ is a smooth wave function of compact support, then $(U_{t} \psi)(x)$ does not have compact support for any $t > 0$.

{\bf{Proof}}

(This result is well-known; we provide a proof here based on the Paley-Wiener theorem).

By the Paley-Wiener theorem, $f \in C_{0}^{\infty}(\mathbb{R}^{n})$ with support in a ball of radius $R < \infty$ if and only if its Fourier transform, $\hat{f}$, satisfies for all $N$:
\begin{quote}

$| \hat{f}(z) | \leq \dfrac{C_{N}e^{R | Im(z) |}}{(1 + |z|)^{N}}$ \hspace{3pt} $\forall z \in C^{n}$.

\end{quote}

For the free-particle evolution we have:
\begin{quote}
$(U_{t}^{\wedge} \psi)(k) = e^{\frac{-ik^{2}t}{2m}} \hat{\psi}(k)$.
\end{quote}
$(U_{t}^{\wedge} \psi)(z)$ cannot satisfy the required condition when $t > 0$ since for fixed $Im(z) > 0$
\begin{quote}
${|(U_{t}^{\wedge} \psi)(z)|} \rightarrow \infty$ as $Re(z) \rightarrow \infty$.
\end{quote}

{\bf{4. Compact Sets and Phase Space Localisation}}

{\bf{3M.2 Definition}} ((RS 4) p. 247)

Let $F > 0$ be a measurable function, then we say $F \rightarrow \infty$ if and only if for every $N > 0$ there is an $R_{N}$ such that $F(x) \geq N$ $\forall$ $| x | \geq R_{N}$.

{\bf{3M. 3 Proposition}}

Let $\omega \subseteq$ unit ball of $L^{2}(\mathbb{R}^{k})$ then the following are equivalent:
\begin{quote}
(i) $\overline{\omega}$ is compact

(ii) $\exists$ $F, G \rightarrow \infty$ such that $\omega$ is contained in the compact set:
\begin{quote}
$\{ \psi \in L^{2} (\mathbb{R}^{k})$ $|$ $\langle \psi, \psi \rangle \leq 1$, $\langle \psi, F(Q)\psi \rangle \leq 1$, $\langle \psi, G(P)\psi \rangle \leq 1 \}$
\end{quote}
where $Q$ and $P$ are the position and momentum operators, and the inner products are to be interpreted as quadratic forms.
\end{quote}

{\bf{Proof}}

(ii) $\Rightarrow$ (i): This is Rellich's criterion for compactness ((RS 4) Theorem XIII.65).

(i) $\Rightarrow$ (ii): For this we use Riesz's criterion for compactness ((RS 4) Theorem XIII.66) to construct the functions $F$ and $G$. Inspection of Riesz's criterion reveals that compact $\overline{S}$ is equivalent to a uniform convergence at infinity in both position and momentum space. It suffices, indeed, to construct just one of the functions, $F$ say for position, as construction of the other is analogous.

From Riesz's criterion we have that for any $\epsilon_{n} > 0$ $\exists$ a bounded set $K_{n} \subset \mathbb{R}^{k}$ such that:
\begin{quote}
$\int_{\mathbb{R}^{k \smallsetminus K_{n}}} |\psi(x)|^{2} dx \leq \epsilon_{n}^{2}$.
\end{quote}
Choose the sequence $\epsilon_{n} = 2^{-n}$ for $n = 1, 2 ...$ and define:
\begin{quote}
$F(x) = 2^{(m - 2)}$ where $m = min \{ n$ $|$ $x \in K_{n} \}$.
\end{quote}
Then, for $\psi \in S$:
\begin{quote}
$\langle \psi, F(Q) \psi \rangle = \int_{K_{1}}F(x) |\psi(x)|^{2} dx + \sum_{n=1}^{\infty} \int_{K_{n+1} \smallsetminus U_{m=1}^{m=n}K_{m}} F(x) | \psi (x) |^{2} dx \leq 2^{-1} \int_{K_{1}} | \psi (x) |^{2} dx + \sum_{n=1}^{\infty} 2^{(n-1)} \int_{\mathbb{R}^{k \smallsetminus K_{n}}} |\psi(x)|^{2} dx \leq 2^{-1} + \sum_{n=1}^{\infty} 2^{(n-1)} 2^{-2n} = 1$
\end{quote}
It is readily seen that $F \rightarrow \infty$ as per Definition 3M.2.

{\bf{3M.4 Proposition}}

Let $F, G \rightarrow \infty$ and let $f, g$ be the quadratic forms on $H = L^{2}(\mathbb{R}^{k})$ given by:
\begin{quote}
$f(\psi) = \int F(x) |\psi(x)|^{2} dx$

$g(\psi) = \int G(k) |\hat{\psi}(k)|^{2} dk$
\end{quote}
where $\hat{}$ denotes the Fourier transform, then $f$ and $g$ are quadratic forms of the operators $F \equiv F(Q)$ and $G \equiv G(P)$ where $Q$ and $P$ are the position and momentum operators and where:
\begin{quote}
(i) $F$ and $G$ are self-adjoint on the Hilbert subspaces $\overline{Quad(f)}$ and $\overline{Quad(g)}$ respectively;

(ii) The quadratic form and operator domains are related by:
\begin{quote}
$Quad(f) = D (F^{\frac{1}{2}})$; $Quad(g) = D(G^{\frac{1}{2}})$.
\end{quote}
\end{quote}

{\bf{Proof}}

It suffices to consider one of the functions, $F$ say. Define:
\begin{quote}
$F'(x) = F(x)$ if $F(x)$ is finite

\hspace{25pt} $= 0$ otherwise.
\end{quote}
By the spectral theorem, $F'(Q)$ is self-adjoint on $H$, hence also self-adjoint on the Hilbert subspace $Ker(F')^{\bot}$. Also, by (Da 2) Theorem 4.12,
\begin{quote}
$D(F'^{\frac{1}{2}}) = Quad (f')$ is dense in $H$
\end{quote}
where $f'$ denotes the quadratic form on $H$ associated to the function $F'$.

As we shall shortly demonstrate, $Quad(f)$ is dense in $Ker(F')^{\bot}$, and we may define $F \equiv F'$ as the unique self-adjoint operator on $\overline{Quad(f)} = Ker(F')^{\bot}$ such that:
\begin{quote}
$f(\psi) = \langle F^{\frac{1}{2}}\psi, F^{\frac{1}{2}}\psi \rangle$ $\forall$ $\psi \in Quad (f)$.
\end{quote}
Evidently, $Quad(f) = D(F^{\frac{1}{2}})$.

To prove that $Quad(f)$ is dense in $Ker(F')^{\bot}$ consider the measurable set in $\mathbb{R}^{k}$ given by:
\begin{quote}
$M = \{ x$ $|$ $F(x)$ is non-finite$\}$.
\end{quote}
We have, by definition that:
\begin{quote}
$f'(x) > 0$ if $x \in \mathbb{R}^{k} \smallsetminus M$

\hspace{25pt} $= 0$ if $x \in M$
\end{quote}
and:
\begin{quote}
$f'(x) = \int f'(x)$ $|\psi(x)|^{2} dx$.
\end{quote}
Hence if $f'(\psi) = 0$ then $\psi$ has support almost everywhere in $M$ and, conversely, if $\psi$ has such support then $f'(\psi) = 0$. Expressed symbolically:
\begin{quote}
$Ker (F') = \{ \psi$ $|$ $\psi$ has support a.e. in $M \}$

\hspace{36pt} = $L^{2}(M)$.
\end{quote}
Hence: $Ker(F')^{\bot} = L^{2}(\mathbb{R}^{k} \smallsetminus M)$.

Now let $\psi \in Quad(f)$, then evidently $\psi \in Quad (f')$ and $\psi$ has support where $F(x)$ is finite. Thus:
\begin{quote}
$Quad (f) \subseteq Quad (f') \cap Ker(F')^{\bot}$.
\end{quote}
Conversely, if $\psi \in Ker(F')^{\bot}$ then we have just seen $\psi$ has support in $\mathbb{R}^{k} \smallsetminus M$ so if also $\psi \in Quad(f')$ the integral
\begin{quote}
$\int F(x) |\psi(x)|^{2} dx$
\end{quote}
converges and $\psi \in Quad (f)$. Overall, therefore:
\begin{quote}
$Quad (f) = Quad (f') \cap Ker(F')^{\bot}$.
\end{quote}
Denseness of $Quad(f')$ now allows us to conclude that $Quad(f)$ is dense in $Ker(F')^{\bot}$.

{\bf{3M.5 Proposition}}

Let $F, G, f$ and $g$ be as in Proposition 3M.4 and suppose further that:
\begin{quote}
$L = \overline{D(F^{\frac{1}{2}})} = \overline{D(G^{\frac{1}{2}})}$,
\end{quote}
and $D(F^{\frac{1}{2}}) \cap D(G^{\frac{1}{2}})$ is dense in $L$, then there exists a unique self-adjoint operator (the quadratic form sum), $F + G$, on $L$ such that:
\begin{quote}
(i) $Quad(F + G) = D (( F + G)^{\frac{1}{2}}) = D(F^{\frac{1}{2}}) \cap D(G^{\frac{1}{2}})$

(ii) $\langle (F + G)^{\frac{1}{2}} \psi, (F + G)^{\frac{1}{2}} \phi \rangle = \langle F^{\frac{1}{2}} \psi, F^{\frac{1}{2}} \phi \rangle + \langle G^{\frac{1}{2}} \psi, G^{\frac{1}{2}} \phi \rangle$ 

$\forall \psi$, $\phi \in D((F + G)^{\frac{1}{2}})$.
\end{quote}

{\bf{Proof}}

With the additional conditions we may apply (Da 2) Corollary 4.13 to deduce the result for the Hilbert space given by $L$.

{\bf{3M.6 Corollary}}

Let $\omega \subseteq$ unit ball of $L^{2}(\mathbb{R}^{k})$ then the following are equivalent:
\begin{quote}
(i) $\overline{\omega}$ is compact

(ii) $\exists$ $F, G \rightarrow \infty$ satisfying the conditions of Proposition 3M.5, and a constant $K < \infty$ such that $\omega$ is contained in the compact set:
\begin{quote}
$\{ \psi \in D (( F + G)^{\frac{1}{2}})$ $|$ $||\psi ||$ $\leq 1$, $||( F + G)^{\frac{1}{2}}\psi || \leq K \}$.
\end{quote}
\end{quote}

{\bf{Proof}}

(ii) $\Rightarrow$ (i) follows from Propositions 3M.5 and 3M.3.

(i) $\Rightarrow$ (ii): Let $\underline{F}$, $\underline{G}$ be functions such that by Proposition 3M.3 $\omega$ is contained in:
\begin{quote}
$\{ \psi$ $|$ $\langle \psi, \psi \rangle \leq 1$, $\langle \psi, \underline{F}(Q)\psi \rangle \leq 1$, $\langle \psi, \underline{G}(P) \psi \rangle \leq 1 \}$.
\end{quote}
Define:
\begin{quote}
$F (x) = min (x^{2}, \underline{F}(x))$

$G(x) = min (x^{2}, \underline{G}(x))$.
\end{quote}
Then $F(Q) + G(P)$ is a densely-defined quadratic form sum and on $D(F^{\frac{1}{2}}) \cap D(G^{\frac{1}{2}})$:
\begin{quote}
$|| (F + G)^{\frac{1}{2}} \psi ||^{2} = \langle F^{\frac{1}{2}} \psi, F^{\frac{1}{2}} \psi \rangle + \langle G^{\frac{1}{2}} \psi, G^{\frac{1}{2}} \psi \rangle \leq 2$.
\end{quote}

{\bf{3M.7 Lemma}}

Let $A \in L(H)$ be a self-adjoint bounded operator then:
\begin{quote}
(i) $Ker (A) = Ran (A)^{\bot}$.

(ii) $A^{-1}$ is a self-adjoint operator on $Ker(A)^{\bot}$.
\end{quote}

{\bf{Proof}}

(i): see e.g. (Ru 1) Theorem 12.10.

(ii) $A$ is a one-to-one mapping in $Ker(A)^{\bot}$:
\begin{quote}
$A$: $Ker(A)^{\bot} \rightarrow Ran(A) \subseteq Ker(A)^{\bot}$
\end{quote}
Thus $A$ is self-adjoint on $Ker(A)^{\bot}$ and $A^{-1}$ is well-defined on $Ran(A)$. $A^{-1}$ is clearly also symmetric $Ran(A)$ and hence closeable on $Ker(A)^{\bot}$. We have that:
\begin{quote}
$Ran(A) = D(A^{-1}) \subseteq D(A^{-1*})$.
\end{quote}
Hence it will suffice to show that:
\begin{quote}
$D(A^{-1*}) \subseteq Ran(A)$.
\end{quote}
To see this let $\psi \in D(A^{-1*})$ then by the definition of an adjoint on $Ker(A)^{\bot}$ we have that $\exists$ $\theta \in Ker(A)^{\bot}$ such that:
\begin{quote}
$\langle \psi, A^{-1} \phi \rangle = \langle \theta, \phi \rangle$ $\forall$ $\phi \in Ran (A).$
\end{quote}
But $A$ is one-to-one so $\exists$! $\Phi \in Ker (A)$ such that $A^{-1} \phi = \Phi$, hence:
\begin{quote}
$\langle \psi, \Phi \rangle = \langle \theta, A \Phi \rangle$ $\forall$ $\Phi \in Ker (A)^{\bot}$

$\Rightarrow \langle \psi, \Phi \rangle = \langle A\theta, \Phi \rangle$ $\forall$ $\Phi \in Ker (A)^{\bot}$.
\end{quote}
Hence, since $Ker(A)^{\bot}$ is a Hilbert space: $\psi = A \theta$ and $\psi \in Ran(A)$ as required.

{\bf{3M.8 Lemma}}

Let $Ran_{K}(A)$ denote the range of an operator $A$ acting on the K-ball of a Hilbert space, then the following are equivalent:
\begin{quote}
(i) $A$ is a compact operator.

(ii) $Ran_{K}(A)$ is a compact set.
\end{quote}

{\bf{Proof}}

(i) $\Rightarrow$ (ii): follows from the well-known facts that a compact operator takes weakly convergent sequences into strongly convergent ones ((RS 1) Theorem VI.11); that any Hilbert space is sequentially weakly compact ((K 1) Chapter 5 Lemma 1.5) and the Bolzano-Weierstrass theorem ((RS 1) Theorem IV.3).

(ii) $\Rightarrow$ (i): follows from the definition of a compact operator as taking bounded sets into precompact sets.

{\bf{3M.9 Theorem}}

Let $\omega \subseteq$ unit ball of $L^{2}(\mathbb{R}^{k})$ then the following are equivalent:
\begin{quote}
(i) $\overline{\omega}$ is compact

(ii) $\exists$ $F, G \rightarrow \infty$ satisfying the conditions of Proposition 3M.5 and $K < \infty$ such that $\omega$ is contained in the compact set:
\begin{quote}
$\{ \psi \in D (( F + G)^{\frac{1}{2}})$ $|$ $||\psi ||$ $\leq 1$ and $||( F + G)^{\frac{1}{2}}\psi || \leq K \}$.
\end{quote}
(iii) $\exists$ $F, G \rightarrow \infty$ satisfying the conditions of Proposition 3M.5 and $K < \infty$ such that $(F + G)^{-\frac{1}{2}}$ defined on  $\overline{D(( F + G)^{\frac{1}{2}})}$ is compact and $\omega$ is contained in the compact set:
\begin{quote}
$Ran_{K}((F + G)^{-\frac{1}{2}})$, acting on the K-ball in $Ker((F + G)^{-\frac{1}{2}})^{\bot}$.
\end{quote}
(iv) $\exists$ a positive compact operator $\Omega$ and $K < \infty$ such that $\omega$ is contained in the compact set:
\begin{quote}
$\{ \psi \in Ran(\Omega)$ $|$ $||\psi ||$ $\leq 1$ and $||\Omega^{-1} \psi || \leq K \}$.
\end{quote}
(v) $\exists$ a positive compact operator $\Omega$ and $K < \infty$ such that $\omega$ is contained in the compact set:
\begin{quote}
$Ran_{K}(\Omega)$ acting on the K-ball in $Ker(\Omega)^{\bot}$.
\end{quote}
\end{quote}

{\bf{Proof}}

By Corollary 3M.6 we have (i) $\Leftrightarrow$ (ii). We shall prove (ii) $\Leftrightarrow$ (iii), (iii) $\Rightarrow$ (v), (iv) $\Leftrightarrow$ (v). (v) $\Rightarrow$ (i) is obvious. (ii) $\Leftrightarrow$  (iii): follows from (RS 4) Theorem XIII.64 and the fact that $F + G$ is a strictly positive operator on $\overline{D((F + G)^{\frac{1}{2}})}$. Compactness in (iii) follows from Lemma 3M.8. Notice that:
\begin{quote}
$\overline{D((F + G)^{\frac{1}{2}})} = Ker((F + G)^{-\frac{1}{2}})^{\bot}$
\end{quote}
and
\begin{quote}
$D((F + G)^{\frac{1}{2}}) = Ran((F + G)^{-\frac{1}{2}})$.
\end{quote}

(iii) $\Rightarrow$ (v): is now trivial.

(iv) $\Leftrightarrow$ (v): follows from Lemma 3M.7 where the relevant Hilbert space is $Ker(\Omega)^{\bot}$. Compactness again follows from Lemma 3M.8.

\newpage

{\bf{3.2 Geometric Bound and Scattering States}}

{\bf{1. Classical Ideas}}

{\bf{3M.10 Proposition}}

Let $B_{\pm}$ and $S_{\pm}$ be defined as in Section 3.2.1, then:
\begin{quote}
(a) $B_{\pm} = \{ \alpha \in S$ $|$ $|| U_{t} \alpha || _{S} < \infty$ $\forall t \in \mathbb{R}^{\pm} \}$

(b) $S_{\pm} = \{ \alpha \in S$ $|$ for each $\beta \in S$ $|| \beta - U_{t} \alpha ||_{S} \rightarrow \infty$ as $t \rightarrow \pm \infty \}$

\hspace{31pt} $= \{ \alpha \in S$ $|$ $\exists$ $\beta$ with $|| \beta - U_{t} \alpha ||_{S} \rightarrow \infty$ as $t \rightarrow \pm \infty \}$.
\end{quote}

{\bf{Proof}}

For (a) we need only notice that for a set $\Omega \subseteq \mathbb{R}^{2n}$:
\begin{quote}
$\Omega$ precompact $\Leftrightarrow \underset{\alpha \in \Omega}{sup}$ $||\alpha||_{S} < \infty$.
\end{quote}
For (b) it suffices to note that if $\alpha$ is a scattering state it permanently escapes from every k-ball in $S$ for large enough time.

{\bf{3M.11 Proposition}}

Let $\tau_{\pm}^{\Omega}(\alpha)$ and $S_{\pm}$ be defined as in Section 3.2.1. Let $U_{t}$ be a Hamiltonian evolution and suppose that the Hamiltonian $h \in C^{\infty}(S)$, then:
\begin{quote}
$S_{\pm} = \{ \alpha \in S$ $|$ $\tau_{\pm}^{\Omega}(\alpha) < \infty$ \hspace{2pt} $\forall$ compact $\Omega \subseteq S \}$.
\end{quote}

{\bf{Proof}}

$\subseteq$: if $\alpha \in S_{+}$ then for each $\Omega$ $\exists T < \infty$ such that $\{U_{t}\alpha \}_{t >T} \cap \Omega = \O$, but then:
\begin{quote}
$\tau_{\pm}^{\Omega}(\alpha) = \int_{0}^{\infty} p^{\Omega}(\alpha, t) dt \leq \int_{0}^{T} p^{\Omega}(\alpha, t) dt < \infty$.
\end{quote}
Similarly for $S_{-}$.

$\supseteq$: We shall show that if $\alpha \in S_{\pm}^{c}$ and the phase-space velocity is bounded on compact sets then $\tau_{\pm}^{\Omega}(\alpha)$ cannot be finite for all compact sets, $\Omega$. Again we prove only for future (+) as past (-) is analogous. Bound states cannot have finite transit times for all compact sets, hence we consider only exceptional states.

Suppose that our claim is false - that is, for some $\alpha \in E_{+}$ $\tau_{+}^{\Omega}(\alpha) < \infty$ for every compact $\Omega$. Since $\alpha \in E_{+}$ then there exists some compact set (which without loss of generality we take to be an R-ball $\Omega_{R}$) and a non-terminating sequence of intervals $\Delta_{n} \subseteq [0, \infty )$ during which the state returns to $\Omega_{R}$. That is:
\begin{quote}
$\tau_{+}^{\Omega_{R}}(\alpha) = \sum_{n} m (\Delta_{n})$
\end{quote}
The transit time is finite, so $\underset{n \rightarrow \infty}{lim} m (\Delta_{n}) = 0$.

Now consider the ball $\Omega_{R + \frac{1}{2}}$. Let $\Delta'_{n}$ denote the interval $\Delta_{n}$ extended to include the time spent inside $\Omega_{R + \frac{1}{2}}$. The phase space distance travelled in each $\Delta'_{n}$ must be greater than or equal to 1. However, this distance is given by:
\begin{quote}
$\int_{\Delta'_{n}}$ $||\dot{\alpha}_{t}||$ $dt$
\end{quote}
where
\begin{quote}
$||\dot{\alpha}_{t}||^{2} = ||\dot{x}_{t}||^{2} + ||\dot{p}_{t}||^{2}$
\end{quote}
the dot denoting time derivative and the norm on the tangent space lifted from phase space.

Supposing that the phase space velocity $\dot{\alpha}_{t}$ is bounded on any compact set, that is
\begin{quote}
$||\dot{\alpha}_{t}|| \leq$ const \hspace{3pt} for $t$ such that \hspace{3pt} $U_{t}\alpha \subseteq$ compact set
\end{quote}
then the phase space distance travelled in each $\Delta'_{n}$ satisfies:
\begin{quote}
$1 \leq \int_{\Delta'_{n}}$ $||\dot{\alpha}_{t}||$ $dt$ $\leq$ const. $m(\Delta'_{n}) \rightarrow 0$.
\end{quote}
This contradiction shows the transit time to be non-finite.

The Proposition follows by noting $h \in C^{\infty}(S)$ implies that the phase space derivatives of $h$ are bounded on any compact set. For a Hamiltonian flow so, therefore, is the phase space velocity.

\vspace{8pt}

{\bf{2. Classical No-Capture Theorem}}

{\bf{3M.12 Proposition}} (Schwarzschild)

Let $U_{t}$ be an evolution such that $U_{t}^{-1} = U_{-t}$ and suppose that $\Omega$ is a phase-space region of finite Liouville measure, then:

For almost every point $\alpha \in \Omega$, if the trajectory $U_{t}\alpha$ will remain in $\Omega$ in the future it must always have been in $\Omega$ in the past. Conversely, if it was always in $\Omega$ in the past it will remain in $\Omega$ in the future.

\vspace{8pt}

{\bf{Proof}}

See (Th 1) Volume 1 Theorem 2.6.14.

\vspace{8pt}
\newpage
{\bf{3M.13 Corollary}}

Let $U_{t}$ be an evolution such that $U_{t}^{-1} = U_{-t}$ and let $B_{\pm}^{\Omega}$, $B_{\pm}$ be defined as in Section 3.2.1, then:
\begin{quote}
(i)  $B_{+}^{\Omega} =  B_{-}^{\Omega}$ a.e.

(ii) $B_{+} = B_{-}$ a.e.
\end{quote}

{\bf{3. General Questions}}

{\bf{3M.14 Proposition}}

Let $U_{t}$ be an evolution. Let $V_{t}$ be another evolution with only scattering states (i.e. $S_{\pm}(V) =$ all of phase space $S$). Let the V-asymptotic states $S_{\pm}(U, V)$ be defined as in Section 3.2.3. Let the bound states $B(U)$ be defined as in Section 3.2.2, then:
\begin{quote}
$S_{\pm}(U, V) \subseteq B(U)^{c}$.
\end{quote}

{\bf{Proof}}

In what follows, convergence is meant in phase space norm.

Let $\alpha \in S_{\pm}(U, V)$ then for some $\alpha' \in S$:
\begin{quote}
$\alpha = \underset{t \rightarrow \pm \infty}{lim}$ ${U_{t}}^{-1} V_{t} \alpha'$.
\end{quote}
Suppose $\alpha$ is in $B(U)$, then $U_{t}\alpha$ is bounded. However $\alpha'$ is in $S_{\pm}(V)$ so $V_{t}\alpha'$ is unbounded. But $U_{t}\alpha - V_{t} \alpha' \rightarrow 0$, hence $\alpha$ cannot be in $B(U)$.

\vspace{8pt}

{\bf{5. Quantum Mechanics - Bound Vectors}}

{\bf{Note}}: If $\Omega$ is a positive compact operator on $H$ then, as in Section 3.1 we understand its ``inverse'' $\Omega^{-1}$ to be the operator on $Ran \Omega$ such that:
\begin{quote}
$\Omega^{-1}\Omega = P(Ker(\Omega)^{\bot})$
\end{quote}
where $P(L)$ denotes the orthogonal projection onto the closed linear manifold $L$. $\Omega \Omega^{-1}$ is defined as the adjoint of $\Omega^{-1} \Omega$ so that
\begin{quote}
$\Omega \Omega^{-1}$ \hspace{0.4pt} $P (Ker(\Omega)^{\bot}) = P (Ker (\Omega)^{\bot})$.
\end{quote}

{\bf{3M.15 Proposition}}

The following sets are the same:
\begin{quote}
(i) $\{ \psi \in H$ $|$ $\exists$ compact $\omega \subseteq H$ with $\{U_{t}\psi \}_{t \in \mathbb{R}^{+} }\subseteq \omega \}$

(ii) $\{ \psi \in H$ $|$ $\exists$ positive compact operator $\Omega$ with $\{U_{t}\psi \}_{t \in \mathbb{R}^{+} }\subseteq Ran(\Omega)$ and $||\Omega^{-1}U_{t}\psi|| \leq 1$ $\forall t \in \mathbb{R}^{\pm} \}$.
\end{quote}

{\bf{Proof}}

(i) $\subseteq$ (ii): $\{ U_{t}\psi \} \subseteq$ compact set $\omega$. In Section 3.1.1 it was shown how to construct a positive compact operator $\Omega_{\omega}$ such that $\omega \subseteq$ range of $\Omega_{\omega}$ acting on the unit ball in $H$.

(ii) $\subseteq$ (i): Choose $\omega$ as the range of $\Omega$ acting on the unit ball.

{\bf{3M. 16 Proposition}}

$B_{\pm}$, defined in Section 3.2.5 are linear manifolds.

\vspace{8pt}

{\bf{Proof}}

Suppose $\psi, \phi \in B_{+}$ then there are compact sets $S$ and $T$, say, such that
\begin{quote}
$U_{t}\psi \in S$ and $U_{t}\phi \in T$ \hspace{2pt} $\forall t \in \mathbb{R}^{+}$
\end{quote}
But then, by the continuity of vector addition, the set
\begin{quote}
$S + T = \{ f + g$ $|$ $f \in S$ and $g \in T \}$
\end{quote}
is compact. From the linearity of $U_{t}$:
\begin{quote}
$U_{t} (\psi + \phi) = U_{t}\psi + U_{t} \phi \in S + T$ \hspace{2pt} $\forall t \in \mathbb{R}^{+}$.
\end{quote}
We conclude that the vector $\psi + \phi$ is in $B_{+}$, which is sufficient to prove the Proposition.

{\bf{3M.17 Proposition (No Capture Theorem)}}

Let $U_{t}$ be an isometry with ${U_{t}}^{-1} = U_{-t}$ $\forall t \in \mathbb{R}$ then:
\begin{quote}
$B_{+} = B_{-}$
\end{quote}

{\bf{Proof}}

Let $\psi \in B_{+}$ and call
\begin{quote}
$S = \{ U_{t} \psi$ $|$ $t \in \mathbb{R}^{+} \}$.
\end{quote}
Then it is easy to see that $\overline{S}$ is a compact set satisfying:
\begin{quote}
$U_{t}(\overline{S}) \subseteq \overline{S}$ \hspace{2pt} $\forall t \in \mathbb{R}^{+}$.
\end{quote}
From the fact that any isometry of a compact metric space to itself which is into is also onto we conclude that the inclusion is an equality. So, by our hypothesis on $U_{t}$:
\begin{quote}
$\overline{S} = U_{-t}(\overline{S})$ \hspace{2pt} $\forall t \in \mathbb{R}^{+}$
\end{quote}
so that $\overline{S} \subseteq B_{-}$ hence $\psi \in B_{-}$. Similarly, we prove that $B_{-} \subseteq B_{+}$.

Note that the method of proof can be used to show that $B_{+}^{\omega} = B_{-}^{\omega}$ and \\ $B_{+}^{\Omega_{R}} = B_{-}^{\Omega_{R}}$.

{\bf{3M.18 Proposition}} (Poincar\'{e} Recurrence Theorem)

Suppose $U_{t}$ is a one-parameter unitary group.

Let $\psi \in B$ then for any $T < \infty$ and any $\epsilon > 0$ $\exists T_{\epsilon} \geq T$ such that:
\begin{quote}
$|| U_{T_{\epsilon}}\psi - \psi ||$ $< \epsilon$.
\end{quote}

{\bf{Proof}}

For $\psi \in B$ and $T < \infty$ consider the sequence $\psi_{n} = U_{nT}\psi$.

Since $\{ U_{t}\psi \}$ lies in a compact set in $H$, and compact metric spaces are also sequentially compact ((Su 2) Ch. 7), then there exists a converging subsequence, say $\{ \psi_{m} \}$. Hence, for any $\epsilon > 0$ $\exists m, m'$ with $m' > m$ such that:
\begin{quote}
$|| U_{m'T}\psi - U_{mT}\psi ||$ $< \epsilon$

$\Rightarrow$ $|| U_{(m'-m)T}\psi - \psi ||$ $< \epsilon$
\end{quote}
so we choose $T_{\epsilon} = (m' - m)T \geq T$.

{\bf{3M.19 Proposition}}

Let $U_{t} = e^{-iht}$ then
\begin{quote}
$clin$ (eigenvectors of $h$) $\subseteq B$.
\end{quote}

{\bf{Proof}}

Eigenvectors of $h$ are obviously bound vectors and so, by Proposition 3M.16, are finite linear combinations of eigenvectors. Suppose $\psi$ is an infinite linear combination of eigenvectors, $\phi_{n}$, of $h$, then:
\begin{quote}
$\psi = \sum_{n=1}^{\infty} \alpha_{n}\phi_{n}$; \hspace{3pt} $\alpha_{n} = \langle \phi_{n}, \psi \rangle$ and

$U_{t} \psi = \sum_{n=1}^{\infty} e^{-i \lambda_{n} t}\alpha_{n} \phi_{n}$; \hspace{3pt} $h\phi_{n} = \lambda_{n} \phi_{n}$.
\end{quote}
Noticing that $\{ e^{-i \lambda_{n} t} \}$ is in $\ell^{\infty}(\mathbb{Z}^{+})$ with norm = 1, consider the mapping:
\begin{quote}
$T$ : $\ell^{\infty}(\mathbb{Z}^{+}) \rightarrow H$; \hspace{2pt} $\{Z_{n}\} \rightarrow T(\{Z_{n}\}) = \sum_{n=1}^{\infty}Z_{n}\alpha_{n}\phi_{n}$
\end{quote}
Since $\psi \in$ range of $T$ acting on the unit ball of $\ell^{\infty}(\mathbb{Z}^{+})$ it will suffice if we can show $T$ to be a compact operator. To see this, notice that $T_{N}$ defined by:
\begin{quote}
$T_{N}(\{Z_{n}\}) = \sum_{n=1}^{N}Z_{n}\alpha_{n}\phi_{n}$, \hspace{3pt} $N < \infty$
\end{quote}
is finite rank and that
\begin{quote}
$||(T - T_{N})(\{Z_{n}\})||^{2} = || \sum_{n = N + 1}^{\infty}Z_{n}\alpha_{n}\phi_{n}||^{2}$

\hspace{87pt} $ = \sum_{n = N + 1}^{\infty}|Z_{n}\alpha_{n}|^{2}$
\end{quote}
Hence, for all $\{Z_{n}\}$ with $||\{Z_{n}\}||_{\infty} \leq 1$ we have:
\begin{quote}
$||(T - T_{N})(\{Z_{n}\})||^{2} \leq  \sum_{n = N + 1}^{\infty} |\alpha_{n}|^{2} \rightarrow 0$ as $N \rightarrow \infty$
\end{quote}
the convergence following from $\psi \in H (\{ \alpha_{n} \} \in \ell^{2}(\mathbb{Z}^{+}))$.

The uniformity of this convergence over the unit ball of $\ell^{\infty}$ allows us to conclude that the operator norm converges:
\begin{quote}
$|| T - T_{N}|| \rightarrow 0$ as $N \rightarrow \infty$.
\end{quote}
Thus $T$ is the norm-limit of a sequence of finite rank operators and is thereby compact.

{\bf{3M.20 Theorem}}

Let $U_{t} = e^{-iht}$ then
\begin{quote}
$B = clin$(eigenvectors of $h$).
\end{quote}

{\bf{Proof}}

By Proposition 3M.19 we need only show $B \subseteq clin$(eigenvectors of $h$). Our first claim is that if $\psi \in B$ then $h$ possesses an eigenvector (has a non-empty point spectrum) in $H$.

So let $\psi \in B$, then the closure $S$ of the set
\begin{quote}
$\{ U_{t} \psi$ $|$ $t \in \mathbb{R} \}$
\end{quote}
is compact. Suppose, without loss, that $|| \psi || = 1$ then by the continuity of the mapping:
\begin{quote}
$j$ : $H \rightarrow J^{+}(H)$; $\psi \rightarrow |\psi \rangle \langle \psi |$
\end{quote}
the set $j(s)$ is compact in trace norm, hence so is its closed convex hull ((Pr 2) Th.4.15) $\overline{co}(j(S))$. Now
\begin{quote}
$\overline{co}(j(S)) \subseteq \{ \rho \in J^{+}(H)$ $|$ $\rho \geq 0$ \hspace{2pt} $Tr[\rho] = 1 \}$.
\end{quote}
$S$ is invariant under $U_{t}$, hence $j(S)$ and $\overline{co}(j(S))$ are invariant under
\begin{quote}
$u_{t}$ : $\rho \rightarrow u_{t}(\rho) = U_{t} \rho {U_{t}}^{*}$.
\end{quote}
So, by the Leray-Schauder-Tychonoff theorem ((RS 1) Th.V.19), $u_{t}$ has a fixed point in $\overline{co}(j(S))$. Denote this fixed point by $\sigma$, then
\begin{quote}
$u_{t}(\sigma) = \sigma$ $\Leftrightarrow$ $U_{t} \sigma = \sigma U_{t}$
\end{quote}
We claim that $\sigma$ can be written in terms of the eigenvectors of $h$:
\begin{quote}
$\sigma = \sum_{n,i} \lambda_{n} | \phi_{n,i} \rangle \langle \phi_{n,i} |$
\end{quote}
where
\begin{quote}
$h\phi_{n,i} = a_{n,i}\phi_{n,i}$ \hspace{3pt} $a_{n,i} \in \mathbb{R}$.
\end{quote}
By the spectral theorem:
\begin{quote}
$\sigma = \sum_{n}\lambda_{n}P_{n}$
\end{quote}
where $P_{n}$ is the orthogonal projection onto the finite dimensional subspace
\begin{quote}
$L_{n} = \{ \psi \in H$ $|$ $\sigma \psi = \lambda_{n}\psi \}$
\end{quote}
and $L_{n} \bot L_{m}$ if $n \not= m$. From this mutual orthogonality and the fact that $U_{t}$ commutes with $\sigma$ we conclude that
\begin{quote}
$U_{t}L_{n} = L_{n}$.
\end{quote}
So, by the spectral theorem:
\begin{quote}
$hL_{n} = L_{n}$
\end{quote}
$L_{n}$ is finite dimensional so we may diagonalise $h$ on $L_{n}$ to obtain eigenvectors $\phi_{n,i}$ of $h$ such that
\begin{quote}
$P_{n} = \sum_{i} | \phi_{n,i} \rangle \langle \phi_{n,i} |$
\end{quote}
which proves our assertion. In particular, we have shown that if there is a non-zero $\psi \in B$ then $h$ possesses an eigenvector, which was our first claim. In fact the conclusion is true if there exists any non-zero invariant compact set.

To complete the proof, let $P_{c}$ denote the projection onto the continuous spectral subspace $H_{c}(h)$ where:
\begin{quote}
$H_{c}(h) = (clin$(eigenvectors of $h))^{\bot}$.
\end{quote}
Consider $P_{c}B$ then, as $U_{t}$ commutes with $P_{c}$, we have:
\begin{quote}
$U_{t}P_{c}B = P_{c}B \subseteq H_{c}(h)$.
\end{quote}
Hence, suppose there is a non-zero $\psi \in P_{c}B$. We may repeat the above argument to show that $h$ must have an eigenvector in $H_{c}(h)$. But this is impossible, so:
\begin{quote}
$P_{c}(B) = 0$

$\Leftrightarrow$ $B \subseteq clin$(eigenvectors $h$).
\end{quote}

{\bf{3M.21 Proposition}}

Let $F$ be any operator such that $P$ $F$ $P$ is bounded, where $P$ is the orthogonal projection onto the pure point spectral subspace (i.e. $B$) of the Hamiltonian $h$ and $U_{t} = e^{-iht}$. Then for $\psi \in B$:
\begin{quote}
(i) $\hat{F} (\psi) \equiv \underset{T \rightarrow \infty}{lim}\frac{1}{2T} \int_{-T}^{T} \langle U_{t} \psi, F U_{t} \psi \rangle dt$ exists 

(ii) $\hat{F}(\psi) = Tr[F \rho]$ where $\rho \in J^{+}(H)$ with:

$\rho = \sum_{n}P_{n}|\psi \times \psi | P_{n}$; $P_{n}$ the projection onto the eigenspace $L_{n}$ of $h$: $L_{n} = \{ \phi$ $|$ $h\phi = a_{n} \phi \}$.
\end{quote}

{\bf{Proof}}

Let $L_{n}$ be the eigenspaces of $h$, i.e., 
\begin{quote}
$h \phi_{n} = a_{n} \phi_{n} \Leftrightarrow \phi_{n} \in L_{n}$
\end{quote}
and let $P_{n}$ denote the projection onto $L_{n}$. If $P$ denotes the projection onto the pure point spectral subspace of $h$ then:
\begin{quote}
$P = \underset{N \rightarrow \infty}{\text{{\em{w-lim}}}} \sum_{n}^{N}P_{n} = \underset{N \rightarrow \infty}{\text{{\em{s-lim}}}} \sum_{n}^{N}P_{n} = \underset{N \rightarrow \infty}{\text{{\em{uw-lim}}}} \sum_{n}^{N}P_{n}$
\end{quote}
where `uw-lim' means ultraweak limit. These results follow from the fact that $\sum_{n}^{N}P_{n}$ is an increasing norm-bounded sequence of positive operators. Call:
\begin{quote}
$\psi_{N} = \sum_{n=1}^{N}P_{n}\psi$
\end{quote}
then:
\begin{quote}
$\langle U_{t} \psi_{N}, FU_{t} \psi_{N} \rangle = \sum_{n,m}^{N}e^{i(a_{n}-a_{m})t}\langle P_{n} \psi, F P_{m} \psi \rangle$
\end{quote}
Hence:
\begin{quote}
$\hat{F}(\psi_{N}) = \sum_{n,m}^{N} \langle P_{n} \psi, F P_{m} \psi \rangle \underset{T \rightarrow \infty}{lim}\frac{1}{2T}\int_{-T}^{T}e^{i(a_{n}-a_{m})t}dt$.
\end{quote}
The cross-terms vanish by the Riemann-Lebesgue Lemma, leaving:
\begin{quote}
$\hat{F}(\psi_{N}) = \sum_{n}^{N} \langle P_{n} \psi, F P_{n} \psi \rangle = Tr[F \rho_{N}]$
\end{quote}
where:
\begin{quote}
$\rho_{N} = \sum_{n}^{N} P_{n} | \psi \rangle \langle \psi | P_{n}$.
\end{quote}
Noting that $P$ $F$ $P$ is norm-bounded and writing for $\phi \in B$:
\begin{quote}
$\langle U_{t} \phi, FU_{t} \phi \rangle = \langle U_{t} \psi_{N}, FU_{t} \psi_{N} \rangle + \langle U_{t} \psi_{N}, FU_{t}(\phi - \psi_{N}) \rangle + \langle U_{t} (\phi - \psi_{N}), FU_{t}\psi_{N} \rangle + \langle U_{t} (\phi - \psi_{N}), FU_{t} (\phi - \psi_{N}) \rangle$ 
\end{quote}
we obtain for $\phi$ such that $\hat{F}(\phi)$ exists:
\begin{quote}
$| \hat{F}(\phi) - \hat{F}(\psi_{N})| \leq$ $||\phi - \psi_{N}||$ $\{Tr[(FPF + F^{*}PF^{*})\rho_{N}] + ||\phi - \psi_{N}||$ $||PFP||$ $\}$
\end{quote}
Choosing $\phi = \psi_{M}$, $M > N$ we see that $\hat{F}(\psi_{N})$ is Cauchy so that $\hat{F}(\psi)$ exists.

The ultraweak convergence of $\sum_{n}^{N}P_{n}$ enables us to conclude that:
\begin{quote}
$\underset{N \rightarrow \infty}{lim} Tr[F \rho_{N}] = Tr[F \rho] = \hat{F}(\psi)$
\end{quote}
where:
\begin{quote}
$\rho = \sum_{n}P_{n} |\psi \rangle \langle \psi | P_{n}$.
\end{quote}
(Note $Tr[\rho] = 1$ and $\rho > 0$).

\vspace{8pt}

{\bf{7. Quantum Mechanics - Average Stays}}

{\bf{3M.22 Proposition}}

Let $\mu^{\Omega}(\psi)$ be defined as in Section 3.2.7. Let $U_{t} = e^{-iht}$ then:
\begin{quote}
(i) $B = \{ \psi \in H$ $|$ $\exists$ positive compact operator with $\mu^{\Omega}(\psi) > 0 \}$

(ii) $B^{\bot} = \{ \psi \in H$ $|$ $\mu^{\Omega}(\psi) = 0$ for all positive compact operators$\}$.
\end{quote}

{\bf{Proof}}

(i) Follows directly from Proposition 3M.21 by choosing $\Omega = F = |\phi_{n} \rangle \langle \phi_{n} |$ where $\phi_{n}$ is an eigenvector of $h$ such that $| \langle \phi_{n}, \psi \rangle | > 0$. Such a $\phi_{n}$ must exist if $B$ is non zero.

(ii) is proved in the RAGE theorem ((RS 3) Th. XI. 115).

\vspace{8pt}

{\bf{8. Quantum Mechanics - Transit Times $\&$ Scattering States}}

{\bf{3M.23 Proposition}}

Let $S$ be defined as in Section 3.2.8 then, for $U_{t} = e^{-iht}$:
\begin{quote}
$\overline{S} = H_{ac}(h)$
\end{quote}
where $H_{ac}(h)$ is the absolutely continuous spectral subspace of $h$.

{\bf{Proof}}

See Lemma 1 and the remarks before it in Section XI.3 of (RS 3).

{\bf{3M.24 Corollary}}

Let $U_{t} = e^{-iht}$. Let $H_{pp}(h)$, $H_{sc}(h)$ and $H_{ac}(h)$ denote the pure point, singular continuous and absolutely continuous spectral subspaces of $h$ ($H = H_{pp}(h) \oplus H_{sc}(h) \oplus H_{ac}(h))$. Let $S, B$ and $E$ be defined as in Section 3.2.8, then:
\begin{quote}
(i) $H_{pp}(h)$

$ =  B \equiv \{ \psi \in H$ $|$ evolution $U_{t}\psi$ is contained in some compact set $\}$

(ii) $H_{sc}(h) \oplus H_{ac}(h)$

$ = \{ \psi \in H$ $|$ average stay $\mu^{\Omega}(\psi)$ w.r.t any compact $\Omega$ is zero $\}$

(iii) $H_{ac}(h)$

$= \overline{S} \equiv \overline{\{ \psi \in H \hspace{1pt} | \hspace{2pt} \text{transit time} \hspace{2pt} \tau^{\Omega}(\psi) \hspace{2pt} \text{w.r.t. any compact} \hspace{1.5pt} \Omega \hspace{2.8pt} \text{is finite} \hspace{1pt} \} }$

(iv) $H_{sc}(h) = E \equiv S^{\bot} \cap B^{\bot}$

\end{quote}

{\bf{Proof}}

From previous results. Note that $\mu^{\Omega}$ and $\tau^{\Omega}$ are for all (past and future) times.

\vspace{8pt}

{\bf{9. Quantum Mechanics - Comparison Dynamics}}

{\bf{3M.25 Lemma}}

Define the sets $D_{\pm}$ as follows:
\begin{quote}
$D_{\pm} = \{ \psi \in H$ $|$ $||\Omega U_{t} \psi ||$ $\rightarrow 0$ as $t \rightarrow \pm \infty$ \hspace{2pt} $\forall$ compact operators $\Omega \}$
\end{quote}
then:
\begin{quote}
(i) $D_{\pm} = \{ \psi \in H$ $|$ $\underset{t \rightarrow \pm \infty}{\text{{\em{w-lim}}}}$ $U_{t} \psi = 0 \}$

(ii) $D_{\pm}$ are closed linear manifolds

(ii) $S_{\pm} \subseteq D_{\pm}$

(iv) $D_{\pm} \bot B$

(v) If $U_{t} = e^{-iht}$ then $H_{ac}(h) \subseteq D_{\pm}$.
\end{quote}

{\bf{Proof}}

(i) $\subseteq$: Use $\Omega = |\phi \rangle \langle \phi |$ for each $\phi \in H$.

\hspace{11pt} $\supseteq$: a compact operator takes weakly convergent sequences into strongly convergent sequences.

(ii) Linearity is obvious; for closure suppose $\psi_{n} \in D_{\pm}$ and $\psi_{n} \rightarrow \psi$. Then:
\begin{quote}
$||\Omega U_{t} \psi ||$ $\leq$ $||\Omega ||$ $|| \psi_{n} - \psi ||$ + $|| \Omega U_{t} \psi_{n} ||$ $\rightarrow 0$.
\end{quote}
(iii) First note that any positive compact operator may be written as $\Omega^{2}$ (or, indeed, as $\Omega^{\frac{1}{2}}$), where $\Omega$ is another positive compact operator. Next note that $||\Omega U_{t} \psi ||$ is a continuous function of $t$. Finally, suppose $\psi \in S_{\pm}$, then for any $\Omega$:
\begin{quote}
$\tau_{\pm}^{\Omega^{2}}(\psi) = \int_{\mathbb{R}^{\pm}}||\Omega U_{t} \psi ||^{2}$ $dt$ $< \infty$
\end{quote}
so $||\Omega U_{t} \psi ||$ is a non-negative square-integrable continuous function. Hence $||\Omega U_{t} \psi || \rightarrow 0$ as $t \rightarrow \pm \infty$.

(iv) Let $\psi \in D_{\pm}$ and $\phi \in B$. Then there exists a positive compact operator $\Omega$ such that $||\Omega^{-1} U_{t} \phi || \leq 1$ $\forall t \in \mathbb{R}$. Hence:
\begin{quote}
$\langle \psi, \phi \rangle = \langle U_{t}\psi, U_{t}\phi \rangle = \langle \Omega U_{t} \psi, \Omega^{-1} U_{t} \phi \rangle$
\end{quote}
and:
\begin{quote}
$|\langle \psi, \phi \rangle|$ $\leq$ $||\Omega U_{t} \psi ||$ $||\Omega^{-1} U_{t} \phi ||$ $\leq$ $||\Omega U_{t} \psi ||$ $\rightarrow 0$.
\end{quote}

(v) Follows directly from (ii), (iii) and Proposition 3M.23.

{\bf{3M.26 Proposition}} (Chernoff)

Let the V-asymptotic vectors $S_{\pm}(U,V)$ of the evolution $U_{t}$ be defined as in Section 3.2.9. Let $V_{t}$ be an evolution with $S_{\pm}(V) = H$ and let $B(U)$ be the bound vectors of $U_{t}$. Then:
\begin{quote}
$S_{\pm}(U,V) \subseteq B(U)^{\bot}$.
\end{quote}

{\bf{Proof}}

Let $\psi \in S_{\pm}(U,V)$. By assumption $\exists \psi'$ such that:
\begin{quote}
$\psi = \underset{t \rightarrow \pm \infty}{lim}{U_{t}}^{*}V_{t}\psi'$.
\end{quote}
Let $\phi \in B(U)$ and consider:
\begin{quote}
$\langle \phi, \psi \rangle = \underset{t \rightarrow \pm \infty}{lim} \langle \phi, {U_{t}}^{*}V_{t}\psi' \rangle = \underset{t \rightarrow \pm \infty}{lim} \langle U_{t} \phi, V_{t}\psi' \rangle$.
\end{quote}
Since $\phi \in B(U)$ then $\{U_{t}\phi\} \subseteq$ compact set. Also, $\psi'$ is in $D_{\pm}(V)$, so by using Lemma 3M.25 (i) in the form:
\begin{quote}
$\langle \phi, V_{t} \psi' \rangle \rightarrow 0$ uniformly for $\phi$ in a compact set,
\end{quote}
we conclude that $\langle \phi, \psi \rangle = 0$ as required.

\chapter{The Relation Between Classical and Quantum Mechanics} \markboth{Chapter 4}{Chapter 4: The Relation Between Classical and Quantum Mechanics}

This Chapter draws on ideas and results presented in previous Chapters to address the key question of the thesis: in what sense are classical and quantum mechanics related?

To place the question in context the Chapter starts with a critique of previous approaches - quantisation and classical limits. It concludes that serious flaws exist in these approaches, most importantly at the highest level of stating the problem to be solved. Accordingly, resort is made to the analysis of inter-theoretic reduction, introduced in Chapter 1, in order to formulate the analytic problem of reduction of classical mechanics to quantum mechanics. This leads to statement of the problem of reduction in the following form:

Given a classical mechanical system, a set of physical circumstances, a set of empirical propositions and an ``acceptable error'', find a quantum mechanical description of a system with predictions indistinguishable, within the acceptable error, from those of the classical propositions.

In this way the onus is placed on quantum mechanics to provide a subtheory weakly equivalent to the classical description of a system.

Choice of a theory within quantum mechanics is constrained by the requirement of identifying symbols in the empirical propositions of the secondary theory (classical mechanics) with symbols in the primary theory (quantum mechanics). The analysis of Chapter 2 provides two such identifications:
\begin{itemize}
\item Identification of basic abstractions in the theory of systems (e.g. pure state, property, expected value).
\item Identification of kinematic properties and propositions arising from a common space-time geometry (e.g. evolution of expected values of kinematic properties).
\end{itemize}
Incorporating these constraints leads to a much more precise statement of the problem in which essentially the only variable is the choice of state in quantum mechanics. Results from geometric quantisation and the work of Hagedorn (Ha 2) are applied to determine suitable states.

Overall, therefore, this Chapter provides a procedure for testing whether a reduction of classical mechanics to quantum mechanics is acceptable. Although particular cases or examples are not examined, these could provide the content of future research.

\newpage

\section{Review}

Ever since the Old Quantum Theory there have been attempts to relate the formalisms of classical and quantum mechanics. This Section reviews some of these attempts, which fall into two broad categories:

\begin{itemize}
\item quantisations: derivation of quantum mechanics from classical mechanics;
\item classical limits: derivation of classical mechanics from quantum mechanics.
\end{itemize}

\vspace{8pt}
{\bf{1) Quantisations}}
\vspace{8pt}

Quantisations date back to the prescriptions of the Old Quantum Theory. In more recent times there have been basically two approaches, quantisation of observables and quantisation of states.

\vspace{8pt}
{\bf{a) Observables}}

Dirac successfully used the analogy between Poisson brackets and commutators to provide quantum mechanical descriptions for the simpler Hamiltonian systems with symmetry. Effort was directed to raising this analogy to the status of a Lie algebra homomorphism between certain functions on phase space (with Poisson bracket as Lie product) and certain operators on Hilbert space (with commutators as Lie product). That this homomorphism does not exist was demonstrated by Van Hove (see AM 1) Section 5.4).

We note four points in conclusion to this ``Dirac problem'':

(i) The analogy depends for its success on the symmetry of the problem. This is not surprising in view of the ambition of Lie algebra homomorphism. For `kinematic observables' the relativity group ensures success (see Chapter 2).

(ii) For Riemannian configuration space manifolds the analogy breaks down even for kinetic energy and momentum (see (AM 1) page 242).

(iii) In the usual phase space $S \approx  \mathbb{R}^{2n}$, the Lie algebra homomorphism fails for polynomials in position and momentum of degree greater than or equal to three (see (AM 1) Theorem 5.49).

(iv) The Weyl-Wigner-Moyal correspondence between the `Moyal bracket' and commutators achieves a modified form of quantisation. See, for example, (AW 1) for some details of this and similar correspondences in the `phase-space' formulations of quantum mechanics.

\vspace{8pt}
{\bf{b) States}}

The Souriau-Kostant program is to construct the quantum dynamics on a Hilbert space from a Hamiltonian flow on a symplectic manifold. It comprises three stages, the first two of which provide a Hilbert space:

(i) Prequantisation - the prequantisation Hilbert space is the space of square-integrable sections of the complex line bundle of a suitable (`quantisable') symplectic manifold.

(ii) Polarisation - selection of a Lagrangian foliation of the symplectic manifold and identification of a quantum Hilbert space as those prequantisation functions which are constant on the leaves of the foliation.

(iii) Dynamics - construction of unitary operators on this Hilbert space corresponding to the classical flow.

Unfortunately there are, in general, unitarily inequivalent polarisations of a symplectic manifold so that the quantum system is not uniquely determined. Moreover, analysis of the dynamics has not been carried out except in special cases. See (Vo 2) Section 6.3 for discussion of these points.

In conclusion, quantisation programmes have failed to provide an abstract connection between classical and quantum mechanics. Where they have succeeded the success is attributable to a common kinematic group structure.

\vspace{8pt}
{\bf{2) Classical Limits}}
\vspace{8pt}

Classical limits also date back to the Old Quantum Theory in the form of the `Correspondence Principle'. Nowadays the classical limit refers to the behaviour of some quantum mechanical object of interest as Planck's constant tends to zero. Before attempting to interpret this limit we review the two principal approaches to classical limits:

\vspace{8pt}
{\bf{a) Asymptotic Expansions}}

This approach involves making asymptotic expansions of relevant mathematical objects in powers of Planck's constant. There are two types of this theory.

\begin{quote}

(i) WKB-Maslov asymptotic solutions to the Schr\"{o}dinger equation. This method provides eigenstates, or solutions to the Cauchy problem, as asymptotic expansions for the time-independent, or time-dependent, Schr\"{o}dinger equation. The zeroth order terms in Planck's constant are the Hamilton-Jacobi equations of classical mechanics. Maslov regularised the traditional WKB method by formulating it in phase space. See (ES 1) for an accessible account of Maslov's work.

(ii) Phase-space formulations of quantum mechanics. In this method, algebras of pseudo-differential operators are applied to the Weyl-Wigner-Moyal formulation of quantum mechanics. Semi-classical states are defined as `asymptotic functionals' on these algebras. See (Vo 1) for further details.

\end{quote}
The asymptotic expansion techniques have enjoyed considerable success, notably in the computation of `semi-classical' eigenvalues which are often in good agreement with experiment. Their theoretical status is not, however, at all clear, not least because error estimates for the expansions are rarely provided. Furthermore, whilst useful for `stationary state' problems the methods are difficult to apply to finite-time evolution problems.

\vspace{8pt}
{\bf{b) Evolution of Coherent States}}

It has long been a folk-lore that coherent states provide the most classical-like quantum states. Dating back to Schr\"{o}dinger's minimum-uncertainty wave packets, coherent states have more recently found application in the area of quantum optics. The extensive literature on coherent states partly arises from the variety of definitions in use, depending on which abstract feature is being emphasised. In this thesis we shall mean Gaussian coherent states (cf. Appendix 4.2).

In the `coherent state' method the basic idea is to approximate the quantum evolution with an evolution generated by a time-dependent quadratic Hamiltonian which is the full Hamiltonian expressed to second order around the classical trajectory.

Amongst those with contributions in this line are Hepp (He 1), Hagedorn (Ha 2) and Heller (He 2, 3, 4).

Hepp's work is discussed in Appendix 4.2 as it exemplifies the problems associated with $\hbar$ (Planck's constant) $\rightarrow$ 0 limit. His scope is the one-dimensional case. Hagedorn treats the multi-dimensional case and provides the relative phase of the approximating evolution (which cancels out in Hepp's approach), and it is this work which provides the basis for our analysis in Section 4.3. Hagedorn also treats scattering theory as does Yajima (Ya 1), although Yajima uses stationary phase methods quite different from Hagedorn. Heller, notably in (He 2), analyses quadratic approximation dynamics for Gaussian wave packets and provides some computations as well as dealing with scattering theory.

Other approaches to the classical limit using coherent states include Davies in (Da 3) and Simon in (Si 1), the latter for statistical mechanical applications.
\vspace{8pt}

{\bf{c) Analysis of the Classical Limit $\hbar \rightarrow$0}}

Everyone agrees that in the real world Planck's constant, $\hbar$, is a fundamental constant; that is, it has a fixed magnitude. Besides devotees of hidden-variable theories everyone also agrees that for certain phenomena quantum mechanics makes different predictions to classical mechanics. Finally it is agreed that the empirically relevant propositions of both theories refer to physical quantities which have magnitude (physical dimensions).

\vspace{8pt}

In light of these principles, $\hbar \rightarrow$0 could mean either that the {\em{magnitude}} of $\hbar \rightarrow$0 or that the {\em{numerical value}} (relative to a family of physical units) of $\hbar \rightarrow$0. Let us examine these in turn:

(i) Magnitude of $\hbar \rightarrow$0. This approach makes statements about a family of possible quantum theories, parameterised by $\hbar$, of which the real world is one. From Chapter 3 we know that existence of a limit is neither necessary nor sufficient for an approximation to hold. In particular, sufficiency additionally requires estimates of convergence in order to determine the error incurred when the asymptotic parameter has a fixed non-zero value. Thus, just because the magnitude of $\hbar$ is small relative to everyday magnitudes we cannot conclude that merely the {\em{existence}} of a limit solves the reduction problem.

(ii) Numerical value of $\hbar \rightarrow$0. This approach aims to show that by rescaling physical quantities the `quantum effects' become relatively, indeed numerically, small. The drawbacks of using limits just discussed apply in a similar way. However, this approach is distinguished from (i) by the ability to go arbitrarily near to the limit by means of suitable scaling. It is therefore important to examine the ideas behind scaling. Mathematical physics typically treats all quantities as dimensionless relative to a fixed choice of physical units. The rescaling of units requires careful attention to the consistent use of symbols. For example, the dilation which transforms position and momentum to the form ((He 1) equation (1.6)):
\begin{quote}
``$p_{h} = \sqrt \hbar p$; $q_{\hbar} = \sqrt \hbar q$ where $p = - i\frac{d}{dx}$ and $q = x$''
\end{quote}
is, as it stands, meaningless in terms of physical quantities. Appendix 4.1 presents a consistent theory of scaling which is applied in Appendix 4.2 to the work of Hepp (He 1) on the classical limit. The conclusion of Appendix 4.2 is that in terms of scaling a classical limit holds for a {\em{family of suitably scaled Hamiltonians}} and for large magnitudes of position and momentum. The interpretation of $\hbar \rightarrow$0 as simply a change of scale is therefore unacceptable as it involves a concurrent change in the form of the Hamiltonian.

\vspace{8pt}

In conclusion, not only is the idea of `classical limit' flawed but additionally the lack of clear concepts can lead to `proofs' which are attacking a different physical problem.

\newpage

\section{Formulation of the Approach}
With neither `quantisations' nor `classical limits' providing a sound basis for relating classical mechanics to quantum mechanics, we return to the principles of intertheoretic reduction already considered from a philosophical angle in Section 1.4.

Identifying the primary theory as quantum mechanics and the secondary theory as classical mechanics it is necessary to specify:
\begin{quote}
(a) Fundamental models - for classical and quantum mechanics.

(b) Empirically relevant propositions - in classical mechanics.

(c) Identifications - between the symbols in the empirically relevant propositions of classical mechanics and certain symbols in quantum mechanics.

(d) Criteria of identity - in order to accept that certain propositions in quantum mechanics are (weakly) equivalent to those identified - using (c) - from the empirically relevant propositions in classical mechanics.

(e) Conditions of deducibility - in quantum mechanics such that propositions satisfying the criteria of identity are true.
\end{quote}
(a) to (e) constitute a solution to the Analytic Problem of Reduction; for this to be acceptable it is also necessary to apply:
\begin{quote}
(f) Co-ordinative definitions - for both classical and quantum mechanics to test the applicability, connectivity and indistinguishability of the reduction.
\end{quote}
To build up a statement of what needs to be proved we examine each of these in turn.

\vspace{8pt}

{\bf{a) Fundamental models}}

These were developed and stated in Chapter 2. With that in mind, the scope of this analysis will be a {\em{single elementary Galilean system in an external field}}. Moreover, spin will be ignored.

\vspace{8pt}
\newpage
{\bf{b) Empirically Relevant Propositions}}

Candidates for the empirically relevant propositions in classical mechanics include:
\begin{quote}
(i) `Infinitesimal' propositions: For a system specified by a Hamiltonian function, $h$, on some set, $\Omega$, in phase space, then any state $\alpha \equiv (\xi, \pi) \in \Omega$ satisfies Hamilton's equations:
\begin{quote}
$\dot{\alpha} \equiv (\dot{\xi}, \dot{\pi})$ = $(\nabla_{\pi}h, - \nabla_{\xi}h)$.
\end{quote}
\end{quote}
(ii) `Finite' propositions: For a system specified by a Hamiltonian function, h, and for some time T, and some set, $\Omega$, in phase space, then:
\begin{quote}
There exists a solution $\alpha(t) \equiv (\xi(t), \pi(t))$ of Hamilton's equations for $0 \leq t \leq T$ and initial data $\alpha (0) \equiv (\xi(0), \pi(0)) \in \Omega$.
\end{quote}

\vspace{8pt}
{\bf{Notes}}:
\begin{enumerate}
\item The infinitesimal propositions could have been phrased in more geometric terms but we choose this form to connect them with Ehrenfest's theorem in Appendix 4.3.
\item In the finite propositions $\alpha(t)$ can be viewed as a `state trajectory' or a `kinematic properties trajectory' or an `expected value trajectory' (of the kinematic properties of a state), these being degenerate in classical mechanics.
\end{enumerate}

The infinitesimal propositions were considered early in the development of quantum theory, a `solution' to the problem of reduction being Ehrenfest's theorem which is presented in Appendix 4.3. From there it is evident that the `solution' is unsatisfactory, and it will not be pursued further. We shall instead concentrate on the finite propositions.

Although the reader may feel that there are features of classical mechanics which have been overlooked, the finite propositions are the core of classical mechanics, being the basic statement of classical particle dynamics. A solution of the Analytic Problem of Reduction will therefore be taken with respect to the finite propositions.

\vspace{8pt}
{\bf{c) Identifications}}

Up to this point the problem of reduction according to Chapter 1 still leaves us free to formulate any sub-theory of quantum mechanics which is weakly equivalent to classical mechanics. It is the identification of symbols in the propositions of classical mechanics with certain symbols in quantum mechanics which now constrains the form of the quantum sub-theory. By their nature, such identifications reflect the common ground of the two theories which we know from Chapter 2 to comprise the abstractions, such as pure state, in the theory of systems and the propositions arising from a common space-time structure.

From the empirically relevant propositions chosen in (b) it is necessary to identify:
\begin{quote}
(i) Classes of objects for (pure) states, properties and expected values.

(ii) Time T.

(iii) Expected values of position and momentum (see Note 2 in (b) above).

(iv) Hamiltonian h.
\end{quote}

For (i), formulation of the fundamental models already provides the identifications. Thus, for example, we are clear from Chapter 2 on what a pure state is in both theories.

Time is straightforward from its status in the Galilei group together with the definition of flow.

For the expected values of position and momentum we again use the Galilei group to link the theories, this time through the kinematic properties as generators of symmetry transformations. If the classical state is $\alpha \equiv (\xi, \pi)$ and the classical kinematic properties of position and momentum are $a^{c} \equiv (q^{c}, p^{c})$, the expected values in the classical case are:
\begin{quote}
$E(a^{c}, \alpha) \equiv (E(q^{c}, \alpha), E(p^{c}, \alpha)) = (\xi, \pi) \equiv \alpha$.
\end{quote}
In the quantum case, with a pure state represented by a unit vector, $\Psi$, in Hilbert space, and $a \equiv (q, p)$ denoting the quantum kinematic properties of position and momentum, we have for suitable $\Psi$:
\begin{quote}
$E(\alpha, \Psi) \equiv (E(q, \Psi)$, $E(p, \Psi)) = (<\Psi, q\Psi>, <\Psi, p\Psi>) \equiv \overline{a}$.
\end{quote}
We therefore choose the identification, $I$, of kinematic propositions in the two theories as the association (denoted by the symbol $\curvearrowright$):
\begin{quote}
$\alpha \equiv (\xi, \pi)$ $\overset{I}{\curvearrowright}$ $(<\Psi, q\Psi>$, $<\Psi, p\Psi>) \equiv \overline{a}$.
\end{quote}
$I$ is {\em{not}} a mapping as the identification merely specifies which categories of object are to be related by the criteria of identity.

Finally we need to consider the Hamiltonian. From Chapter 2, Section 2.7, Galilean space-time leads in both classical and quantum mechanics to generators of evolution (Hamiltonians) of the form:
\begin{quote}
$\frac{1}{2m} (p - A)^{2} + V$.
\end{quote}
The identification we therefore choose for Hamiltonians is an identical specification of the `free variables' - the vector potential, A, and the scalar potential, V.

\vspace{8pt}

{\bf{(d) Criteria of Identity}}

Following the discussion of Chapter 3, Section 3.1, the criteria of identity are that propositions must agree to within an acceptable error, $\epsilon$ say, which is specified for each required reduction. The criterion for identity is therefore the condition:
\begin{quote}
$|$$\alpha(t) - \overline{a(t)}$$|$ $<$ $\epsilon$.
\end{quote}

\vspace{8pt}
{\bf{(e) Conditions of Deducibility}}

These are conditions, in solely quantum-mechanical terms, for which the reduction can be proved. Note that to solve the Analytic Problem of Reduction only one such condition, subject to the identifications, need be found even though others may exist.

The only remaining free quantum `parameter' is the choice of state, so our final phrasing of the Analytic Problem of Reduction is therefore:
\begin{quote}
The Analytic Problem of Reduction is solved relative to the classical parameters $(\Omega, h, T)$ and acceptable error $\epsilon$ if for each $\alpha(0) \in \Omega$ and for each $t \in [0, T]$ there exists $\Psi \in \overline{\Omega}$, where $\overline{\Omega}$ is a set of quantum states, such that $|$$\alpha(t) - \overline{a(t)}$$|$ $<$ $\epsilon$.
\end{quote}

\vspace{8pt}
{\bf{(f) Co-ordinative definitions}}

The co-ordinative definitions will not trouble us much as we are analysing the behaviour of expected values of kinematic properties under non-relativistic dynamics, and the significance of these is guaranteed by the representation of space-time structure. Note, however, that co-ordinative definitions could be a problem if the empirically relevant propositions were at a lower level of abstraction, such as the analysis of a particular experiment. In general, propositions involving exotic `observables' would be difficult to interpret by themselves. However, as argued in Chapter 2, we take the view that `observables' in an experiment arise from the more fundamental dynamical behaviour (albeit in a non-trivial way!).

\newpage

\section{A Solution to the Analytic Problem of Reduction}
Our problem is to find quantum states such that the magnitude $|\alpha(t) - \overline{a(t)}|$ is less than an asserted acceptable error $\epsilon$. Note that $\alpha$, $a$ and $\epsilon$ are 6-dimensional vectors corresponding to position and momentum.

Our overall strategy in solving this problem will be to
\begin{itemize}
\item Find an `approximating evolution' $W(t,0)$ and state $\psi$ such that:
\begin{quote}
$\alpha(t) = \langle \psi, W(t,0)^{*} a W(t,0) \psi \rangle$
\end{quote}
\item Use a `comparator' operator $\Omega$ to control the possibly erratic behaviour of quantum states outside the region of physical significance. This operator acts as a sort of phase-space projection and converts the unbounded operators, $a$, into a more friendly form.
\end{itemize}

{\bf{1. Abstract Estimate of the Error}}

For the first part of the development assume that an `approximating evolution' $W(t,0)$ and state $\psi$ have been found. With these given the argument is based on the following formal result:

\vspace{8pt}
{\bf{4.1 Proposition}} (Formal)

Suppose there exists a propagator $W (t,0)$ and a state $\psi$ such that:
\begin{quote}
$\alpha (t) = \langle \psi, W(t,0)^{*} a W(t,0) \psi \rangle$.
\end{quote} 
Then formally:
\begin{quote}
$|\alpha(t) - \overline{a(t)}|$ $\leq$ $||(W(t,0) - U(t))\psi||$ . $||a(W(t,0) + U(t))\psi||.$
\end{quote}

{\bf{Proof}}

Use the following formula applicable to bounded operators $A$, $B$, $C$:
\begin{quote}
$2(B^{*}AB - C^{*}AC) = (B^{*} - C^{*}) A (B + C) + (B^{*} + C^{*}) A (B - C)$.
\end{quote}
Then set $A = a$, $B = W(t,0)$ and $C = U(t)$ and apply the Schwartz inequality to:
\begin{quote}
$|\alpha(t) - \overline{a(t)}|$ $=$ $|\langle \psi, W(t,0)^{*} a W(t,0) \psi \rangle - \langle \psi, U(t)^{*} a U(t) \psi \rangle |$.
\end{quote}
An interpretation of Proposition 4.1 is that the difference between the classical and quantum evolutions of expected values is dominated by the product of two terms:
\begin{quote}
(i) The norm difference $||(W-U)\psi||$ between the approximating and full quantum evolutions of the state, AND

(ii) The approximate phase-space position $||a(W+U)\psi||$ of the state.
\end{quote}
Most work on the relation between classical and quantum mechanics has hitherto focussed on estimating (i) so our immediate objectives are to:
\begin{itemize}
\item Make the formal Proposition 4.1 rigorous.

\item Estimate the approximate phase-space position arising from the rigorous version of Proposition 4.1.
\end{itemize}
Throughout, we assume the classical evolution $\alpha (t)$ is given.

\vspace{8pt}

{\bf{4.2 Proposition}}

Suppose there exists a propagator $W (t,0)$ and a state vector $\psi$ in a Hilbert space $H \approx L^{2} (\mathbb{R}^{3})$ such that:
\begin{quote}
$\alpha(t) = \langle \psi, W(t,0)^{*} a W(t,0) \psi \rangle$.
\end{quote}
Suppose further that $U(t) \psi \in D(a)$ for all $t \in [0, T]$.

Let $\Omega$ be a bounded operator such that $\Omega^{*}a\Omega$ are bounded operators, then:
\begin{quote}
$|\alpha(t) - \overline{a(t)}|$ $\leq$ $||(W(t,0) - U(t))\psi||$ . 2$||\Omega^{*}a\Omega ||$ 

\hspace{55pt} + $| \langle W(t,0)\psi, (a - \Omega^{*}a\Omega)W(t,0)\psi \rangle |$ 

\hspace{55pt} + $| \langle U(t)\psi, (a - \Omega^{*}a\Omega)U(t)\psi \rangle |$.
\end{quote}

\vspace{8pt}

{\bf{Proof}}

Apply the result quoted in the proof of Proposition 4.1 and the Schwartz inequality to:
\begin{quote}
$|\langle \psi, W^{*}aW\psi \rangle - \langle \psi, W^{*}\Omega^{*}a\Omega W \psi \rangle$  

$+ \langle \psi, W^{*}\Omega^{*}a\Omega W \psi \rangle -  \langle \psi, U^{*}\Omega^{*}a\Omega U \psi \rangle$

$+ \langle \psi, U^{*}\Omega^{*}a\Omega U \psi \rangle - \langle \psi, U^{*}a U \psi \rangle |$.

\end{quote}

\vspace{8pt}

{\bf{4.3 Remarks}}

Notice that by this Proposition the term representing approximate phase-space position has changed to $|| \Omega^{*} a \Omega ||$ which depends on the `comparator' operator $\Omega$. Dependency on the state has transferred to the two terms $\langle W (t,0), (a - \Omega^{*}a\Omega) W(t,0)\psi \rangle$ and $\langle U (t) \psi, (a - \Omega^{*}a\Omega) U(t)\psi \rangle$ which represent the error in restricting $a$ to $\Omega^{*}a\Omega$.

\vspace{8pt}

{\bf{4.4 Definition}}

Define, for a compact self-adjoint operator $\Omega$, the {\em{set of states within magnitude}} $E$ by:
\begin{quote}
$\Omega_{E} \equiv \{ \psi \in Ran (\Omega)$$|$ $||\Omega^{-1}\psi ||$ $\leq$ $E$ $\}$. 
\end{quote}

\vspace{8pt}

{\bf{4.5 Proposition}}

Let $B$ be a self-adjoint operator. Let $\Omega$ be a positive compact operator with dense range such that $B$ is $\Omega^{-1}$-bounded, then for any $\psi \in \Omega_{E}$:
\begin{quote}
$| \langle \psi, (B - \Omega B \Omega) \psi \rangle|$ $\leq$ $(E + 1)$ $||B\Omega ||$ $||(1 - \Omega)\psi ||$. 
\end{quote}

\vspace{8pt}

{\bf{Proof}}

Recall first that if $B$ is $A$-bounded for some operator $A$ then $D(A) \subset D(B)$ and there exist constants $a,b \geq 0$ such that:
\begin{quote}
$||B\psi ||$ $\leq$ $a ||A \psi ||$ $+ b ||\psi ||$ \hspace{2pt} $\forall \psi \in D(A)$
\end{quote}
Since $D(\Omega^{-1}) = Ran (\Omega)$ then $\Omega_{E} \subset D(B)$. Also, we have that $B\Omega$ is bounded since:
\begin{quote}
$||B \Omega \phi ||$ $\leq$ $a ||\Omega^{-1} \Omega \phi ||$ $+ b ||\Omega \phi ||$.
\end{quote}
These results justify the manipulations in the following argument:
\begin{quote}
$| \langle \psi, (B - \Omega B \Omega) \psi \rangle |$ $\leq$ $||B(1 + \Omega) \psi ||$ $||(1 - \Omega)\psi||$

\hspace{92pt} $\leq$ $||B \Omega (\Omega^{-1} + 1) \psi ||$ $||(1 - \Omega)\psi||$

\hspace{92pt} $\leq$ $||(\Omega^{-1} + 1) \psi ||$ $||B\Omega ||$ $||(1 - \Omega)\psi ||$.

\end{quote}

The result follows from noting that $\psi \in \Omega_{E}$.

\vspace{8pt}

{\bf{4.6 Remarks}}

Proposition 4.5 expresses the idea that the difference in expected value between an operator $B$ and its $\Omega$-restricted form $\Omega B \Omega$ is given by the product of three terms:
\begin{itemize}
\item The order of magnitude threshold.

\item The bound of the operator when dominated by the bounding inverse $\Omega^{-1}$ of the comparator.

\item The error in `projecting' with $\Omega$; that is, the difference between $\Omega$ and {\em{the identity}} as far as $\psi$ is concerned.

\end{itemize}
As shown by the Proposition, the operator $\Omega$ fulfils two principal functions:
\begin{itemize}
\item As an appropriate comparator for the physics of interest. Typically, $\Omega$ might be chosen as a measure of  the energy range applicable to a problem.
\item Providing an approximate identity for states of interest, effectively acting as a phase-space projection even though no true phase-space projection operators exist. The phase-space aspect follows from the analysis of compact operators in Chapter 3.
\end{itemize}
Finally, we note that the condition that $Ran (\Omega)$ is dense is only included to tie up with the usual definition of relative boundedness.

\vspace{8pt}

{\bf{4.7 Theorem}}

Suppose there exists a propagator $W(t,0)$ and a state vector $\psi$ in a Hilbert space $H \approx L^{2} (\mathbb{R}^{3}$) such that:
\begin{quote}
$\alpha (t) = \langle \psi, W(t,0)^{*} a W(t,0) \psi \rangle$.
\end{quote}
Let $\Omega$ be a positive compact operator such that the operators in $a$ are each $\Omega^{-1}$-bounded.

Suppose finally that $U(t)\psi \in \Omega_{E}$ and $W(t,0)\psi \in \Omega_{E}$ for all $t \in [0, T]$, then:
\begin{quote}
$|\alpha(t) - \overline{a(t)}|$ $\leq$

$||a\Omega ||$ $\{ \{ 2$$||\Omega ||$ $+ (E + 1)$ $||1-\Omega ||$ $\}$ $||(W(t,0) - U(t))\psi ||$ 

$+ 2(E + 1)$ $|| (1 - \Omega) W(t,0) \psi ||$ $\}$.
\end{quote}

\vspace{8pt}

{\bf{Proof}}

The result follows from Propositions 4.2 and 4.5.

\vspace{8pt}

{\bf{4.8 Remarks}}

(1) The right-hand side of the inequality in the theorem resembles that in Proposition 4.1 with the additional terms deriving from the `comparator' operator $\Omega$. Specifically, the conclusion of the Theorem has the form:
\begin{quote}
$|\alpha(t) - \overline{a(t)}|$ $\leq$ $\omega \{m_{1}\Delta_{1} + m_{2} \Delta_{2} \}$
\end{quote}
where:
\begin{quote}
$\omega \equiv$ $||a\Omega||$ is the approximate phase-space position.

$m_{1} \equiv$ 2$||\Omega ||$ $+ (E + 1)$ $||1-\Omega ||$ is a fixed magnitude determined by the operator $\Omega$ and set of states $\Omega_{E}$.

$\Delta_{1} \equiv$ $||(W(t,0) - U(t))\psi ||$ is the difference between the approximating and full quantum evolutions of the particular state.

$m_{2} \equiv$ $2(E + 1)$ is another fixed magnitude determined by the set of states $\Omega_{E}$.

$\Delta_{2} \equiv$ $|| (1 - \Omega) W(t,0) \psi ||$ represents the difference between the comparator $\Omega$ and the identity for the particular state.
\end{quote}

(2) Concerning the conditions:
\begin{itemize}
\item The supposition that $W(t,0)$ and $\psi$ exist, and an example choice for the operator $\Omega$ as well as the requirement that $W(t,0) \in \Omega_{E}$ will shortly be examined.
\item The major problem with the theorem is the requirement that $U(t)\psi \in \Omega_{E}$. We are {\em{unable to offer a satisfactory solution to this problem}} in the thesis. The aim is to determine conditions on Hamiltonian $h$, rather than the evolution $U(t)$, so that a suitable $\Omega$ could be found for some $E, t \in [0, T]$. For instance, would a condition based on $h$ being $\Omega^{-1}$-bounded be appropriate? Apart from this notable deficiency, Theorem 4.7 solves the Analytic Problem of Reduction in a manner we shall now make clear.
\end{itemize}

With Theorem 4.7 the first part of our development is complete. It remains to find suitable $\Omega$, $W(t,0)$ and $\psi$ and then estimate the right-hand side of the inequality in the theorem. As we shall see, however, the equations whose solution is needed for the estimates are very complicated and a closed-form estimate is unrealistic. So, furthering the call-and-response approach already adopted, our aim will {\em{not}} be an explicit error estimate but rather a procedure within which numerical methods may be applied to evaluate the error for particular circumstances.

Overall, therefore, our solution to the Analytic Problem of Reduction will be a method for determining whether the reduction holds. If the reader doubts that this is a solution let him provide particular classical circumstances and an acceptable error. Although we cannot provide the answer we can show the reader how to go about determining an answer.

We look first at the form of the approximating evolution $W(t,0)$ and the choice of state $\psi$, and then at the comparator operator $\Omega$.

\vspace{8pt}

{\bf{2. The Approximating Evolution and Choice of State}}

The first task is to find a $\psi$ and $W(t,0)$ such that they yield the classical evolution $\alpha (t)$ in the form:
\begin{quote}
$\alpha (t) = \langle \psi, W(t,0)^{*} a W(t,0) \psi \rangle$.
\end{quote}

The most obvious candidate is a unitary automorphism $T_{\alpha}$ of the Weyl algebra $\{ a, 1\} \equiv \{Q, P, 1 \}$ such that:
\begin{quote}
$T_{\alpha}(a) = a + \alpha$;  \hspace{3pt} $a = \begin{pmatrix} Q \\ P \end{pmatrix} $; \hspace{3pt} $\alpha = \begin{pmatrix} \xi \\ \pi \end{pmatrix} $   
\end{quote}
and $T_{\alpha}(a) = U(\alpha)^{*}aU(\alpha)$ for some unitary operator $U(\alpha)$.

The solution to this problem is well-known (see, for instance, (Vo 1)) and provided by the Weyl operators. The next Lemma collects together some pertinent features of the Weyl operators which we shall need.

\vspace{8pt}

{\bf{4.8 Lemma}}

Let $U(\alpha) \equiv e^{-i \omega(\alpha, a)} \equiv U(\xi, \pi)$, where $\omega(\alpha, a) \equiv \xi.P - \pi.Q$, then:
\begin{quote}
(i) $(U(\alpha)\psi)(x) = exp(\frac{i}{2} \pi \xi)$ $exp(i \pi. (x - \xi))$ $\psi(x - \xi).$

(ii) $U(\alpha)U(\beta) = exp(-\frac{i}{2} \omega(\alpha, \beta))$ $U(\alpha + \beta)$.

(iii) $U(\alpha) a U(\alpha)^{*} = a - \alpha$.
\end{quote}

\vspace{8pt}

{\bf{Proof}}

For (i) see (Da 1) Equation (5.1). (ii) and (iii) may be proved by direct computation from (i).

Of particular interest to us is the case where $\alpha (t)$ is a continuous trajectory in the classical phase-space. The next Lemma looks at $U(\alpha (t))$ as a propagator:

\vspace{8pt}
{\bf{4.9 Lemma}}

Let $\alpha (t) \in C^{1} (\mathbb{R}, \mathbb{R}^{6})$ then the strong derivative of $U(\alpha (t))$ is given by:
\begin{quote}
$i \frac{d}{dt} U(\alpha (t)) = \{ \omega (\dot{\alpha}(t), a) - \frac{1}{2}\omega(\dot{\alpha}(t), \alpha(t)) \}$ $U(\alpha (t))$.
\end{quote}

\vspace{8pt}
{\bf{Proof}}

We give only a formal proof - for a rigorous treatment of domain questions see (GV 1).
\begin{quote}
$\frac{d}{dt} U(\alpha(t)) = \underset{\delta t \rightarrow 0}{lim} \frac{1}{\delta t}$ $\{U(\alpha(t + \delta t)) - U(\alpha(t)) \}$

\hspace{47pt} $ = \underset{\delta t \rightarrow 0}{lim} \frac{1}{\delta t}$ $\{U(\alpha(t + \delta t)) U(\alpha (t))^{*} - 1 \}$ $U(\alpha(t))$

\hspace{47pt} $ = \underset{\delta t \rightarrow 0}{lim} \frac{1}{\delta t}$ $\{U(\alpha(t + \delta t) - \alpha (t))$ . $exp(\frac{i}{2} \omega (\alpha (t + \delta t), \alpha(t))) - 1 \}$ $U(\alpha (t))$

\hspace{47pt} $= \{ - i\omega(\dot{\alpha}(t), a) + \frac{i}{2} \omega (\dot{\alpha}(t), \alpha(t)) \}$ $U(\alpha(t))$.  
\end{quote}

\vspace{8pt}

{\bf{4.10 Corollary}}

Let $\alpha (t)$ be Hamiltonian; that is, there exists a Hamiltonian function $h (\alpha)$ such that:
\begin{quote}
$\dot{\alpha}(t) = J.h^{(1)}(\alpha(t))$

where $J \equiv \begin{pmatrix} 0 & 1 \\ -1 & 0 \end{pmatrix}$ ; \hspace{4pt} $h^{(1)}(\alpha(t)) \equiv  {\begin{pmatrix} \delta_{\xi}h \\ \delta_{\pi}h \end{pmatrix}}_{|\alpha(t)}$.
\end{quote}
Then:
\begin{quote}
$i \frac{d}{dt} U(\alpha(t)) = \{ \langle h^{(1)}(\alpha(t)), a \rangle - \frac{1}{2} \langle h^{(1)}(\alpha(t)), \alpha(t) \rangle \}$ $U(\alpha(t))$
\end{quote}
where $\langle \cdot, \cdot \rangle$ denotes the inner product on $\mathbb{R}^{6}$.

\vspace{8pt}

{\bf{Proof}}

Follows from Lemma 4.9 if we notice that $\omega (\alpha, \beta) = \langle \alpha, J. \beta \rangle$.

Now, to satisfy the initial condition $\langle \psi, a\psi \rangle = \alpha(0)$ we may choose \\ $\psi \equiv U(\alpha(0))\Gamma$ where the state vector $\Gamma$ satisfies $\langle \Gamma, a\Gamma \rangle = 0$.

It follows that any $W(t,0)$ of the form
\begin{quote}
$U(\alpha(t)) V(t, 0) U(\alpha(0))^{*}$
\end{quote}
where $V(t,0)$ is some propagator, satisfies:
\begin{quote}
$\langle \psi, W(t,0)^{*} a W(t,0) \psi \rangle = \langle V(t,0)\Gamma, aV(t,0)\Gamma \rangle + \alpha(t)$.
\end{quote}
So, provided $\langle V(t,0)\Gamma, aV(t,0)\Gamma \rangle = 0$ we have found a suitable $W(t,0)$ and state $\psi \equiv U(\alpha(0))\Gamma$.

We round-off our preliminary results by a set of notational definitions and a Proposition providing a general propagator.

\vspace{8pt}
\newpage
{\bf{4.11 Definition}}

(1) Let $h(a)$ denote the self-adjoint quantum Hamiltonian operator corresponding to the classical Hamiltonian function $h(\alpha)$.

(2) $h_{0}(t) \equiv h(\alpha(t))$

\hspace{14pt} $h_{1}(t) \equiv \langle h^{(1)}(\alpha(t)), a \rangle$

\hspace{14pt} $h_{2}(t) \equiv \frac{1}{2} \langle a,  h^{(2)}(\alpha(t)).a \rangle$

\hspace{14pt} $h^{(1)}(\alpha(t)) \equiv$ ${\begin{pmatrix} \delta_{\xi}h \\ \delta_{\pi}h \end{pmatrix}}_{|\alpha(t)}$

\hspace{14pt} $h^{(2)}(\alpha(t)) \equiv {\begin{pmatrix} \delta_{\xi \xi}^{2}h & \delta_{\xi \pi}^{2}h \\ \delta_{\pi \xi}^{2}h & \delta_{\pi \pi}^{2}h \end{pmatrix}}_{|\alpha(t)}$.

\vspace{7pt} 

(3) $h_{quad}(t) \equiv h_{0}(t) + h_{1}(t) + h_{2}(t)$.

(4) Let $U(t)$ denote the one-parameter unitary group generated by $h(a)$.

(5) Let $X(t,0)$ denote the propagator generated by:
\begin{quote}
$h(\alpha(t)) - \frac{1}{2} \langle h^{(1)}(\alpha(t)), \alpha(t) \rangle$.
\end{quote}

\vspace{8pt}

{\bf{4.12 Proposition}}

Introduce an operator $f(t)$ which has the useful property that any manipulation it is used in is valid.

Let $W(t,0)$ be the propagator generated by:
\begin{quote}
$U(\alpha(t)) g(t) U(\alpha(t))^{*}$
\end{quote}
where $g(t) \equiv h_{0}(t) + h_{1}(t) + f(t)$.

Let $Z(t,0)$ be the propagator generated by $f(t)$, then:
\begin{quote}
$W(t,0) = X(t,0) U(\alpha(t)) Z(t,0) U(\alpha(0))^{*}$
\end{quote}

\vspace{8pt}

{\bf{Proof}}

Again ignoring domain questions (which are somewhat irrelevant given the magical power of $f(t)$!), the result can be verified by obtaining the generator of the right-hand-side. Differentiation gives:
\begin{quote}
$h_{0}(t) - \frac{1}{2} \langle h^{(1)}(\alpha(t)), \alpha(t) \rangle$ + $\langle h^{(1)}(\alpha(t)), a \rangle$ $- \frac{1}{2} \langle h^{(1)}(\alpha(t)), \alpha(t) \rangle$ + $U(\alpha(t))f(t)U(\alpha(t))^{*}$

$= U(\alpha(t))$ $\{ h_{0}(t) + h_{1}(t) + f(t) \}$ $U(\alpha(t))^{*}$

$= U(\alpha(t)) g(t) U(\alpha(t))^{*}$.
\end{quote}
This generator together with the correct value at $t=0$ provides the result.

Clearly this $W(t,0)$ has the required form since $X(t,0)$ is only a phase. It remains to find $Z(t,0)$ - that is to say, our magical operator $f(t)$ - and a state $\Gamma$ such that:
\begin{quote}
$\langle Z (t,0) \Gamma, a Z(t,0)\Gamma \rangle = 0$.
\end{quote}

Before doing this let us suppose that such $Z(t,0)$ and $\Gamma$ exist and, following Hepp (He 1), estimate the difference in state evolution $||(W-U)\psi ||$ using the Duhamel formula:

\vspace{8pt}

{\bf{4.13 Proposition}}

Let $U, W, X, Z$ be as in Definition 4.11 and Proposition 4.12. Suppose $Z(t,0)$ and $\Gamma \equiv U(\alpha(0))^{*}\psi$ exist such that:
\begin{quote}
$\langle Z (t,0) \Gamma, a Z(t,0)\Gamma \rangle = 0$ \hspace{2pt} $\forall t \in [0,T]$.
\end{quote}
Define a `remainder' operator $R(s)$ as the difference between the quantum Hamiltonian centred around the classical trajectory and the generator of the approximating evolution:
\begin{quote}
$R(t) \equiv h(\alpha(t) + a) - g(t)$.
\end{quote}
Provided $U(t - s) W(s,0)\psi$ is strongly differentiable in $s$, then:
\begin{quote}
$|| (W(t,0) - U(t))\psi ||$ $\leq$ $\int_{0}^{t} ds$ $||R(s)Z(s,0)\Gamma ||$.
\end{quote}

\vspace{8pt}

{\bf{Proof}}

Given that the strong derivative of $U(t - s) W(s,0)\psi$ exists then the Duhamel formula is valid:
\begin{quote}
$(W(t,0) - U(t))\psi = \int_{0}^{t} ds \frac{d}{ds}U(t - s) W(s,0) \psi$.
\end{quote}
Now:
\begin{quote}
$\frac{d}{ds}U(t - s) W(s,0) \psi = U (t - s)$ $\{ ih(a) - iU(\alpha(s))g(s)U(\alpha(s))^{*}\}$ $W(s,0)\psi$.
\end{quote}

So, taking the norm:
\begin{quote}
$|| (W(t,0) - U(t))\psi ||$ $\leq$ $\int_{0}^{t} ds$ $|| \{ h(a) - U(\alpha(s))g(s)U(\alpha(s))^{*}\}$ $W(s,0)\psi ||$

$= \int_{0}^{t}ds$ $|| \{h(a + \alpha(s)) - g(s) \}$ $Z(s,0)\Gamma ||$. 
\end{quote}

\vspace{8pt}
\newpage
{\bf{Remarks}}

(1) To meet our aim of providing a means to compute the error it will be necessary to find expressions for $R(s)$, $Z(s,0)$, $\Gamma$ and, hopefully, the state $Z(s,0)\Gamma$.

(2) In the case where the Hamiltonian is $h(a) = \frac{1}{2m}P^{2} + V(Q)$ and $V$ is twice-differentiable the `remainder' R(t) takes the form (using Taylor's theorem):
\begin{quote}
$R(t) = \frac{P^{2}}{2m} + \int_{0}^{Q}(Q-y)V^{(2)}(\alpha_{t} + y)$ $dy -f(t)$.
\end{quote}
It would seem appropriate that $f(t)$ should at least include a term to cancel the quantum kinetic energy $\frac{P^{2}}{2m}$.

The form we choose for the propagator $Z(t,0)$ derives from the following abstract group theoretical result concerning the metaplectic group:

\vspace{8pt}

{\bf{4.15 Proposition}}

Let $g \in Sp(n)$ denote the symplectic group and let $sp(n)$ denote its Lie algebra, then: 

(1) There exists a projective representation, $U$, of $Sp(n)$ in the Hilbert space $L^{2}(\mathbb{R}^{n})$ generated by quadratic operators of the form $\langle a, Ga \rangle$ where $G \in sp(n)$. This representation is known as the {\em{metaplectic}} representation of $Sp(n)$ and is a faithful realisation of the metaplectic group $Mp(n)$.

(2) Each $U(g)$ generates an automorphism of the Weyl algebra $\{ a, 1 \}$ according to:
\begin{quote}
$U(g) a U(g)^{*} = g \circ a$.
\end{quote}
For example, if
\begin{quote}
$g = \begin{pmatrix} \alpha &\beta \\ \gamma & \delta \end{pmatrix}$ \hspace{3pt} then $g \circ a = \begin{pmatrix} \alpha Q + \beta P \\ \gamma Q + \delta P \end{pmatrix}$
\end{quote}

\vspace{8pt}

{\bf{Proof}}

See, for example, Section 4 of (Vo 2).

This result tells us a great deal if we choose $f(t)$ to be quadratic of the form $\langle a, Ga \rangle$, because then $Z(t,0) = U(g)$ for some $g \in Sp(n)$ and:
\begin{quote}
$\langle \psi, a \psi \rangle = 0$

$\Rightarrow \langle \psi, Z(t,0) a Z(t,0) \psi \rangle$

$= g \circ \langle \psi, a \psi \rangle = 0$.
\end{quote}
That is, {\em{any}} such $Z(t,0)$ meets our requirement. Recalling Definition 4.11 it makes obvious sense to choose the generator of $f(t)$ as $h_{2}(t)$.
Hence, for a Hamiltonian of the form:
\begin{quote}
$h(a) = \frac{1}{2m}P^{2} + V(q)$
\end{quote}
we may apply Taylor's theorem to determine the remainder as:
\begin{quote}
$R(t) = \frac{1}{2} \int_{0}^{Q} (Q - y)^{2} V^{(3)}(\alpha_{t} + y) dy$.
\end{quote}
This leaves the state $\Gamma \equiv U(\alpha(0))^{*}\psi$ as the only `unknown' in the error term $||R(t)Z(t,0)\Gamma ||$. For this error we can anticipate a dependency on the `dispersion' of $\Gamma$. For example, if $\Gamma$ is widely spread over space the remainder term threatens to be large. As we shall see, there is a play-off between the position-space and momentum-space dispersions in choosing a suitable $\Gamma$.

The following result provides an appropriate class of states together with the equations necessary to determine their evolution under $Z(t,0)$.

\vspace{8pt}

{\bf{4.16 Proposition}}

Let $g = \begin{pmatrix} \alpha &\beta \\ \gamma & \delta \end{pmatrix} \in Sp(n)$ and define in $L^{2}(\mathbb{R}^{n})$:
\begin{quote}
$\Gamma(x) = \pi^{\frac{-n}{4}}exp(-\frac{1}{2} \langle x, x \rangle)$.
\end{quote}
Let $U(g)$ be the metaplectic representative of $g$, then:
\begin{quote}
$U(g)\Gamma \equiv \Gamma^{M(g)}$
\end{quote}
where:
\begin{quote}
$M(g) = B(g)^{-1}A(g)$

$A(g) = \alpha + i\gamma$

$B(g) = \delta -i\beta$

$\Gamma^{M(g)}(x) = \pi^{\frac{-n}{4}}|B(g)|^{-\frac{1}{2}}exp(-\frac{1}{2} \langle x, M(g)x \rangle)$.
\end{quote}

Moreover, if $g(t)$ is the symplectic transformation generated by $h^{(2)}(\alpha(t))$ and $Z(t,0)$ is the corresponding metaplectic transformation generated by $h_{2}(t)$ (see Definition 4.11), then:
\begin{quote}
$Z(t,s)\Gamma^{M(s)} = \Gamma^{M(t)}$
\end{quote}
where the equation of motion for $M$ is:
\begin{quote}
$\dot{M} = i \delta_{\xi \xi}^{2}h - (M.\delta_{\xi \pi}^{2}h + \delta_{\pi \xi}^{2}h.M) - iM.\delta_{\pi \pi}^{2}h.M$.
\end{quote}

\vspace{8pt}

{\bf{Proof}}

The first part of the proof follows Hepp's analysis in (He 1). It is easy to see that:
\begin{quote}
$(Q + iP)\Gamma = 0$.
\end{quote}
Hence:
\begin{quote}
$U(g)(Q + iP) U(g)^{*} U(g) \Gamma = 0$.
\end{quote}
That is, if $A = \alpha + i\gamma$, $B = \delta - i\beta$ then:
\begin{quote}
$(A.Q + iB.P)U(g)\Gamma = 0$.
\end{quote}
It is then elementary to show that:
\begin{quote}
$U(g)\Gamma \equiv \Gamma^{M(g)} = K$ $exp(-\frac{1}{2} \langle x, B^{-1} A x \rangle )$
\end{quote}
satisfies this equation (note that $M = B^{-1}A$ is symmetric). The phase $|B(g)^{-\frac{1}{2}}|$ is chosen in accord with the second part of the proof, to which we now turn. 

From the first result we know that
\begin{quote}
$Z(t,0)\Gamma = \Gamma^{M(t)}$
\end{quote}
for some symmetric complex matrix $M$.

Taking the time derivative of the left-hand side gives:
\begin{quote}
$i \frac{d}{dt}Z(t,0)\Gamma = \frac{1}{2} \{Tr[\delta_{\pi \pi}^{2}h.M - i \delta_{\pi \xi}^{2}h]$ $+$ \\ $\langle x, (\delta_{\xi \xi}^{2}h + i(M.\delta_{\xi \pi}^{2}h + \delta_{\pi \xi}^{2}h.M) - M.\delta_{\xi \xi}^{2}h) x \rangle \}$. $Z(t,0)\Gamma$.
\end{quote}
Taking the time derivative of the right-hand side and noting that a determinant $|B| = exp (Tr$ $\ell_{n} B)$ gives:
\begin{quote}
$- \frac{i}{2} Tr [\delta_{\pi \xi}^{2}h + iM.\delta_{\pi \pi}^{2}h] - \frac{i}{2} \langle x, \dot{M}x \rangle$.
\end{quote}

Equating terms provides the required result.

\vspace{8pt}

{\bf{4.17 Remarks}}

(1) These results are well-known in a variety of guises. See, for example, Hagedorn in (Ha 1) who also investigates the abstract behaviour of the matrices $A$, $B$.

(2) The state $\Gamma^{M}$ is a Gaussian coherent state and it is evident from the Proposition that we are free to choose {\em{any}} such $\Gamma^{M}$ as our initial state. Thus, for instance we may choose a dilated family of Gaussians and see how they affect our error.

(3) To tie up to previously derived results (e.g. (Ha 1)) notice that the equation of motion for $M$ is compatible with the coupled first-order matrix equations:
\begin{quote}
$\dot{A} = iB.\delta_{\xi \xi}^{2}h - A.\delta_{\xi \pi}^{2}h$

$\dot{B} = B.\delta_{\pi \xi}^{2}h + iA.\delta_{\pi \pi}^{2}h$.
\end{quote}

(4) Our analysis has not needed to concentrate on rigorous consideration of operator domains as the results are essentially group-theoretical.

(5) It may well be asked if the approach can be extended beyond quadratic generators of an approximating evolution. To answer this question notice that we have relied upon Lie algebras - first the Weyl algebra, then the metaplectic Lie algebra. It is easy to see that any power of a greater than or equal to 3 does not lead to an algebra of finite order. It would seem, then, that using the metaplectic group is as far as one can go in providing approximating evolutions along the lines adopted in this thesis.

To summarise the results on the approximating evolution we present a straightforward Corollary:

\vspace{8pt}

{\bf{4.18 Corollary}}

Let $W(t,0)$ be an approximating evolution generated by $h_{quad}(t)$. Let $\Gamma^{M}(\alpha) \equiv U(\alpha(0))\Gamma^{M}$ where $\Gamma^{M}$ is as given in Proposition 4.16, then:
\begin{quote}
$||W(t,0) - U(t))$ $\Gamma^{M}(\alpha)||$ $\leq$ $\int_{0}^{t}$ $ds$ $||R(s)\Gamma^{M(s)}||$
\end{quote}
where:
\begin{quote}
$R(s) \equiv h(\alpha(s) + a) -h_{quad}(s)$, and 

$M(s)$ is a solution of the differential equation of motion in Proposition 4.16.
\end{quote}

\vspace{16pt}

{\bf{3. The Comparator Operator}}

In this Section we look at a particular choice of comparator operator both to show that such objects exist and also provide some estimates for the error terms in Theorem 4.7.

The choice we make is the family of compact operators $\Omega_{S}$ described in the following Lemma:

\vspace{8pt}
\newpage
{\bf{4.19 Lemma}}

Let $A$ denote the `annihilation' operator $\frac{1}{\sqrt{2}}(Q + iP)$.

Let $\Gamma (\alpha) \equiv U(\alpha)\Gamma$ be a coherent state.

Define:
\begin{quote}
$\tilde{\Omega_{s}} = \sigma_{s} e^{-sA^{*}A}$; \hspace{2pt} $s \in \mathbb{R}^{+}$; \hspace{2pt} $\sigma_{s} = 1 - e^{-s}$; \hspace{2pt} $\Omega_{s} = \sigma_{s}^{-1}\tilde{\Omega_{s}}$
\end{quote}
then:
\begin{quote}
(i) $0 < \tilde{\Omega_{s}} < 1$

(ii) $|| \tilde{\Omega_{s}} ||$ $=$ $\sigma_{s}$

(iii) $Tr[\tilde{\Omega_{s}}] = 1$

(iv) $\tilde{\Omega_{s}}^{-1}$ is well-defined.

(v) $\tilde{\Omega_{s}} = \lambda_{s} \int \frac{d^{2}\alpha}{\pi}e^{- \lambda_{s}|\alpha |^{2}}$ $|\Gamma (\alpha) \rangle \langle \Gamma (\alpha)|$ 

\hspace{7pt} $\tilde{\Omega_{s}}^{-1} = \frac{1}{\lambda_{s}} \int \frac{d^{2}\alpha}{\pi}e^{\sigma_{s}|\alpha |^{2}}$ $|\Gamma (\alpha) \rangle \langle \Gamma (\alpha)|$ 

where $\lambda_{s} = e^{s} - 1$ and the integrals converge weakly in the sense of distributions (see (AW 1)).

(vi) $\langle \Gamma (\alpha), \tilde{\Omega_{s}} \Gamma (\alpha) \rangle = \sigma_{s} e^{- \sigma_{s}|\alpha |^{2}}$

(vii) $|| \tilde{\Omega_{s}}^{-1} \Gamma (\alpha)||^{2} = {\sigma_{s}}^{-2} e^{\lambda_{2s}|\alpha |^{2}}$

(viii) $|| (1 - \tilde{\Omega_{s}})$ $\Gamma (\alpha)||$ $\leq$ $1 - \sigma_{s} e^{- \sigma_{s}|\alpha |^{2}}$ 

(ix) $|| a$ $\tilde{\Omega_{s}}||^{2}$ $\leq$ $\frac{1}{s}\sigma_{s}^{2}e^{s-1}$ where $a$ is any one of the operators $Q, P$.
\end{quote}

\vspace{8pt}

{\bf{Proof}}

We use the well-known properties of the `number operator' 

$A^{*}A = \frac{1}{2}(Q^{2} + P^{2} - 1)$.

(i) $\&$ (ii): The function $f(n) = \sigma_{s}e^{-sn}$ for $n = 0, 1, 2, ....$ has range in $(0, \sigma_{s}]$.

(iii): Use the orthonormal eigenvectors of $A^{*}A$ in the trace:
\begin{quote}
$Tr[\tilde{\Omega_{s}}] = \sum_{n=0}^{\infty} \sigma_{s}e^{-sn} = 1$.

(iv) Use the spectral theorem.

(v) See (AW 1).

(vi) See (AW 1).

(vii) Since $\langle \Gamma (\alpha), \tilde{\Omega_{s}}^{-1} \Gamma (\alpha) \rangle$ = $\frac{1}{\sigma_{s}}e^{\lambda_{s}|\alpha |^{2}}$ (see (AW 1)) then:
\begin{quote}
$||\tilde{\Omega_{s}}^{-1} \Gamma (\alpha)||^{2} = \frac{\sigma_{2s}}{\sigma_{s}^{2}}$ $\langle \Gamma(\alpha), {\tilde{\Omega_{2s}}}^{-1} \Gamma (\alpha) \rangle$

\hspace{51pt} $= \sigma_{s}^{-2} e^{\lambda_{2s}|\alpha |^{2}}$.
\end{quote}
(viii) $|| (1 - \tilde{\Omega_{s}}) \Gamma(\alpha)||$ $\leq$ $|| (1 - \tilde{\Omega_{s}})^{\frac{1}{2}}||$ $||(1 - \tilde{\Omega_{s}})^{\frac{1}{2}} \Gamma(\alpha)||$

and so the result follows from (vi). 

(ix) For any $a$, $a^{*}a \leq 2A^{*}A + 1$, hence if $\psi_{n}$ denote the eigenfunctions of $A^{*}A$:
\begin{quote}
$|| a$ $\tilde{\Omega_{s}}||^{2}$ $=$ $\underset{n}{sup}$ $\langle \tilde{\Omega_{s}} \psi_{n}, a^{*}a \tilde{\Omega}\psi_{n} \rangle$

$\leq$ $\underset{n}{sup}$ $\langle \tilde{\Omega_{s}} \psi_{n}, (2A^{*}A + 1) \tilde{\Omega_{s}}\psi_{n} \rangle$ 
\end{quote}
But $\tilde{\Omega_{s}}\psi_{n} = \sigma_{s}e^{-sn}\psi_{n}$ and $A^{*}A \psi_{n} = n$, so:
\begin{quote}
$|| a$ $\tilde{\Omega_{s}}||^{2}$ $\leq$ $\underset{n}{sup} (2n + 1) {\sigma_{s}}^{2}e^{-2sn}$.

Now the function $x \rightarrow (2x + 1) e^{-bx}$ has a maximum at $\frac{1}{b} - \frac{1}{2}$, hence for $b = 2s$ we find:
\begin{quote}
$|| a$ $\tilde{\Omega_{s}}||^{2}$ $\leq$ $\frac{1}{s}{\sigma_{s}}^{2}e^{s-1}$.
\end{quote}
\end{quote}
\end{quote}
\vspace{8pt}

{\bf{4.20 Remarks}}

We provide here some heuristic remarks on the choice of $\Omega_{s} = {\sigma_{s}}^{-1}\tilde{\Omega_{s}}$. 

(1) The following `operator' is a projection:
\begin{quote}
$\int \frac{d^{2}\alpha}{\pi} \delta(\alpha - \beta)$ $|\Gamma(\alpha) \rangle \langle \Gamma(\alpha)|$ $=$ $|\Gamma(\beta) \rangle \langle \Gamma(\beta)|$.
\end{quote}

(2) Consider a family $\{ f_{\mu} (\alpha) \}$ of functions such that:
\begin{quote}
$f_{\mu} (\alpha) \rightarrow \delta(\alpha - \beta)$ as $\mu \rightarrow \infty$.
\end{quote}
Such a family is:
\begin{quote}
$f_{\mu} (\alpha) = \mu e^{-\mu |\alpha|^{2}}$
\end{quote}
where for a test function, $F$, the following limit exists:
\begin{quote}
$\underset{\mu \rightarrow \infty}{lim} \int \frac{d^{2} \alpha}{\pi}f_{\mu}(\alpha)F(\alpha) = F(0)$.
\end{quote}

(3) Note that the family:
\begin{quote}
$g_{\mu} (\alpha) = e^{-\frac{1}{\mu} |\alpha|^{2}}$
\end{quote}
satisfies:
\begin{quote}
$\underset{\mu \rightarrow \infty}{lim} \int \frac{d^{2} \alpha}{\pi}g_{\mu}(\alpha)F(\alpha) = \int \frac{d^{2} \alpha}{\pi} F(\alpha) $.
\end{quote}

(4) Hence as $\mu \rightarrow \infty$:
\begin{quote}
$\int \frac{d^{2} \alpha}{\pi}f_{\mu}(\alpha)$ $|\Gamma(\alpha) \rangle \langle \Gamma(\alpha)|$ $\rightarrow$ $|\Gamma(0) \rangle \langle \Gamma(0)|$

$\int \frac{d^{2} \alpha}{\pi}g_{\mu}(\alpha)$ $|\Gamma(\alpha) \rangle \langle \Gamma(\alpha)|$ $\rightarrow 1$.
\end{quote}

(5) For our operator $\Omega_{s}$ where
\begin{quote}
$\tilde{\Omega_{s}} = \int \frac{d^{2} \alpha}{\pi} \lambda_{s} e^{-\lambda_{s}|\alpha|^{2}}$ $|\Gamma(\alpha) \rangle \langle \Gamma(\alpha)|$
\end{quote}
we have, since $\lambda_{s} \rightarrow \infty$ as $s \rightarrow \infty$ that:
\begin{quote}
$\Omega_{s} \equiv {\sigma_{s}}^{-1} \tilde{\Omega_{s}} \begin{cases} \rightarrow |\Gamma(0) \rangle \langle \Gamma(0)| \hspace{8pt} \text{as} \hspace{3pt} s \rightarrow \infty \\ \rightarrow 1 \hspace{58pt} \text{as} \hspace{3pt} s \rightarrow 0 \end{cases}$
\end{quote}

(6) Thus, the chosen comparator $\Omega_{s}$ is not only a compact operator representing a phase-space localisation (see Chapter 3) but additionally acts, for small $s$, as a coherent-state projection.

\vspace{8pt}

{\bf{4.21 Proposition}}

With the assumptions of Theorem 4.7 and choosing $\psi$ as $\Gamma(\alpha)$ and $\Omega$ as $\Omega_{s}$ we have:
\begin{quote}
$|\alpha(t) - \overline{\alpha(t)}|$ $\leq$ $(\frac{e^{s}}{se})^{\frac{1}{2}} \{ (E + 3) ||(W(t,0) - U(t))\Gamma(\alpha)|| + 2(E + 1) || (1 - \Omega_{s})W(t,0)\Gamma (\alpha)|| \}$.
\end{quote}

\vspace{8pt}

{\bf{Proof}}

Use the results in Lemma 4.19.

\vspace{8pt} 

{\bf{4. Warning Example}}

In the case where the Hamiltonian 
\begin{quote}
$h(a) = \frac{1}{2m}P^{2} + V(Q)$
\end{quote}
the estimate for  $||(W(t,0) - U(t))\Gamma^{M}(\alpha)||$  in Corollary 4.18 shows that the difference between approximate and true quantum state evolution can be made arbitrarily small by concentrating $\Gamma^{M}(\alpha)$ around the classical position. (This behaviour should be compared to the critique of Hepp's paper in our Appendix 4.2).

So it would appear from this that the problem is solved simply by making $\Gamma^{M}(\alpha)$ as concentrated as need be in position-space.

However, as illustrated by Proposition 4.21 this fails to take into account the comparator $\Omega$, since as $\Gamma^{M}$ gets more concentrated in position-space so it disperses in momentum-space, accordingly making the term
\begin{quote}
$|| (1 - \Omega)W(t,0)\Gamma^{M}||$
\end{quote}
increasingly significant.

\vspace{8pt}

{\bf{4.22 Remarks}}

The centering, and the value of $s$, in the definition of the comparator operator $\Omega_{s}$ may be chosen to minimise the right-hand side of the inequality in Proposition 4.21. Note that we have, for simplicity, only considered the case where $\Omega_{s}$ is centred around the origin in phase-space.

Were it not for the term $(\frac{e^{s}}{se})^{\frac{1}{2}}$ - the approximate radius of the phase-space region - it would be best to set $s = 0$. The size of the error terms depends on the dynamics - the Hamiltonian, the initial state, and the interval of time under consideration.

\newpage

\addcontentsline{toc}{section}{Appendices to Chapter 4}
\section*{Appendices to Chapter 4}
\subsection*{Appendix 4.1}

\hspace{145pt} {\bf{Theory of Scaling}}

The {\em{magnitude}} of a physical quantity is independent of the choice of units - thus, Plack's constant, $\hbar$, has the magnitude:
\begin{quote}
$6.625 \times 10^{-34}$ $js$ = $6.625 \times 10^{-27}$ $erg$ $s$.
\end{quote}
The numerical {\em{value}} of a physical quantity does, however, depend on the units in which that quantity is measured.

Let [ ] denote a choice of units. In particular we shall be interested in:
\begin{quote}
[M] - unit of mass

[L] - unit of length

[T] - unit of time.

\end{quote}

Let ( ) denote the magnitude of a physical quantity.

Let [ ] denote the units (or physical dimensions) of a physical quantity.

If $g$ is a physical quantity we have the following equation relating the magnitude to the value of $g$:
\begin{quote}
$(g) = g[g]$
\end{quote}
where $g$ is the numerical value in the units [ ]. If [ ]' is another choice of units, the invariance of magnitude is expressed by:
\begin{quote}
$g' [g]' = (g) = g [g]$
\end{quote}
where $g'$ is the numerical value of $g$ in the units [ ]'.

In Table 4.A1.1 we give the physical dimensions of some physical quantities and the symbols by which we shall denote them in these appendices. Throughout we refer to a mass, length, time system of units.

Of particular interest to us will be a change in units (a `scaling') in which the numerical value of Planck's constant gets smaller.

To this end we introduce the parameter $\lambda \in \mathbb{R}^{+}$ and consider the systems of units [ ]$_{\lambda}$. To specify this family of systems of units, let $\hbar_{\lambda}$ denote the numerical value of Planck's constant in [ ]$_{\lambda}$ units. We require:
\begin{quote}
$\hbar_{\lambda} = \lambda \hbar_{1}$

$\Leftrightarrow [\hbar]_{\lambda} = \lambda^{-1}[\hbar]_{1}$

$\Leftrightarrow [M]_{\lambda} {[L]_{\lambda}}^{2} {[T]_{\lambda}}^{-1} = \lambda^{-1} [M] [L]^{2} [T]^{-1}$.
\end{quote}
We choose the following additional conditions as an example:

(a) Fix mass and time units once and for all:
\begin{quote}
$[M]_{\lambda} = [M]$; \hspace{2pt} $[T]_{\lambda} = [T]$
\end{quote}
so that mass and time numerical values remain proportional to their physical magnitudes irrespective of the scaling chosen.

(b) Choose $[M]_{1}$, $[L]_{1}$, $[T]_{1}$ such that 
\begin{quote}
$\hbar_{1} = 1$
\end{quote}
that is, the numerical value of Planck's constant is chosen as one in the case $\lambda = 1$.

Overall, therefore we have
\begin{quote}
$[L]_{\lambda} = \lambda^{-\frac{1}{2}}[L]$

and $\hbar_{\lambda} = \lambda$.
\end{quote}

In Table 4.A1.2 the effect on the numerical values of our physical quantities under a $\lambda$-scaling (i.e. in these [ ]$_{\lambda}$ units) is summarised.

\vspace{7pt}
\newpage
{\bf{Table 4.A1.1 - Physical dimensions of Physical quantities}}

\begin{center}
\begin{tabular}{l c l}
{\bf{Quantity}} & {\bf{Symbol}} & {\bf{Dimensions}} \\ 
Position & $\xi$ & $[\xi] = [L]$ \\

Momentum & $\pi$ & $[\pi]$ = $[M]$ $[L]$ $[T]^{-1}$ \\

Time & $t$ & $[t] = [T]$ \\

Mass & $m$ & $[m] = [M]$ \\

Energy & $h$  or $V$ & $[V]$ = $[M]$ $[L]^{2}$ $[T]^{-2}$ \\

Planck's Constant & $\hbar$ & $[\hbar]$ = $[M]$ $[L]^{2}$ $[T]^{-1}$ \\

\end{tabular}
\end{center}

\vspace{9pt}

{\bf{Table 4.A1.2 - Effect on numerical values under a $\lambda$-scaling}}

\begin{center}
\begin{tabular}{l c l}
{\bf{Quantity}} & {\bf{Value in [ ]$_{1}$ units}} & {\bf{Value in [ ]$_{\lambda}$ units}} \\ 
Position & $\xi_{1} \equiv \xi$ & $\xi_{\lambda} = \lambda^{\frac{1}{2}} \xi$ \\

Momentum & $\pi_{1} \equiv \pi$ & $\pi_{\lambda} = \lambda^{\frac{1}{2}} \pi$ \\

Time & $t_{1} \equiv t$ & $t_{\lambda} = t$ \\

Mass & $m_{1} \equiv m$ & $m_{\lambda} = m$ \\

Energy & $V_{1} \equiv V$ & $V_{\lambda} = \lambda V$ \\

Planck's Constant & $\hbar_{1} \equiv 1$ & $\hbar_{\lambda}$ = $\lambda$ \\

\end{tabular}
\end{center}

\vspace{24pt}

{\bf{Representation of Physical Quantities as Functions}}

Let $(g)$ denote a physical quantity (e.g. energy) which takes magnitudes for various values of position, ($\xi$), and momentum, ($\pi$).

Let [ ] be a choice of units. If $(g)$ denotes the mapping of physical position and momentum magnitudes to the $g$-magnitude:
\begin{quote}
$(g) \equiv (g) ((\xi), (\pi))$
\end{quote} 
we introduce the {\em{numerical function}} $g$ as:
\begin{quote}
$(g) ((\xi), (\pi)) = g(\xi, \pi) [g]$
\end{quote}
where $g$ is a function of the numerical values $\xi_{1}$ and $\pi$ in the [ ] units.

Now let [ ]$_{\lambda}$ be another choice of units. Again we introduce a numerical function, the $\lambda$-scaled function $g_{\lambda}$ as:
\begin{quote}
$(g) ((\xi), (\pi)) = g_{\lambda}(\xi_{\lambda}, \pi{_\lambda}) [g]_{\lambda}$.
\end{quote}
Since magnitudes are independent of scaling, we have:
\begin{quote}
$g_{\lambda}(\xi_{\lambda}, \pi_{\lambda})[g]_{\lambda} = g(\xi, \pi)[g]$.
\end{quote}
As an example, let us take $g$ as the energy $h$ and choose $[ ] \equiv [ ]_{1}$ and $[ ]_{\lambda}$ as before.

Using Table 4.A1.2 we immediately conclude that the $\lambda$-scaled energy function, $h_{\lambda}$, is given as:
\begin{quote}
$h_{\lambda}(\lambda^{\frac{1}{2}}\xi, \lambda^{\frac{1}{2}}\pi)\lambda^{-1}[h] = h(\xi, \pi) [h]$
\end{quote}
or:
\begin{quote}
$h_{\lambda}(x,k) = \lambda h(\lambda^{-\frac{1}{2}}x, \lambda^{-\frac{1}{2}}k)$.
\end{quote}

\newpage

\subsection*{Appendix 4.2}

\hspace{115pt} {\bf{Hepp's Analysis of the Classical Limit}}

In this Appendix we describe the work of Hepp in his paper (He 1) on the classical limit of quantum mechanics. In accord with our analysis in Section 4.1, his results may be interpreted as describing either:
\begin{quote}
(a) A family of quantum theories with decreasing magnitude of Planck's constant.

(b) A family of evolutions with respect to fixed numerical energy and position/momentum under the scaling described in Appendix 4.1.
\end{quote}
Option (a) is rejected and option (b) evaluated.

To introduce the methods used we consider the case of no time evolution and explicate Hepp's equations (1.8) and (1.9):

\vspace{8pt}

{\bf{1. No time evolution}}

We introduce the vacuum vector $\Gamma^{\hbar}(0)$ as:
\begin{quote}
$(\Gamma^{\hbar}(0)) (x) = (\hbar \pi)^{-\frac{n}{4}}exp(\frac{-x^{2}}{2 \hbar})$
\end{quote}
and the coherent state $\Gamma^{\hbar}(\alpha)$ as:
\begin{quote}
$\Gamma^{\hbar}(\alpha) = U^{\hbar}(\alpha) \Gamma^{\hbar}(0)$
\end{quote}
where: $U^{\hbar}(\alpha) \equiv exp(\frac{-i \omega(\alpha, a^{\hbar})}{\hbar})$

\vspace{4pt}

$\alpha \equiv \begin{pmatrix} \xi \\ \pi \end{pmatrix}$; $a^{\hbar} \equiv \begin{pmatrix} q^{\hbar} \\ p^{\hbar} \end{pmatrix}$ $\equiv \begin{pmatrix} x \\ -i \hbar \frac{d}{dx} \end{pmatrix}$; $\omega (\alpha, a^{\hbar}) \equiv \xi p^{\hbar} - \pi q^{\hbar}$.

\vspace{4pt}

Apart from our notation for coherent states, the notation used is essentially Hepp's. Note, however, that our $a \not= \frac{(q + ip)}{\sqrt{2}}$ but is to be viewed as a vector of operators.

The principal object implicit in the theory is the dilation operator $D(\hbar)$ defined by:
\begin{quote}
$(D(\hbar)\psi) (x) = \hbar^{\frac{n}{4}} \psi (\hbar^{\frac{1}{2}}x)$.
\end{quote}
It is readily seen that:
\begin{quote}
$D(\hbar) \Gamma^{\hbar}(0) = \Gamma(0)$ where $(\Gamma(0))(x) = (\pi)^{\frac{-n}{4}}$ $exp(\frac{-x^{2}}{2})$

$D(\hbar)a^{\hbar}D(\hbar)^{*} = a_{\hbar} = \hbar^{\frac{1}{2}}a$

\end{quote}

where $a_{\hbar} \equiv \begin{pmatrix} q_{\hbar} \\ p_{\hbar} \end{pmatrix} \equiv \hbar^{\frac{1}{2}} \begin{pmatrix} x \\ -i \frac{d}{dx} \end{pmatrix} \equiv \hbar^{\frac{1}{2}} \begin{pmatrix} q \\ p \end{pmatrix} \equiv \hbar^{\frac{1}{2}}a$.

Thus the dilation removes the $\hbar$-dependence from the vacuum vector and generates the `symmetric' representation of the CCR (Hepp's equation (1.6)).

A little calculation gives us: 
\begin{quote}
$D(\hbar) \Gamma^{\hbar}(\alpha) = U(\hbar^{-\frac{1}{2}}\alpha)\Gamma(0) \equiv \Gamma (\hbar^{-\frac{1}{2}}\alpha) \equiv \Gamma_{\hbar}(\alpha)$
\end{quote}
where
\begin{quote}
$U (\hbar^{-\frac{1}{2}}\alpha) \equiv exp(-i \omega (\hbar^{-\frac{1}{2}} \alpha, a))$.
\end{quote}
We also have that
\begin{quote}
$U(\alpha)aU(\alpha)^{*} = a - \alpha$
\end{quote}
The object of interest in its `full' version is the expectation:
\begin{quote}
$\langle \Gamma^{\hbar}(\alpha), (a^{\hbar} - \alpha)\Gamma^{\hbar}(\alpha) \rangle$

$= \langle \Gamma_{\hbar}(\alpha), (a_{\hbar} - \alpha)\Gamma_{\hbar}(\alpha) \rangle$

$= \hbar^{\frac{1}{2}} \langle \Gamma(0), a\Gamma(0) \rangle$

$= \langle \Gamma^{\hbar}(0), a^{\hbar}\Gamma^{\hbar}(0) \rangle$.
\end{quote}
These equations include Hepp's equations (1.8) and (1.9). The argument is then that as $\hbar \rightarrow 0$ we have
\begin{quote}
$\langle \Gamma^{\hbar}(\alpha), a^{\hbar}\Gamma^{\hbar}(\alpha) \rangle \rightarrow \alpha$.
\end{quote}
Thus, in a family of quantum theories with decreasing magnitude of $\hbar$, the expectation value of position/momentum in coherent states tends to the coherent state parameters (interpreted as classical position/momentum).

Let us now rephrase this result in terms of the scaling theory of Appendix 4.1:

(i) We treat the vacuum vector as an invariant under scale changes, noting that what was `-$\frac{x^{2}}{2\hbar}$' in $\Gamma^{\hbar}(0)$ must be invariant (has no physical dimensions). Thus we choose:
\begin{quote}
$(\Gamma(0))(x) = \pi^{\frac{-n}{4}}$ $exp(\frac{-x^{2}}{2})$
\end{quote}
so that `$x$' is also treated as an invariant.

(ii) We consider the following physical magnitude:
\begin{quote}
$(\overline{a}) \equiv \langle \Gamma ((\alpha)), (a) \Gamma((\alpha)) \rangle$
\end{quote}
in various scales, but {\em{keep the numerical value of the classical position/momentum vector}}, $\alpha$, {\em{constant}}.

In magnitude terms we are interested in:
\begin{quote}
$\langle \Gamma ((\alpha)), (a) \Gamma((\alpha)) \rangle = \langle \Gamma (0), (a) \Gamma(0) \rangle + (\alpha)$.
\end{quote}
Let [ ] be a system of units. In this system this equation may be written in numerical values as:
\begin{quote}
$\overline{a} = \langle \Gamma (\alpha), a \Gamma (\alpha) \rangle = \langle \Gamma (0), a \Gamma(0) \rangle + \alpha$.
\end{quote}

Choose [ ] $\equiv$ [ ]$_{1}$ and consider the system of units [ ]$_{\lambda}$ as in Appendix 1.

In the new scale, the magnitude $(\overline{a})$ has numerical values:
\begin{quote}
$\overline{a}_{\lambda} = \langle \Gamma (\alpha_{\lambda}), a_{\lambda} \Gamma (\alpha_{\lambda}) \rangle$.
\end{quote}

Now $a$ transforms under the scale change as a position and momentum numerical value, hence
\begin{quote}
$a_{\lambda} = \lambda^{\frac{1}{2}}a$
\end{quote}
and:
\begin{quote}
$\overline{a}_{\lambda} = \lambda^{\frac{1}{2}}\langle \Gamma (0), a \Gamma (0) \rangle + \alpha_{\lambda}$.
\end{quote}
By fixing the numerical value, $\alpha_{\lambda}$, as the scale changes we see that as $\lambda \rightarrow 0$:
\begin{quote}
$\langle \Gamma (\alpha_{\lambda}), a_{\lambda} \Gamma (\alpha_{\lambda}) \rangle \rightarrow \alpha_{\lambda}$.
\end{quote}
This result may be equivalently expressed by saying that the relative error between the expectation and the coherent state parameters tends to zero as these parameters get large in a {\em{fixed scale}}. Hardly a remarkable result in view of the equation for $\overline{a}$. However, we note the following changes in numerical values under $\lambda$-scaling:
\begin{quote}
$a \rightarrow a_{\lambda} = \lambda^{\frac{1}{2}}a$

$U(\alpha) \rightarrow U_{\lambda}(\alpha_{\lambda}) = exp(\frac{-iw(\alpha_{\lambda}, a_{\lambda})}{\hbar})$

\hspace{79pt} $ = exp(-iw(\lambda^{-\frac{1}{2}} \alpha_{\lambda}, a))$

\hspace{79pt} $\equiv U(\lambda^{-\frac{1}{2}}  \alpha_{\lambda})$

$\Gamma(\alpha) \rightarrow \Gamma_{\lambda}(\alpha_{\lambda}) = U_{\lambda}(\alpha_{\lambda})\Gamma(0)$

\hspace{74pt} $ = \Gamma(\lambda^{-\frac{1}{2}} \alpha_{\lambda})$

\end{quote}
which may be directly compared to the transformation of the $\hbar$-dependent formulae under the dilation $D(\hbar)$.

\vspace{8pt}

{\bf{2. Time Evolution - Hepp's Version}}

Hepp's stated aim is equation (1.11) (for $t$ in a compact set [0, T]). In the form presented it is somewhat confusing, but may be written in our notation as:
\begin{quote}
$\underset{\hbar \rightarrow 0}{lim} \hspace{3pt} \hbar^{- \frac{1}{2}} \langle \Gamma^{\hbar}(\alpha),  (U^{\hbar}(t)^{*} a^{\hbar} U^{\hbar}(t) - \alpha(t)) \Gamma^{\hbar}(\alpha) \rangle$

$ = \hbar^{- \frac{1}{2}} \langle \Gamma^{\hbar}(0),  W^{\hbar}(t,0)^{*} a^{\hbar} W^{\hbar}(t,0) \Gamma^{\hbar}(0) \rangle$

$ = \langle \Gamma (0),  W(t,0)^{*} a W(t,0) \Gamma (0) \rangle$
\end{quote}
where:
\begin{quote}
$\alpha(t)$ are solutions of classical equations of motion $(\alpha \equiv \alpha(0))$.

$U^{\hbar}(t) \equiv exp(\frac{-ih^{\hbar}t}{\hbar})$; $h^{\hbar} = \frac{(p^{\hbar})^{2}}{2m} + V (q^{\hbar})$.

$W^{\hbar}(t,0)$ is the propagator with generator given by:
\begin{quote}
$i \hbar \frac{d}{dt} W^{\hbar}(t,0) = \frac{(p^{\hbar})^{2}}{2m} + V^{(2)} (\xi (t)) \frac{(q^{\hbar})^{2}}{2}$
\end{quote}
$W(t,0)$ is the propagator with generator given by:
\begin{quote}
$i \frac{d}{dt} W(t,0) = \frac{p^{2}}{2m} + V^{(2)} (\xi (t)) \frac{q^{2}}{2}$.
\end{quote}
\end{quote}
Transforming using the dilation $D(\hbar)$ as before, this equation then takes Hepp's form (1.11):
\begin{quote}
$\underset{\hbar \rightarrow 0}{lim} \hspace{3pt} \hbar^{- \frac{1}{2}} \langle \Gamma(\hbar^{- \frac{1}{2}} \alpha), (U_{\hbar}(t)^{*} a_{\hbar} U_{\hbar}(t) - \alpha(t)) \Gamma(\hbar^{- \frac{1}{2}} \alpha) \rangle$

$ = \langle \Gamma (0), W(t,0)^{*}aW(t,0) \Gamma(0) \rangle$
\end{quote}
where:
\begin{quote}
$U_{\hbar}(t) \equiv exp(\frac{-ih_{\hbar}t}{\hbar})$; \hspace{2pt} $h_{\hbar} = \frac{{(p_{\hbar})}^{2}}{2m} + V(q_{\hbar})$.
\end{quote}
Noting that:
\begin{quote}
$U^{\hbar}(\alpha(t))a^{\hbar}U^{\hbar}(\alpha(t))^{*} = a^{\hbar} - \alpha(t)$
\end{quote}
we may write the stated aim in the form:
\begin{quote}
$\underset{\hbar \rightarrow 0}{lim} \hspace{3pt} \langle \Gamma(0),  (V_{\hbar}(t,0)^{*} a V_{\hbar}(t,0) - W(t,0)^{*} a W(t,0)) \Gamma(0) \rangle = 0$
\end{quote}
where:
\begin{quote}
$V_{\hbar}(t, 0) \equiv U_{\hbar} (\alpha(t))^{*} U_{\hbar}(t) U_{\hbar}(\alpha(0))$

$ = U(\hbar^{- \frac{1}{2}} a(t))^{*} U_{\hbar}(t)U(\hbar^{- \frac{1}{2}} \alpha(0))$.
\end{quote}
Apart from a phase, our $V_{\hbar}(t, 0)$ is the $W_{\hbar}(t, 0)$ defined by equation (2.10) in Hepp's paper.

In order to avoid domain questions it is convenient to use the Weyl operator:
\begin{quote}
$U(\eta) = e^{-i \omega (\eta, a)}$; \hspace{2pt} $\eta \equiv \begin{pmatrix} -s \\ r \end{pmatrix}$
\end{quote}
instead of $a$. In this form the stated aim finally becomes:
\begin{quote}
$\underset{\hbar \rightarrow 0}{lim} \hspace{3pt} \langle \Gamma(0),  (V_{\hbar}(t,0)^{*} U(\eta) V_{\hbar}(t,0) - W(t,0)^{*} U(\eta) W(t,0)) \Gamma(0) \rangle = 0$.
\end{quote}
What is actually proved is considerably more, namely equation (2.1) which in our notation is:
\begin{quote}
$\underset{\hbar \rightarrow 0}{\text{\em{s-lim}}} \hspace{3pt} V_{\hbar}(t,0)^{*} U(\eta)V_{\hbar}(t,0) = W(t,0)^{*} U(\eta) W(t,0)$.
\end{quote}
In other words, Hepp proves the stated aim not just for the vacuum vector, but for {\em{every vector}} in Hilbert space!

To see what is happening here choose, as Hepp does, the dense set of Gaussians (one dimensional case):
\begin{quote}
$\psi_{a}(x) = \pi^{-\frac{1}{4}}exp(-\frac{(x-a)^{2}}{2})$ \hspace{4pt} $\forall a \in \mathbb{R}$.
\end{quote}
Now for any $\psi \in L^{2}(\mathbb{R})$:
\begin{quote}
$[\psi^{\hbar}(\alpha)](x) = [U^{\hbar}(\alpha)D(\hbar)^{*}\psi](x)$

$ = exp(\frac{i \pi(x - \frac{1}{2}\xi)}{\hbar})\hbar^{-\frac{1}{4}}\psi(\hbar^{-\frac{1}{2}}(x - \xi))$.
\end{quote}
Hence:
\begin{quote}
$|\psi_{a}^{\hbar}(\alpha)|^{2}(x) = (\pi \hbar)^{-\frac{1}{2}} exp(-\frac{(x - \xi -\hbar^{\frac{1}{2}}a)^{2}}{\hbar})$  
\end{quote}
so:
\begin{quote}
$|\psi_{a}^{\hbar}(\alpha)|^{2} \rightarrow |\Gamma^{\hbar}(\alpha)|^{2}$ for each $a$ as $\hbar \rightarrow 0$.
\end{quote}
Thus, irrespective of the vector, the $\hbar$-dependence guarantees localisation around the classical trajectory as $\hbar \rightarrow 0$. The reason why each vector is in this way `sucked into' a neighbourhood of the classical trajectory is the $\hbar$-dependence caused by the dilation $D(\hbar)$. The effect of this dilation is also seen in the time evolution. For a small time, $\delta t$:
\begin{quote}
$U^{\hbar}(\delta t) \thicksim 1-i\delta t(-\frac{\hbar^{2}}{2m} \frac{d^{2}}{dx^{2}} + \frac{1}{\hbar}V(x))$.
\end{quote}
This affects $\Gamma^{\hbar}(\alpha)$ in the following way:
\begin{quote}
$U^{\hbar}(\delta t) \Gamma^{\hbar}(\alpha) \thicksim (1-i\delta t(\frac{1}{\hbar}(\frac{\pi^{2}}{2m} + V(\xi)) + \frac{1}{2}V^{(2)}(\xi) + 0 (\hbar)))\Gamma^{\hbar}(\alpha)$
\end{quote}
where we have calculated the Gaussian integrals in $\langle \Gamma^{\hbar}(\alpha), U^{\hbar}(\delta t) \Gamma^{\hbar}(\alpha) \rangle$ and assumed $V(x)$ can be written as a Taylor series about $\xi$:
\begin{quote}
$V(x) = \sum_{n=0}^{\infty}V^{(n)}(\xi)\frac{(x-\xi)^{n}}{n!}$.
\end{quote}
Here we see that the evolution picks up the classical evolution plus a quadratic correction. The asymptotic formula here holds primarily because of the $\hbar$-dependence in the vacuum state:
\begin{quote}

$[\Gamma^{\hbar}(0)](x) = [D(\hbar)^{*}\Gamma(0)](x) = (\pi \hbar)^{-\frac{1}{4}} exp(-\frac{x^{2}}{2\hbar})$.
\end{quote}
In the `dilated' form of the quantities used in Hepp's proof of Theorem 2.1 the evolution again picks out the classical term plus a quadratic correction - both of which are eliminated by a comparison evolution. This time, however, we view from a fixed vector as $\hbar \rightarrow 0$. The circumstances are depicted in Figure 4.A2.1, which shows how, as $\hbar \rightarrow 0$:
\begin{quote}
(i) A neighbourhood ($\propto \hbar^{-\frac{1}{2}}$) of the classical trajectory expands to encompass any vector $(\psi_{a})$. This neighbourhood represents a region of fixed continuity of the potential energy function $V$, hence:

(ii) The potential energy dilates so that the region of applicability of the quadratic approximation gets larger. The vector being fixed means that the approximation thereby gets better.
\end{quote}

I believe we can draw two conclusions from this analysis of Hepp's result (Theorem 2.1):
\begin{quote}
(a) In a family of theories parameterised by the magnitude of $\hbar$, Planck's constant, the quadratic (classical) approximation gets better as $\hbar \rightarrow 0$. This is achieved by holding the mass and potential energy parameters fixed and `condensing' the vectors around the classical trajectories.

(b) For a given magnitude of $\hbar$, a coherent state with the same position/momentum parameters as used in the quadratic approximation provides a better approximation than a vector delocalised away from the classical trajectory.
\end{quote}

The arguments of Section 4.1 lead us to reject the `family of theories' parameterised by the magnitude of $\hbar$, since the latter is fixed and not at our disposal to vary. We can, however, vary the value of $\hbar$ by changing the units (scaling) - to this we turn shortly and it will be seen that Hepp's type of result may be obtained by changing the magnitude of the evolution parameters as the scale changes.

The difference between Hepp's claim and his result - namely that the limit holds for all vectors as a strong limit - can be attributed to the dilation $D(\hbar)$. By enabling a strong limit to be concluded it is apparent that the problem has been poorly phrased as what is needed is some estimate of how `classical' a quantum state is for a fixed magnitude of $\hbar$.

\vspace{8pt}

{\bf{3. Time evolution - in terms of scaling}}

As in the case of no time evolution (see 1. above) the quantum state is taken as an invariant under scale changes - in particular, we choose a fixed representation of a vector $\psi$ as $\psi(x)$ so that `$x$' is an invariant and {\em{not}} to be viewed as position space. Again we consider the scale change associated with the change in units from [ ] $\equiv$ [ ]$_{1}$ to [ ]$_{\lambda}$ introduced in Appendix 4.1 and used in the no time evolution case above.

The plan is to reproduce the formulae used by Hepp in his Theorem 2.1 but with $\hbar$ replaced by the scaling parameter $\lambda$. We already have the transformations of the operators:
\begin{quote}
$a \rightarrow a_{\lambda} = \lambda^{\frac{1}{2}}a$

$U(\alpha) \rightarrow U_{\lambda} (\alpha_{\lambda})= U(\lambda^{-\frac{1}{2}} \alpha_{\lambda})$
\end{quote}
where everything is treated as a numerical quantity.

As before, we are therefore interested in the behaviour of the formulae for {\em{fixed numerical values}} of position and momentum.

Let ($h$) be the Hamiltonian physical quantity. We shall consider it both as a function of classical position/momentum quantities and as a function of quantum position/momentum operator quantities. Let $h (\xi, \pi)$ be the value of $h$ as a function of position and momentum in the scale [ ]. In the scale [ ]$_{\lambda}$ we have, from Appendix 4.1, that:
\begin{quote}
$h_{\lambda}(\xi_{\lambda}, \pi_{\lambda}) = \lambda h(\xi, \pi)$.
\end{quote}
Consider now the parameterised family of functions:
\begin{quote}
$g^{\lambda}(\xi, \pi) = \lambda^{-1} h(\xi_{\lambda}, \pi_{\lambda})$
\end{quote}
then
\begin{quote}
$g_{\lambda}^{\lambda}(\xi_{\lambda}, \pi_{\lambda}) = h(\xi_{\lambda}, \pi_{\lambda})$.
\end{quote}
That is, in the $\lambda$-scale $g^{\lambda}$ has the same numerical value for the fixed numerical value of position and momentum. Thus $g^{\lambda}$ provides a family of Hamiltonians such that under the $\lambda$-scaling $g^{\lambda}$ is the same function of $\xi_{\lambda}$ and $\pi_{\lambda}$ as $h$ was of $\xi$ and $\pi$.

\vspace{24pt}


{\bf{Figure 4.A2.1: Scaling of the Potential energy in Hepp's Proof}}

\hspace{-1.1cm} {\includegraphics[scale=0.54]{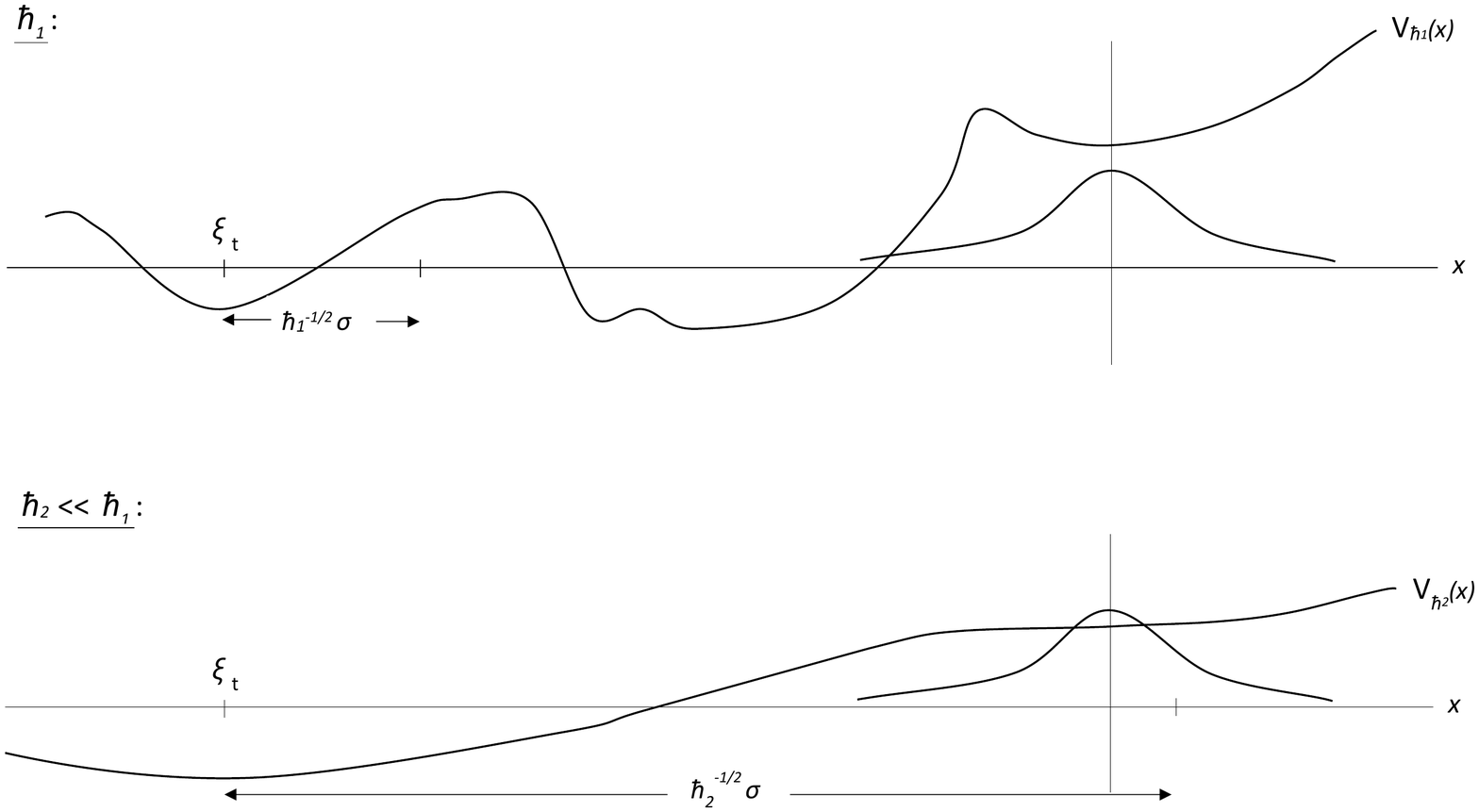}}


where $V_{\hbar}(x) \equiv V(\xi_{t} + {\hbar}^{-\frac{1}{2}} x)$

\hspace{36pt} $\psi_{a}(x) \equiv \pi^{-\frac{1}{4}}exp (- (x - a)^{2}/2)$.\\

This diagram should be compared to the equation (2.18) in Hepp's paper (He~{1}).  Note that $V(\xi_{t} + x)$ is $C^{2 + \delta}$ for all $|x| \leq \sigma$.

\newpage

It is, however, only the classical values of position and momentum which we wish to keep fixed. In terms of the quantum operators:
\begin{quote}
$g_{\lambda}^{\lambda}(q_{\lambda}, p_{\lambda}) = \frac{(p_{\lambda})^{2}}{2m} + V(q_{\lambda})$

\hspace{49pt} $ = - \frac{\lambda}{2m} \frac{d^{2}}{dx^{2}} + V (\lambda^{\frac{1}{2}}x)$

and $ U_{\lambda}(t) = exp (- \frac{i g_{\lambda}^{\lambda}t}{\lambda})$.
\end{quote}
These are equivalent to Hepp's formulae for $H_{\hbar}$ and $U_{\hbar}(t)$. We immediately conclude Hepp's results (2.1) and (2.2) noting that the linearised Hamiltonian has the form:
\begin{quote}
$H(t) = \frac{(p_{\lambda})^{2}}{2m} + V^{(2)}(\xi_{\lambda}(t)) \frac{(q_{\lambda})^{2}}{2} = \lambda(\frac{p^{2}}{2m} + V^{(2)} (\xi_{\lambda}(t)) \frac{q^{2}}{2})$
\end{quote}
the $\lambda$ cancelling in the evolution generated by $\frac{H(t)}{\lambda}$. (This is thereby analogous to Hepp's equation (2.3)). All the other $\hbar$-dependence in Theorem 2.1 and its proof may be similarly derived as $\lambda$-dependence.

There are a number of ways of expressing this result:

(1) For fixed numerical values of classical position and momentum, and fixed value of the Hamiltonian (energy) as a function of these values, the quadratic approximation of the evolution gets better as the units get larger in magnitude.

(1)' The quadratic approximation gets relatively better as the magnitudes of position and momentum get larger, provided that the Hamiltonian of the evolution is altered as:
\begin{quote}
$g^{\lambda}(q, p) = \lambda^{-1} h (q_{\lambda}, p_{\lambda})$

\hspace{38pt} $ = \frac{p^{2}}{2m} + \frac{1}{\lambda} V(\lambda^{\frac{1}{2}}q)$
\end{quote}
where $\xi$ and $\pi$ increase as $\lambda^{- \frac{1}{2}}$. (All in a fixed scale).

The formulation in (1)' corresponds most closely to the expression of Hepp's result in Theorem 2.1.

In the `scaling theory' form of Hepp's approach we can see more clearly why the result holds - such as, for example, the `expansion' of the potential to encompass any vector. We must ask, however, if the result is useful. It would be, provided we could use it to give criteria on quantum states and Hamiltonians such that the quadratic (classical) approximation is `good'. Or, conversely, for a given state and Hamiltonian estimate the  error incurred in making the quadratic approximation. Hepp's theory, as it stands, fulfils neither of these objectives. The pedagogical goal of this critique has been to demonstrate that Hepp's theory is {\em{not}} `so simple that it could belong to an elementary course on quantum mechanics'!

\newpage

\subsection*{Appendix 4.3}

\hspace{155pt} {\bf{Ehrenfest's Theorem}}

What is usually called Ehrenfest's Theorem is really just a statement of the Heisenberg equations of motion. We shall derive these in a sequence of Lemmas below. However, everything in this Appendix is formal in that we do not discuss existence or domain questions at all. Only sketch proofs are given.

\vspace{8pt}

{\bf{3A.1 Lemma}} (formal)

Let
\begin{quote}
$h = \frac{1}{2m}(\pi - A)^{2} + V$
\end{quote}
be a time-independent Hamiltonian function on $\mathbb{R}^{6}$, with the vector potential $A$ and scalar potential $V$ both functions of position $\xi$ only. The momentum is denoted by $\pi$. Then:
\begin{quote}
(i) $\dot{\pi} = \frac{1}{m}((\pi - A) \wedge (\nabla \wedge A) + ((\pi - A). \nabla)(A)) - \nabla(V)$

\hspace{23pt} $ = \frac{1}{m} \sum_{i = 1}^{3} \nabla(A_{i})\pi_{i} - \frac{1}{2m} \nabla(A^{2}) - \nabla (V)$

(ii) $\dot{\xi} = \frac{1}{m}(\pi - A)$

(iii) $m \ddot{\xi} = \frac{1}{m}(\pi - A) \wedge (\nabla \wedge A) - \nabla(V)$

\hspace{38pt} $ = \dot{\xi} \wedge B - \nabla(V)$
\end{quote}
where the magnetic potential $B$ is given by $\nabla \wedge A$.

\vspace{8pt}

{\bf{Proof}}

For (i) and (ii) use Hamilton's equations and vector identities. (iii) follows from (i) and (ii) and vector identities.

\vspace{8pt}
{\bf{3A.2 Lemma}} (Formal)

Let
\begin{quote}
$h = \frac{1}{2m}(p - A)^{2} + V$
\end{quote}
be the time-independent Hamiltonian operator on $L^{2}(\mathbb{R}^{3})$, with $p \equiv - i \hbar \nabla$ and $A$, V both operator functions of the position operator $q \equiv x$. In the Coulomb gauge ($\nabla . A = 0$) we have:
\begin{quote}
(i) $\dot{p} = \frac{1}{m} \sum_{i = 1}^{3} \nabla(A_{i}) p_{i} - \frac{1}{2m} \nabla (A^{2}) - \nabla (V)$

(ii) $\dot{q} = \frac{1}{m} (p - A)$

(iii) $m \ddot{q} = \frac{1}{2} (\dot{q} \wedge B - B \wedge \dot{q}) - \nabla (V)$.

\end{quote}

\vspace{8pt}

{\bf{Proof}}

Use the Heisenberg formula:
\begin{quote}
$\dot{\Omega}_{t} = \frac{i}{\hbar} [h, \Omega_{t}] + \delta_{t} \Omega_{t}$
\end{quote}
for the evolution of an operator $\Omega_{t} = e^{\frac{i h t}{\hbar}} \Omega e^{\frac{- i h t}{\hbar}}$.

\vspace{8pt}

{\bf{3A.3 Lemma}} (Formal Ehrenfest Theorem)

Let
\begin{quote}
$h = \frac{1}{2m}(p - A)^{2} + V$
\end{quote}
as in Lemma 3A.2. Let $\psi \in L^{2}(\mathbb{R}^{3})$ and 
\begin{quote}
$\psi_{t} \equiv exp^{\frac{-i h t}{\hbar}}\psi$.
\end{quote}
For any operator $\Omega$ let $\overline{\Omega}$ denote $\langle \psi_{t}, \Omega \psi_{t} \rangle$, then:
\begin{quote}
(i) $\dot{\overline{p}} = \frac{1}{m} \sum_{i = 1}^{3} \overline{\nabla(A_{i}) p_{i}} - \frac{1}{2m} \overline{\nabla (A^{2})} - \overline{\nabla (V)}$

(ii) $\dot{\overline{q}} = \frac{1}{m} \overline{(p - A)}$

(iii) $m \ddot{\overline{q}} = \frac{1}{2} (\overline{\dot{q} \wedge B} - \overline{B \wedge \dot{q}}) - \overline{\nabla (V)}$
\end{quote}
where $\dot{q} \equiv \frac{1}{m} (p - A)$.

\vspace{8pt}

{\bf{3A.4 Remarks}}

The equations of Lemmas 3A.1 and 3A.3 should be compared to yield the spirit of Ehrenfest's theorem - namely that expectation values of the quantum operators satisfy the classical equations of motion. We are, however, far from proving this since, for example:
\begin{quote}
$\overline{\nabla (V)} \not= \nabla(\overline{V})$
\end{quote}
indeed, we cannot even make sense of the second $\nabla$!

Note, for equation (iii), that:
\begin{quote}
$\frac{1}{2} (\dot{q} \wedge B - B \wedge \dot{q}) = \dot{q} \wedge B + \frac{i \hbar}{2} \nabla \wedge B = \dot{q} \wedge B - \frac{i \hbar}{2} \Delta (A)$.
\end{quote}
\singlespacing

\newpage 
\addcontentsline{toc}{chapter}{References} \markboth{References}{References}

\section*{References}
\begin{quote}
(Ac 1) Achinstein, P. (1968)

Concepts of Science,

Johns Hopkins Press.

\vspace{9pt}

(AJS) Amrein, W.O., Jauch, J.M. $\&$ Sinha, K.B. (1977)

Scattering Theory in Quantum Mechanics,

Benjamin.

\vspace{9pt}

(AM 1) Abraham, R. $\&$ Marsden, J.E. (1978).

Foundations of Mechanics (Second Edition),

Benjamin/Cummings.

\vspace{9pt}

(AW 1) Agarwal, G.S. $\&$ Wolf, E. (1970).

Calculus for Functions of Noncommuting Operators

and General Phase-Space Methods in Quantum

Mechanics I, II, $\&$ III,

Phys. Rev. D \underline{2}, 2161-2225.

\vspace{9pt}

(B $\&$ C 1) Beltrametti, E. G. $\&$ Cassinelli, G. (1981)

The Logic of Quantum Mechanics,

Addison-Wesley.

\vspace{9pt}

(Be 1) Bez, H. (1976).

Some Applications of Group Theory to Classical

and Quantum Mechanics,

D.Phil Thesis, Oxford University (unpublished).

\vspace{9pt}

\newpage

(Ch 1) Chernoff, P. R. (1977)

The quantum n-body problem and a theorem of Littlewood,

Pacific J. Math.

\vspace{9pt}

(Da 1) Davies, E.B. (1976)

Quantum Theory of Open Systems,

Academic Press.

\vspace{9pt}

(Da 2) Davies, E. B. (1980)

One-Parameter Semigroups,

Academic Press.

\vspace{9pt}

(Da 3) Davies, E.B. (1976)

The classical limit for quantum dynamical semigroups,

Commun. Math. Phys. \underline{49}, 113-129.

\vspace{9pt}

(ES 1) Eckmann, J. P. $\&$ Seneor, R. (1976)

Arch. Rational Mechanics \underline{61} p.153.

\vspace{9pt}

(Gr 1) Griffith, J. S. (1971)

The theory of Transition-Metal Ions,

Cambridge University Press.

\vspace{9pt}

(GS 1) Guillemin, V. $\&$ Sternberg, S. (1977)

Geometric Asymptotics,

American Mathematical Society.

\vspace{9pt}

\newpage

(GV 1) Ginibre, J. $\&$ Velo, G. (1979)

The Classical Field Limit of Scattering Theory for Non-Relativistic Boson Systems,

Commun. Math. Phys. \underline{66}, 37-76.

\vspace{9pt}

(Ha 1) Halmos, P. R. (1950)

Meausure Theory,

Springer.

\vspace{9pt}

(Ha 2) Hagedorn, G. A. (1980)

Semiclassical quantum mechanics. 1. The $\hbar \rightarrow$ 0

limit for coherent states,

Commun. Math. Phys. \underline{71}, 77-93.

\vspace{9pt}

(He 1) Hepp, K. (1974)

The classical limit for quantum mechanical 

correlation functions,

Commun. Math. Phys. \underline{35}, 265-277.

\vspace{9pt}

(He 2) Heller, E. J. (1975)

Time-dependent approach to semiclassical dynamics,

J. Chem. Phys. \underline{62}, 1544-1555.

\vspace{9pt}

(He 3) Heller, E. J. (1976)

Classical S-matrix limit of wave packet dynamics,

J. Chem. Phys. \underline{65}, 4979-4989.

\vspace{9pt}

(He 4) Heller, E.J. (1977)

Generalised theory of semiclassical amplitudes,

J. Chem. Phys. \underline{66}, 5777-5785.

\vspace{9pt}

\newpage

(Ja 1) Jauch, J. M. (1968)

Foundations of Quantum Mechanics,

Addison-Wesley.

\vspace{9pt}

(Jo 1) Jost, R. (1976)

Measures on the finite-dimensional subspaces of a

Hilbert Space: remarks to a theorem by A. M. Gleason,

In `Studies in Mathematical Physics', Editors

E. H. Lieb, B. Simon $\&$ A. S. Wightman,

Princeton University Press.

\vspace{9pt}

(K 1) Kato, T. (1976)

Perturbation Theory for Linear Operators,

Springer.

\vspace{9pt}

(Kh 1) Khinchin, A. I. (1949)

Mathematical Foundations of Statistical Mechanics,

Dover.

\vspace{9pt}

(LL 1) Levy-Leblond, J. M. (1976)

One More derivation of the Lorentz transformation,

Amer. J. Phys. \underline{44}, 271-277.

\vspace{9pt}

(LL 2) Levy-Leblond, J. M. (1971)

Galilei group and Galilei invariance,

in `Group Theory and its Applications. Volume II',

Editor E. M. Loebl,

Academic Press.

\vspace{9pt}

\newpage

(LS 1) Loomis, L. H. $\&$ Sternberg, S. (1968)

Advanced Calculus,

Addison-Wesley.

\vspace{9pt}

(Ma 1) Mackey, G. W. (1963)

Mathematical Foundations of Quantum Mechanics,

Benjamin-Cummings.

\vspace{9pt}

(Na 1) Nagel, E. (1961)

The Structure of Science,

Harcourt, Brace, and World.

\vspace{9pt}

(Pe 1) Perry, R. A. (1983)

Scattering Theory by the Enss Method,

Math. Reports Vol. 1 Part 1.

\vspace{9pt}

(Po 1) Popper, K. (1959)

The Logic of Scientific Discovery

Hutchinson

\vspace{9pt}

(Pr 1) Primas, H. (1975)

Pattern recognition in molecular quantum mechanics.

1. Background dependence of molecular states,

Theor. Chim. Acta \underline{39}, 127-148.

\vspace{9pt}

(Pr 2) Primas, H. (1981)

Chemistry, Quantum Mechanics and Reductionism,

Lecture Notes in Chemistry \underline{24},

Springer-Verlag.

\vspace{9pt}

\newpage

(Pry 1) Pryce, J. D. (1973)

Basic Methods of Linear Functional Analysis,

Hutchinson.

\vspace{9pt}

(Ro 1) Roxburgh, I. W. (1977)

Is space curved?,

in `The Encyclopaedia of Ignorance', Editors

R. Duncan $\&$ M. Weston-Smith.

\vspace{9pt}

(RS 1) Reed, M. $\&$ Simon, B. (1972)

Methods of Modern Mathematical Physics.

Volume I: Functional Analysis,

Academic Press.

\vspace{9pt}

(RS 2) Reed, M. $\&$ Simon, B. (1975)

Methods of Modern Mathematical Physics.

Volume II: Fourier Analysis, Self-Adjointness, 

Academic Press.

\vspace{9pt}

(RS 3) Reed, M. $\&$ Simon, B. (1979)

Methods of Modern Mathematical Physics.

Volume III: Scattering theory,

Academic Press.

\vspace{9pt}

(RS 4) Reed, M. $\&$ Simon, B. (1978)

Methods of Modern Mathematical Physics.

Volume IV: Analysis of Operators,

Academic Press.

\vspace{9pt}

\newpage

(Ru 1) Rudin, W. (1973)

Functional Analysis,

McGraw-Hill.

\vspace{9pt}

(Sc 1) Schiebe, E. (1973),

The Logical Analysis of Quantum Mechanics,

Pergamon.

\vspace{9pt}

(Si 1) Simon, B. (1980)

The classical limit of quantum partition functions,

Commun. Math. Phys. \underline{71}, 247-276.

\vspace{9pt}

(Su 1) Suppe, F. (1974)

The Structure of Scientific Theories,

University of Illinois Press.

\vspace{9pt}

(Su 2) Sutherland, W. A. (1975)

Introduction to Metric and Topological Spaces,

Oxford University Press.

\vspace{9pt}

(Th 1) Thirring, W. (1978)

A Course in Mathematical Physics. 1. Classical

Dynamical Systems,

Springer-Verlag.

\vspace{9pt}

(Va 1) Varadarajan, V. S. (1968)

Geometry of Quantum Theory. Volume 1,

Van Nostrand.

\vspace{9pt}

\newpage

(Va 2) Varadarajan, V. S.  (1970)

Geometry of Quantum Theory. Volume 2,

Van Nostrand.

\vspace{9pt}

(Vo 1) Voros, A. (1976)

Semi-classical approximations,

Ann. Inst. Henri Poincar\'{e} \underline{24A}, 31-90.

\vspace{9pt}

(Vo 2) Voros, A. (1977)

Asymptotic $\hbar$-expansions of stationary quantum states,

Ann. Inst. Henri Poincar\'{e} \underline{26A}, 343-403.

\vspace{9pt}

(Ya 1) Yajima, K. (1979)

The quasi-classical limit of quantum scattering theory,

Commun. Math. Phys. \underline{69}, 101-130.

\end{quote}

\end{document}